%% file: Sp-copula_v10.tex
\documentclass[11pt, a4paper]{article}
\usepackage[top=1in, bottom=1in, left=1.25in, right=1.25in]{geometry}
\usepackage{booktabs}
\usepackage{multirow}
\usepackage{setspace}
\usepackage{natbib}
\usepackage{graphicx}
\usepackage{amsmath}
\usepackage{amsthm}
\usepackage{mathtools}
\usepackage{amsfonts}
\usepackage{ascmac}
\usepackage{amssymb}
\usepackage{bm}
\usepackage{color}
\usepackage{enumitem}
\usepackage{mathrsfs}
\usepackage[ 
	colorlinks  = true, 
	anchorcolor = black,
	linkcolor   = blue,
	citecolor   = blue, 
	filecolor   = black, 
	urlcolor    = blue
]{hyperref}
\usepackage[hyphenbreaks]{breakurl}
\usepackage{comment}
\usepackage{float}
\usepackage{enumitem}
\usepackage{longtable}
\usepackage{tabularx}
\usepackage{colortab}
\usepackage{colortbl}
\usepackage{arydshln}
\usepackage{subcaption}  

\newcommand{\biblist}{\begin{list}{}
{\listparindent 0.0cm \leftmargin 0.50cm \itemindent -0.50 cm
\labelwidth 0 cm \labelsep 0.50 cm
\usecounter{list}}\clubpenalty4000\widowpenalty4000}
\newcommand{\ebiblist}{\end{list}}

\def\Corr{{\rm Corr}}

\def\S{\mathcal{S}}
\def\C{\mathcal{C}}
\def\N{\mathcal{N}}
\def\D{\mathcal{D}}

\newcommand{\indep}{\perp \!\!\! \perp}

\title{{\bf Semiparametric Copula Estimation for Spatially Correlated Multivariate Mixed Outcomes:  Analyzing Visual Sightings of Fin Whales from a Line Transect Survey}}
\date{}

\begin{document}


\maketitle
\doublespacing

\vspace{-1.5cm}
\begin{center}
Tomotaka Momozaki$^1$, Tomoyuki Nakagawa$^{2,5}$, Shonosuke Sugasawa$^3$\\
and Hiroko Kato Solvang$^4$
\end{center}

\noindent
$^1$Department of Information Sciences, Tokyo University of Science\\
$^2$School of Data Science, Meisei University\\
$^3$Faculty of Economics, Keio University\\
$^4$Marine Mammals Research Group, Institute of Marine Research\\
$^5$Statistical Mathematics Unit, RIKEN Center for Brain Science\\

\medskip
\noindent

\singlespacing 

\medskip
\begin{center}
{\bf \large Abstract}
\end{center}
For marine biologists, ascertaining the dependence structures between marine species and marine environments, such as sea surface temperature and ocean depth, is imperative for defining ecosystem functioning and providing insights into the dynamics of marine ecosystems. 
However, obtained data include not only continuous but also discrete data, such as binaries and counts (referred to as mixed outcomes), as well as spatial correlations, both of which make conventional multivariate analysis tools impractical. 
To solve this issue, we propose semiparametric Bayesian inference and develop an efficient algorithm for computing the posterior of the dependence structure based on the rank likelihood under a latent multivariate spatial Gaussian process using the Markov chain Monte Carlo method. 
To alleviate the computational intractability caused by the Gaussian process, we also provide a scalable implementation that leverages the nearest-neighbor Gaussian process. 
Extensive numerical experiments reveal that the proposed method reliably infers the dependence structures of spatially correlated mixed outcomes. 
Finally, we apply the proposed method to a dataset collected during an international synoptic krill survey in the Scotia Sea of the Antarctic Peninsula to infer the dependence structure between fin whales ({\it Balaenoptera physalus}), krill biomass, and relevant oceanographic data. 


\vspace{0.5cm}
\section{Introduction}
\subsection{Background and motivation}
Many marine ecosystems are characterized by several levels of interactions between marine species such as populations of fish, birds, and sea mammals, and the marine environment \citep{mann2005dynamics}. 
Understanding the nature of these interactions is imperative because they define ecosystem functioning and provide insights into the effects of climate change on the dynamics of marine ecosystems \citep{tett2013framework}. 
The data used to investigate these interactions are usually obtained through scientific surveys, and the outcomes of abiotic and biotic observations are mixed with continuous (e.g., acoustic registration, biomass, abundance) and discrete numbers (e.g., counted and sighted data). 
In addition, the data are spatially correlated. 
Therefore, it is difficult to estimate the interactions between marine species and environmental factors in marine ecosystems accurately using these data. 

To model the interactions among variables, multivariate analysis is a ubiquitous tool, with many methodologies developed, such as factor analysis, principal component analysis \citep[e.g.,][]{manly2016multivariate}, graphical modeling \citep[e.g.,][]{jordan2004graphical}, and copulas \citep[e.g.,][]{joe2014dependence}. 
In particular, copulas are attractive because they decouple the dependence structure from the marginal distributions, allowing each component to be modeled separately \citep[e.g.,][]{joe2014dependence}.
However, inference on copula parameters generally requires assumptions or estimates for the marginal distributions.
To address this issue, \cite{hoff2007extending} proposed the extended rank likelihood, which enables inference on copula parameters without explicit parametric assumptions on the marginals and without requiring their estimation.
This approach can be successfully applied even when outcomes are mixed (i.e., outcomes include both continuous and discrete variables).
Following \cite{hoff2007extending}, we refer to this framework as ``semiparametric'' because the dependence structure is modeled parametrically through the Gaussian copula, while the marginal distributions are left completely unspecified.
Although these methods are useful for modeling the dependence structure, they assume that the multivariate observations are independent. 
However, beginning with the example of marine ecosystems mentioned above, in many applications such as epidemiology, climatology, medicine, and sociology, multivariate data with location information are often available, and spatial correlation must be appropriately considered; otherwise, the estimation of the dependence structure among outcomes could be severely biased. 

\begin{figure}[htb!]
    \centering
    \includegraphics[width=\columnwidth]{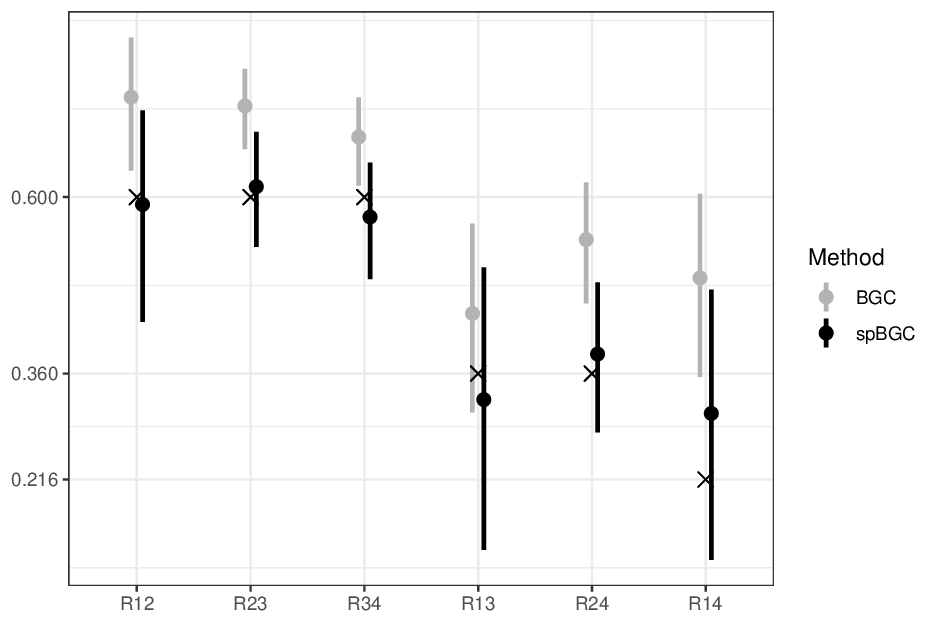}
    \caption{The 95\% credible interval and the posterior median of each correlation coefficient by \cite{hoff2007extending}'s (BGC) and our proposed methods (spBGC): Cross marks denote true values.
    }
    \label{fg:toy}
\end{figure}

To demonstrate the potential effects of spatial correlation in dependence modeling, we consider simulated multivariate data with 300 samples and four mixed outcomes, the detailed settings of which are explained in Section \ref{sec:vshoff}. 
Figure \ref{fg:toy} shows the estimation results of the method in \cite{hoff2007extending} (denoted as BGC) when applied to the simulated data.
The 95\% credible intervals of the BGC do not include most of the true values, indicating that the inference of the dependence structure is severely biased without considering the spatial correlation. 
However, the 95\% credible intervals based on the proposed method (spBGC) presented in Section \ref{sec:spbgc} can suitably capture the true correlation. 
Therefore, for multivariate data equipped with location information, it is essential to adequately consider the spatial correlation to make a correct inference on the dependence structure. 

Our proposal defines a spatially correlated hierarchical model combined with extended rank likelihood. 
By doing so, we can provide a semiparametric copula inference that considers the spatial correlation between multivariate observations. 
To this end, we consider a spatial hierarchical model that embeds a Gaussian copula for the dependence among mixed outcomes and a Gaussian process for spatial correlation.
We refer to it as a {\it spatial Gaussian copula}. 
For Bayesian inference, we develop an efficient algorithm to calculate the posterior of the dependence structure based on the extended rank likelihood using the Markov chain Monte Carlo (MCMC) method. 
Because the Gaussian process is computationally intractable for large datasets, for example, in terms of matrix operation cost and memory \citep{heaton2019case, liu2020gaussian}, we scale our algorithm using the nearest-neighbor Gaussian process \citep{datta2016hierarchical}. 
Extensive numerical experiments reveal that our proposed procedure successfully accounts for the spatial correlation and correctly infers the dependence structure among outcomes. 
Remarkably, as the spatial correlation among observations becomes stronger, our proposed method outperforms \cite{hoff2007extending}'s, which does not consider spatial correlation. 

Related work in spatial modeling includes several studies on modeling multivariate spatial data. 
For example, \cite{dey2022graphical} and \cite{krock2023modeling} proposed efficient modeling (relatively high-dimensional) of multivariate spatial data for Gaussian outcomes, whereas \cite{feng2012joint} and \cite{torabi2014spatial} considered multivariate models based on generalized linear mixed models. 
Furthermore, attempts have been made to use copulas in multivariate spatial modeling, such as \cite{musafer2017nonlinear}, \cite{krupskii2018factor}, \cite{krupskii2019copula}, and \cite{gong2022flexible}. 
Because the above methods consider the joint estimation of both marginal and dependence structures, one may lose the efficiency of the estimation by a large number of nuisance parameters in the marginal distributions, while the estimation of the dependence could be biased owing to potential misspecification of the marginal distributions. 
Hence, it is more reasonable to model the dependence structure directly, as in the proposed method. 

The remainder of this paper is organized as follows.
Section \ref{sec:bgc} introduces the extended rank likelihood and semiparametric Bayesian inference for the dependence structure proposed by \cite{hoff2007extending}.
Section \ref{sec:spbgc} proposes semiparametric Bayesian inference for the dependence structure in spatially correlated mixed outcomes and develops an efficient algorithm for computing the posterior using MCMC.
The algorithm is then extended to a more scalable implementation using the nearest-neighbor Gaussian process.
Section \ref{sec:vshoff} presents the simulation results, including a comparison with the method of \cite{hoff2007extending}, to validate the usefulness of the proposed method.
Section \ref{sec:real} presents the application of the proposed method using the fin whale sighting data described in Section \ref{sec:data}.
Section \ref{sec:end} provides directions for future research.
The \texttt{R} code for implementing the proposed method is available in the GitHub repository (\burl{https://github.com/t-momozaki/spBGC}).

\subsection{Motivating application: Fin whale sighting data} \label{sec:data}
To motivate the proposed methodology, we describe a dataset for characterizing the Sourthern Ocean ecosystem.
Antarctic krill ({\it Euphausia superba}) is a characteristic species of the Soruthern Ocean and exists within a narrow band of cold temperatures, as this species is known as cold adaptive species \citep{cui2025}. 
This is a major prey item for a diverse suite of predators including whales, penguins, seals and fish and is an important fishery resource \citep{krafft2019report}. 
However, recent ocean warming has impacted population dynamics, with changes in sea ice dynamics-including a reduction-being observed \citep{Kawaguchi2023}.

The data was collected during the 2019 Area 48 Survey for Antarctic krill in the Scotia Sea of the Antarctic Peninsula \citep{krafft2021}.
During the krill survey period, visual sightings of fin whales ({\it Balaenoptera physalus}) were carried out along survey transects from three of the six participating vessels: R/V Kronprins Haakon (KPH), F/V Cabo de Hornos (CDH), and RRS Discovery (DIS), following the observation protocols detailed in \cite{biuw2024estimated}.
The observed data in the survey consisted of the number of fin whale sightings (groups and individuals) and krill acoustic data (recalculated to krill biomass; \citep[details in the Supplementary Material of][]{krafft2021}) from vessels operating along the transects.
The observation transects were split into 1-nm long segments, resulting in 3,833 segments based on 42 transects.
For each segment, these data were summarized.

In addition, four environmental variables were included to investigate the association of environmental data with the biological community: sea surface temperature (SST, in Celsius), water depth (Depth, in meters), slope of the depth (Slope), and gradient of surface temperature (SST.grd).
Water depth data were obtained from the ETOPO 1 bathymetric dataset (\burl{https://www.ncei.noaa.gov/products/etopo-global-relief-model}), and SST data were obtained from the OISST dataset (\burl{https://www.ncei.noaa.gov/products/optimum-interpolation-sst}), both extracted for the middle position of each 1-nm segment.
The slope of the depth is the maximum rate of change in depth from that cell to its neighbors, calculated using the {\it Slope} tool of the Surface toolset in ArcGIS (\burl{https://desktop.arcgis.com/en/arcmap/10.3/tools/spatial-analyst-toolbox/an-overview-of-the-surface-tools.htm}).
The lower the slope value is, the flatter is the terrain, and vice versa.
The gradient surface temperature (SST.grd) was calculated using the {\it Slope} tool applied to the SST.

\begin{figure}[H]
    \begin{tabular}{cc}
    \begin{minipage}[b]{0.49\linewidth}
        \centering
        \includegraphics[width=\columnwidth]{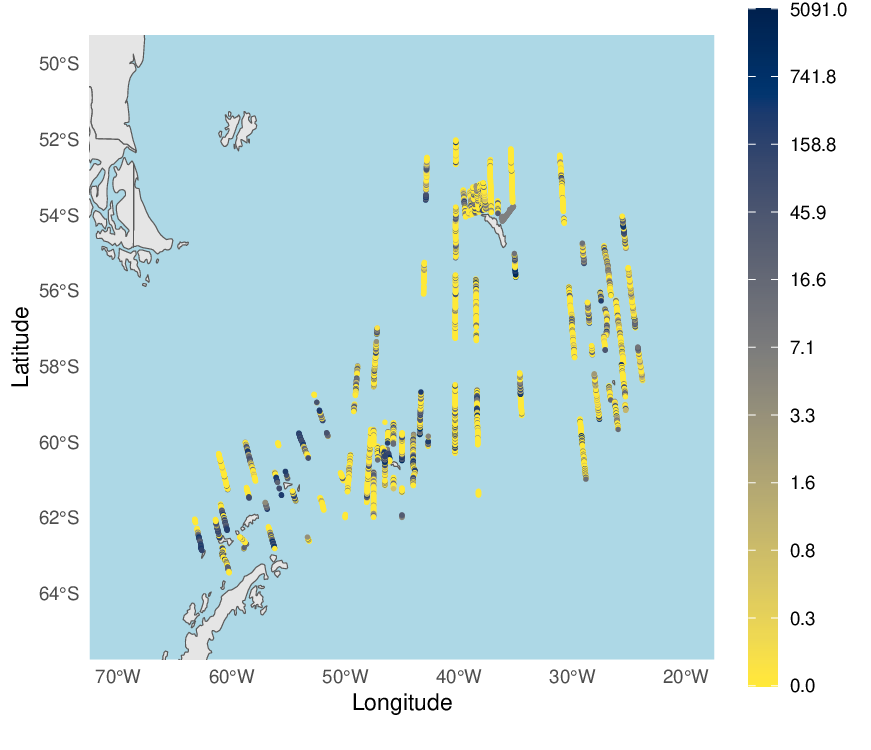}
        \subcaption{Krill}
        \label{fg:geo_krill}
    \end{minipage} &
    \begin{minipage}[b]{0.49\linewidth}
        \centering
        \includegraphics[width=\columnwidth]{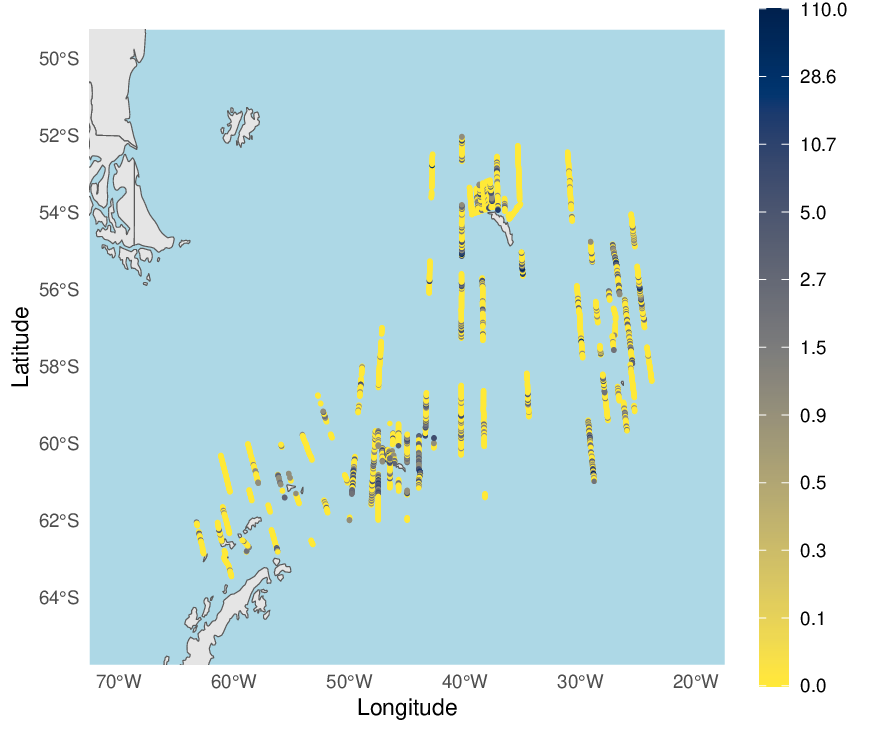}
        \subcaption{Whale}
        \label{fg:geo_whale}
    \end{minipage} \\
    \begin{minipage}[b]{0.49\linewidth}
        \centering
        \includegraphics[width=\columnwidth]{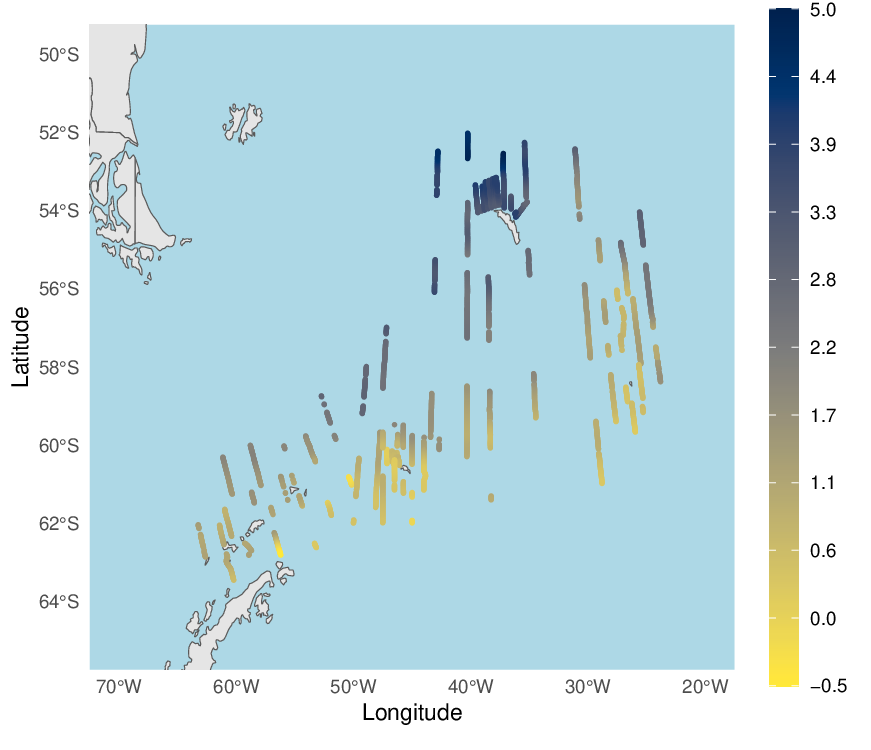}
        \subcaption{SST}
        \label{fg:geo_sst}
    \end{minipage} &
    \begin{minipage}[b]{0.49\linewidth}
        \centering
        \includegraphics[width=\columnwidth]{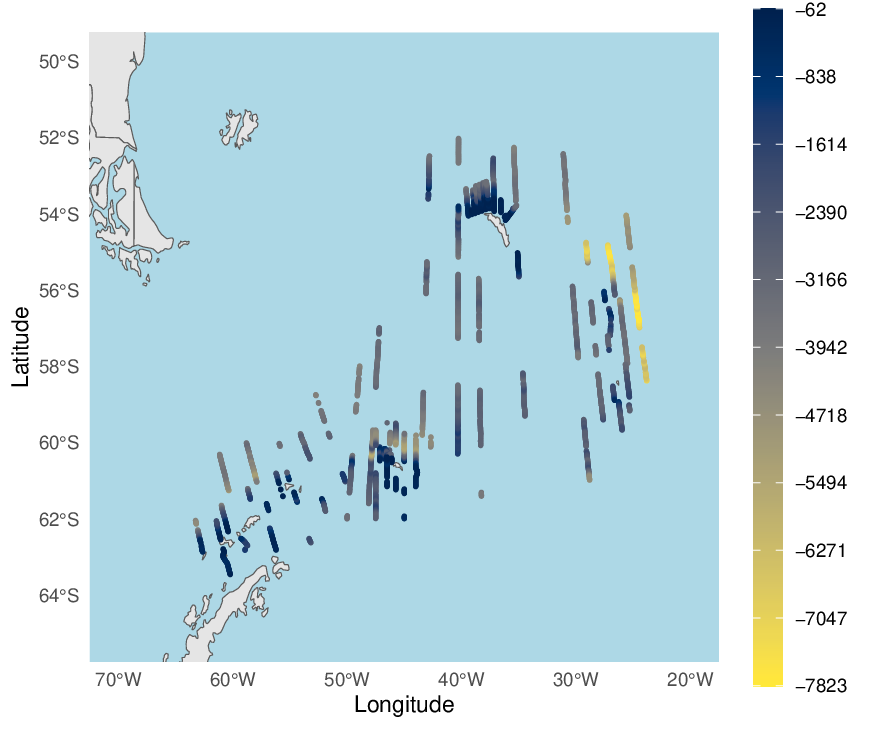}
        \subcaption{Depth}
        \label{fg:geo_depth}
    \end{minipage} \\
    \begin{minipage}[b]{0.49\linewidth}
        \centering
        \includegraphics[width=\columnwidth]{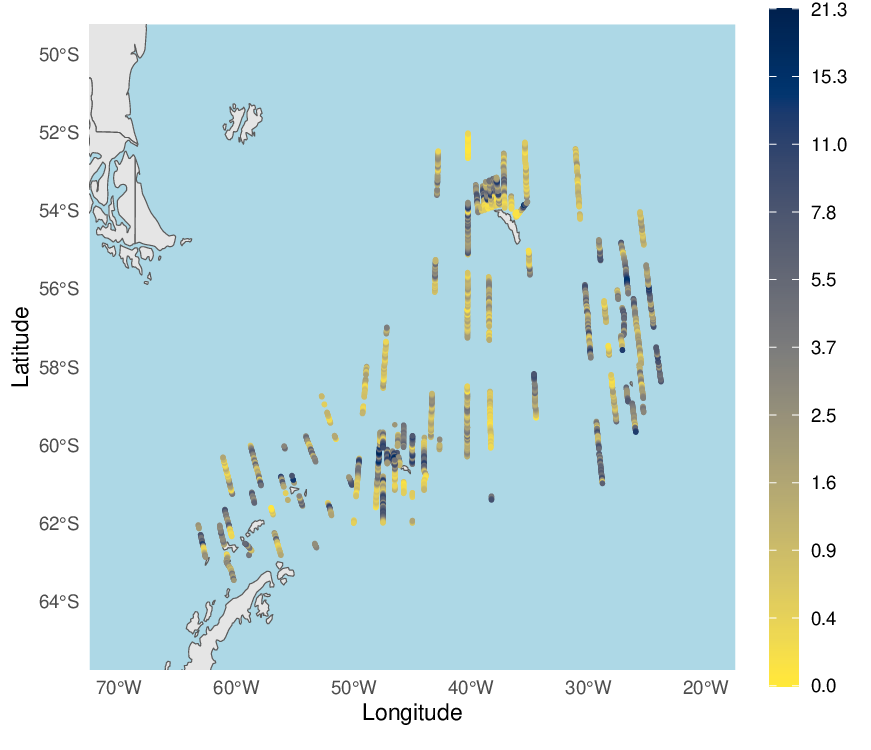}
        \subcaption{Slope}
        \label{fg:geo_slope}
    \end{minipage} &
    \begin{minipage}[b]{0.49\linewidth}
        \centering
        \includegraphics[width=\columnwidth]{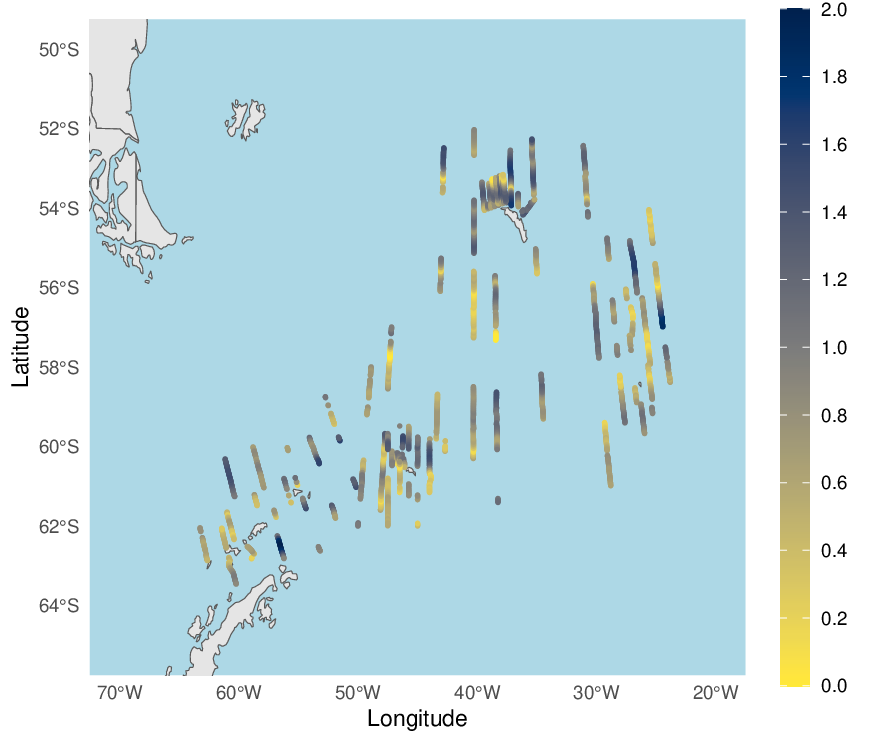}
        \subcaption{SST.grd}
        \label{fg:geo_sstgrd}
    \end{minipage}
    \end{tabular}
    \caption{Spatial distribution of the six variables in the fin whale sighting dataset: krill biomass (Krill), fin whale sightings (Whale), sea surface temperature (SST), water depth (Depth), slope of the depth (Slope), and gradient of surface temperature (SST.grd). 
    In each panel, darker colors correspond to higher values of the variable shown (for Depth, higher values indicate shallower water); the figure is displayed in color in the online version.}
    \label{fg:geo_da}
\end{figure}

Figure \ref{fg:geo_da} shows the spatial plots of all six variables across the study region.
The spatial plots reveal substantial spatial heterogeneity in both biological and environmental variables.
To quantify the spatial correlation in the data, we compute Moran's $I$ statistic \citep{moran1950notes}, a widely used measure of spatial autocorrelation where values close to $+1$ indicate strong positive spatial correlation.
Table~\ref{tab:moran_obs} shows that all six variables exhibit positive spatial autocorrelation, with environmental variables (SST, Depth, SST.grd) showing particularly strong spatial correlation ($I > 0.86$), while biological variables (Krill, Whale) exhibit moderate correlation ($I \approx 0.4$).
This spatial correlation, combined with the mixed nature of the outcomes, makes conventional multivariate analysis methods inappropriate for inferring the dependence structure among these variables.

\begin{table}[H]
\centering
\caption{Moran's $I$ statistic for observed data. Values close to $+1$ indicate strong positive spatial autocorrelation.}
\label{tab:moran_obs}
\begin{tabular}{lcccccc}
\hline
 & Krill & Whale & SST & Depth & SST.grd & Slope \\
\hline
$I_{\text{obs}}$ & 0.410 & 0.442 & 0.999 & 0.989 & 0.981 & 0.868 \\
\hline
\end{tabular}
\end{table}

Understanding the relationships between fin whales, their prey (krill), and environmental factors is essential for assessing the impacts of climate change on Antarctic marine ecosystems.
However, this task presents several statistical challenges.
First, the observed data consist of mixed outcomes: the number of fin whale sightings (count data) and krill biomass (continuous data).
Second, the data exhibit strong spatial correlation due to the spatial distribution of both species and oceanographic features.

Our goal is to infer the dependence structure between krill biomass and fin whale sightings, as well as their relationships with environmental factors, while properly accounting for the spatial correlation in the data.
This application demonstrates the need for the semiparametric spatial copula approach developed in this paper.

\section{Extended rank likelihood for semiparametric copula estimation} \label{sec:bgc}
Suppose that we observe $p$-dimensional observations of mixed outcomes $\bm{y}_i=(y_{i1},\ldots,y_{ip})^\top$ for $i=1,2,\ldots,n$, where each variable could be a variety of outcomes, such as continuous, count, and ordered variables. 
Here, we are interested in the dependence structure of $p$ outcomes. 
The marginal distributions of each component in $\bm{y}_i$ are not of primary interest. 
Let $\bm{z}_i$ be a $p$-dimensional continuous latent variable for $\bm{y}_i$. 
Then, we consider the following Gaussian copula model: 
\begin{equation}
\label{eq:bgc}
\begin{gathered}
\bm{z} = (\bm{z}_1^\top, \bm{z}_2^\top, \ldots, \bm{z}_n^\top)^\top \sim N_{pn}(\bm{0}, \bm{I}_n \otimes \bm{R}), \\
y_{ij} = F_j^{-1}[\Phi(z_{ij})], 
\end{gathered}
\end{equation}
where $\bm{R}$ is the $p\times p$ correlation matrix and $\bm{I}_n$, $\otimes$, $F_j^{-1}$, and $\Phi(\cdot)$ denote the $n \times n$ identity matrix, Kronecker product, inverse of an unknown univariate cumulative distribution function, and cumulative distribution function of the standard normal distribution, respectively.
The second line of \eqref{eq:bgc} represents the relationship between the latent variable $z_{ij}$ and the observed outcome $y_{ij}$, following the notation of \cite{hoff2007extending}.

Here, we focus on correlation matrix $\bm{R}$ as the dependence structure.
Although the Gaussian copula model \eqref{eq:bgc} separates the dependence structure $\bm{R}$ from the marginal distributions $F_1, \ldots, F_p$, inference on $\bm{R}$ generally requires estimation or specification of these marginals.
To overcome this difficulty, \cite{hoff2007extending} exploits the ordering constraints induced by the observed data.
Since $F_j$ is a nondecreasing function, observing $y_{ij} < y_{i'j}$ for any $i\neq i'$ implies that $z_{ij} < z_{i'j}$, regardless of the specific form of $F_j$.
For discrete outcomes with tied observations (i.e., $y_{ij} = y_{i'j}$), no ordering constraint is imposed between $z_{ij}$ and $z_{i'j}$.
Therefore, observing $\bm{y}=(\bm{y}_1^\top, \bm{y}_2^\top, \ldots, \bm{y}_n^\top)^\top$ indicates that $\bm{z}$ must lie within the set
\begin{equation*}
D = \{\bm{z}\in\mathbb{R}^{pn}: \max(z_{kj}: y_{kj}<y_{ij}) < z_{ij} < \min(z_{kj}: y_{ij}<y_{kj})\}.
\end{equation*}
Since the set $D$ is determined solely by the ordering of the observed data, it does not depend on the specific form or continuity of the marginal distributions $F_1, \ldots, F_p$.
Based on this observation, \cite{hoff2007extending} defines the following extended rank likelihood:
\begin{equation}
\label{eq:erl_bgc}
L(\bm{R}) = \Pr(\bm{z}\in D|\bm{R}) = \int_D \phi_{pn}(\bm{z};\bm{0},\bm{I}_n\otimes\bm{R}) d\bm{z},
\end{equation}
where $\phi_k(\bm{x}; \bm{\mu}, \bm{\Sigma})$ is the $k$-dimensional multivariate normal density with a mean vector $\bm{\mu}$ and a variance-covariance matrix $\bm{\Sigma}$.

Note that the extended rank likelihood \eqref{eq:erl_bgc} depends only on the parameter of interest, $\bm{R}$, and not on the marginal distributions $\bm{F}=(F_1,F_2,\ldots,F_p)^\top$.
Although the marginal distributions implicitly influence the observed data $\bm{y}$ through the transformation in \eqref{eq:bgc}, the likelihood itself does not require their explicit specification or estimation.
This also means that the discontinuity of $F_j$ for discrete outcomes does not affect inference on the dependence structure $\bm{R}$.
Following \cite{hoff2007extending}, we refer to this framework as ``semiparametric'' because the dependence structure is modeled parametrically through the Gaussian copula, while the marginal distributions are left completely unspecified.

The extended rank likelihood can be regarded as a type of marginal likelihood and can be derived in terms of the decomposition theorem on sufficient statistics.
That is,
\begin{align*}
p(\bm{y}|\bm{R},\bm{F}) &= p(\bm{z}\in D, \bm{y}|\bm{R},\bm{F}) \\
&= \Pr(\bm{z}\in D|\bm{R}) p(\bm{y}|\bm{z}\in D, \bm{R},\bm{F})
\end{align*}
and to estimate $\bm{R}$, it is sufficient to use only $\Pr(\bm{z}\in D|\bm{R})$.
The extended rank likelihood can also be seen as a multivariate version of the rank likelihood \citep{pettitt1982inference, heller2001pairwise}.
For a more detailed discussion on the extended rank likelihood and sufficient statistics, see \cite{hoff2007extending}, Section 5. 

\cite{hoff2007extending} further developed an algorithm to compute the posterior of $\bm{R}$ using the Gibbs sampler with the parameter expansion technique \citep{liu1999parameter}. 
This algorithm can also be used when observations are missing at random. 
The algorithm can be easily implemented in the \texttt{R} programming language using the \texttt{sbgcop.mcmc} function of the {\bf sbgcop} package.

\section{Copula estimation under spatial correlation} \label{sec:spbgc}

\subsection{Latent models with spatial correlation}
Suppose that we obtain observations of mixed outcomes $\bm{y}_{\S}=(\bm{y}(\bm{s}_1)^\top, \bm{y}(\bm{s}_2)^\top, \ldots, \bm{y}(\bm{s}_n)^\top)^\top$ equipped with location information $\S=\{\bm{s}_1,\bm{s}_2,\ldots\bm{s}_n\}$. 
Here, $\bm{y}(\bm{s}_i)=(y_1(\bm{s}_i),y_2(\bm{s}_i),\ldots,y_p(\bm{s}_i))^\top$ and $\bm{s}_i$ denotes a two-dimensional vector of longitude and latitude. 
These observations are characterized by spatial dependence and correlation, and nearby observations have similar properties. 
In such data, spatial correlations should be adequately considered; otherwise, the estimation of the dependence structure among outcomes would be severely biased. 
Therefore, we consider a hierarchical spatial model and latent multivariate Gaussian process. 
We define the following {\it spatial Gaussian copula} as a Gaussian copula model combined with a spatial Gaussian process to consider spatial correlation: 
\begin{equation}
\label{eq:spbgc}
\bm{z}_{\S} = (\bm{z}(\bm{s}_1)^\top, \bm{z}(\bm{s}_2)^\top, \ldots, \bm{z}(\bm{s}_n)^\top)^\top \sim N_{pn}(\bm{0}, \bm{H}(\phi)\otimes\bm{R}), 
\end{equation}
where $\bm{z}(\bm{s}_i)$ denotes a latent variable equipped with location information $\bm{s}_i$ and $\bm{H}(\phi)$ is a $n\times n$ matrix whose $(i,i')$-element is a valid correlation function $\rho(\|\bm{s}_i-\bm{s}_{i'}\|;\phi)$, such as exponential correlation function $\exp(-\|\bm{s}_i-\bm{s}_{i'}\|/\phi)$, with spatial range parameter $\phi$. 
$\bm{H}(\phi)$ is also interpreted as the correlation matrix of $\bm{z}_{(j)}^{\S} = (z_j(\bm{s}_1),z_j(\bm{s}_2),\ldots,z_j(\bm{s}_n))^\top$ with $\Corr(z_j(\bm{s}_i),z_j(\bm{s}_{i'}))=\rho(\|\bm{s}_i-\bm{s}_{i'}\|;\phi)$. 
Hereafter, we write $\bm{H}(\phi)$ as $\bm{H}$, where the dependence on $\phi$ is implicit, with similar notation for all spatial correlation matrices. 
The spatial Gaussian copula is identical to that of a multivariate Gaussian process with a correlation structure
\begin{equation*}
\Corr(\bm{z}(\bm{s}_i), \bm{z}(\bm{s}_{i'})) = \rho(\|\bm{s}_i-\bm{s}_{i'}\|; \phi) \cdot \bm{R},
\end{equation*} 
and its correlation structure is known as a separable correlation \citep[e.g., Chapter 9 in][]{banerjee2003hierarchical}. 

We consider the extended rank likelihood for the inference of the dependence structure in spatially correlated mixed outcomes, that is, the correlation matrix $\bm{R}$ in the spatial Gaussian copula \eqref{eq:spbgc}. 
Observation $\bm{y}_{\S}$ provides us information about the latent variable $\bm{z}_{\S}$ such that, for $j=1,2,\ldots,p$, $z_{j}(\bm{s}_i)<z_{j}(\bm{s}_{i'})$ holds when $y_{j}(\bm{s}_i)<y_{j}(\bm{s}_{i'})$ for an arbitrary pair $z_{j}(\bm{s}_i)$ and $z_{j}(\bm{s}_{i'})$ with $i\neq i'$, that is, $\bm{z}_{(j)}^{\S}$ must lie in the set 
\begin{equation*}
D_j = \{\bm{z}_{(j)}^{\S} \in \mathbb{R}^n: \max[z_{j}(\bm{s}_k): y_{j}(\bm{s}_k)<y_{j}(\bm{s}_i)] < z_{j}(\bm{s}_i) < \min[z_{j}(\bm{s}_k): y_{j}(\bm{s}_i)<y_{j}(\bm{s}_k)] \}. 
\end{equation*}
Then, the extended rank likelihood of $\bm{R}$ including the spatial range parameter $\phi$ under spatial Gaussian copula \eqref{eq:spbgc} is given by 
\begin{equation}
\label{eq:erl_spbgc}
L(\bm{R}, \phi) = \int_{D_1} \int_{D_2} \cdots \int_{D_p} \phi_{pn}(\bm{z}_{\S}; \bm{0},\bm{H}\otimes\bm{R}) d\bm{z}_{\S}. 
\end{equation}
Note that this extended rank likelihood considers spatially varying ranks that are not captured by the likelihood in \cite{hoff2007extending}. 

\subsection{Posterior computation} \label{subsec:post}
Extended rank likelihood \eqref{eq:erl_spbgc} allows us to develop an effective algorithm for calculating the posteriors of $\bm{R}$ and $\phi$ using MCMC. 
The joint posterior of $\bm{R}$, $\phi$, and $\bm{z}_{\S}$ given $\bm{y}_{\S}$ based on the extended rank likelihood \eqref{eq:erl_spbgc}, can be expressed as 
\begin{equation}
p(\bm{R},\phi,\bm{z}_{\S}|\bm{y}_{\S}) \propto p(\bm{R}) p(\phi) \phi_{pn}(\bm{z}_{\S};\bm{0},\bm{H}\otimes\bm{R}), ~~ \bm{z}_{\S} \in D_1 \times D_2 \times \cdots \times D_p. 
\end{equation}
Because it is difficult to construct a semi-conjugate prior distribution for $\bm{R}$ due to correlation matrix $\bm{R}$, we can consider a semi-conjugate prior distribution for the covariance matrix $\bm{V}$ using parameter expansion \citep{liu1999parameter}, as in \cite{hoff2007extending}. 
In other words, let $\bm{V}$ follow the inverse-Wishart prior distribution, $IW(v_0,v_0\bm{V}_0)$, and use the fact that $\bm{R}$ is equal to the distribution of the correlation matrix, where each element is $\bm{V}_{ij} / \sqrt{ \bm{V}_i \bm{V}_j }$. 
The MCMC algorithm for generating the posterior samples of $\bm{R}$, $\phi$, and $\bm{z}(\bm{s_i})$ for $i=1,2,\ldots,n$ is expressed as follows: 

\begin{itemize}
    \item Sampling of $\bm{z}(\bm{s}_i)$: 
    Generate $\bm{z}(\bm{s}_i)$ from a $p$-dimensional truncated normal distribution, $TN_p(\bm{\mu}_{\bm{s}_i},\bm{\Sigma}_{\bm{s}_i}; \bm{\ell}, \bm{u})$, where
    \begin{equation*}
        \bm{\mu}_{\bm{s}_i} = (\bm{H}_{\bm{s}_i,\S_{-i}} \bm{H}_{\S_{-i}}^{-1} \otimes \bm{I}_p) \bm{z}_{\S_{-i}}, \quad 
        \bm{\Sigma}_{\bm{s}_i} = (1 - \bm{H}_{\bm{s}_i,\S_{-i}} \bm{H}_{\S_{-i}}^{-1} \bm{H}_{\bm{s}_i,\S_{-i}}^\top) \bm{R}, 
    \end{equation*}
    $\S_{-i}=\{\bm{s}_{i'}|i\neq i', i'\in\mathbb{N}\}$, 
    $\bm{H}_{\bm{s}_i,\S_{-i}}$ is the $(n-1)$-dimensional cross-correlation row vector between $\bm{z}(\bm{s}_i)$ and $\bm{z}_{\S_{-i}}$, 
    $\bm{H}_{\S_{-i}}$ is the $(n-1)\times(n-1)$ correlation matrix of $\bm{z}_{\S_{-i}}$, 
    and the $j$-th element of $p$-dimensional vectors $\bm{\ell}$ and $\bm{u}$ are 
    \begin{equation*}
        \ell_j = \max[z_{j}(\bm{s}_k): y_{j}(\bm{s}_k)<y_{j}(\bm{s}_i)] ~~ \mbox{and} ~~ u_j = \min[z_{j}(\bm{s}_k): y_{j}(\bm{s}_i)<y_{j}(\bm{s}_k)], 
    \end{equation*}
    denoting the lower and upper bounds of the truncated normal distribution in each dimension. 
    
    \item Sampling $\bm{R}$: 
    Generate $\bm{V}$ from $IW\left(v_0+n, v_0\bm{V}_0 + \sum_{i=1}^n h_i^{-1}[\bm{z}(\bm{s}_i)-\bar{\bm{z}}_i][\bm{z}(\bm{s}_i)-\bar{\bm{z}}_i]^\top \right)$, where 
    \begin{equation*}
        h_i = 1 - \bm{H}_{\bm{s}_i,\C_i} \bm{H}_{\C_i}^{-1} \bm{H}_{\bm{s}_i, \C_i}^\top, \quad 
        \bar{\bm{z}}_i = ( \bm{H}_{\bm{s}_i,\C_i} \bm{H}_{\C_i}^{-1} \otimes \bm{I}_p ) \bm{z}_{\C_i}, 
    \end{equation*}
    $\C_i=\{\bm{s}_{i'}|i'<i, i'\in\mathbb{N}\}$, 
    $\bm{H}_{\bm{s}_i,\C_i}$ is the $c_i$-dimensional cross-correlation row vector between $\bm{z}(\bm{s}_i)$ and $\bm{z}_{\C_i}$ with $c_i=|\C_i|$, 
    and $\bm{H}_{\C_i}$ is the $c_i\times c_i$ correlation matrix of $\bm{z}_{\C_i}$, 
    and transform $\bm{R}_{ij} = \bm{V}_{ij} / \sqrt{ \bm{V}_i \bm{V}_j }$. 
    
    \item Sampling $\phi$: 
    The full conditional of $\phi$ is proportional to 
    \begin{equation*}
        |\bm{H}|^{-p/2} \exp\left\{ -\frac{1}{2} \sum_{i=1}^n h_i^{-1}[\bm{z}(\bm{s}_i)-\bar{\bm{z}}_i]^\top \bm{V}^{-1} [\bm{z}(\bm{s}_i)-\bar{\bm{z}}_i] \right\}. 
    \end{equation*}
    A random-walk Metropolis-Hastings is used to sample from this distribution. 
\end{itemize}

The posterior computation above involves sampling from a $p$-dimensional truncated normal (tMVN) distribution. 
For scalable posterior computation, it is important to discuss sampling from tMVN. 
Several approaches exist for sampling from a tMVN distribution, including sequential sampling from conditionally univariate truncated normal distributions \citep{geweke1991efficient, kotecha1999gibbs, damien2001sampling}, the Hamiltonian Markov chain algorithm \citep{pakman2014exact}, the minimax tilting accept-reject algorithm \citep{botev2017normal}, and an approximate sampling algorithm based on the cumulative distribution function of the standard logistic distribution \citep{souris2018soft}. 
Although the details are omitted here, we performed simulations to identify the most efficient tMVN sampling algorithm for our posterior computation, except for \cite{pakman2014exact}, which required tuning. 
Consequently, because there were no notable changes in the mixing properties, we adopted the algorithm of \cite{botev2017normal}, which is easier to implement using the \texttt{R} package \texttt{TruncatedNormal} \citep{botev2021truncatednormal}. 
However, when the number of dimensions exceeds 100, the acceptance probability decreases, and the algorithm slows down. 
In such cases, we should consider using other algorithms, such as an approximate sampling algorithm that uses the probit function for sampling from the tMVN. 

\subsection{Scalable posterior computation under large spatial data } \label{sec:nngp}
The Gaussian process has gained considerable popularity in spatial modeling owing to its excellent properties; however, it suffers from computational intractability, for example, in matrix operations and memory related to the covariance matrix (or correlation matrix) \citep{heaton2019case, liu2020gaussian}. 
The inference algorithm proposed in the previous section is no exception. 
In each iteration of the MCMC, the inverse of the spatial correlation matrix $\bm{H}$ in the Gaussian process and related matrix operations are required for every update of $\bm{z}(\bm{s}_i)$, $\bm{R}$ and $\phi$. 
Therefore, in the current algorithm, even with a dataset of a few thousand observations, it would take more than one day of computation time with a standard laptop and, in some cases, memory explosion would occur, rendering the analysis impossible. 

To address this problem, we employ the nearest-neighbor Gaussian process (NNGP) proposed by \cite{datta2016hierarchical} for $\bm{z}_\S$ with a multivariate normal distribution and sparse precision matrix defined as 
\begin{equation*}
p(\bm{z}_{\S}) = \prod_{i=1}^n \phi_p(\bm{z}(\bm{s}_i); \bm{B}_{\bm{s}_i}\bm{z}_{\N_i}, F_{\bm{s}_i} \bm{R} ), 
\end{equation*}
where
\begin{equation*}
\bm{B}_{\bm{s}_i} = \bm{H}_{\bm{s}_i,\N_i} \bm{H}_{\N_i}^{-1} \otimes \bm{I}_p, \quad F_{\bm{s}_i} = 1 - \bm{H}_{\bm{s}_i,\N_i} \bm{H}_{\N_i}^{-1} \bm{H}_{\bm{s}_i, \N_i}^\top, 
\end{equation*}
$\N_i = \{\bm{s}_{i'} | \mbox{$i'$ is an index of $m$-nearest neighbor of $\bm{s}_i$} \}$, $\bm{H}_{\bm{s}_i,\N_i}$ is the $n_i$-dimensional cross-correlation row vector between $\bm{z}(\bm{s}_i)$ and $\bm{z}_{\N_i}$ with $n_i=|\N_i| (\leq m)$, and $\bm{H}_{\N_i}$ is the $n_i\times n_i$ correlation matrix of $\bm{z}_{\N_i}$. 
Note that when $m = n - 1$, each observation has all other observations as its nearest neighbors, that is, $\N_i = \S_{-i}$, which recovers the full Gaussian process described in Section \ref{subsec:post}. 
Thus, the NNGP framework naturally encompasses the full GP through the single parameter $m$. 

Under NNGP, which is a type of Vecchia approximation \citep{vecchia1988estimation}, the conditional distribution of $\bm{z}(\bm{s}_i)$ can be expressed as 
\begin{equation*}
p(\bm{z}(\bm{s}_i) | \bm{z}_{\S_{-i}}) = p(\bm{z}(\bm{s}_i) | \bm{z}_{\D_i}), 
\end{equation*}
where $\D_i = \{\bm{s}_{i'} | \bm{s}_{i} \in \N_{i'} \} \cup \N_i$ since $(\bm{z}(\bm{s}_i), \bm{z}_{\D_i}) \indep \bm{z}_{\S_{-i} \setminus \D_i}$. 
Then, the MCMC algorithm for generating posterior samples of $\bm{R}$, $\phi$, and $\bm{z}(\bm{s}_i)$ is provided as follows: 

\begin{itemize}
    \item Sampling of $\bm{z}(\bm{s}_i)$: 
    Generate $\bm{z}(\bm{s}_i)$ from a $p$-dimensional truncated normal distribution, $TN_p(\tilde{\bm{\mu}}_{\bm{s}_i},\tilde{\bm{\Sigma}}_{\bm{s}_i}; \bm{\ell}, \bm{u})$, where 
    \begin{equation*}
        \tilde{\bm{\mu}}_{\bm{s}_i} = (\bm{H}_{\bm{s}_i,\D_i} \bm{H}_{\D_i}^{-1} \otimes \bm{I}_p) \bm{z}_{\D_i}, \quad 
        \tilde{\bm{\Sigma}}_{\bm{s}_i} = (1 - \bm{H}_{\bm{s}_i,\D_i} \bm{H}_{\D_i}^{-1} \bm{H}_{\bm{s}_i,\D_i}^\top) \bm{R}, 
    \end{equation*}
    $\bm{H}_{\bm{s}_i,\D_i}$ is the $d_i$-dimensional cross-correlation row vector between $\bm{z}(\bm{s}_i)$ and $\bm{z}_{\D_i}$ with $d_i=|\D_i|$, 
    $\bm{H}_{\D_i}$ is the $d_i\times d_i$ correlation matrix of $\bm{z}_{\D_i}$. 
    
    \item Sampling $\bm{R}$: 
    Generate $\bm{V}$ from 
    \begin{equation*}
        IW\left(v_0+n, v_0\bm{V}_0 + \sum_{i=1}^n F_{\bm{s}_i}^{-1}[\bm{z}(\bm{s}_i)-\bm{B}_{\bm{s}_i}\bm{z}_{\N_i}][\bm{z}(\bm{s}_i)-\bm{B}_{\bm{s}_i}\bm{z}_{\N_i}]^\top \right), 
    \end{equation*}
    and transform $\bm{R}_{ij} = \bm{V}_{ij} / \sqrt{ \bm{V}_i \bm{V}_j }$. 
    
    \item Sampling $\phi$: 
    The full conditional of $\phi$ is proportional to 
    \begin{equation*}
        |\bm{H}|^{-p/2} \exp\left\{ -\frac{1}{2} \sum_{i=1}^n F_{\bm{s}_i}^{-1}[\bm{z}(\bm{s}_i)-\bm{B}_{\bm{s}_i}\bm{z}_{\N_i}]^\top \bm{V}^{-1} [\bm{z}(\bm{s}_i)-\bm{B}_{\bm{s}_i}\bm{z}_{\N_i}] \right\}. 
    \end{equation*}
    A random-walk Metropolis-Hastings is used to sample from this distribution. 
\end{itemize}
The full conditional of $\bm{z}(\bm{s}_i)$ then requires only $d_i$- rather than $(n-1)$-dimensional matrix operations (Section~\ref{subsec:post}), and those of $\bm{R}$ and $\phi$ only $m$ ($\ll n$)-dimensional ones.
Specifically, the computational complexity of the NNGP is $O(nm^3)$ per MCMC iteration, compared to $O(n^3)$ for the full GP \citep{datta2016hierarchical, finley2019efficient}.
With respect to the number of outcomes, the matrix operations are inexpensive, $O(p^3)$ once per iteration for the inverse-Wishart update of $\bm{R}$ and $O(np^2)$ for the conditional moments, whereas the cost of the $n$ draws from $p$-variate truncated normals is less predictable: each needs an $O(p^3)$ factorization and an optimization for the minimax tilting \citep{botev2017normal}, its number of accept--reject proposals depends on the truncation region, and the rank-based truncation limits must be found for all $np$ latent variables.
A benchmark with $p\le 50$ (Section~\ref{sec:p_scalability} of the Supplementary Material) shows that the truncated-normal draws and the search for their truncation limits take $87$--$97\%$ of the time per iteration, which grows only about $4.5$-fold from $p=6$ to $50$.

Although the value of $m$ needs to be chosen based on the available computational resources, in many situations, relatively small values of $m$ (e.g., $m < 100$) can achieve high accuracy. 
For a detailed discussion, see, for example, \cite{katzfuss2021general}. 
In practice, in our method, even with a sample from a few hundreds to thousands of observations, the estimation accuracy remains almost the same, whereas the computation time is significantly reduced compared to the full Gaussian process.
More specifically, our sensitivity analyses across a range of spatial correlation strengths (Section~\ref{sec:nngp_sensitivity} of the Supplementary Material) and under the actual, non-uniform survey configuration (Section~\ref{sec:real_locations}) confirm that $m$ between $10$ and $15$ (or $n/10$) attains accuracy and coverage close to those of the full Gaussian process, with only $m=5$ showing mild undercoverage at the largest sample sizes.

\section{Simulation study} \label{sec:vshoff}
Using a simulation study, this section demonstrates that the proposed approach successfully accounts for spatial correlation and correctly infers the dependence structure among spatially correlated multivariate mixed outcomes, including a comparison with \cite{hoff2007extending}. 
To this end, we consider spatial Gaussian copula \eqref{eq:spbgc} with $n \in \{50, 500, 1000\}$ and $p \in \{6, 9\}$, where $\phi \in \{0.05, 0.25, 0.5\}$ and $\bm{R}_{12}=0.5$, $\bm{R}_{14}=0.3$, $\bm{R}_{15}=0.2$, $\bm{R}_{23}=-0.2$, $\bm{R}_{24}=-0.3$, $\bm{R}_{35}=0.4$, $\bm{R}_{45}=-0.5$, and all others to zero for $j<j'$. 
Location information $\bm{s}_i = (s_{i1}, s_{i2})^{\top}$ are generated independently, where both coordinates $s_{i1}$ and $s_{i2}$ are drawn from a uniform distribution on the unit square $[0, 1] \times [0, 1]$, which represents a common scenario in spatial statistics where observations are randomly distributed across a study region.
To examine the sensitivity of our results to the spatial configuration, we also conduct simulations using the actual spatial locations from the real data application; see Section~\ref{sec:real_locations} of the Supplementary Material.

Using the spatial Gaussian copula, we generate a simulated dataset based on \cite{smith2021implicit}: 
\begin{enumerate}
    \item
    Calculate spatial correlation matrix $\bm{H}$ with location information $\bm{s}_i = (s_{i1},s_{i2})^\top$, using the exponential correlation function for $\rho(\cdot; \phi)$ in \eqref{eq:spbgc}. 

    \item 
    Generate latent variables $\bm{z}_{\S}$ from multivariate Gaussian process \eqref{eq:spbgc}. 
    
    \item 
    Calculate $y_j(\bm{s}_i) = F_j^{-1}[\Phi(z_j(\bm{s}_i))]$, where $F_j$ is a specified cumulative distribution function for $j=1,2,\ldots,p$. 
\end{enumerate}
In our simulation setting, $y_1(\bm{s}_i) \sim {\rm Bernoulli}(0.5)$, $y_2(\bm{s}_i) \sim {\rm Poi}(15)$, $y_3(\bm{s}_i) \sim {\rm Poi}(5)$, $y_4(\bm{s}_i) \sim {\rm Ordered Categorical}(0.3,0.15,0.1,0.25,0.2)$, and all other $y_j(\bm{s}_i)$s are generated from normal distributions. 

For the simulated dataset, we apply the proposed and \cite{hoff2007extending} methods, denoted as spBGC and BGC, respectively. 
In doing so, we use 2000 draws for the posterior computation after discarding the first 1000 draws as burn-in, and we set $IW(p+2, (p+2)\bm{I})$ prior for $\bm{V}$ as a prior for $\bm{V}$. 
We compute posterior medians $\{\hat{\bm{R}}_{jj'}\}_{j<j'}$ as point estimates of $\bm{R}_{jj'}$ for $j<j'$ and evaluate their performance by the mean squared error (MSE), defined as $q^{-1} \sum_{j<j'} (\hat{\bm{R}}_{jj'} - \bm{R}_{jj'})^2$ where $q=p(p-1)/2$. 
We also compute 95\% credible intervals and calculate coverage probabilities (CP) and average lengths (AL), defined as $q^{-1} \sum_{j<j'} I(\bm{R}_{jj'} \in {\rm CI}_{jj'})$ and $q^{-1} \sum_{j<j'} |{\rm CI}_{jj'}|$, respectively, where $I(\cdot)$ is the indicator function and ${\rm CI}_{jj'}$ is the 95\% credible interval of $\bm{R}_{jj'}$. 
These values are averaged over 300 replications of the simulated datasets. 

Furthermore, we determine the advantages of using NNGP in the spBGC. 
Hereafter, we denote spBGCNNGP to clarify its use. 
As described below, although spBGCNNGP can considerably reduce the computation time compared to spBGC, even for a sample size of 500, its estimation accuracy remains almost the same. 
Because spBGC requires considerable time per simulation when the sample size is 1000, we evaluate the performance of spBGC with only 10 replications in this setting; consequently, the spBGC results at $n=1000$ are subject to larger Monte Carlo error than those of spBGCNNGP, which use 300 replications.
This simulation study is conducted on an Apple Mac Studio with an Apple M1 Ultra chip, using programming language \texttt{R} \citep{R2024}. 

\begin{figure}[H]
    \centering
    \begin{minipage}{\textwidth}
        \centering
        \includegraphics[width=\textwidth]{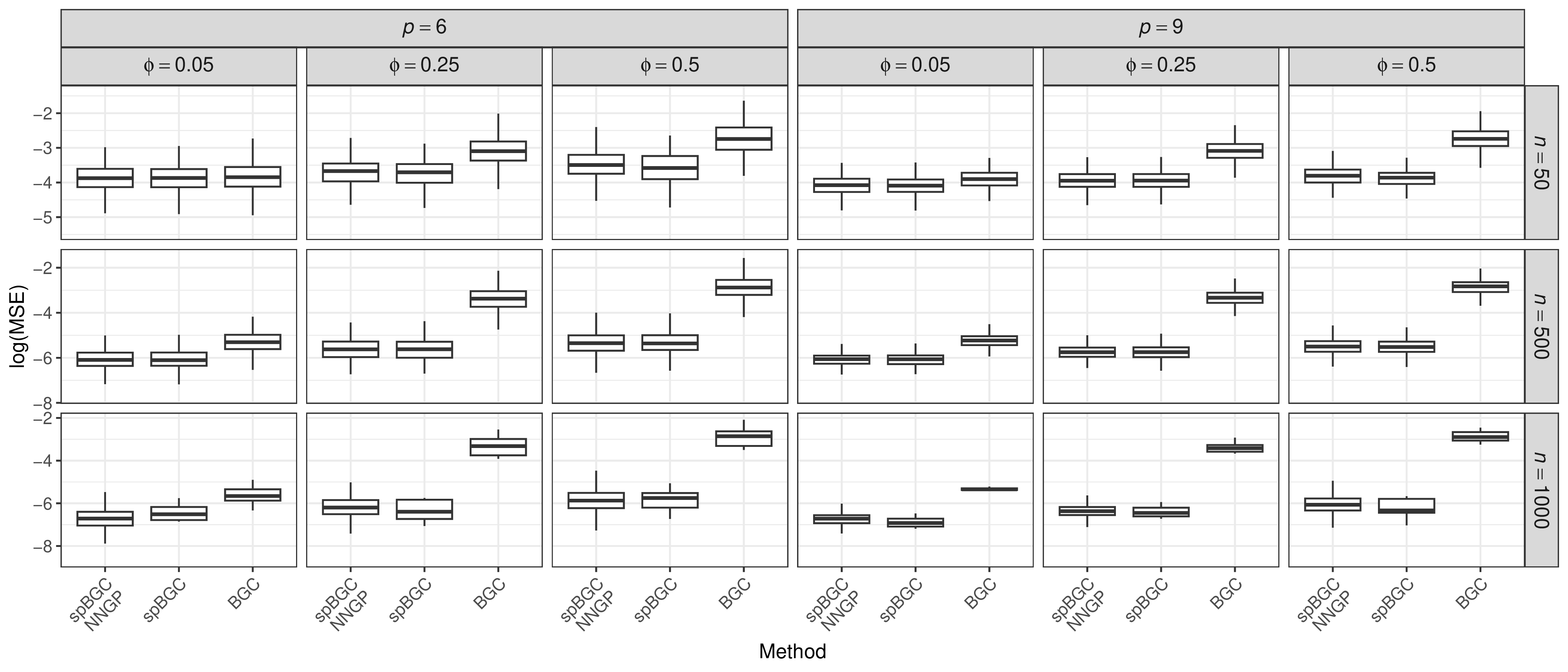}
        \subcaption{The logarithm of the MSEs for $p=6$ (left) and $p=9$ (right)}
    \end{minipage}\\
    \begin{minipage}{\textwidth}
        \centering
        \includegraphics[width=\textwidth]{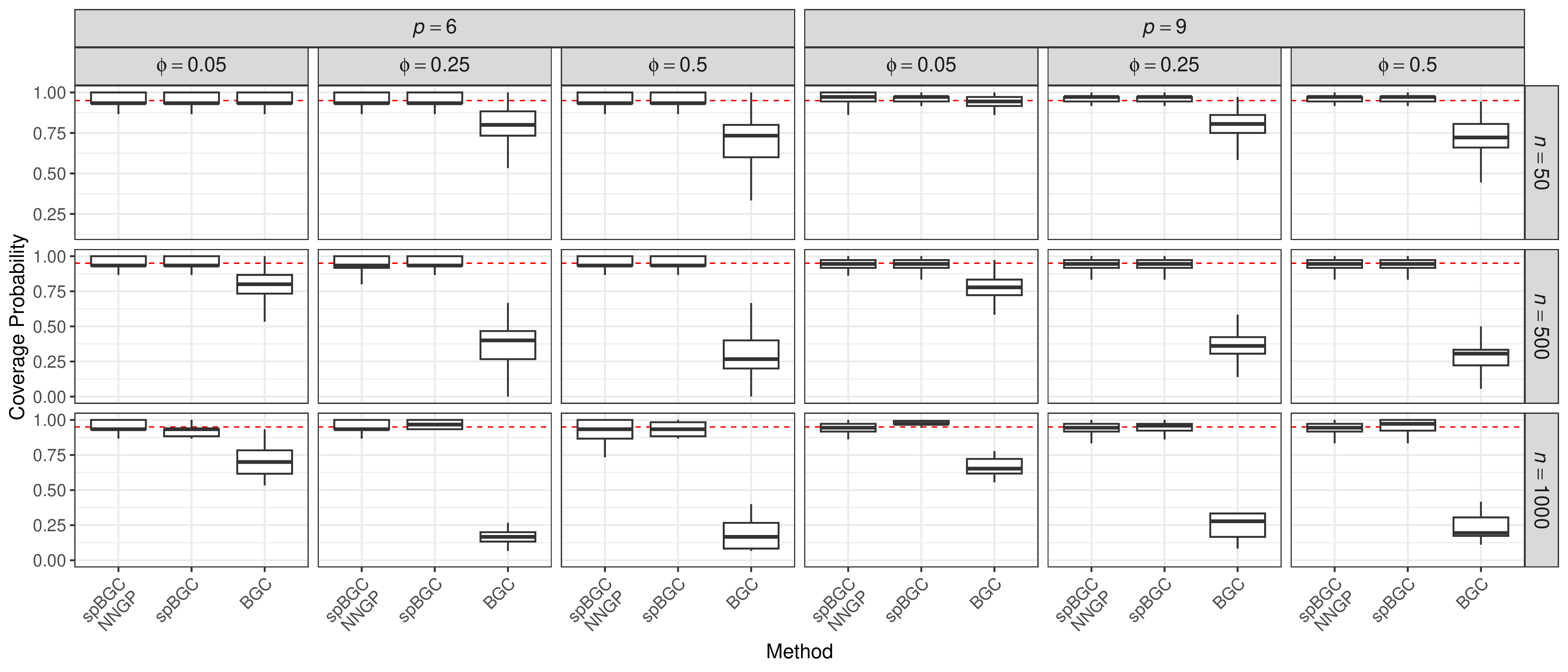}
        \subcaption{The CPs for $p=6$ (left) and $p=9$ (right)}
    \end{minipage}\\
    \begin{minipage}{\textwidth}
        \centering
        \includegraphics[width=\textwidth]{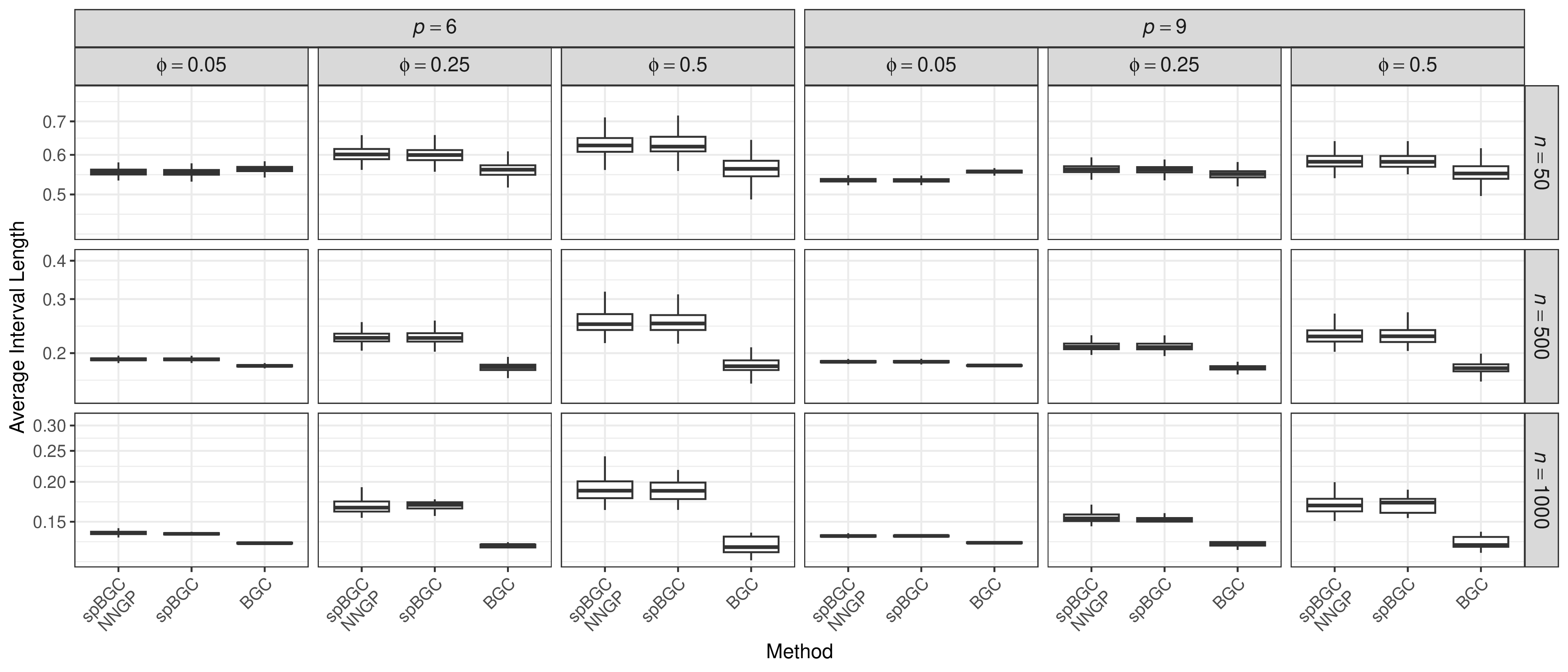}
        \subcaption{The ALs for $p=6$ (left) and $p=9$ (right)}
    \end{minipage}
    \caption{
        Comparisons of the logarithm of the MSEs (a), CPs (b), and ALs (c) for the spBGC, spBGCNNGP, and BGC under varying sample sizes ($n \in \{50, 500, 1000\}$) and spatial range parameters ($\phi \in \{0.05, 0.25, 0.5\}$).
        The dashed line in the CP graphs indicates the target coverage probability of 0.95.
        Each figure shows results for the number of outcomes $p=6$ on the left and $p=9$ on the right.
    }
    \label{fig:ne}
\end{figure}

The simulation results presented in Figure \ref{fig:ne} and Tables \ref{tb:mse}--\ref{tb:time} clearly demonstrate the superior performance of our proposed methods, spBGC, and its computationally efficient version, spBGCNNGP, compared with the existing BGC method. 
Figure \ref{fig:ne}(a) and Table \ref{tb:mse} show that both spBGC and spBGCNNGP achieve significantly lower MSEs than BGC across all scenarios. 
This difference becomes more pronounced as spatial correlation ($\phi$) increases, highlighting the ability of the proposed method to effectively account for spatial dependence. 
The CPs displayed in Figure \ref{fig:ne}(b) and Table \ref{tb:cp} reveal that spBGC and spBGCNNGP consistently maintain CPs close to the nominal 95\% level. 
By contrast, BGC coverage deteriorates substantially as the spatial correlation strengthens, often falling well below the desired level. 
This underscores the reliability of the proposed method for providing accurate credible intervals. 
The ALs of the credible intervals, as shown in Figure \ref{fig:ne}(c) and Table \ref{tb:al}, indicate that spBGC and spBGCNNGP achieve high coverage probabilities, while maintaining reasonably narrow interval widths. 
Balancing coverage and precision is crucial for practical applications. 
Most notably, Table \ref{tb:time} highlights the computational efficiency of spBGCNNGP. 
As the sample size increases, the time savings offered by spBGCNNGP become increasingly significant. 
For instance, for $n = 1000$, spBGCNNGP completes the computation in approximately one-ninth of the time required by the spBGC, while maintaining a comparable estimation accuracy. 

These results collectively demonstrate that the proposed methods, particularly the spBGCNNGP, offer a powerful combination of estimation accuracy and computational efficiency for analyzing spatially correlated multivariate mixed outcomes. 
The ability of spBGCNNGP to significantly reduce computation time without sacrificing estimation quality is particularly valuable for large-scale spatial datasets, which are becoming increasingly common in various research areas.
By employing the spBGCNNGP, researchers and practitioners can perform high-precision estimations of spatial dependence structures on datasets that were previously computationally infeasible. 
This opens new possibilities for in-depth spatial analyses across a wide range of applications, from environmental science to epidemiology. 
In conclusion, the spBGCNNGP is a robust and efficient tool for analyzing spatially correlated multivariate mixed outcomes, offering an optimal balance between accuracy and computational efficiency. 
Its ability to handle large-scale spatial datasets with high precision renders it an invaluable asset for researchers studying complex spatial datasets.
To assess the robustness of our findings to the choice of spatial correlation function, copula model, and the structure and magnitude of the correlation matrix $\bm{R}$, we also conduct simulations using Mat\'ern 3/2 and Mat\'ern 5/2 correlation functions, the $t$-copula model, and dense and small-magnitude correlation matrices; see Sections~\ref{sec:correlation_misspec}--\ref{sec:R_structure} of the Supplementary Material.

\begin{table}[H]
\centering
\caption{Comparisons of the logarithm of the MSEs for the spBGC, spBGCNNGP (with $m=n/10$), and BGC under varying sample sizes and spatial range parameters.
The values represent average log(MSE)s from 300 calculations (10 for spBGC at $n=1000$), with standard errors in parentheses.}
\label{tb:mse}
\begin{minipage}{.48\linewidth}
  \centering
  \subcaption{Number of outcomes $p=6$}
  \begin{tabularx}{\linewidth}{l *{3}{>{\centering\arraybackslash}X}}
  \hline
  & \multicolumn{3}{c}{$\phi$} \\
  \cline{2-4}
   & 0.05 & 0.25 & 0.50  \\
  \hline
  \multicolumn{4}{l}{$n=50$} \\
  spBGC & $-3.884$ (0.023) & $-3.729$ (0.021) & $-3.567$ (0.024) \\
  spBGCNNGP & $-3.876$ (0.023) & $-3.687$ (0.023) & $-3.479$ (0.025) \\
  BGC & $-3.853$ (0.024) & $-3.106$ (0.028) & $-2.729$ (0.027) \\
  \hline
  \multicolumn{4}{l}{$n=500$} \\
  spBGC & $-6.085$ (0.026) & $-5.639$ (0.028) & $-5.322$ (0.033) \\
  spBGCNNGP & $-6.089$ (0.026) & $-5.636$ (0.028) & $-5.320$ (0.032) \\
  BGC & $-5.326$ (0.026) & $-3.395$ (0.030) & $-2.888$ (0.030) \\
  \hline
  \multicolumn{4}{l}{$n=1000$} \\
  spBGC & $-6.538$ (0.172) & $-6.343$ (0.162) & $-5.776$ (0.326) \\
  spBGCNNGP & $-6.734$ (0.028) & $-6.189$ (0.027) & $-5.847$ (0.035) \\
  BGC & $-5.602$ (0.026) & $-3.333$ (0.027) & $-2.874$ (0.029) \\
  \hline
  \end{tabularx}
\end{minipage}%
\hfill
\begin{minipage}{.48\linewidth}
  \centering
  \subcaption{Number of outcomes $p=9$}
  \begin{tabularx}{\linewidth}{l *{3}{>{\centering\arraybackslash}X}}
  \hline
  & \multicolumn{3}{c}{$\phi$} \\
  \cline{2-4}
   & 0.05 & 0.25 & 0.50  \\
  \hline
  \multicolumn{4}{l}{$n=50$} \\
  spBGC & $-4.089$ (0.016) & $-3.952$ (0.015) & $-3.864$ (0.015) \\
  spBGCNNGP & $-4.087$ (0.016) & $-3.943$ (0.016) & $-3.790$ (0.018) \\
  BGC & $-3.916$ (0.016) & $-3.084$ (0.017) & $-2.739$ (0.019) \\
  \hline
  \multicolumn{4}{l}{$n=500$} \\
  spBGC & $-6.086$ (0.015) & $-5.751$ (0.018) & $-5.493$ (0.023) \\
  spBGCNNGP & $-6.085$ (0.015) & $-5.749$ (0.018) & $-5.481$ (0.023) \\
  BGC & $-5.242$ (0.017) & $-3.337$ (0.020) & $-2.851$ (0.019) \\
  \hline
  \multicolumn{4}{l}{$n=1000$} \\
  spBGC & $-6.886$ (0.080) & $-6.393$ (0.091) & $-6.238$ (0.145) \\
  spBGCNNGP & $-6.736$ (0.016) & $-6.352$ (0.018) & $-6.015$ (0.027) \\
  BGC & $-5.406$ (0.011) & $-3.376$ (0.015) & $-2.860$ (0.016) \\
  \hline
  \end{tabularx}
\end{minipage}
\end{table}

\begin{table}[H]
\centering
\caption{Comparisons of the coverage probabilities for the spBGC, spBGCNNGP (with $m=n/10$), and BGC under varying sample sizes and spatial range parameters.
The values represent average coverage probabilities from 300 calculations (10 for spBGC at $n=1000$), with standard errors in parentheses.}
\label{tb:cp}
\begin{minipage}{.48\linewidth}
  \centering
  \subcaption{Number of outcomes $p=6$}
  \begin{tabularx}{\linewidth}{l *{3}{>{\centering\arraybackslash}X}}
  \hline
  & \multicolumn{3}{c}{$\phi$} \\
  \cline{2-4}
   & 0.05 & 0.25 & 0.50  \\
  \hline
  \multicolumn{4}{l}{$n=50$} \\
  spBGC & $0.945$ (0.004) & $0.954$ (0.003) & $0.942$ (0.004) \\
  spBGCNNGP & $0.946$ (0.004) & $0.948$ (0.003) & $0.936$ (0.004) \\
  BGC & $0.942$ (0.004) & $0.802$ (0.008) & $0.712$ (0.009) \\
  \hline
  \multicolumn{4}{l}{$n=500$} \\
  spBGC & $0.946$ (0.003) & $0.940$ (0.004) & $0.937$ (0.004) \\
  spBGCNNGP & $0.947$ (0.003) & $0.938$ (0.004) & $0.940$ (0.004) \\
  BGC & $0.786$ (0.007) & $0.368$ (0.008) & $0.299$ (0.008) \\
  \hline
  \multicolumn{4}{l}{$n=1000$} \\
  spBGC & $0.920$ (0.019) & $0.967$ (0.011) & $0.933$ (0.017) \\
  spBGCNNGP & $0.945$ (0.004) & $0.937$ (0.004) & $0.932$ (0.004) \\
  BGC & $0.680$ (0.010) & $0.173$ (0.006) & $0.193$ (0.007) \\
  \hline
  \end{tabularx}
\end{minipage}%
\hfill
\begin{minipage}{.48\linewidth}
  \centering
  \subcaption{Number of outcomes $p=9$}
  \begin{tabularx}{\linewidth}{l *{3}{>{\centering\arraybackslash}X}}
  \hline
  & \multicolumn{3}{c}{$\phi$} \\
  \cline{2-4}
   & 0.05 & 0.25 & 0.50  \\
  \hline
  \multicolumn{4}{l}{$n=50$} \\
  spBGC & $0.960$ (0.002) & $0.961$ (0.002) & $0.962$ (0.002) \\
  spBGCNNGP & $0.961$ (0.002) & $0.959$ (0.002) & $0.958$ (0.002) \\
  BGC & $0.949$ (0.002) & $0.798$ (0.005) & $0.716$ (0.006) \\
  \hline
  \multicolumn{4}{l}{$n=500$} \\
  spBGC & $0.947$ (0.002) & $0.943$ (0.002) & $0.945$ (0.002) \\
  spBGCNNGP & $0.949$ (0.002) & $0.943$ (0.002) & $0.941$ (0.002) \\
  BGC & $0.780$ (0.004) & $0.363$ (0.005) & $0.289$ (0.005) \\
  \hline
  \multicolumn{4}{l}{$n=1000$} \\
  spBGC & $0.972$ (0.008) & $0.944$ (0.014) & $0.956$ (0.017) \\
  spBGCNNGP & $0.950$ (0.002) & $0.944$ (0.002) & $0.941$ (0.002) \\
  BGC & $0.664$ (0.005) & $0.250$ (0.006) & $0.233$ (0.006) \\
  \hline
  \end{tabularx}
\end{minipage}
\end{table}

\begin{table}[H]
\centering
\caption{Comparisons of the average credible interval lengths for the spBGC, spBGCNNGP (with $m=n/10$), and BGC under varying sample sizes and spatial range parameters.
The values represent average interval lengths from 300 calculations (10 for spBGC at $n=1000$), with standard errors in parentheses.}
\label{tb:al}
\begin{minipage}{.48\linewidth}
  \centering
  \subcaption{Number of outcomes $p=6$}
  \begin{tabularx}{\linewidth}{l *{3}{>{\centering\arraybackslash}X}}
  \hline
  & \multicolumn{3}{c}{$\phi$} \\
  \cline{2-4}
   & 0.05 & 0.25 & 0.50  \\
  \hline
  \multicolumn{4}{l}{$n=50$} \\
  spBGC & $0.554$ (0.001) & $0.603$ (0.001) & $0.632$ (0.002) \\
  spBGCNNGP & $0.555$ (0.001) & $0.605$ (0.001) & $0.632$ (0.002) \\
  BGC & $0.562$ (0.001) & $0.561$ (0.002) & $0.568$ (0.003) \\
  \hline
  \multicolumn{4}{l}{$n=500$} \\
  spBGC & $0.191$ (0.000) & $0.226$ (0.001) & $0.256$ (0.002) \\
  spBGCNNGP & $0.191$ (0.000) & $0.226$ (0.001) & $0.256$ (0.002) \\
  BGC & $0.182$ (0.000) & $0.180$ (0.000) & $0.188$ (0.002) \\
  \hline
  \multicolumn{4}{l}{$n=1000$} \\
  spBGC & $0.138$ (0.000) & $0.169$ (0.003) & $0.196$ (0.011) \\
  spBGCNNGP & $0.138$ (0.000) & $0.168$ (0.001) & $0.194$ (0.001) \\
  BGC & $0.129$ (0.000) & $0.127$ (0.000) & $0.141$ (0.003) \\
  \hline
  \end{tabularx}
\end{minipage}%
\hfill
\begin{minipage}{.48\linewidth}
  \centering
  \subcaption{Number of outcomes $p=9$}
  \begin{tabularx}{\linewidth}{l *{3}{>{\centering\arraybackslash}X}}
  \hline
  & \multicolumn{3}{c}{$\phi$} \\
  \cline{2-4}
   & 0.05 & 0.25 & 0.50  \\
  \hline
  \multicolumn{4}{l}{$n=50$} \\
  spBGC & $0.534$ (0.000) & $0.561$ (0.001) & $0.585$ (0.001) \\
  spBGCNNGP & $0.534$ (0.000) & $0.563$ (0.001) & $0.584$ (0.001) \\
  BGC & $0.555$ (0.000) & $0.550$ (0.001) & $0.555$ (0.002) \\
  \hline
  \multicolumn{4}{l}{$n=500$} \\
  spBGC & $0.188$ (0.000) & $0.211$ (0.000) & $0.231$ (0.001) \\
  spBGCNNGP & $0.188$ (0.000) & $0.211$ (0.000) & $0.230$ (0.001) \\
  BGC & $0.182$ (0.000) & $0.180$ (0.000) & $0.182$ (0.001) \\
  \hline
  \multicolumn{4}{l}{$n=1000$} \\
  spBGC & $0.135$ (0.000) & $0.153$ (0.002) & $0.170$ (0.004) \\
  spBGCNNGP & $0.135$ (0.000) & $0.155$ (0.000) & $0.172$ (0.001) \\
  BGC & $0.129$ (0.000) & $0.128$ (0.000) & $0.130$ (0.000) \\
  \hline
  \end{tabularx}
\end{minipage}
\end{table}

\begin{table}[H]
\centering
\caption{Comparison of execution times for the spBGC and spBGCNNGP under varying sample sizes and spatial range parameters.
The values represent average execution times from 300 calculations (10 for spBGC at $n=1000$), with standard errors in parentheses.}
\label{tb:time}
\begin{minipage}{.48\linewidth}
  \centering
  \subcaption{Number of outcomes $p=6$}
  \begin{tabularx}{\linewidth}{l *{3}{>{\centering\arraybackslash}X}}
  \hline
  & \multicolumn{3}{c}{$\phi$} \\
  \cline{2-4}
   & 0.05 & 0.25 & 0.50  \\
  \hline
  \multicolumn{4}{l}{$n=50$} \\
  spBGC  & 111.766 & 109.170 & 112.211 \\
  & (0.569) & (0.847) & (0.564) \\
  spBGC  & 110.811 & 110.377 & 112.377 \\
  NNGP & (0.637) & (0.651) & (0.404) \\
  \hline
  \multicolumn{4}{l}{$n=500$} \\
  spBGC  & 2895.816 & 2817.129 & 2904.990 \\
  & (19.365) & (21.468) & (19.360) \\
  spBGC  & 1106.890 & 1094.114 & 1075.157 \\
  NNGP & (7.495) & (7.380) & (7.944) \\
  \hline
  \multicolumn{4}{l}{$n=1000$} \\
  spBGC  & 23516.638 & 23486.437 & 23267.765 \\
  & (18.329) & (21.710) & (184.699) \\
  spBGC  & 2674.612 & 2667.926 & 2639.537 \\
  NNGP & (13.036) & (9.235) & (15.207) \\
  \hline
  \end{tabularx}
\end{minipage}%
\hfill
\begin{minipage}{.48\linewidth}
  \centering
  \subcaption{Number of outcomes $p=9$}
  \begin{tabularx}{\linewidth}{l *{3}{>{\centering\arraybackslash}X}}
  \hline
  & \multicolumn{3}{c}{$\phi$} \\
  \cline{2-4}
   & 0.05 & 0.25 & 0.50  \\
  \hline
  \multicolumn{4}{l}{$n=50$} \\
  spBGC  & 123.568 & 123.158 & 126.839 \\
  & (1.038) & (1.038) & (0.843) \\
  spBGC  & 126.917 & 127.625 & 121.738 \\
  NNGP & (0.752) & (0.588) & (0.817) \\
  \hline
  \multicolumn{4}{l}{$n=500$} \\
  spBGC  & 3083.322 & 3046.740 & 3004.888 \\
  & (22.758) & (24.516) & (21.437) \\
  spBGC  & 1272.467 & 1256.959 & 1291.827 \\
  NNGP & (10.080) & (9.477) & (6.836) \\
  \hline
  \multicolumn{4}{l}{$n=1000$} \\
  spBGC  & 23999.193 & 23957.520 & 23921.606 \\
  & (38.057) & (32.902) & (103.828) \\
  spBGC  & 3168.153 & 3101.358 & 3123.644 \\
  NNGP & (12.227) & (18.245) & (14.450) \\
  \hline
  \end{tabularx}
\end{minipage}
\end{table}

\section{Application to fin whale sighting data} \label{sec:real}
We apply our method to the data set comprising krill biomass data, visual sighting data of fin whales from line transect surveys in the Southern Ocean, and four environmental variables.
The detailed description of the data is provided in Section \ref{sec:data}.
We are interested in the dependence structure of krill biomass and sighting data of fin whales, and that of each of these outcomes and environmental factors (SST, Depth, Slope, and SST.grd). 
Therefore, this application focuses on the posterior inference of these dependence structures. 
We compare the results of the proposed method (spBGC) with those of \cite{hoff2007extending}'s method (BGC), which does not consider spatial correlation. 
Before the comparison, we examine the mixing properties of spBGC. 
Using an $IW(p+2, (p+2)\bm{I})$ prior, that is, $IW(8, 8\bm{I})$ for $\bm{V}$, we conduct MCMC with 25,000 iterations, discarding the first 5,000 as burn-in and retaining every 10th sample from the remaining chain, yielding 2,000 samples for posterior calculations.
For the spatial correlation function $\rho(\cdot; \phi)$ in \eqref{eq:spbgc}, we use the exponential correlation function.
To assess the sensitivity of our results to this choice, we also conduct analyses using Mat\'ern 3/2 and Mat\'ern 5/2 correlation functions; see Section B of the Supplementary Material.

Figure \ref{fg:da_mixing} presents the mixing and autocorrelation results for the correlation coefficients of interest. 
From these figures, the mixing properties of our spBGC algorithm are quite satisfactory, that is, convergence to stationarity appears and the autocorrelation at lag-20 is close to zero for most elements. 

\begin{figure}[H]
    \begin{minipage}[b]{0.49\linewidth}
        \centering
        \includegraphics[width=\columnwidth]{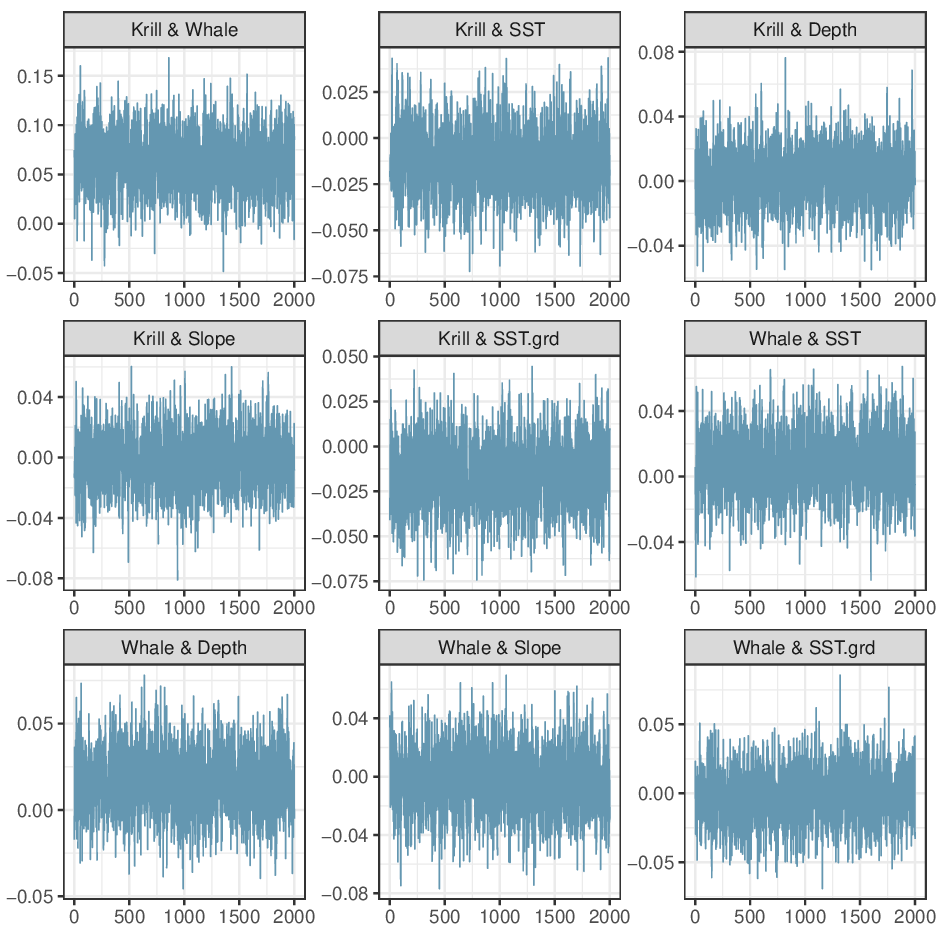}
    \end{minipage}
    \begin{minipage}[b]{0.49\linewidth}
        \centering
        \includegraphics[width=\columnwidth]{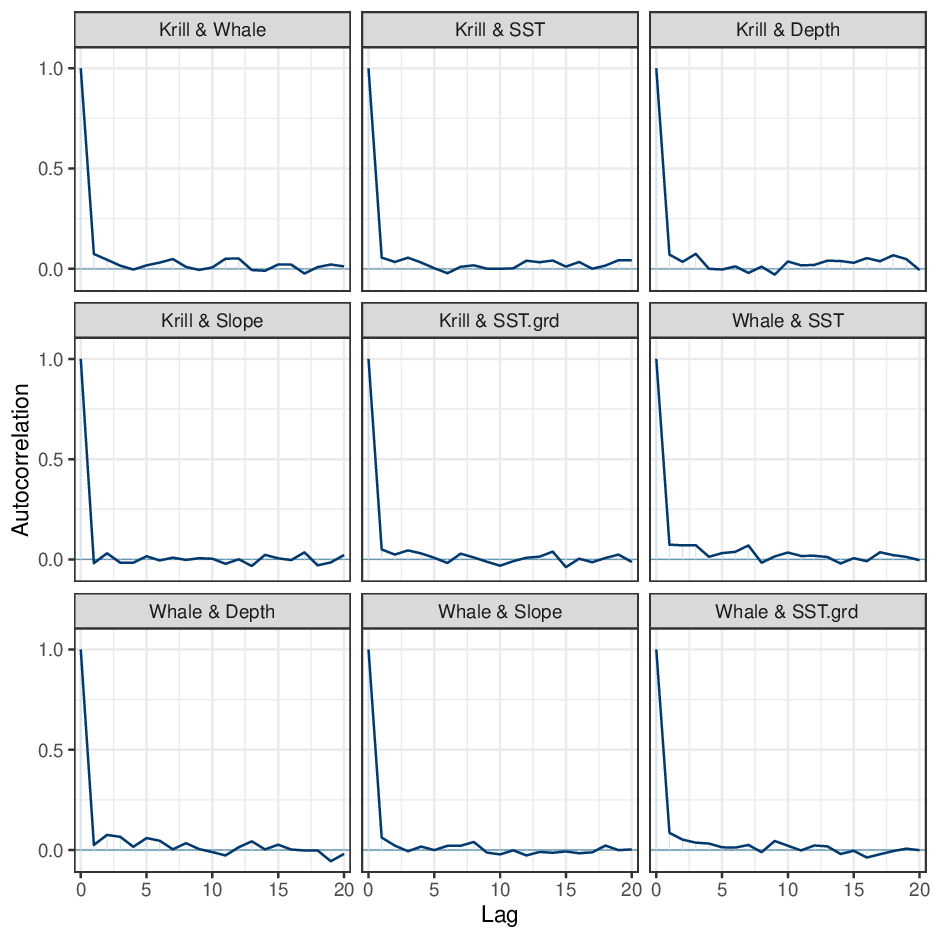}
    \end{minipage}
    \caption{Trace plots and autocorrelation of posterior draws of the correlation coefficients based on the spBGC algorithm.}
    \label{fg:da_mixing}
\end{figure}

\begin{figure}[H]
    \centering
    \includegraphics[width=\columnwidth]{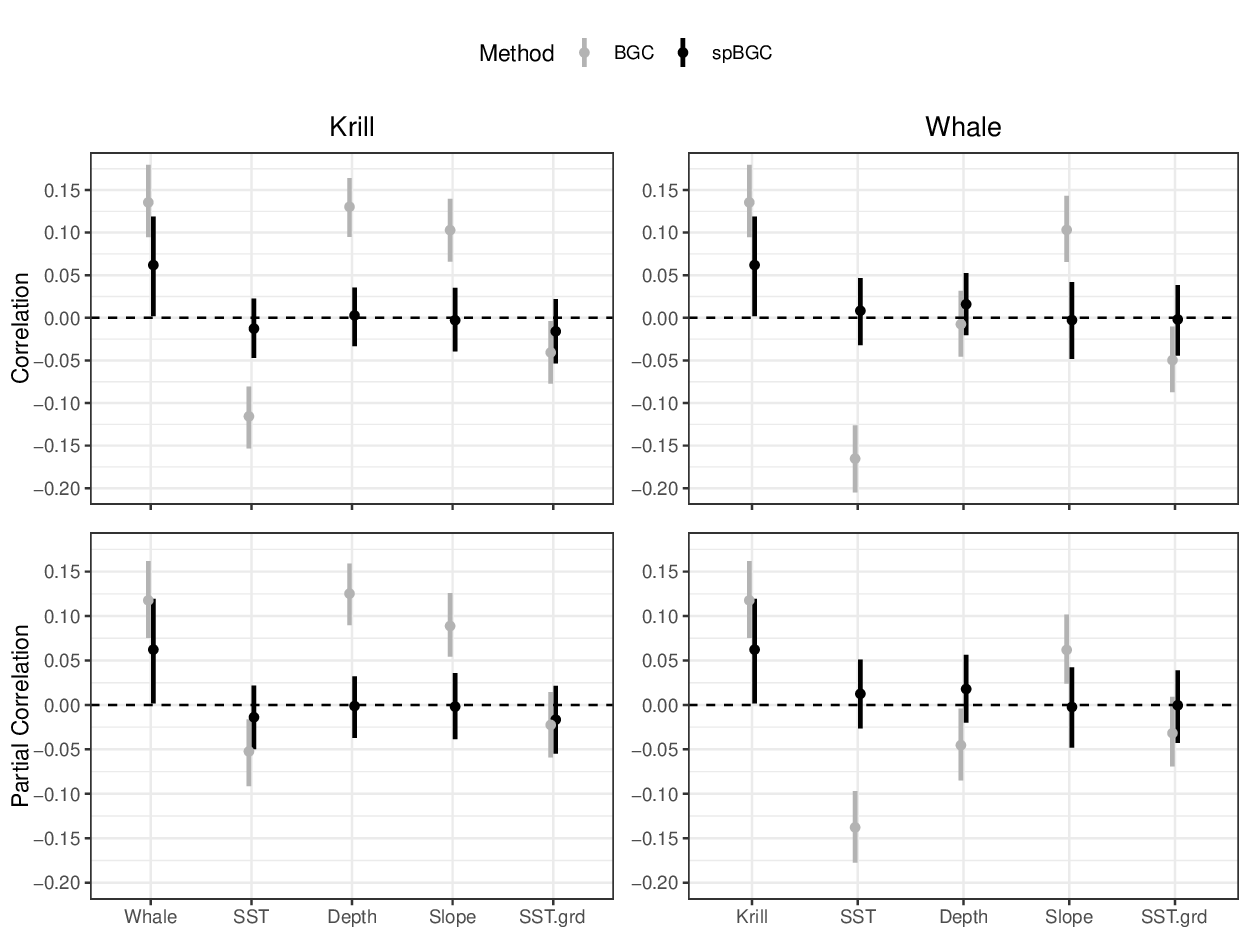}
    \caption{95\% credible intervals with posterior medians ($\bullet$) of the correlation coefficients (upper) and partial correlations (bottom) based on the proposed method, spBGC (black), and \cite{hoff2007extending}'s one, BGC (grey).}
    \label{fg:da_r_pcor}
\end{figure}

Based on 2,000 posterior draws, we compute the posterior medians and 95\% credible intervals of the correlations between the variables, as shown in Figure \ref{fg:da_r_pcor} and Tables \ref{tb:da_spBGC} and \ref{tb:da_BGC}. 
The results reveal that spBGC and BGC yield markedly different estimates for several correlations. 
In particular, spBGC intervals suggest that Krill has no strong positive or negative correlation with outcomes other than Whale. 
By contrast, BGC intervals exclude zero for all Krill- and Whale-related outcomes except Depth. 
Turning to the structure of conditional independence, we also examine partial correlations. 
Figure \ref{fg:da_r_pcor} and Tables \ref{tb:da_spBGC} and \ref{tb:da_BGC} also show that these partial correlation estimates differ substantially between spBGC and BGC. 
Notably, the only spBGC 95\% credible interval for a partial correlation that excludes zero is that for Krill and Whale, whereas the BGC identifies additional pairs with non-zero partial correlations.

Under the separable covariance $\bm{H}(\phi)\otimes\bm{R}$, $\bm{R}$ is the same-location latent cross-outcome correlation, not a residual left after removing a shared spatial effect, so the strong spatial autocorrelation of the outcomes (Table~\ref{tab:moran_obs}) does not by itself explain why most entries are small.
Accounting for spatial dependence explains instead the contrast with BGC, whose estimates are highly variable when the dependent sites are treated as independent replicates while its intervals stay narrow (cf.\ its undercoverage in Section~\ref{sec:vshoff}).
The magnitudes are estimates under a separable model with a common $\phi$ that fits some outcomes better than others (see the posterior predictive check below), and they are sensitive to this specification: the coregionalization models in Section~\ref{sec:inla_comparison} of the Supplementary Material give Krill--Whale correlations of $0.094$ and $0.353$ (vs.\ $0.062$) and Krill--Depth correlations near $0.3$; only the positive Krill--Whale association is stable across the three models.

As a related analysis, \cite{solvang2024} applied standard generalized linear models to the same dataset.
The results showed a significant positive association between Krill and Whale. 
Furthermore, the estimated coefficients on SST were significantly negative, while those on Slope were positive for both Krill and Whale. 
These results are similar to those obtained for the BGC, possibly because neither BGC nor the generalized linear models account for spatial correlation. 
In contrast, by using the proposed spBGC that considers spatial correlation, we identify relationships that slightly differ from previous findings, with the exception of SST and SST.grd, which showed negative for Krill.

\begin{table}[H]
    \caption{2.5\%, 50\%, and 97.5\% posterior quantiles of the correlation coefficients and partial correlations based on the proposed method, spBGC.}
    \label{tb:da_spBGC}
    \begin{minipage}[b]{0.49\linewidth}
        \centering
        \subcaption{Krill}
        \scalebox{0.80}{
        \begin{tabular}{r rrr} 
            \multicolumn{1}{l}{}&\multicolumn{1}{c}{2.5\%}&\multicolumn{1}{c}{Median}&\multicolumn{1}{c}{97.5\%}\tabularnewline \hline
            {\bf Correlation} & & & \\ 
            Whale&$ 0.0019$&$ 0.0619$&$0.1187$\tabularnewline
            SST&$-0.0472$&$-0.0126$&$0.0225$\tabularnewline
            Depth&$-0.0332$&$ 0.0028$&$0.0356$\tabularnewline
            Slope&$-0.0393$&$-0.0029$&$0.0352$\tabularnewline
            SST.grd&$-0.0536$&$-0.0159$&$0.0219$\tabularnewline
            &&& \\ 
            {\bf Partial correlation} & & & \\ 
            Whale&$ 0.0018$&$ 0.0623$&$0.1193$\tabularnewline
            SST&$-0.0497$&$-0.0139$&$0.0217$\tabularnewline
            Depth&$-0.0371$&$-0.0012$&$0.0320$\tabularnewline
            Slope&$-0.0385$&$-0.0019$&$0.0359$\tabularnewline
            SST.grd&$-0.0547$&$-0.0166$&$0.0214$\tabularnewline \hline
        \end{tabular}
        }
        \label{tb:krill_spBGC}
    \end{minipage}
    \begin{minipage}[b]{0.49\linewidth}
        \centering
        \subcaption{Whale}
        \scalebox{0.80}{
        \begin{tabular}{r rrr} 
            \multicolumn{1}{l}{}&\multicolumn{1}{c}{2.5\%}&\multicolumn{1}{c}{Median}&\multicolumn{1}{c}{97.5\%}\tabularnewline \hline
            {\bf Correlation} & & & \\ 
            Krill&$ 0.0019$&$ 0.0619$&$0.1187$\tabularnewline
            SST&$-0.0322$&$ 0.0083$&$0.0467$\tabularnewline
            Depth&$-0.0202$&$ 0.0159$&$0.0526$\tabularnewline
            Slope&$-0.0481$&$-0.0027$&$0.0420$\tabularnewline
            SST.grd&$-0.0443$&$-0.0019$&$0.0383$\tabularnewline
            &&& \\ 
            {\bf Partial correlation} & & & \\ 
            Krill&$ 0.0018$&$ 0.0623$&$0.1193$\tabularnewline
            SST&$-0.0264$&$ 0.0125$&$0.0511$\tabularnewline
            Depth&$-0.0198$&$ 0.0179$&$0.0563$\tabularnewline
            Slope&$-0.0480$&$-0.0023$&$0.0424$\tabularnewline
            SST.grd&$-0.0427$&$-0.0004$&$0.0389$\tabularnewline \hline
        \end{tabular}
        }
        \label{tb:whale_spBGC}
    \end{minipage}
\end{table}

\begin{table}[H]
    \caption{2.5\%, 50\%, and 97.5\% posterior quantiles of the correlation coefficients and partial correlations based on the method of \cite{hoff2007extending}, BGC.}
    \label{tb:da_BGC}
    \begin{minipage}[b]{0.49\linewidth}
        \centering
        \subcaption{Krill}
        \scalebox{0.80}{
        \begin{tabular}{r rrr} 
            \multicolumn{1}{l}{}&\multicolumn{1}{c}{2.5\%}&\multicolumn{1}{c}{Median}&\multicolumn{1}{c}{97.5\%}\tabularnewline \hline
            {\bf Correlation} & & & \\ 
            Whale&$ 0.0945$&$ 0.1355$&$ 0.1794$\tabularnewline
            SST&$-0.1531$&$-0.1156$&$-0.0806$\tabularnewline
            Depth&$ 0.0951$&$ 0.1303$&$ 0.1640$\tabularnewline
            Slope&$ 0.0660$&$ 0.1029$&$ 0.1396$\tabularnewline
            SST.grd&$-0.0772$&$-0.0406$&$-0.0040$\tabularnewline
            &&& \\ 
            {\bf Partial correlation} & & & \\ 
            Whale&$ 0.0755$&$ 0.1177$&$ 0.1617$\tabularnewline
            SST&$-0.0913$&$-0.0523$&$-0.0162$\tabularnewline
            Depth&$ 0.0898$&$ 0.1253$&$ 0.1591$\tabularnewline
            Slope&$ 0.0542$&$ 0.0888$&$ 0.1258$\tabularnewline
            SST.grd&$-0.0590$&$-0.0225$&$ 0.0145$\tabularnewline \hline
        \end{tabular}
        }
        \label{tb:krill_BGC}
    \end{minipage}
    \begin{minipage}[b]{0.49\linewidth}
        \centering
        \subcaption{Whale}
        \scalebox{0.80}{
        \begin{tabular}{r rrr} 
            \multicolumn{1}{l}{}&\multicolumn{1}{c}{2.5\%}&\multicolumn{1}{c}{Median}&\multicolumn{1}{c}{97.5\%}\tabularnewline \hline
            {\bf Correlation} & & & \\ 
            Krill&$ 0.0945$&$ 0.1355$&$ 0.1794$\tabularnewline
            SST&$-0.2047$&$-0.1653$&$-0.1263$\tabularnewline
            Depth&$-0.0455$&$-0.0075$&$ 0.0315$\tabularnewline
            Slope&$ 0.0656$&$ 0.1032$&$ 0.1431$\tabularnewline
            SST.grd&$-0.0873$&$-0.0497$&$-0.0104$\tabularnewline
            &&& \\ 
            {\bf Partial correlation} & & & \\ 
            Krill&$ 0.0755$&$ 0.1177$&$ 0.1617$\tabularnewline
            SST&$-0.1775$&$-0.1378$&$-0.0970$\tabularnewline
            Depth&$-0.0850$&$-0.0453$&$-0.0042$\tabularnewline
            Slope&$ 0.0239$&$ 0.0619$&$ 0.1018$\tabularnewline
            SST.grd&$-0.0691$&$-0.0318$&$ 0.0090$\tabularnewline \hline
        \end{tabular}
        }
        \label{tb:whale_BGC}
    \end{minipage}
\end{table}

To validate the spatial correlation assumptions underlying each model, we conduct a posterior predictive check (PPC), which assesses model fit by comparing summary statistics of observed data with those of data generated from the posterior predictive distribution.
We use Moran's $I$ statistic (Table~\ref{tab:moran_obs}) as the summary statistic.
The PPC procedure is as follows: 
(i) compute Moran's $I$ for the observed ranks of each variable; 
(ii) for each posterior sample $(\bm{R}^{(s)}, \phi^{(s)})$, generate replicated latent Gaussian data from \eqref{eq:spbgc}, i.e., $\bm{z}^{(s)} \sim N_{pn}(\bm{0}, \bm{H}(\phi^{(s)})\otimes\bm{R}^{(s)})$, and compute the ranks; 
(iii) compute Moran's $I$ for each replicated rank data; 
(iv) compare the observed Moran's $I$ with the distribution of the replicated values.

The results reveal a stark contrast between the two models.
BGC generates data with virtually no spatial correlation ($I \approx 0$), completely failing to reproduce the strong spatial patterns observed in the data (Table~\ref{tab:moran_obs}).
In contrast, spBGC generates spatially correlated data with $I \approx 0.98$.
For Depth and SST.grd, the observed Moran's $I$ falls within the range of replicated values from spBGC, indicating adequate model fit. 
However, for Krill, Whale and Slope, spBGC overestimates spatial correlation, markedly so for Krill and Whale (observed $I\approx 0.4$), while for SST, it slightly underestimates.
This suggests that a single spatial range parameter $\phi$ shared across all variables may not fully capture the varying degrees of spatial dependence, particularly for biological variables (Krill) that exhibit weaker spatial correlation than environmental variables.
We also conduct PPC using Mat\'ern 3/2 and Mat\'ern 5/2 correlation functions, with similar results reported in Supplementary Material Section B.

\section{Discussion} \label{sec:end}
We discuss here the limitations of our method based on its application in Section \ref{sec:real}. 
As the proposed method, we employ a Gaussian process with a covariance matrix that assumes isotropy and common spatial range parameter $\phi$ across all outcomes. 
Although such a simple Gaussian process facilitates efficient inference algorithms, it may struggle to fit data collected along transect lines or outcomes in which the degree of spatial correlation decay varies with distance, as illustrated by Figure \ref{fg:geo_da}. 
The assumption of isotropy implies that all locations equidistant from a given point have the same spatial correlation, which leads to identical spatial correlations between data points on the same transect line and those on different transect lines. 
One way to mitigate these issues is to employ Gaussian processes with more complex covariance structures, such as covariance functions that incorporate transect lines, anisotropic and nonstationary covariance functions \citep{paciorek2006spatial}, spatiotemporal covariance functions \citep{cressie1999classes, stein2005space}, and multivariate cross-covariance functions \citep{gneiting2010matern, apanasovich2012valid}. 
However, the use of such complex Gaussian processes increases the computational burden on inference algorithms, highlighting the need for more scalable algorithmic developments. 
In addition, the benchmark in Section~\ref{sec:p_scalability} of the Supplementary Material shows that, as $p$ grows, the nonzero correlations are increasingly shrunk toward zero; a sparsity-inducing prior for $\bm{R}$, which would require a new sampler, is left for future work.

The proposed method can be applied and extended in several ways. 
The first is a Gaussian process formulation for the latent variables, from which we can obtain the posterior predictive distribution of the latent variables in arbitrary spatial locations (including non-sampled locations). 
Then, using the marginal distribution of each outcome estimated in a parametric or non-parametric manner, we can perform spatial predictions in non-sampled locations. 
Second, although this study focuses on the latent Gaussian process, the proposed method may be conceptually extended to other copula models, such as elliptical copulas and skew elliptical copulas \citep{smith2021implicit}.
For example, if one is interested in the tail dependence of a latent process, it may be preferable to consider a $t$-copula model.
Such heavy-tailed copulas could be particularly useful when dealing with zero-inflated count data, as is common in ecological applications, where extreme co-occurrences may exhibit stronger dependence than what a Gaussian copula can capture.
However, posterior computation of additional parameters using such copula models may not be feasible.
In this case, an MCMC sampler based on the Metropolis-Hastings algorithm or more efficient computational algorithms may need to be developed. 
Third, we extend our method to spatiotemporal data. 
Multivariate data may contain time and location information. 
For example, the data used in the application example were observed in the Southern Ocean over different periods. 
In the analysis of the dependence structure of such data, it may be necessary to consider both spatial and temporal correlations. 
To this end, we consider a hierarchical Bayesian spatiotemporal model based on extended rank likelihood. 
However, the development of posterior computation for the model is challenging, thus, a subject to address in future work. 
Finally, although the proposed methodology is for point-referenced data, it is possible to develop a similar method to estimate the copula for multivariate data observed on a graph.

\section*{Acknowledgements}
This research was supported by JSPS KAKENHI (grant numbers 20H00080, 21H00699, and 	24K23870).

\vspace{0.5cm}
\bibliographystyle{chicago}
\bibliography{References}

\newpage
\appendix  

\setcounter{page}{1}
\renewcommand{\thepage}{S\arabic{page}}
\setcounter{section}{0}
\renewcommand{\thesection}{S\arabic{section}}
\setcounter{figure}{0}
\renewcommand{\thefigure}{S\arabic{figure}}
\setcounter{table}{0}
\renewcommand{\thetable}{S\arabic{table}}
\setcounter{equation}{0}
\renewcommand{\theequation}{S\arabic{equation}}

\doublespacing 

\input{Supplementary_Material_body_v3}

\end{document}

%% file: Supplementary_Material_body_v3.tex
\begin{center}
  {\Large \textbf{Supplementary Material of ``Semiparametric Copula Estimation for Spatially Correlated Multivariate Mixed Outcomes: Analyzing Visual Sightings of Fin Whales from a Line Transect Survey''}}\\[1.5em]
  {\large Tomotaka Momozaki$^1$, Tomoyuki Nakagawa$^2$, Shonosuke Sugasawa$^3$, Hiroko Kato Solvang$^4$}\\[1em]
  {\small $^1$Department of Information Sciences, Tokyo University of Science\\
  $^2$School of Data Science, Meisei University\\
  $^3$Faculty of Economics, Keio University\\
  $^4$Marine Mammals Research Group, Institute of Marine Research, Bergen, Norway}\\[0.5em]
  {\small Last update: \today}
\end{center}
\vspace{1.5em}

\singlespacing 

\appendix

\renewcommand{\thetable}{\Alph{section}.\arabic{table}}
\renewcommand{\thefigure}{\Alph{section}.\arabic{figure}}
\setcounter{table}{0}
\setcounter{figure}{0}

\section{Additional Simulation Studies}

This section presents additional simulation studies to further evaluate the robustness of the proposed methods.
Section~\ref{sec:nngp_sensitivity} examines the sensitivity of spBGCNNGP to the choice of the number of neighbors $m$.
Section~\ref{sec:real_locations} evaluates performance under the realistic spatial configurations from line transect survey data.
Section~\ref{sec:correlation_misspec} investigates the robustness to misspecification of the spatial correlation function.
Section~\ref{sec:copula_misspec} evaluates the robustness to misspecification of the copula family.
Section~\ref{sec:R_structure} examines the robustness to the structure and magnitude of the correlation matrix $\bm{R}$, considering dense and small-magnitude designs.
Section~\ref{sec:p_scalability} benchmarks the runtime and accuracy for up to $p=50$ outcomes.

\subsection{Sensitivity to Number of Neighbors $m$ in spBGCNNGP}
\label{sec:nngp_sensitivity}

We investigate the sensitivity of the spBGCNNGP method to the choice of the number of neighbors $m$.
We compare the performance of spBGCNNGP with $m \in \{5, 10, 15, n/10\}$ against spBGC and the baseline BGC method.
The data generating process follows the same setup as in Section 4 of the main manuscript.

\begin{figure}[H]
\centering
\includegraphics[width=\columnwidth]{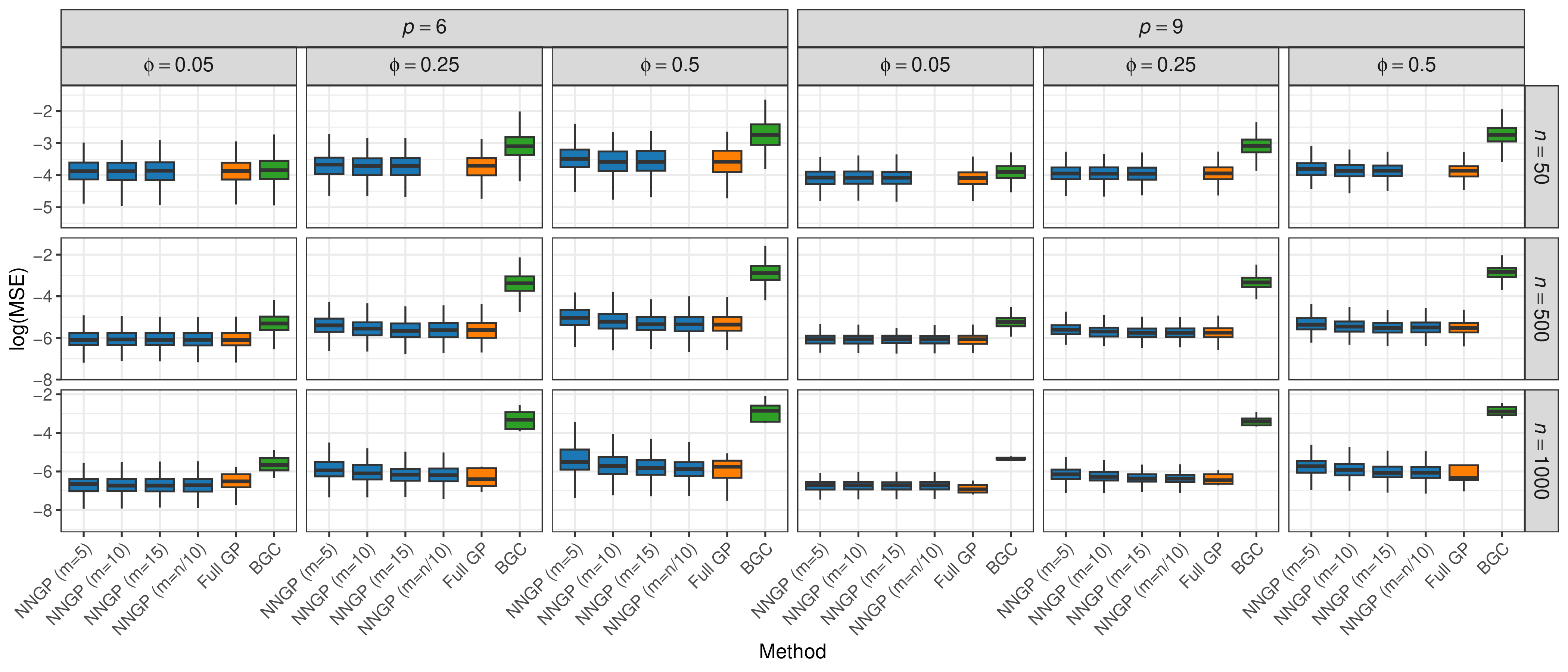}
\caption{Sensitivity analysis for the number of neighbors $m$ in spBGCNNGP: logarithm of MSEs.}
\label{fig:nngp_mse_sensitivity}
\end{figure}

\begin{figure}[H]
\centering
\includegraphics[width=\columnwidth]{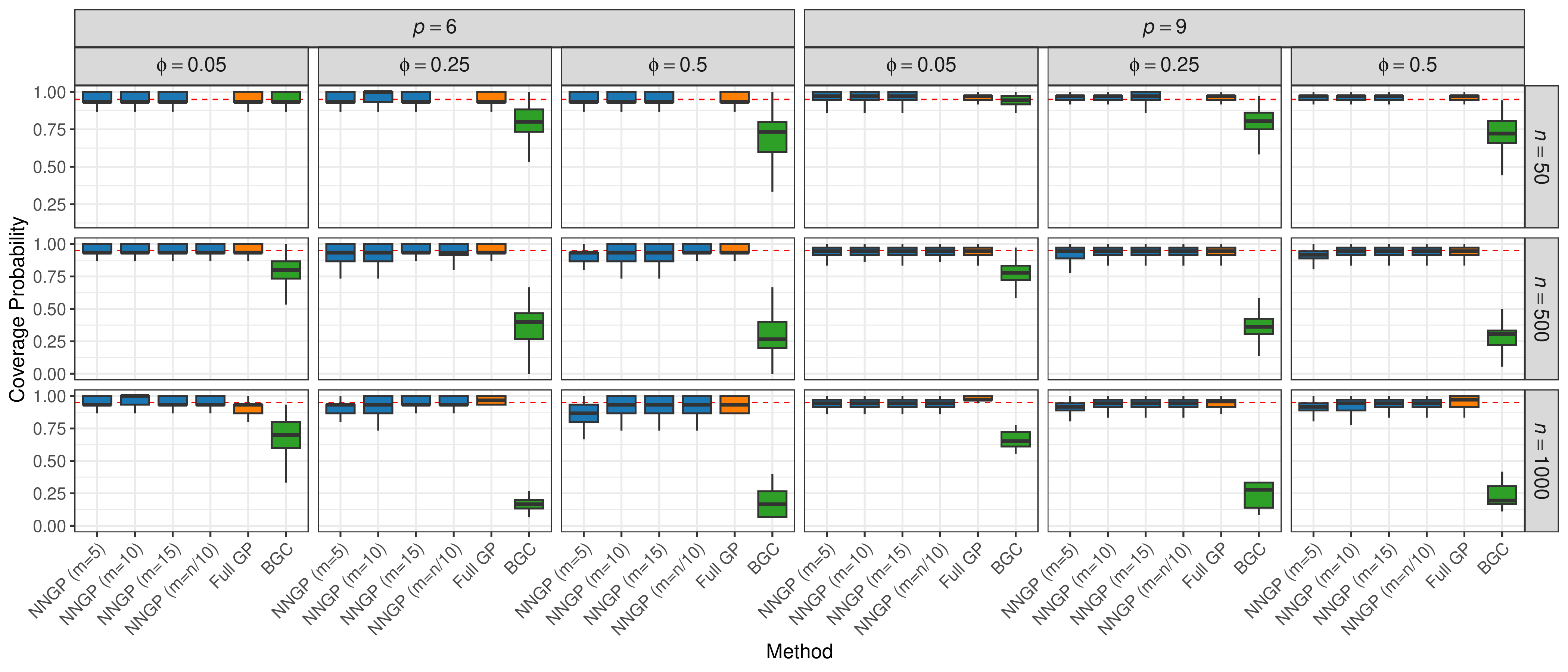}
\caption{Sensitivity analysis for the number of neighbors $m$ in spBGCNNGP: coverage probabilities.}
\label{fig:nngp_cp_sensitivity}
\end{figure}

\begin{figure}[H]
\centering
\includegraphics[width=\columnwidth]{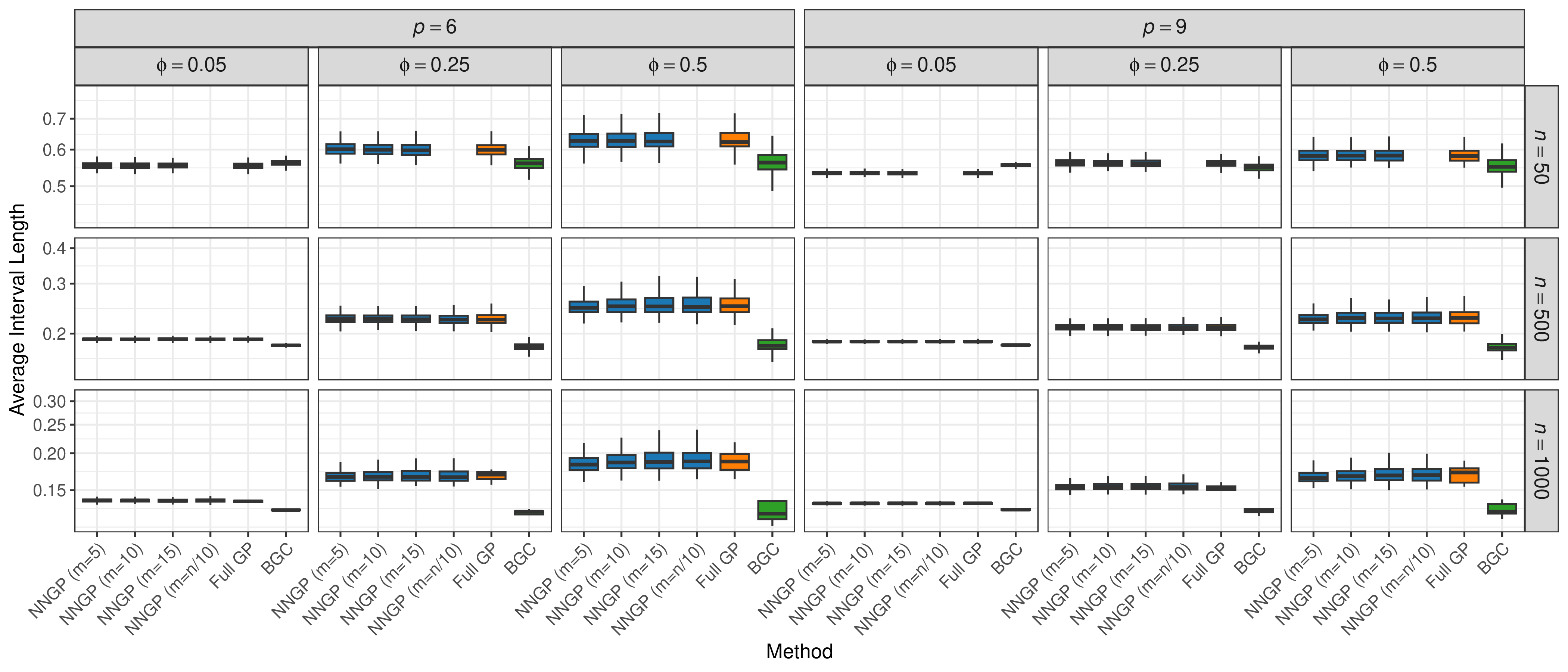}
\caption{Sensitivity analysis for the number of neighbors $m$ in spBGCNNGP: average credible interval lengths.}
\label{fig:nngp_len_sensitivity}
\end{figure}

Tables~\ref{tb:mse_sensitivity_random},~\ref{tb:cp_sensitivity_random}, and~\ref{tb:avl_sensitivity_random} present the logarithm of mean squared errors (MSEs), coverage probabilities, and average credible interval lengths, respectively, for different values of $m$.
Figures~\ref{fig:nngp_mse_sensitivity},~\ref{fig:nngp_cp_sensitivity}, and~\ref{fig:nngp_len_sensitivity} provide visual comparisons of these metrics across different neighbor sizes.

The results reveal several important findings regarding the sensitivity of spBGCNNGP to the choice of $m$.
First, for small sample sizes ($n=50$), the choice of $m$ has negligible impact on estimation accuracy; the differences in log(MSE) across all values of $m$ are within the standard errors.
This is intuitive because when the sample size is small, even the smallest neighbor set ($m=5$) captures a substantial portion of the spatial dependence structure.

As the sample size increases to $n=500$ and $n=1000$, the effect of $m$ becomes more pronounced, particularly for stronger spatial correlations ($\phi = 0.25$ and $0.50$).
With $m=5$, the log(MSE) is approximately 0.1--0.4 higher than spBGC for $\phi \geq 0.25$, whereas $m \geq 10$ yields results nearly indistinguishable from spBGC.
For instance, at $n=1000$ and $\phi=0.50$ with $p=6$, the log(MSE) improves from $-5.407$ ($m=5$) to $-5.847$ ($m=n/10$), approaching spBGC value of $-5.776$.
These findings suggest that $m=10$ to $15$ provides a reasonable balance between computational efficiency and estimation accuracy in practice.
In contrast, the BGC method, which ignores spatial correlations, exhibits substantially worse performance across all settings, with the performance gap widening dramatically as spatial correlation strength increases.
For example, at $n=1000$ and $\phi=0.50$, the log(MSE) for BGC is $-2.874$ for $p=6$ and $-2.860$ for $p=9$, representing a degradation of approximately 2.9--3.4 log units compared to spBGC.

Regarding coverage probability, the spBGCNNGP method maintains near-nominal coverage (around 0.95) across most settings when $m \geq 10$.
However, with $m=5$, coverage can drop to approximately 0.88--0.91 under strong spatial correlations ($\phi=0.50$) at $n=1000$.
This slight undercoverage with small $m$ is attributable to the NNGP approximation not fully capturing the long-range spatial dependencies.
In contrast, the BGC method, which ignores spatial correlations entirely, exhibits severe undercoverage that deteriorates dramatically with increasing sample size and spatial correlation strength.
For example, at $n=1000$ and $\phi=0.50$, BGC coverage falls to approximately 0.19 for $p=6$ and 0.23 for $p=9$, rendering its uncertainty quantification unreliable.

The average credible interval lengths are nearly identical between spBGCNNGP and spBGC across all values of $m$, indicating that the NNGP approximation appropriately quantifies uncertainty.
The BGC method produces shorter intervals because it fails to account for the additional uncertainty induced by spatial correlations; this explains its substantial undercoverage.

From a computational perspective, the NNGP approximation reduces the complexity from $O(n^3)$ to $O(nm^2)$, making it feasible for large datasets.
Given that $m=10$ to $15$ provides accuracy comparable to spBGC while maintaining computational tractability, we recommend using $m=10$ to $15$ as a default choice in practice.
For applications where computational resources permit, $m=n/10$ (or larger) may provide marginal improvements, particularly under strong spatial correlations.

\begin{table}[H]
\centering
\caption{Sensitivity analysis for the number of neighbors $m$ in spBGCNNGP: comparisons of the logarithm of the MSEs for the spBGCNNGP, spBGC, and BGC under varying neighbor sizes and spatial range parameters.
The values represent average log(MSE)s from 300 calculations (10 for spBGC at $n=1000$), with standard errors in parentheses.}
\label{tb:mse_sensitivity_random}
\begin{minipage}{.48\linewidth}
  \centering
  \subcaption{Number of outcomes $p=6$}
  \begin{tabularx}{\linewidth}{l c *{3}{>{\centering\arraybackslash}X}}
  \hline
  & & \multicolumn{3}{c}{$\phi$} \\
  \cline{3-5}
  Method & $m$ & 0.05 & 0.25 & 0.50 \\
  \hline
  \multicolumn{5}{l}{\textbf{$n=50$}} \\
  spBGC & $5$ & $-3.876$ & $-3.876$ & $-3.687$ \\
  NNGP &  & (0.023) & (0.023) & (0.023) \\
   & $10$ & $-3.882$ & $-3.724$ & $-3.561$ \\
   &  & (0.023) & (0.021) & (0.024) \\
   & $15$ & $-3.882$ & $-3.720$ & $-3.566$ \\
   &  & (0.023) & (0.021) & (0.024) \\
  spBGC & -- & $-3.884$ & $-3.884$ & $-3.729$ \\
   &  & (0.023) & (0.023) & (0.021) \\
  BGC & -- & $-3.853$ & $-3.853$ & $-3.106$ \\
   &  & (0.024) & (0.024) & (0.028) \\
  \hline
  \multicolumn{5}{l}{\textbf{$n=500$}} \\
  spBGC & $5$ & $-6.077$ & $-5.407$ & $-5.012$ \\
  NNGP &  & (0.026) & (0.028) & (0.032) \\
   & $10$ & $-6.084$ & $-5.561$ & $-5.216$ \\
   &  & (0.026) & (0.028) & (0.032) \\
   & $15$ & $-6.082$ & $-5.639$ & $-5.314$ \\
   &  & (0.026) & (0.028) & (0.030) \\
   & $n/10$ & $-6.089$ & $-5.636$ & $-5.320$ \\
   &  & (0.026) & (0.028) & (0.032) \\
  spBGC & -- & $-6.085$ & $-5.639$ & $-5.322$ \\
   &  & (0.026) & (0.028) & (0.033) \\
  BGC & -- & $-5.326$ & $-3.395$ & $-2.888$ \\
   &  & (0.026) & (0.030) & (0.030) \\
  \hline
  \multicolumn{5}{l}{\textbf{$n=1000$}} \\
  spBGC & $5$ & $-6.713$ & $-5.876$ & $-5.407$ \\
  NNGP &  & (0.027) & (0.032) & (0.041) \\
   & $10$ & $-6.731$ & $-6.049$ & $-5.657$ \\
   &  & (0.028) & (0.028) & (0.038) \\
   & $15$ & $-6.736$ & $-6.161$ & $-5.799$ \\
   &  & (0.028) & (0.027) & (0.036) \\
   & $n/10$ & $-6.734$ & $-6.189$ & $-5.847$ \\
   &  & (0.028) & (0.027) & (0.035) \\
  spBGC & -- & $-6.538$ & $-6.343$ & $-5.776$ \\
   &  & (0.172) & (0.162) & (0.326) \\
  BGC & -- & $-5.602$ & $-3.333$ & $-2.874$ \\
   &  & (0.026) & (0.027) & (0.029) \\
  \hline
  \end{tabularx}
\end{minipage}%
\hfill
\begin{minipage}{.48\linewidth}
  \centering
  \subcaption{Number of outcomes $p=9$}
  \begin{tabularx}{\linewidth}{l c *{3}{>{\centering\arraybackslash}X}}
  \hline
  & & \multicolumn{3}{c}{$\phi$} \\
  \cline{3-5}
  Method & $m$ & 0.05 & 0.25 & 0.50 \\
  \hline
  \multicolumn{5}{l}{\textbf{$n=50$}} \\
  spBGC & $5$ & $-4.087$ & $-4.087$ & $-3.943$ \\
  NNGP &  & (0.016) & (0.016) & (0.016) \\
   & $10$ & $-4.087$ & $-3.953$ & $-3.865$ \\
   &  & (0.016) & (0.015) & (0.016) \\
   & $15$ & $-4.093$ & $-3.957$ & $-3.868$ \\
   &  & (0.016) & (0.016) & (0.015) \\
  spBGC & -- & $-4.089$ & $-4.089$ & $-3.952$ \\
   &  & (0.016) & (0.016) & (0.015) \\
  BGC & -- & $-3.916$ & $-3.916$ & $-3.084$ \\
   &  & (0.016) & (0.016) & (0.017) \\
  \hline
  \multicolumn{5}{l}{\textbf{$n=500$}} \\
  spBGC & $5$ & $-6.079$ & $-5.595$ & $-5.325$ \\
  NNGP &  & (0.015) & (0.019) & (0.022) \\
   & $10$ & $-6.081$ & $-5.707$ & $-5.446$ \\
   &  & (0.015) & (0.019) & (0.022) \\
   & $15$ & $-6.085$ & $-5.746$ & $-5.499$ \\
   &  & (0.015) & (0.018) & (0.022) \\
   & $n/10$ & $-6.085$ & $-5.749$ & $-5.481$ \\
   &  & (0.015) & (0.018) & (0.023) \\
  spBGC & -- & $-6.086$ & $-5.751$ & $-5.493$ \\
   &  & (0.015) & (0.018) & (0.023) \\
  BGC & -- & $-5.242$ & $-3.337$ & $-2.851$ \\
   &  & (0.017) & (0.020) & (0.019) \\
  \hline
  \multicolumn{5}{l}{\textbf{$n=1000$}} \\
  spBGC & $5$ & $-6.728$ & $-6.137$ & $-5.716$ \\
  NNGP &  & (0.016) & (0.022) & (0.029) \\
   & $10$ & $-6.733$ & $-6.252$ & $-5.870$ \\
   &  & (0.016) & (0.019) & (0.027) \\
   & $15$ & $-6.734$ & $-6.338$ & $-5.998$ \\
   &  & (0.016) & (0.018) & (0.026) \\
   & $n/10$ & $-6.736$ & $-6.352$ & $-6.015$ \\
   &  & (0.016) & (0.018) & (0.027) \\
  spBGC & -- & $-6.886$ & $-6.393$ & $-6.238$ \\
   &  & (0.080) & (0.091) & (0.145) \\
  BGC & -- & $-5.406$ & $-3.376$ & $-2.860$ \\
   &  & (0.011) & (0.015) & (0.016) \\
  \hline
  \end{tabularx}
\end{minipage}
\end{table}

\begin{table}[H]
\centering
\caption{Sensitivity analysis for the number of neighbors $m$ in spBGCNNGP: comparisons of the coverage probabilities for the spBGCNNGP, spBGC, and BGC under varying neighbor sizes and spatial range parameters.
The values represent average coverage probabilities from 300 calculations (10 for spBGC at $n=1000$), with standard errors in parentheses.}
\label{tb:cp_sensitivity_random}
\begin{minipage}{.48\linewidth}
  \centering
  \subcaption{Number of outcomes $p=6$}
  \begin{tabularx}{\linewidth}{l c *{3}{>{\centering\arraybackslash}X}}
  \hline
  & & \multicolumn{3}{c}{$\phi$} \\
  \cline{3-5}
  Method & $m$ & 0.05 & 0.25 & 0.50 \\
  \hline
  \multicolumn{5}{l}{\textbf{$n=50$}} \\
  spBGC & $5$ & $0.946$ & $0.946$ & $0.948$ \\
  NNGP &  & (0.004) & (0.004) & (0.003) \\
   & $10$ & $0.944$ & $0.955$ & $0.945$ \\
   &  & (0.004) & (0.003) & (0.004) \\
   & $15$ & $0.946$ & $0.954$ & $0.942$ \\
   &  & (0.004) & (0.003) & (0.004) \\
  spBGC & -- & $0.945$ & $0.954$ & $0.942$ \\
   &  & (0.004) & (0.003) & (0.004) \\
  BGC & -- & $0.942$ & $0.802$ & $0.712$ \\
   &  & (0.004) & (0.008) & (0.009) \\
  \hline
  \multicolumn{5}{l}{\textbf{$n=500$}} \\
  spBGC & $5$ & $0.944$ & $0.918$ & $0.896$ \\
  NNGP &  & (0.004) & (0.004) & (0.005) \\
   & $10$ & $0.944$ & $0.933$ & $0.926$ \\
   &  & (0.004) & (0.004) & (0.004) \\
   & $15$ & $0.947$ & $0.938$ & $0.937$ \\
   &  & (0.004) & (0.004) & (0.004) \\
   & $n/10$ & $0.947$ & $0.938$ & $0.940$ \\
   &  & (0.003) & (0.004) & (0.004) \\
  spBGC & -- & $0.946$ & $0.940$ & $0.937$ \\
   &  & (0.003) & (0.004) & (0.004) \\
  BGC & -- & $0.786$ & $0.368$ & $0.299$ \\
   &  & (0.007) & (0.008) & (0.008) \\
  \hline
  \multicolumn{5}{l}{\textbf{$n=1000$}} \\
  spBGC & $5$ & $0.941$ & $0.906$ & $0.880$ \\
  NNGP &  & (0.004) & (0.004) & (0.005) \\
   & $10$ & $0.948$ & $0.927$ & $0.909$ \\
   &  & (0.004) & (0.004) & (0.004) \\
   & $15$ & $0.946$ & $0.935$ & $0.926$ \\
   &  & (0.004) & (0.004) & (0.004) \\
   & $n/10$ & $0.945$ & $0.937$ & $0.932$ \\
   &  & (0.004) & (0.004) & (0.004) \\
  spBGC & -- & $0.920$ & $0.967$ & $0.933$ \\
   &  & (0.019) & (0.011) & (0.017) \\
  BGC & -- & $0.680$ & $0.173$ & $0.193$ \\
   &  & (0.010) & (0.006) & (0.007) \\
  \hline
  \end{tabularx}
\end{minipage}%
\hfill
\begin{minipage}{.48\linewidth}
  \centering
  \subcaption{Number of outcomes $p=9$}
  \begin{tabularx}{\linewidth}{l c *{3}{>{\centering\arraybackslash}X}}
  \hline
  & & \multicolumn{3}{c}{$\phi$} \\
  \cline{3-5}
  Method & $m$ & 0.05 & 0.25 & 0.50 \\
  \hline
  \multicolumn{5}{l}{\textbf{$n=50$}} \\
  spBGC & $5$ & $0.961$ & $0.961$ & $0.959$ \\
  NNGP &  & (0.002) & (0.002) & (0.002) \\
   & $10$ & $0.960$ & $0.960$ & $0.963$ \\
   &  & (0.002) & (0.002) & (0.002) \\
   & $15$ & $0.961$ & $0.962$ & $0.964$ \\
   &  & (0.002) & (0.002) & (0.002) \\
  spBGC & -- & $0.960$ & $0.961$ & $0.962$ \\
   &  & (0.002) & (0.002) & (0.002) \\
  BGC & -- & $0.949$ & $0.798$ & $0.716$ \\
   &  & (0.002) & (0.005) & (0.006) \\
  \hline
  \multicolumn{5}{l}{\textbf{$n=500$}} \\
  spBGC & $5$ & $0.946$ & $0.931$ & $0.923$ \\
  NNGP &  & (0.002) & (0.003) & (0.003) \\
   & $10$ & $0.946$ & $0.940$ & $0.939$ \\
   &  & (0.002) & (0.002) & (0.002) \\
   & $15$ & $0.947$ & $0.941$ & $0.943$ \\
   &  & (0.002) & (0.003) & (0.002) \\
   & $n/10$ & $0.949$ & $0.943$ & $0.941$ \\
   &  & (0.002) & (0.002) & (0.002) \\
  spBGC & -- & $0.947$ & $0.943$ & $0.945$ \\
   &  & (0.002) & (0.002) & (0.002) \\
  BGC & -- & $0.780$ & $0.363$ & $0.289$ \\
   &  & (0.004) & (0.005) & (0.005) \\
  \hline
  \multicolumn{5}{l}{\textbf{$n=1000$}} \\
  spBGC & $5$ & $0.950$ & $0.925$ & $0.911$ \\
  NNGP &  & (0.002) & (0.002) & (0.003) \\
   & $10$ & $0.950$ & $0.934$ & $0.927$ \\
   &  & (0.002) & (0.002) & (0.003) \\
   & $15$ & $0.951$ & $0.942$ & $0.938$ \\
   &  & (0.002) & (0.002) & (0.003) \\
   & $n/10$ & $0.950$ & $0.944$ & $0.941$ \\
   &  & (0.002) & (0.002) & (0.002) \\
  spBGC & -- & $0.972$ & $0.944$ & $0.956$ \\
   &  & (0.008) & (0.014) & (0.017) \\
  BGC & -- & $0.664$ & $0.250$ & $0.233$ \\
   &  & (0.005) & (0.006) & (0.006) \\
  \hline
  \end{tabularx}
\end{minipage}
\end{table}

\begin{table}[H]
\centering
\caption{Sensitivity analysis for the number of neighbors $m$ in spBGCNNGP: comparisons of the average credible interval lengths for the spBGCNNGP, spBGC, and BGC under varying neighbor sizes and spatial range parameters.
The values represent average interval lengths from 300 calculations (10 for spBGC at $n=1000$), with standard errors in parentheses.}
\label{tb:avl_sensitivity_random}
\begin{minipage}{.48\linewidth}
  \centering
  \subcaption{Number of outcomes $p=6$}
  \begin{tabularx}{\linewidth}{l c *{3}{>{\centering\arraybackslash}X}}
  \hline
  & & \multicolumn{3}{c}{$\phi$} \\
  \cline{3-5}
  Method & $m$ & 0.05 & 0.25 & 0.50 \\
  \hline
  \multicolumn{5}{l}{\textbf{$n=50$}} \\
  spBGC & $5$ & $0.555$ & $0.555$ & $0.605$ \\
  NNGP &  & (0.001) & (0.001) & (0.001) \\
   & $10$ & $0.554$ & $0.603$ & $0.633$ \\
   &  & (0.001) & (0.001) & (0.002) \\
   & $15$ & $0.555$ & $0.602$ & $0.633$ \\
   &  & (0.001) & (0.001) & (0.002) \\
  spBGC & -- & $0.554$ & $0.603$ & $0.632$ \\
   &  & (0.001) & (0.001) & (0.002) \\
  BGC & -- & $0.562$ & $0.561$ & $0.568$ \\
   &  & (0.001) & (0.002) & (0.003) \\
  \hline
  \multicolumn{5}{l}{\textbf{$n=500$}} \\
  spBGC & $5$ & $0.191$ & $0.226$ & $0.250$ \\
  NNGP &  & (0.000) & (0.001) & (0.001) \\
   & $10$ & $0.191$ & $0.227$ & $0.254$ \\
   &  & (0.000) & (0.001) & (0.001) \\
   & $15$ & $0.191$ & $0.226$ & $0.256$ \\
   &  & (0.000) & (0.001) & (0.002) \\
   & $n/10$ & $0.191$ & $0.226$ & $0.256$ \\
   &  & (0.000) & (0.001) & (0.002) \\
  spBGC & -- & $0.191$ & $0.226$ & $0.256$ \\
   &  & (0.000) & (0.001) & (0.002) \\
  BGC & -- & $0.182$ & $0.180$ & $0.188$ \\
   &  & (0.000) & (0.000) & (0.002) \\
  \hline
  \multicolumn{5}{l}{\textbf{$n=1000$}} \\
  spBGC & $5$ & $0.138$ & $0.167$ & $0.187$ \\
  NNGP &  & (0.000) & (0.000) & (0.001) \\
   & $10$ & $0.138$ & $0.168$ & $0.191$ \\
   &  & (0.000) & (0.001) & (0.001) \\
   & $15$ & $0.138$ & $0.168$ & $0.193$ \\
   &  & (0.000) & (0.001) & (0.001) \\
   & $n/10$ & $0.138$ & $0.168$ & $0.194$ \\
   &  & (0.000) & (0.001) & (0.001) \\
  spBGC & -- & $0.138$ & $0.169$ & $0.196$ \\
   &  & (0.000) & (0.003) & (0.011) \\
  BGC & -- & $0.129$ & $0.127$ & $0.141$ \\
   &  & (0.000) & (0.000) & (0.003) \\
  \hline
  \end{tabularx}
\end{minipage}%
\hfill
\begin{minipage}{.48\linewidth}
  \centering
  \subcaption{Number of outcomes $p=9$}
  \begin{tabularx}{\linewidth}{l c *{3}{>{\centering\arraybackslash}X}}
  \hline
  & & \multicolumn{3}{c}{$\phi$} \\
  \cline{3-5}
  Method & $m$ & 0.05 & 0.25 & 0.50 \\
  \hline
  \multicolumn{5}{l}{\textbf{$n=50$}} \\
  spBGC & $5$ & $0.534$ & $0.534$ & $0.563$ \\
  NNGP &  & (0.000) & (0.000) & (0.001) \\
   & $10$ & $0.534$ & $0.561$ & $0.585$ \\
   &  & (0.000) & (0.001) & (0.001) \\
   & $15$ & $0.533$ & $0.561$ & $0.584$ \\
   &  & (0.000) & (0.001) & (0.001) \\
  spBGC & -- & $0.534$ & $0.561$ & $0.585$ \\
   &  & (0.000) & (0.001) & (0.001) \\
  BGC & -- & $0.555$ & $0.550$ & $0.555$ \\
   &  & (0.000) & (0.001) & (0.002) \\
  \hline
  \multicolumn{5}{l}{\textbf{$n=500$}} \\
  spBGC & $5$ & $0.187$ & $0.211$ & $0.226$ \\
  NNGP &  & (0.000) & (0.000) & (0.001) \\
   & $10$ & $0.187$ & $0.211$ & $0.229$ \\
   &  & (0.000) & (0.000) & (0.001) \\
   & $15$ & $0.188$ & $0.211$ & $0.230$ \\
   &  & (0.000) & (0.000) & (0.001) \\
   & $n/10$ & $0.188$ & $0.211$ & $0.230$ \\
   &  & (0.000) & (0.000) & (0.001) \\
  spBGC & -- & $0.188$ & $0.211$ & $0.231$ \\
   &  & (0.000) & (0.000) & (0.001) \\
  BGC & -- & $0.182$ & $0.180$ & $0.182$ \\
   &  & (0.000) & (0.000) & (0.001) \\
  \hline
  \multicolumn{5}{l}{\textbf{$n=1000$}} \\
  spBGC & $5$ & $0.135$ & $0.154$ & $0.167$ \\
  NNGP &  & (0.000) & (0.000) & (0.001) \\
   & $10$ & $0.135$ & $0.155$ & $0.170$ \\
   &  & (0.000) & (0.000) & (0.001) \\
   & $15$ & $0.135$ & $0.155$ & $0.171$ \\
   &  & (0.000) & (0.000) & (0.001) \\
   & $n/10$ & $0.135$ & $0.155$ & $0.172$ \\
   &  & (0.000) & (0.000) & (0.001) \\
  spBGC & -- & $0.135$ & $0.153$ & $0.170$ \\
   &  & (0.000) & (0.002) & (0.004) \\
  BGC & -- & $0.129$ & $0.128$ & $0.130$ \\
   &  & (0.000) & (0.000) & (0.000) \\
  \hline
  \end{tabularx}
\end{minipage}
\end{table}

\subsection{Simulation with Real Data Spatial Locations}
\label{sec:real_locations}

To evaluate the robustness of our method under realistic spatial configurations, we conduct simulations using the actual spatial locations from the fin whale sighting data.
The spatial configuration of line transect surveys often exhibits characteristics such as elongated sampling patterns along transect lines and varying inter-point distances.
This allows us to assess performance under the non-uniform spatial sampling patterns typical of line transect surveys.

\begin{figure}[H]
\centering
\includegraphics[width=\columnwidth]{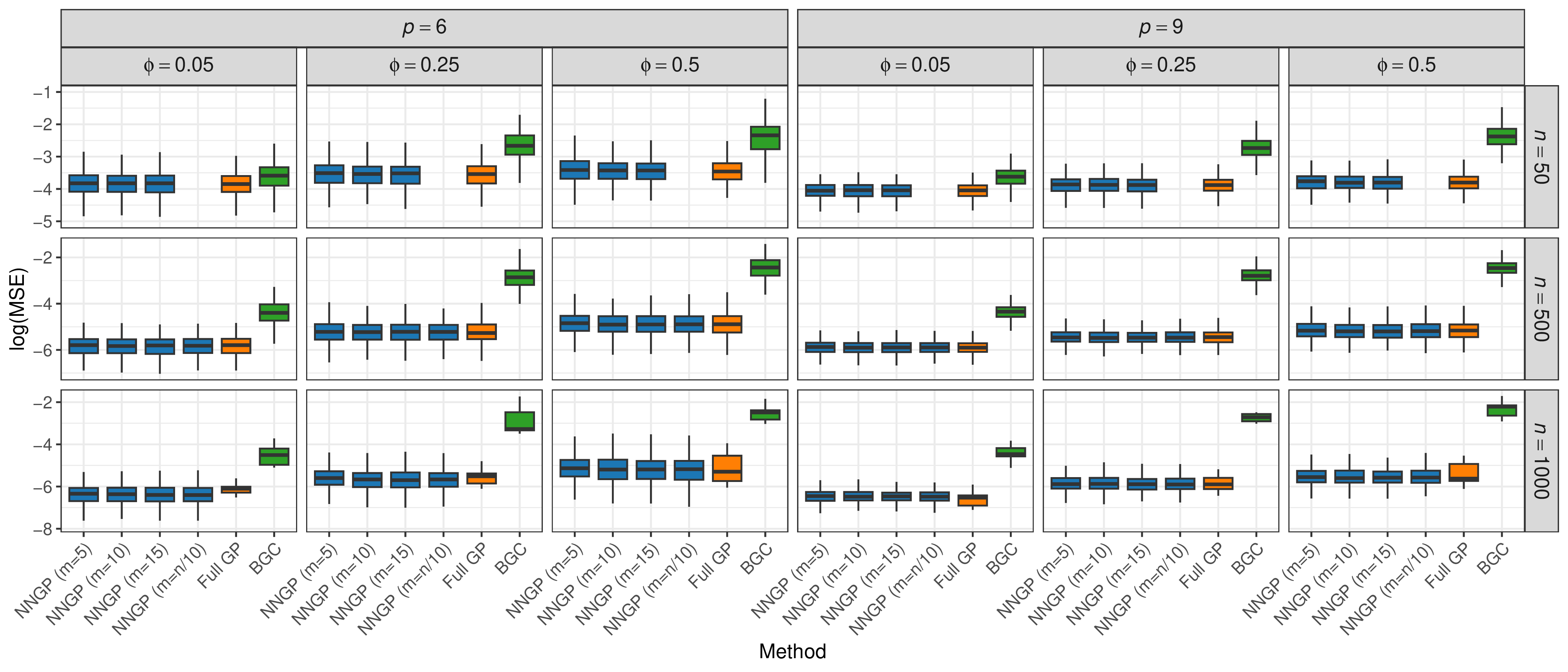}
\caption{Sensitivity analysis using real data spatial locations: logarithm of MSEs.}
\label{fig:real_mse_sensitivity}
\end{figure}

\begin{figure}[H]
\centering
\includegraphics[width=\columnwidth]{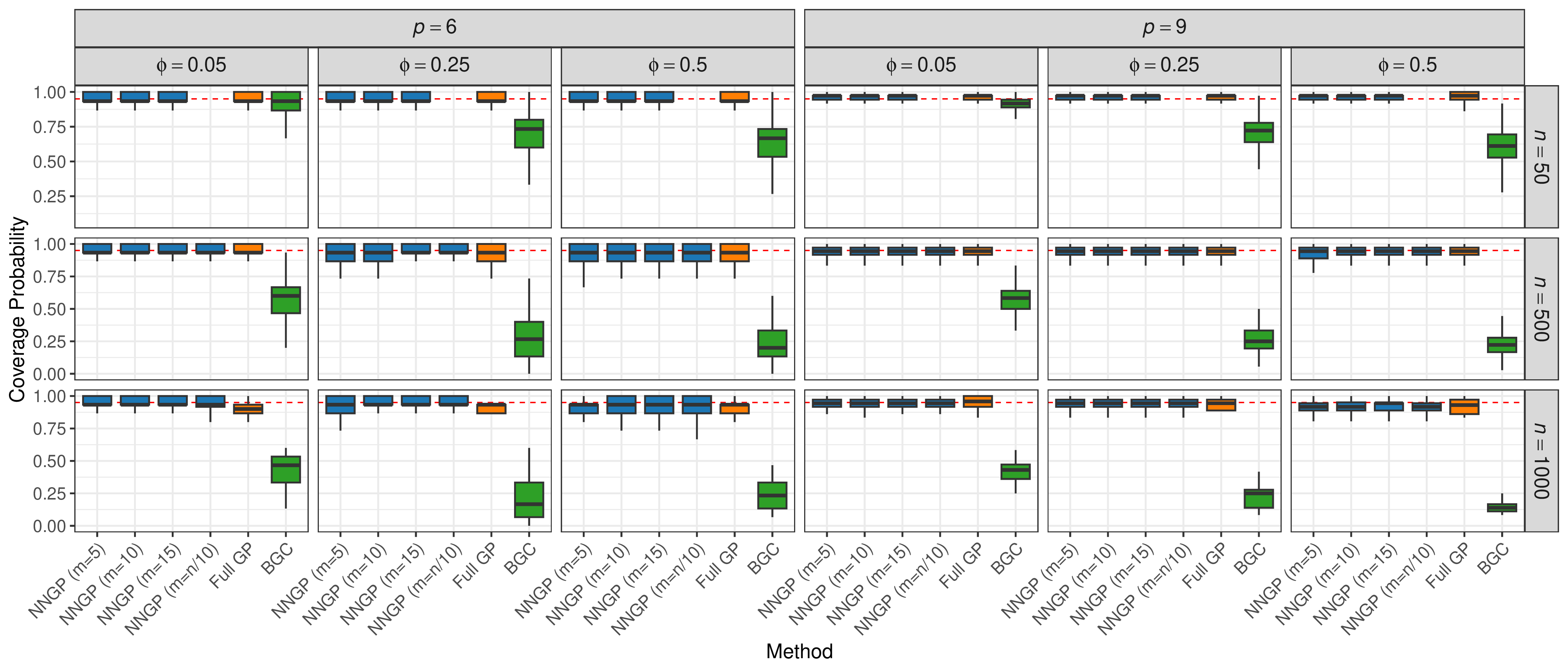}
\caption{Sensitivity analysis using real data spatial locations: coverage probabilities.}
\label{fig:real_cp_sensitivity}
\end{figure}

\begin{figure}[H]
\centering
\includegraphics[width=\columnwidth]{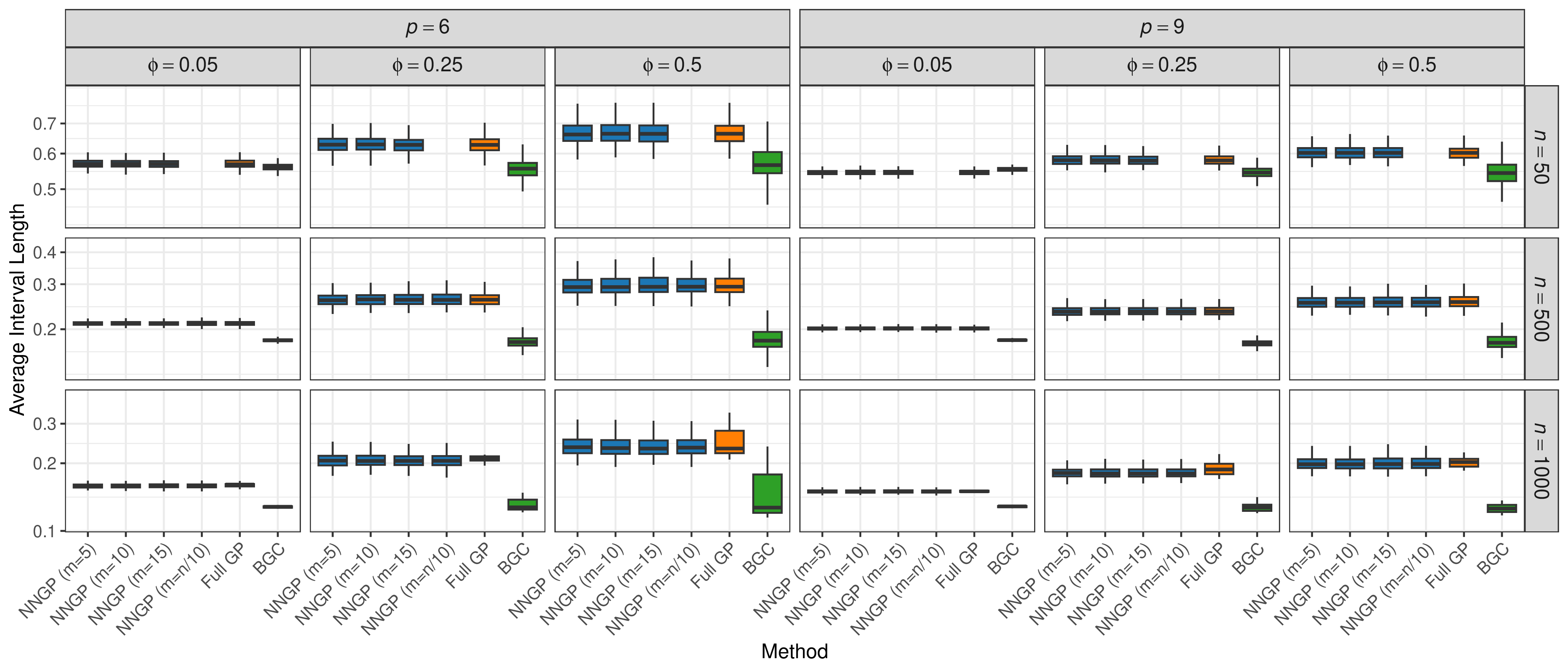}
\caption{Sensitivity analysis using real data spatial locations: average credible interval lengths.}
\label{fig:real_len_sensitivity}
\end{figure}

Tables~\ref{tb:mse_sensitivity_real},~\ref{tb:cp_sensitivity_real}, and~\ref{tb:avl_sensitivity_real} present the simulation results for MSE, coverage probability, and average credible interval lengths under real data spatial configurations.
Figures~\ref{fig:real_mse_sensitivity},~\ref{fig:real_cp_sensitivity}, and~\ref{fig:real_len_sensitivity} provide visual comparisons of these results.

The results demonstrate that spBGCNNGP maintains strong performance even under the non-uniform spatial sampling patterns.
Comparing with the results from random spatial locations (Section~\ref{sec:nngp_sensitivity}), we observe similar overall patterns: the choice of $m$ has minimal impact for small sample sizes, while $m \geq 10$ provides results comparable to spBGC for larger samples.
However, the log(MSE) values are generally slightly higher (indicating somewhat lower precision) under real data locations compared to random locations, particularly for stronger spatial correlations.
For instance, at $n=1000$, $\phi=0.50$, and $p=6$, the log(MSE) for spBGCNNGP with $m=n/10$ is $-5.181$ under real data locations versus $-5.847$ under random locations.
This difference reflects the challenges posed by the elongated and non-uniform spatial distribution along transect lines typical of line transect surveys.

Despite this reduction in precision, the coverage probabilities remain close to the nominal level.
The spBGCNNGP method achieves coverage probabilities in the range of 0.91--0.95 for most settings with $m \geq 10$, demonstrating appropriate uncertainty quantification even under non-uniform spatial sampling patterns.

The BGC method continues to exhibit severe undercoverage under real data spatial configurations, with coverage probabilities falling as low as 0.15--0.25 at $n=1000$ and $\phi=0.50$.
This underscores that accounting for spatial correlations is essential regardless of the specific spatial configuration of the data.

The credible interval lengths produced by spBGCNNGP are comparable to those from spBGC, indicating that the NNGP approximation appropriately captures the uncertainty structure.
Notably, the interval lengths under real data locations are somewhat longer than under random locations (e.g., 0.240 vs.\ 0.194 at $n=1000$, $\phi=0.50$, $p=6$), reflecting the additional uncertainty associated with non-uniform spatial sampling patterns.

These findings provide practical assurance that the proposed spBGCNNGP method can be reliably applied to real-world line transect survey data, where spatial configurations are dictated by survey design and practical constraints rather than statistical optimality.

\begin{table}[H]
\centering
\caption{Sensitivity analysis using real data spatial locations: comparisons of the logarithm of the MSEs for the spBGCNNGP, spBGC, and BGC under varying neighbor sizes and spatial range parameters.
The values represent average log(MSE)s from 300 calculations (10 for spBGC at $n=1000$), with standard errors in parentheses.}
\label{tb:mse_sensitivity_real}
\begin{minipage}{.48\linewidth}
  \centering
  \subcaption{Number of outcomes $p=6$}
  \begin{tabularx}{\linewidth}{l c *{3}{>{\centering\arraybackslash}X}}
  \hline
  & & \multicolumn{3}{c}{$\phi$} \\
  \cline{3-5}
  Method & $m$ & 0.05 & 0.25 & 0.50 \\
  \hline
  \multicolumn{5}{l}{\textbf{$n=50$}} \\
  spBGC & $5$ & $-3.845$ & $-3.538$ & $-3.406$ \\
  NNGP &  & (0.023) & (0.023) & (0.024) \\
   & $10$ & $-3.835$ & $-3.562$ & $-3.427$ \\
   &  & (0.023) & (0.024) & (0.023) \\
   & $15$ & $-3.839$ & $-3.568$ & $-3.431$ \\
   &  & (0.023) & (0.023) & (0.023) \\
  spBGC & -- & $-3.845$ & $-3.564$ & $-3.435$ \\
   &  & (0.023) & (0.023) & (0.023) \\
  BGC & -- & $-3.613$ & $-2.672$ & $-2.401$ \\
   &  & (0.026) & (0.026) & (0.029) \\
  \hline
  \multicolumn{5}{l}{\textbf{$n=500$}} \\
  spBGC & $5$ & $-5.835$ & $-5.218$ & $-4.849$ \\
  NNGP &  & (0.025) & (0.028) & (0.030) \\
   & $10$ & $-5.852$ & $-5.230$ & $-4.878$ \\
   &  & (0.026) & (0.027) & (0.031) \\
   & $15$ & $-5.839$ & $-5.243$ & $-4.870$ \\
   &  & (0.025) & (0.027) & (0.030) \\
   & $n/10$ & $-5.845$ & $-5.237$ & $-4.858$ \\
   &  & (0.025) & (0.028) & (0.031) \\
  spBGC & -- & $-5.840$ & $-5.245$ & $-4.865$ \\
   &  & (0.026) & (0.028) & (0.032) \\
  BGC & -- & $-4.412$ & $-2.881$ & $-2.441$ \\
   &  & (0.030) & (0.029) & (0.028) \\
  \hline
  \multicolumn{5}{l}{\textbf{$n=1000$}} \\
  spBGC & $5$ & $-6.398$ & $-5.578$ & $-5.100$ \\
  NNGP &  & (0.027) & (0.029) & (0.037) \\
   & $10$ & $-6.405$ & $-5.679$ & $-5.160$ \\
   &  & (0.027) & (0.030) & (0.040) \\
   & $15$ & $-6.408$ & $-5.680$ & $-5.162$ \\
   &  & (0.027) & (0.030) & (0.040) \\
   & $n/10$ & $-6.406$ & $-5.685$ & $-5.181$ \\
   &  & (0.027) & (0.029) & (0.040) \\
  spBGC & -- & $-6.117$ & $-5.560$ & $-5.140$ \\
   &  & (0.077) & (0.126) & (0.236) \\
  BGC & -- & $-4.510$ & $-2.973$ & $-2.631$ \\
   &  & (0.025) & (0.033) & (0.034) \\
  \hline
  \end{tabularx}
\end{minipage}%
\hfill
\begin{minipage}{.48\linewidth}
  \centering
  \subcaption{Number of outcomes $p=9$}
  \begin{tabularx}{\linewidth}{l c *{3}{>{\centering\arraybackslash}X}}
  \hline
  & & \multicolumn{3}{c}{$\phi$} \\
  \cline{3-5}
  Method & $m$ & 0.05 & 0.25 & 0.50 \\
  \hline
  \multicolumn{5}{l}{\textbf{$n=50$}} \\
  spBGC & $5$ & $-4.051$ & $-3.876$ & $-3.795$ \\
  NNGP &  & (0.014) & (0.016) & (0.016) \\
   & $10$ & $-4.053$ & $-3.881$ & $-3.802$ \\
   &  & (0.014) & (0.016) & (0.015) \\
   & $15$ & $-4.052$ & $-3.882$ & $-3.810$ \\
   &  & (0.014) & (0.016) & (0.016) \\
  spBGC & -- & $-4.053$ & $-3.883$ & $-3.807$ \\
   &  & (0.014) & (0.016) & (0.015) \\
  BGC & -- & $-3.631$ & $-2.722$ & $-2.385$ \\
   &  & (0.017) & (0.019) & (0.021) \\
  \hline
  \multicolumn{5}{l}{\textbf{$n=500$}} \\
  spBGC & $5$ & $-5.892$ & $-5.428$ & $-5.124$ \\
  NNGP &  & (0.017) & (0.019) & (0.024) \\
   & $10$ & $-5.903$ & $-5.451$ & $-5.154$ \\
   &  & (0.017) & (0.018) & (0.023) \\
   & $15$ & $-5.900$ & $-5.454$ & $-5.154$ \\
   &  & (0.017) & (0.018) & (0.024) \\
   & $n/10$ & $-5.902$ & $-5.443$ & $-5.141$ \\
   &  & (0.016) & (0.018) & (0.024) \\
  spBGC & -- & $-5.901$ & $-5.443$ & $-5.139$ \\
   &  & (0.017) & (0.019) & (0.024) \\
  BGC & -- & $-4.362$ & $-2.791$ & $-2.470$ \\
   &  & (0.018) & (0.019) & (0.019) \\
  \hline
  \multicolumn{5}{l}{\textbf{$n=1000$}} \\
  spBGC & $5$ & $-6.469$ & $-5.858$ & $-5.492$ \\
  NNGP &  & (0.017) & (0.021) & (0.026) \\
   & $10$ & $-6.472$ & $-5.856$ & $-5.514$ \\
   &  & (0.017) & (0.022) & (0.027) \\
   & $15$ & $-6.473$ & $-5.872$ & $-5.529$ \\
   &  & (0.017) & (0.022) & (0.026) \\
   & $n/10$ & $-6.478$ & $-5.859$ & $-5.507$ \\
   &  & (0.017) & (0.023) & (0.027) \\
  spBGC & -- & $-6.602$ & $-5.871$ & $-5.439$ \\
   &  & (0.115) & (0.126) & (0.159) \\
  BGC & -- & $-4.408$ & $-2.794$ & $-2.335$ \\
   &  & (0.021) & (0.019) & (0.021) \\
  \hline
  \end{tabularx}
\end{minipage}
\end{table}

\begin{table}[H]
\centering
\caption{Sensitivity analysis using real data spatial locations: comparisons of the coverage probabilities for the spBGCNNGP, spBGC, and BGC under varying neighbor sizes and spatial range parameters.
The values represent average coverage probabilities from 300 calculations (10 for spBGC at $n=1000$), with standard errors in parentheses.}
\label{tb:cp_sensitivity_real}
\begin{minipage}{.48\linewidth}
  \centering
  \subcaption{Number of outcomes $p=6$}
  \begin{tabularx}{\linewidth}{l c *{3}{>{\centering\arraybackslash}X}}
  \hline
  & & \multicolumn{3}{c}{$\phi$} \\
  \cline{3-5}
  Method & $m$ & 0.05 & 0.25 & 0.50 \\
  \hline
  \multicolumn{5}{l}{\textbf{$n=50$}} \\
  spBGC & $5$ & $0.946$ & $0.946$ & $0.943$ \\
  NNGP &  & (0.004) & (0.003) & (0.003) \\
   & $10$ & $0.946$ & $0.947$ & $0.942$ \\
   &  & (0.004) & (0.003) & (0.004) \\
   & $15$ & $0.945$ & $0.946$ & $0.942$ \\
   &  & (0.004) & (0.003) & (0.004) \\
  spBGC & -- & $0.947$ & $0.942$ & $0.944$ \\
   &  & (0.004) & (0.003) & (0.004) \\
  BGC & -- & $0.905$ & $0.703$ & $0.635$ \\
   &  & (0.005) & (0.008) & (0.010) \\
  \hline
  \multicolumn{5}{l}{\textbf{$n=500$}} \\
  spBGC & $5$ & $0.945$ & $0.930$ & $0.921$ \\
  NNGP &  & (0.004) & (0.004) & (0.004) \\
   & $10$ & $0.947$ & $0.933$ & $0.924$ \\
   &  & (0.004) & (0.004) & (0.004) \\
   & $15$ & $0.947$ & $0.934$ & $0.927$ \\
   &  & (0.004) & (0.004) & (0.004) \\
   & $n/10$ & $0.946$ & $0.936$ & $0.925$ \\
   &  & (0.004) & (0.004) & (0.004) \\
  spBGC & -- & $0.944$ & $0.936$ & $0.923$ \\
   &  & (0.004) & (0.004) & (0.004) \\
  BGC & -- & $0.571$ & $0.268$ & $0.234$ \\
   &  & (0.009) & (0.008) & (0.007) \\
  \hline
  \multicolumn{5}{l}{\textbf{$n=1000$}} \\
  spBGC & $5$ & $0.940$ & $0.928$ & $0.907$ \\
  NNGP &  & (0.004) & (0.004) & (0.004) \\
   & $10$ & $0.943$ & $0.937$ & $0.914$ \\
   &  & (0.004) & (0.003) & (0.004) \\
   & $15$ & $0.942$ & $0.936$ & $0.913$ \\
   &  & (0.004) & (0.004) & (0.004) \\
   & $n/10$ & $0.941$ & $0.938$ & $0.914$ \\
   &  & (0.004) & (0.004) & (0.004) \\
  spBGC & -- & $0.900$ & $0.913$ & $0.907$ \\
   &  & (0.016) & (0.011) & (0.027) \\
  BGC & -- & $0.420$ & $0.220$ & $0.247$ \\
   &  & (0.008) & (0.011) & (0.008) \\
  \hline
  \end{tabularx}
\end{minipage}%
\hfill
\begin{minipage}{.48\linewidth}
  \centering
  \subcaption{Number of outcomes $p=9$}
  \begin{tabularx}{\linewidth}{l c *{3}{>{\centering\arraybackslash}X}}
  \hline
  & & \multicolumn{3}{c}{$\phi$} \\
  \cline{3-5}
  Method & $m$ & 0.05 & 0.25 & 0.50 \\
  \hline
  \multicolumn{5}{l}{\textbf{$n=50$}} \\
  spBGC & $5$ & $0.963$ & $0.959$ & $0.961$ \\
  NNGP &  & (0.002) & (0.002) & (0.002) \\
   & $10$ & $0.961$ & $0.960$ & $0.962$ \\
   &  & (0.002) & (0.002) & (0.002) \\
   & $15$ & $0.962$ & $0.961$ & $0.963$ \\
   &  & (0.002) & (0.002) & (0.002) \\
  spBGC & -- & $0.963$ & $0.962$ & $0.964$ \\
   &  & (0.002) & (0.002) & (0.002) \\
  BGC & -- & $0.909$ & $0.707$ & $0.614$ \\
   &  & (0.003) & (0.006) & (0.007) \\
  \hline
  \multicolumn{5}{l}{\textbf{$n=500$}} \\
  spBGC & $5$ & $0.944$ & $0.935$ & $0.930$ \\
  NNGP &  & (0.002) & (0.003) & (0.002) \\
   & $10$ & $0.948$ & $0.940$ & $0.934$ \\
   &  & (0.002) & (0.002) & (0.003) \\
   & $15$ & $0.946$ & $0.940$ & $0.935$ \\
   &  & (0.002) & (0.002) & (0.002) \\
   & $n/10$ & $0.945$ & $0.939$ & $0.935$ \\
   &  & (0.002) & (0.002) & (0.002) \\
  spBGC & -- & $0.944$ & $0.939$ & $0.934$ \\
   &  & (0.002) & (0.002) & (0.002) \\
  BGC & -- & $0.569$ & $0.264$ & $0.235$ \\
   &  & (0.006) & (0.005) & (0.005) \\
  \hline
  \multicolumn{5}{l}{\textbf{$n=1000$}} \\
  spBGC & $5$ & $0.945$ & $0.933$ & $0.924$ \\
  NNGP &  & (0.002) & (0.002) & (0.003) \\
   & $10$ & $0.944$ & $0.936$ & $0.926$ \\
   &  & (0.002) & (0.002) & (0.003) \\
   & $15$ & $0.945$ & $0.940$ & $0.928$ \\
   &  & (0.002) & (0.002) & (0.002) \\
   & $n/10$ & $0.948$ & $0.938$ & $0.925$ \\
   &  & (0.002) & (0.002) & (0.002) \\
  spBGC & -- & $0.947$ & $0.939$ & $0.925$ \\
   &  & (0.016) & (0.011) & (0.022) \\
  BGC & -- & $0.428$ & $0.222$ & $0.150$ \\
   &  & (0.006) & (0.006) & (0.003) \\
  \hline
  \end{tabularx}
\end{minipage}
\end{table}

\begin{table}[H]
\centering
\caption{Sensitivity analysis using real data spatial locations: comparisons of the average credible interval lengths for the spBGCNNGP, spBGC, and BGC under varying neighbor sizes and spatial range parameters.
The values represent average interval lengths from 300 calculations (10 for spBGC at $n=1000$), with standard errors in parentheses.}
\label{tb:avl_sensitivity_real}
\begin{minipage}{.48\linewidth}
  \centering
  \subcaption{Number of outcomes $p=6$}
  \begin{tabularx}{\linewidth}{l c *{3}{>{\centering\arraybackslash}X}}
  \hline
  & & \multicolumn{3}{c}{$\phi$} \\
  \cline{3-5}
  Method & $m$ & 0.05 & 0.25 & 0.50 \\
  \hline
  \multicolumn{5}{l}{\textbf{$n=50$}} \\
  spBGC & $5$ & $0.571$ & $0.631$ & $0.668$ \\
  NNGP &  & (0.001) & (0.002) & (0.002) \\
   & $10$ & $0.570$ & $0.632$ & $0.669$ \\
   &  & (0.001) & (0.002) & (0.002) \\
   & $15$ & $0.570$ & $0.631$ & $0.668$ \\
   &  & (0.001) & (0.002) & (0.002) \\
  spBGC & -- & $0.569$ & $0.630$ & $0.669$ \\
   &  & (0.001) & (0.002) & (0.002) \\
  BGC & -- & $0.560$ & $0.557$ & $0.576$ \\
   &  & (0.001) & (0.002) & (0.003) \\
  \hline
  \multicolumn{5}{l}{\textbf{$n=500$}} \\
  spBGC & $5$ & $0.211$ & $0.263$ & $0.299$ \\
  NNGP &  & (0.000) & (0.001) & (0.002) \\
   & $10$ & $0.211$ & $0.264$ & $0.299$ \\
   &  & (0.000) & (0.001) & (0.002) \\
   & $15$ & $0.211$ & $0.264$ & $0.301$ \\
   &  & (0.000) & (0.001) & (0.002) \\
   & $n/10$ & $0.211$ & $0.265$ & $0.300$ \\
   &  & (0.000) & (0.001) & (0.002) \\
  spBGC & -- & $0.211$ & $0.264$ & $0.300$ \\
   &  & (0.000) & (0.001) & (0.002) \\
  BGC & -- & $0.181$ & $0.180$ & $0.189$ \\
   &  & (0.000) & (0.001) & (0.002) \\
  \hline
  \multicolumn{5}{l}{\textbf{$n=1000$}} \\
  spBGC & $5$ & $0.159$ & $0.207$ & $0.241$ \\
  NNGP &  & (0.000) & (0.001) & (0.002) \\
   & $10$ & $0.159$ & $0.208$ & $0.240$ \\
   &  & (0.000) & (0.001) & (0.002) \\
   & $15$ & $0.159$ & $0.207$ & $0.239$ \\
   &  & (0.000) & (0.001) & (0.002) \\
   & $n/10$ & $0.159$ & $0.207$ & $0.240$ \\
   &  & (0.000) & (0.001) & (0.002) \\
  spBGC & -- & $0.160$ & $0.210$ & $0.250$ \\
   &  & (0.000) & (0.005) & (0.011) \\
  BGC & -- & $0.128$ & $0.129$ & $0.165$ \\
   &  & (0.000) & (0.001) & (0.005) \\
  \hline
  \end{tabularx}
\end{minipage}%
\hfill
\begin{minipage}{.48\linewidth}
  \centering
  \subcaption{Number of outcomes $p=9$}
  \begin{tabularx}{\linewidth}{l c *{3}{>{\centering\arraybackslash}X}}
  \hline
  & & \multicolumn{3}{c}{$\phi$} \\
  \cline{3-5}
  Method & $m$ & 0.05 & 0.25 & 0.50 \\
  \hline
  \multicolumn{5}{l}{\textbf{$n=50$}} \\
  spBGC & $5$ & $0.545$ & $0.582$ & $0.603$ \\
  NNGP &  & (0.000) & (0.001) & (0.001) \\
   & $10$ & $0.545$ & $0.582$ & $0.603$ \\
   &  & (0.000) & (0.001) & (0.001) \\
   & $15$ & $0.545$ & $0.582$ & $0.603$ \\
   &  & (0.000) & (0.001) & (0.001) \\
  spBGC & -- & $0.545$ & $0.582$ & $0.603$ \\
   &  & (0.000) & (0.001) & (0.001) \\
  BGC & -- & $0.554$ & $0.547$ & $0.547$ \\
   &  & (0.000) & (0.001) & (0.002) \\
  \hline
  \multicolumn{5}{l}{\textbf{$n=500$}} \\
  spBGC & $5$ & $0.201$ & $0.236$ & $0.257$ \\
  NNGP &  & (0.000) & (0.001) & (0.001) \\
   & $10$ & $0.201$ & $0.236$ & $0.258$ \\
   &  & (0.000) & (0.001) & (0.001) \\
   & $15$ & $0.202$ & $0.237$ & $0.258$ \\
   &  & (0.000) & (0.001) & (0.001) \\
   & $n/10$ & $0.202$ & $0.236$ & $0.258$ \\
   &  & (0.000) & (0.001) & (0.001) \\
  spBGC & -- & $0.201$ & $0.237$ & $0.259$ \\
   &  & (0.000) & (0.001) & (0.001) \\
  BGC & -- & $0.181$ & $0.176$ & $0.183$ \\
   &  & (0.000) & (0.000) & (0.001) \\
  \hline
  \multicolumn{5}{l}{\textbf{$n=1000$}} \\
  spBGC & $5$ & $0.150$ & $0.182$ & $0.202$ \\
  NNGP &  & (0.000) & (0.001) & (0.001) \\
   & $10$ & $0.150$ & $0.182$ & $0.201$ \\
   &  & (0.000) & (0.001) & (0.001) \\
   & $15$ & $0.150$ & $0.182$ & $0.202$ \\
   &  & (0.000) & (0.001) & (0.001) \\
   & $n/10$ & $0.150$ & $0.182$ & $0.202$ \\
   &  & (0.000) & (0.001) & (0.001) \\
  spBGC & -- & $0.150$ & $0.190$ & $0.206$ \\
   &  & (0.000) & (0.005) & (0.005) \\
  BGC & -- & $0.128$ & $0.128$ & $0.127$ \\
   &  & (0.000) & (0.000) & (0.000) \\
  \hline
  \end{tabularx}
\end{minipage}
\end{table}

\subsection{Sensitivity to Correlation Function Specification}
\label{sec:correlation_misspec}

We investigate the robustness of the proposed method to misspecification of the spatial correlation function.
The simulation setup follows the same configuration as Section 4 of the main manuscript.

The key difference is in the spatial correlation function used for data generation.
Data are generated using Gaussian copulas with Mat\'{e}rn correlation functions with smoothness parameters $\nu = 3/2$ and $\nu = 5/2$.
The Mat\'{e}rn correlation function is given by
\[
\rho(d; \phi, \nu) = \frac{2^{1-\nu}}{\Gamma(\nu)}\left(\frac{\sqrt{2\nu}d}{\phi}\right)^\nu K_\nu\left(\frac{\sqrt{2\nu}d}{\phi}\right),
\]
where $d = \|\bm{s}_i - \bm{s}_{i'}\|$ is the Euclidean distance between locations $\bm{s}_i$ and $\bm{s}_{i'}$, and $K_\nu$ is the modified Bessel function of the second kind.
For $\nu = 3/2$ and $\nu = 5/2$, this simplifies to
\begin{align*}
\rho(d; \phi, \nu=3/2) &= \left(1 + \frac{\sqrt{3}d}{\phi}\right)\exp\left(-\frac{\sqrt{3}d}{\phi}\right), \\
\rho(d; \phi, \nu=5/2) &= \left(1 + \frac{\sqrt{5}d}{\phi} + \frac{5d^2}{3\phi^2}\right)\exp\left(-\frac{\sqrt{5}d}{\phi}\right),
\end{align*}
respectively.
However, the estimation is performed assuming the exponential correlation function $\rho(d; \phi) = \exp(-d/\phi)$ (equivalent to Mat\'{e}rn with $\nu = 1/2$).
This creates a scenario where the smoothness of the true spatial correlation is underestimated during estimation.

\subsubsection{Mat\'{e}rn 3/2 Data Generating Process}

\begin{figure}[H]
\centering
\includegraphics[width=\columnwidth]{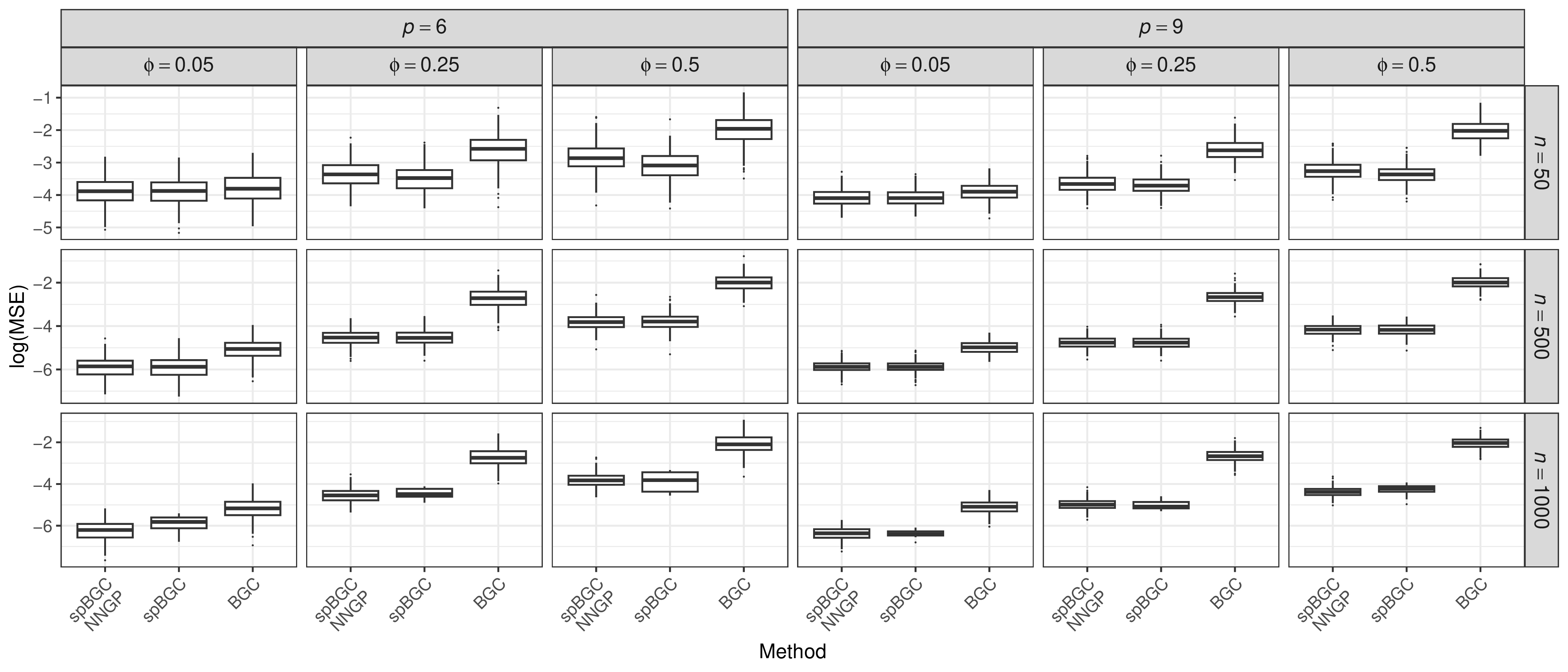}
\caption{Comparisons under Gaussian Copula with Mat\'{e}rn 3/2 kernel: logarithm of MSEs.}
\label{fig:matern3_mse}
\end{figure}

\begin{figure}[H]
\centering
\includegraphics[width=\columnwidth]{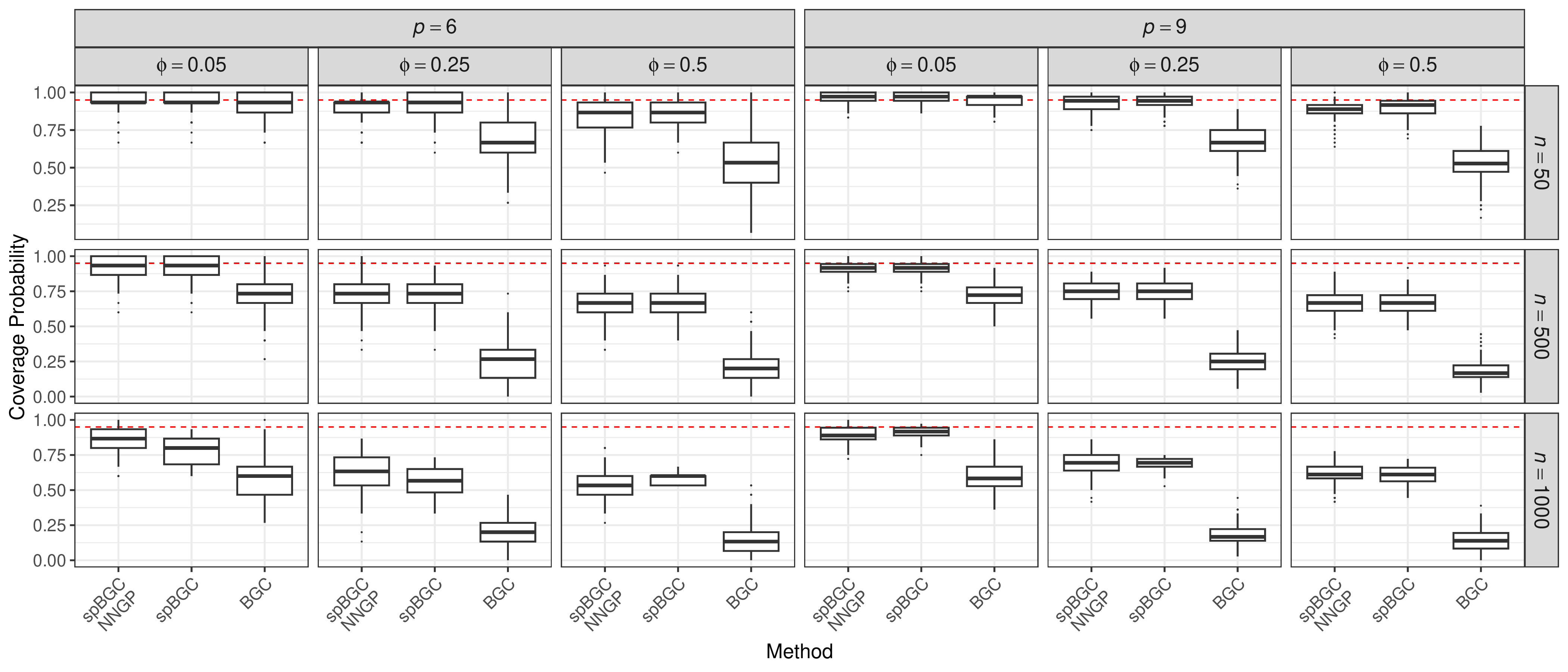}
\caption{Comparisons under Gaussian Copula with Mat\'{e}rn 3/2 kernel: coverage probabilities.}
\label{fig:matern3_cp}
\end{figure}

\begin{figure}[H]
\centering
\includegraphics[width=\columnwidth]{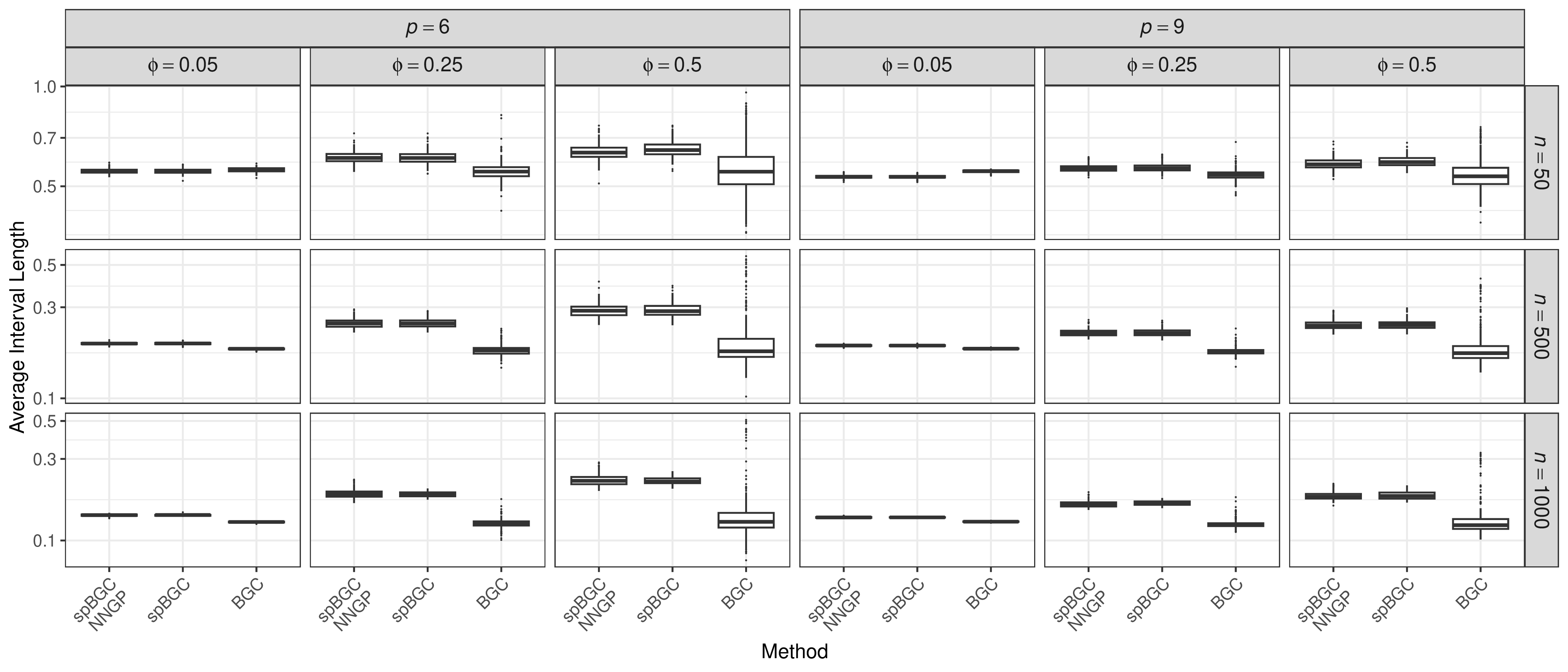}
\caption{Comparisons under Gaussian Copula with Mat\'{e}rn 3/2 kernel: average credible interval lengths.}
\label{fig:matern3_avl}
\end{figure}

Tables~\ref{tb:mse_gaussian_copula_matern3}--\ref{tb:avl_gaussian_copula_matern3} and Figures~\ref{fig:matern3_mse}--\ref{fig:matern3_avl} present the results for parameter estimation under Mat\'{e}rn 3/2.
The results demonstrate that spBGC and spBGCNNGP maintain substantially better performance than BGC despite the correlation function misspecification.
The logarithm of MSEs remain comparable across different spatial range parameters $\phi$, with only modest increases compared to the correctly specified exponential kernel case.
Coverage probabilities remain near the nominal 95\% level for small $\phi$ and small sample sizes; however, as $\phi$ and $n$ increase, moderate to substantial undercoverage emerges (e.g., around 0.55--0.65 at $n \geq 500$ and $\phi \geq 0.25$), reflecting the impact of misspecification on uncertainty quantification.
Importantly, spBGC and spBGCNNGP still achieve considerably higher coverage probabilities than BGC across all scenarios.
The average credible interval lengths of spBGC and spBGCNNGP are appropriately wider than those of BGC, reflecting proper accounting for spatial dependence; in contrast, BGC produces overly narrow intervals that fail to achieve adequate coverage.

\begin{table}[H]
\centering
\caption{Comparisons of the logarithm of the MSEs for the spBGC, spBGCNNGP (with $m=n/10$), and BGC under Gaussian Copula with Mat\'{e}rn 3/2 kernel.
The values represent average log(MSE)s from 300 calculations (10 for spBGC at $n=1000$), with standard errors in parentheses.}
\label{tb:mse_gaussian_copula_matern3}
\begin{minipage}{.48\linewidth}
  \centering
  \subcaption{Number of outcomes $p=6$}
  \begin{tabularx}{\linewidth}{l *{3}{>{\centering\arraybackslash}X}}
  \hline
  & \multicolumn{3}{c}{$\phi$} \\
  \cline{2-4}
   & 0.05 & 0.25 & 0.50 \\
  \hline
  \multicolumn{4}{l}{$n=50$} \\
  spBGC & $-3.807$ (0.024) & $-3.396$ (0.024) & $-3.010$ (0.026) \\
  spBGCNNGP & $-3.807$ (0.024) & $-3.283$ (0.024) & $-2.742$ (0.028) \\
  BGC & $-3.700$ (0.026) & $-2.521$ (0.027) & $-1.901$ (0.026) \\
  \hline
  \multicolumn{4}{l}{$n=500$} \\
  spBGC & $-5.804$ (0.027) & $-4.477$ (0.021) & $-3.728$ (0.023) \\
  spBGCNNGP & $-5.803$ (0.027) & $-4.477$ (0.021) & $-3.748$ (0.023) \\
  BGC & $-4.962$ (0.027) & $-2.642$ (0.026) & $-1.935$ (0.022) \\
  \hline
  \multicolumn{4}{l}{$n=1000$} \\
  spBGC & $-5.833$ (0.123) & $-4.425$ (0.078) & $-3.782$ (0.149) \\
  spBGCNNGP & $-6.164$ (0.024) & $-4.491$ (0.020) & $-3.769$ (0.021) \\
  BGC & $-5.088$ (0.026) & $-2.635$ (0.026) & $-1.970$ (0.026) \\
  \hline
  \end{tabularx}
\end{minipage}
\hfill
\begin{minipage}{.48\linewidth}
  \centering
  \subcaption{Number of outcomes $p=9$}
  \begin{tabularx}{\linewidth}{l *{3}{>{\centering\arraybackslash}X}}
  \hline
  & \multicolumn{3}{c}{$\phi$} \\
  \cline{2-4}
   & 0.05 & 0.25 & 0.50 \\
  \hline
  \multicolumn{4}{l}{$n=50$} \\
  spBGC & $-4.046$ (0.016) & $-3.667$ (0.016) & $-3.330$ (0.017) \\
  spBGCNNGP & $-4.042$ (0.016) & $-3.604$ (0.018) & $-3.203$ (0.019) \\
  BGC & $-3.853$ (0.016) & $-2.559$ (0.018) & $-1.975$ (0.019) \\
  \hline
  \multicolumn{4}{l}{$n=500$} \\
  spBGC & $-5.841$ (0.015) & $-4.719$ (0.016) & $-4.134$ (0.015) \\
  spBGCNNGP & $-5.840$ (0.015) & $-4.729$ (0.015) & $-4.137$ (0.015) \\
  BGC & $-4.952$ (0.016) & $-2.616$ (0.018) & $-1.951$ (0.016) \\
  \hline
  \multicolumn{4}{l}{$n=1000$} \\
  spBGC & $-6.376$ (0.061) & $-4.978$ (0.082) & $-4.264$ (0.090) \\
  spBGCNNGP & $-6.332$ (0.016) & $-4.953$ (0.015) & $-4.357$ (0.014) \\
  BGC & $-5.060$ (0.017) & $-2.630$ (0.017) & $-2.004$ (0.016) \\
  \hline
  \end{tabularx}
\end{minipage}
\end{table}

\begin{table}[H]
\centering
\caption{Comparisons of the coverage probabilities for the spBGC, spBGCNNGP (with $m=n/10$), and BGC under Gaussian Copula with Mat\'{e}rn 3/2 kernel.
The values represent average coverage probabilities from 300 calculations (10 for spBGC at $n=1000$), with standard errors in parentheses.}
\label{tb:cp_gaussian_copula_matern3}
\begin{minipage}{.48\linewidth}
  \centering
  \subcaption{Number of outcomes $p=6$}
  \begin{tabularx}{\linewidth}{l *{3}{>{\centering\arraybackslash}X}}
  \hline
  & \multicolumn{3}{c}{$\phi$} \\
  \cline{2-4}
   & 0.05 & 0.25 & 0.50 \\
  \hline
  \multicolumn{4}{l}{$n=50$} \\
  spBGC & $0.943$ (0.004) & $0.920$ (0.005) & $0.879$ (0.005) \\
  spBGCNNGP & $0.942$ (0.004) & $0.908$ (0.005) & $0.828$ (0.007) \\
  BGC & $0.931$ (0.004) & $0.678$ (0.009) & $0.546$ (0.010) \\
  \hline
  \multicolumn{4}{l}{$n=500$} \\
  spBGC & $0.926$ (0.004) & $0.742$ (0.006) & $0.652$ (0.006) \\
  spBGCNNGP & $0.926$ (0.004) & $0.743$ (0.006) & $0.650$ (0.006) \\
  BGC & $0.727$ (0.008) & $0.258$ (0.007) & $0.215$ (0.007) \\
  \hline
  \multicolumn{4}{l}{$n=1000$} \\
  spBGC & $0.780$ (0.036) & $0.560$ (0.037) & $0.593$ (0.017) \\
  spBGCNNGP & $0.874$ (0.005) & $0.620$ (0.007) & $0.560$ (0.006) \\
  BGC & $0.590$ (0.008) & $0.188$ (0.006) & $0.155$ (0.006) \\
  \hline
  \end{tabularx}
\end{minipage}
\hfill
\begin{minipage}{.48\linewidth}
  \centering
  \subcaption{Number of outcomes $p=9$}
  \begin{tabularx}{\linewidth}{l *{3}{>{\centering\arraybackslash}X}}
  \hline
  & \multicolumn{3}{c}{$\phi$} \\
  \cline{2-4}
   & 0.05 & 0.25 & 0.50 \\
  \hline
  \multicolumn{4}{l}{$n=50$} \\
  spBGC & $0.962$ (0.002) & $0.933$ (0.003) & $0.896$ (0.004) \\
  spBGCNNGP & $0.963$ (0.002) & $0.926$ (0.003) & $0.882$ (0.004) \\
  BGC & $0.949$ (0.002) & $0.677$ (0.006) & $0.530$ (0.007) \\
  \hline
  \multicolumn{4}{l}{$n=500$} \\
  spBGC & $0.920$ (0.003) & $0.756$ (0.004) & $0.670$ (0.004) \\
  spBGCNNGP & $0.920$ (0.003) & $0.759$ (0.004) & $0.668$ (0.004) \\
  BGC & $0.725$ (0.005) & $0.250$ (0.004) & $0.186$ (0.005) \\
  \hline
  \multicolumn{4}{l}{$n=1000$} \\
  spBGC & $0.897$ (0.022) & $0.675$ (0.022) & $0.603$ (0.025) \\
  spBGCNNGP & $0.892$ (0.003) & $0.686$ (0.004) & $0.613$ (0.004) \\
  BGC & $0.586$ (0.006) & $0.183$ (0.004) & $0.139$ (0.004) \\
  \hline
  \end{tabularx}
\end{minipage}
\end{table}

\begin{table}[H]
\centering
\caption{Comparisons of the average credible interval lengths for the spBGC, spBGCNNGP (with $m=n/10$), and BGC under Gaussian Copula with Mat\'{e}rn 3/2 kernel.
The values represent average interval lengths from 300 calculations (10 for spBGC at $n=1000$), with standard errors in parentheses.}
\label{tb:avl_gaussian_copula_matern3}
\begin{minipage}{.48\linewidth}
  \centering
  \subcaption{Number of outcomes $p=6$}
  \begin{tabularx}{\linewidth}{l *{3}{>{\centering\arraybackslash}X}}
  \hline
  & \multicolumn{3}{c}{$\phi$} \\
  \cline{2-4}
   & 0.05 & 0.25 & 0.50 \\
  \hline
  \multicolumn{4}{l}{$n=50$} \\
  spBGC & $0.555$ (0.001) & $0.611$ (0.001) & $0.648$ (0.002) \\
  spBGCNNGP & $0.556$ (0.001) & $0.611$ (0.001) & $0.635$ (0.002) \\
  BGC & $0.561$ (0.001) & $0.555$ (0.002) & $0.575$ (0.006) \\
  \hline
  \multicolumn{4}{l}{$n=500$} \\
  spBGC & $0.194$ (0.000) & $0.247$ (0.001) & $0.291$ (0.001) \\
  spBGCNNGP & $0.194$ (0.000) & $0.247$ (0.001) & $0.290$ (0.001) \\
  BGC & $0.182$ (0.000) & $0.179$ (0.001) & $0.205$ (0.005) \\
  \hline
  \multicolumn{4}{l}{$n=1000$} \\
  spBGC & $0.141$ (0.001) & $0.186$ (0.003) & $0.225$ (0.006) \\
  spBGCNNGP & $0.141$ (0.000) & $0.188$ (0.001) & $0.225$ (0.001) \\
  BGC & $0.128$ (0.000) & $0.126$ (0.000) & $0.146$ (0.004) \\
  \hline
  \end{tabularx}
\end{minipage}
\hfill
\begin{minipage}{.48\linewidth}
  \centering
  \subcaption{Number of outcomes $p=9$}
  \begin{tabularx}{\linewidth}{l *{3}{>{\centering\arraybackslash}X}}
  \hline
  & \multicolumn{3}{c}{$\phi$} \\
  \cline{2-4}
   & 0.05 & 0.25 & 0.50 \\
  \hline
  \multicolumn{4}{l}{$n=50$} \\
  spBGC & $0.534$ (0.000) & $0.569$ (0.001) & $0.594$ (0.001) \\
  spBGCNNGP & $0.534$ (0.000) & $0.567$ (0.001) & $0.584$ (0.001) \\
  BGC & $0.555$ (0.000) & $0.542$ (0.001) & $0.549$ (0.004) \\
  \hline
  \multicolumn{4}{l}{$n=500$} \\
  spBGC & $0.189$ (0.000) & $0.221$ (0.001) & $0.243$ (0.001) \\
  spBGCNNGP & $0.189$ (0.000) & $0.221$ (0.001) & $0.242$ (0.001) \\
  BGC & $0.182$ (0.000) & $0.176$ (0.000) & $0.188$ (0.003) \\
  \hline
  \multicolumn{4}{l}{$n=1000$} \\
  spBGC & $0.136$ (0.000) & $0.165$ (0.002) & $0.184$ (0.004) \\
  spBGCNNGP & $0.136$ (0.000) & $0.163$ (0.000) & $0.182$ (0.001) \\
  BGC & $0.129$ (0.000) & $0.125$ (0.000) & $0.135$ (0.002) \\
  \hline
  \end{tabularx}
\end{minipage}
\end{table}

\begin{figure}[H]
\centering
\includegraphics[width=\columnwidth]{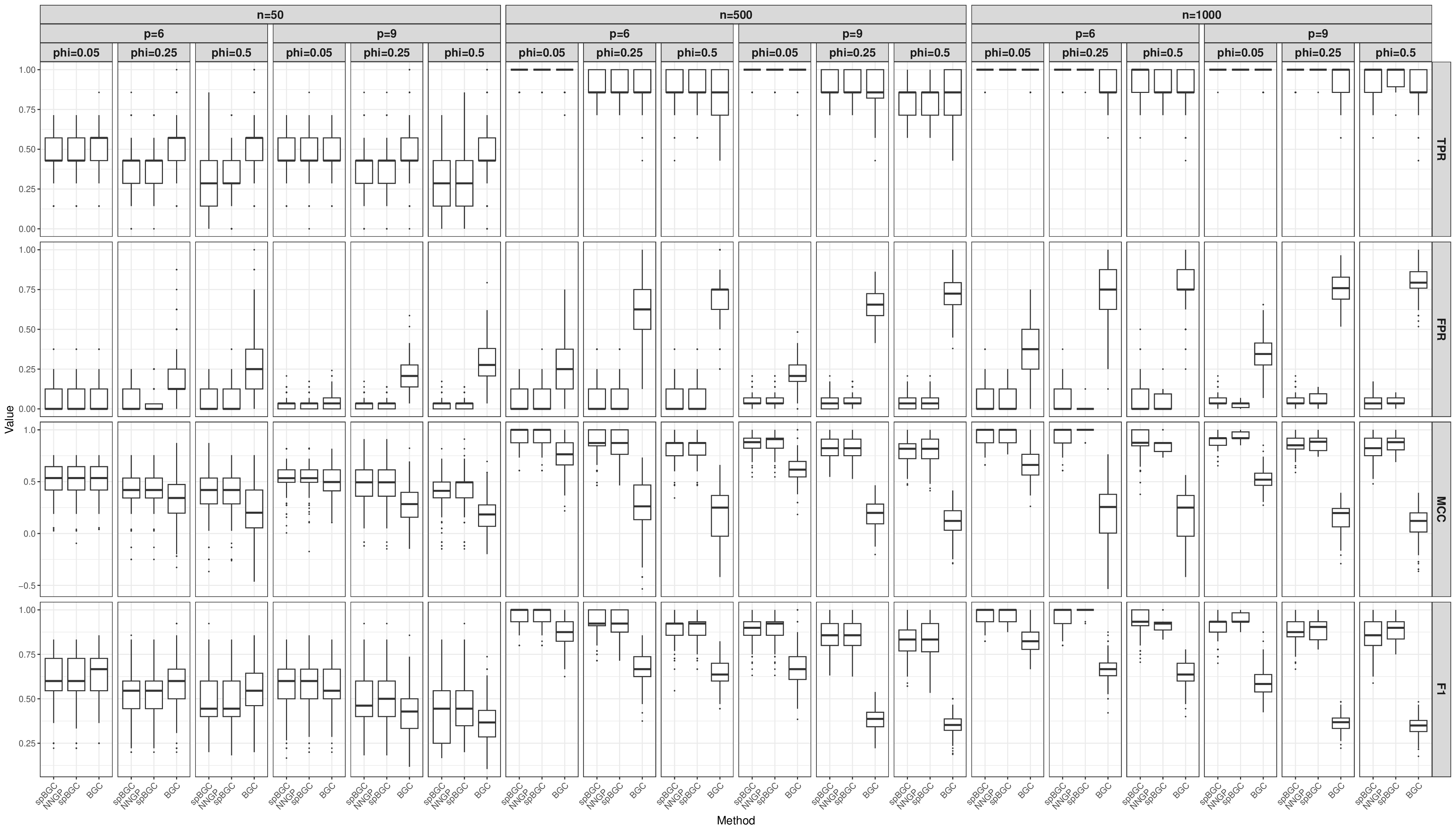}
\caption{Edge selection performance under Gaussian Copula with Exponential kernel (correctly specified): TPR, FPR, MCC, and F1 score across different sample sizes and spatial range parameters.}
\label{fig:edge_metrics_exp_iso}
\end{figure}

\begin{figure}[H]
\centering
\includegraphics[width=\columnwidth]{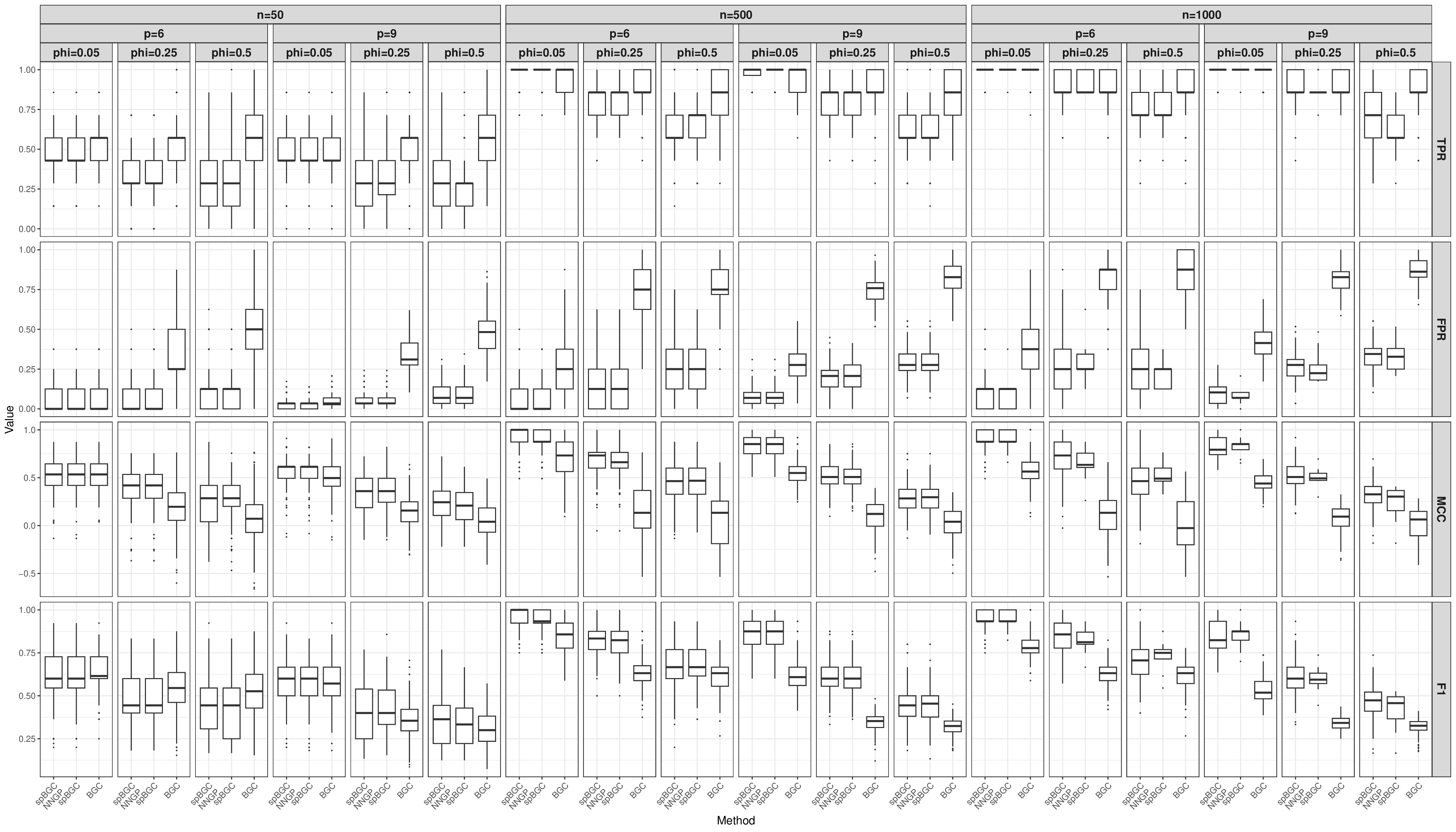}
\caption{Edge selection performance under Gaussian Copula with Mat\'{e}rn 3/2 kernel: TPR, FPR, MCC, and F1 score across different sample sizes and spatial range parameters.}
\label{fig:matern3_edge_metrics}
\end{figure}

For edge selection comparison with the misspecified case, we present the results under the correctly specified exponential correlation function in Tables~\ref{tb:edge_metrics_gaussian_copula_exp_iso_part1}--\ref{tb:edge_metrics_gaussian_copula_exp_iso_part2} and Figure~\ref{fig:edge_metrics_exp_iso}.
Tables~\ref{tb:edge_metrics_gaussian_copula_matern3_part1}--\ref{tb:edge_metrics_gaussian_copula_matern3_part2} and Figure~\ref{fig:matern3_edge_metrics} present the edge selection results under Mat\'{e}rn 3/2.
For edge selection under Mat\'{e}rn 3/2, comparing with the baseline exponential kernel results reveals that spBGC and spBGCNNGP maintain substantially better performance than BGC despite the misspecification.
For small $\phi$ values, TPR remains high and FPR stays low (around 0.05--0.07); as $\phi$ increases, FPR rises to moderate levels (around 0.15--0.30) and MCC decreases, reflecting the impact of misspecification.
Nevertheless, the overall performance metrics (MCC and F1 scores) remain substantially higher than those of BGC across all scenarios.
In contrast, BGC shows dramatic performance deterioration, particularly with very high FPR (exceeding 0.75 at large $\phi$) and near-zero MCC, highlighting the critical importance of incorporating spatial structure even under model misspecification.

\begin{table}[H]
\centering
\small
\caption{Edge selection metrics (TPR, FPR, MCC, F1) for spBGC, spBGCNNGP, and BGC under Gaussian copula with exponential kernel (correctly specified) for $n=50$ and $n=500$.
The values represent averages from 300 calculations with standard errors in parentheses.}
\label{tb:edge_metrics_gaussian_copula_exp_iso_part1}
\begin{minipage}{.48\linewidth}
  \centering
  \subcaption{Number of outcomes $p=6$}
  \begin{tabularx}{\linewidth}{l *{4}{>{\centering\arraybackslash}X}}
  \hline
  & TPR & FPR & MCC & F1 \\
  \hline
  \multicolumn{5}{l}{$n=50$, $\phi=0.05$} \\
  spBGC & $0.477$ (0.008) & $0.045$ (0.004) & $0.505$ (0.009) & $0.615$ (0.008) \\
  spBGCNNGP & $0.472$ (0.008) & $0.045$ (0.004) & $0.501$ (0.009) & $0.609$ (0.008) \\
  BGC & $0.502$ (0.008) & $0.053$ (0.005) & $0.517$ (0.009) & $0.635$ (0.007) \\
  \multicolumn{5}{l}{$n=50$, $\phi=0.25$} \\
  spBGC & $0.387$ (0.008) & $0.037$ (0.004) & $0.440$ (0.009) & $0.529$ (0.009) \\
  spBGCNNGP & $0.378$ (0.008) & $0.038$ (0.004) & $0.431$ (0.009) & $0.522$ (0.008) \\
  BGC & $0.504$ (0.009) & $0.196$ (0.010) & $0.338$ (0.013) & $0.577$ (0.008) \\
  \multicolumn{5}{l}{$n=50$, $\phi=0.50$} \\
  spBGC & $0.334$ (0.009) & $0.040$ (0.004) & $0.396$ (0.010) & $0.490$ (0.009) \\
  spBGCNNGP & $0.325$ (0.009) & $0.044$ (0.004) & $0.380$ (0.011) & $0.474$ (0.009) \\
  BGC & $0.508$ (0.010) & $0.291$ (0.011) & $0.237$ (0.013) & $0.544$ (0.008) \\
  \hline
  \multicolumn{5}{l}{$n=500$, $\phi=0.05$} \\
  spBGC & $0.975$ (0.003) & $0.051$ (0.004) & $0.926$ (0.005) & $0.960$ (0.003) \\
  spBGCNNGP & $0.976$ (0.003) & $0.047$ (0.004) & $0.931$ (0.005) & $0.962$ (0.003) \\
  BGC & $0.965$ (0.004) & $0.217$ (0.009) & $0.761$ (0.009) & $0.877$ (0.005) \\
  \multicolumn{5}{l}{$n=500$, $\phi=0.25$} \\
  spBGC & $0.917$ (0.005) & $0.059$ (0.005) & $0.864$ (0.007) & $0.924$ (0.004) \\
  spBGCNNGP & $0.920$ (0.005) & $0.059$ (0.005) & $0.867$ (0.007) & $0.926$ (0.004) \\
  BGC & $0.870$ (0.006) & $0.641$ (0.010) & $0.261$ (0.014) & $0.671$ (0.005) \\
  \multicolumn{5}{l}{$n=500$, $\phi=0.50$} \\
  spBGC & $0.870$ (0.006) & $0.055$ (0.005) & $0.826$ (0.007) & $0.899$ (0.004) \\
  spBGCNNGP & $0.866$ (0.006) & $0.054$ (0.005) & $0.825$ (0.007) & $0.897$ (0.004) \\
  BGC & $0.850$ (0.007) & $0.699$ (0.010) & $0.182$ (0.013) & $0.642$ (0.004) \\
  \hline
  \end{tabularx}
\end{minipage}
\hfill
\begin{minipage}{.48\linewidth}
  \centering
  \subcaption{Number of outcomes $p=9$}
  \begin{tabularx}{\linewidth}{l *{4}{>{\centering\arraybackslash}X}}
  \hline
  & TPR & FPR & MCC & F1 \\
  \hline
  \multicolumn{5}{l}{$n=50$, $\phi=0.05$} \\
  spBGC & $0.455$ (0.007) & $0.031$ (0.002) & $0.538$ (0.008) & $0.571$ (0.007) \\
  spBGCNNGP & $0.456$ (0.007) & $0.031$ (0.002) & $0.538$ (0.008) & $0.570$ (0.007) \\
  BGC & $0.473$ (0.007) & $0.047$ (0.003) & $0.514$ (0.008) & $0.565$ (0.007) \\
  \multicolumn{5}{l}{$n=50$, $\phi=0.25$} \\
  spBGC & $0.370$ (0.008) & $0.029$ (0.002) & $0.466$ (0.009) & $0.492$ (0.008) \\
  spBGCNNGP & $0.354$ (0.008) & $0.030$ (0.002) & $0.449$ (0.010) & $0.479$ (0.008) \\
  BGC & $0.494$ (0.009) & $0.202$ (0.005) & $0.274$ (0.010) & $0.427$ (0.007) \\
  \multicolumn{5}{l}{$n=50$, $\phi=0.50$} \\
  spBGC & $0.310$ (0.009) & $0.025$ (0.002) & $0.422$ (0.010) & $0.445$ (0.009) \\
  spBGCNNGP & $0.286$ (0.009) & $0.026$ (0.002) & $0.398$ (0.011) & $0.432$ (0.009) \\
  BGC & $0.486$ (0.010) & $0.285$ (0.007) & $0.176$ (0.009) & $0.368$ (0.006) \\
  \hline
  \multicolumn{5}{l}{$n=500$, $\phi=0.05$} \\
  spBGC & $0.975$ (0.003) & $0.051$ (0.002) & $0.875$ (0.005) & $0.896$ (0.004) \\
  spBGCNNGP & $0.977$ (0.003) & $0.052$ (0.002) & $0.874$ (0.005) & $0.895$ (0.004) \\
  BGC & $0.964$ (0.004) & $0.224$ (0.005) & $0.618$ (0.006) & $0.676$ (0.005) \\
  \multicolumn{5}{l}{$n=500$, $\phi=0.25$} \\
  spBGC & $0.903$ (0.005) & $0.051$ (0.003) & $0.826$ (0.006) & $0.858$ (0.005) \\
  spBGCNNGP & $0.902$ (0.005) & $0.052$ (0.003) & $0.824$ (0.006) & $0.856$ (0.005) \\
  BGC & $0.862$ (0.007) & $0.642$ (0.005) & $0.188$ (0.008) & $0.383$ (0.003) \\
  \multicolumn{5}{l}{$n=500$, $\phi=0.50$} \\
  spBGC & $0.837$ (0.006) & $0.047$ (0.003) & $0.788$ (0.007) & $0.826$ (0.006) \\
  spBGCNNGP & $0.840$ (0.006) & $0.047$ (0.002) & $0.790$ (0.007) & $0.827$ (0.005) \\
  BGC & $0.843$ (0.007) & $0.710$ (0.006) & $0.120$ (0.008) & $0.353$ (0.003) \\
  \hline
  \end{tabularx}
\end{minipage}
\end{table}

\begin{table}[H]
\centering
\small
\caption{Edge selection metrics (TPR, FPR, MCC, F1) for spBGC, spBGCNNGP, and BGC under Gaussian copula with exponential kernel (correctly specified) for $n=1000$.
The values represent averages from 300 calculations with standard errors in parentheses.}
\label{tb:edge_metrics_gaussian_copula_exp_iso_part2}
\begin{minipage}{.48\linewidth}
  \centering
  \subcaption{Number of outcomes $p=6$}
  \begin{tabularx}{\linewidth}{l *{4}{>{\centering\arraybackslash}X}}
  \hline
  & TPR & FPR & MCC & F1 \\
  \hline
  \multicolumn{5}{l}{$n=1000$, $\phi=0.05$} \\
  spBGC & $1.000$ (0.000) & $0.062$ (0.028) & $0.939$ (0.027) & $0.968$ (0.014) \\
  spBGCNNGP & $0.998$ (0.001) & $0.054$ (0.005) & $0.946$ (0.005) & $0.971$ (0.003) \\
  BGC & $0.989$ (0.002) & $0.329$ (0.010) & $0.691$ (0.009) & $0.842$ (0.004) \\
  \multicolumn{5}{l}{$n=1000$, $\phi=0.25$} \\
  spBGC & $0.986$ (0.014) & $0.013$ (0.012) & $0.975$ (0.017) & $0.986$ (0.010) \\
  spBGCNNGP & $0.972$ (0.003) & $0.054$ (0.005) & $0.921$ (0.005) & $0.957$ (0.003) \\
  BGC & $0.914$ (0.005) & $0.740$ (0.010) & $0.224$ (0.014) & $0.664$ (0.004) \\
  \multicolumn{5}{l}{$n=1000$, $\phi=0.50$} \\
  spBGC & $0.900$ (0.030) & $0.050$ (0.028) & $0.862$ (0.029) & $0.919$ (0.017) \\
  spBGCNNGP & $0.932$ (0.005) & $0.060$ (0.005) & $0.879$ (0.007) & $0.932$ (0.004) \\
  BGC & $0.899$ (0.006) & $0.789$ (0.009) & $0.155$ (0.014) & $0.642$ (0.004) \\
  \hline
  \end{tabularx}
\end{minipage}
\hfill
\begin{minipage}{.48\linewidth}
  \centering
  \subcaption{Number of outcomes $p=9$}
  \begin{tabularx}{\linewidth}{l *{4}{>{\centering\arraybackslash}X}}
  \hline
  & TPR & FPR & MCC & F1 \\
  \hline
  \multicolumn{5}{l}{$n=1000$, $\phi=0.05$} \\
  spBGC & $1.000$ (0.000) & $0.028$ (0.007) & $0.937$ (0.015) & $0.948$ (0.013) \\
  spBGCNNGP & $1.000$ (0.000) & $0.048$ (0.002) & $0.898$ (0.004) & $0.914$ (0.004) \\
  BGC & $0.990$ (0.002) & $0.337$ (0.006) & $0.526$ (0.005) & $0.593$ (0.004) \\
  \multicolumn{5}{l}{$n=1000$, $\phi=0.25$} \\
  spBGC & $0.986$ (0.014) & $0.059$ (0.016) & $0.870$ (0.030) & $0.891$ (0.026) \\
  spBGCNNGP & $0.972$ (0.003) & $0.054$ (0.002) & $0.869$ (0.005) & $0.891$ (0.004) \\
  BGC & $0.915$ (0.006) & $0.753$ (0.005) & $0.154$ (0.007) & $0.365$ (0.002) \\
  \multicolumn{5}{l}{$n=1000$, $\phi=0.50$} \\
  spBGC & $0.943$ (0.032) & $0.048$ (0.012) & $0.857$ (0.031) & $0.882$ (0.025) \\
  spBGCNNGP & $0.918$ (0.005) & $0.051$ (0.003) & $0.836$ (0.006) & $0.865$ (0.005) \\
  BGC & $0.894$ (0.007) & $0.797$ (0.005) & $0.097$ (0.008) & $0.345$ (0.003) \\
  \hline
  \end{tabularx}
\end{minipage}
\end{table}

\begin{table}[H]
\centering
\small
\caption{Edge selection metrics (TPR, FPR, MCC, F1) for spBGC, spBGCNNGP, and BGC under Gaussian copula with Mat\'{e}rn 3/2 kernel for $n=50$ and $n=500$.
The values represent averages from 300 calculations with standard errors in parentheses.}
\label{tb:edge_metrics_gaussian_copula_matern3_part1}
\begin{minipage}{.48\linewidth}
  \centering
  \subcaption{Number of outcomes $p=6$}
  \begin{tabularx}{\linewidth}{l *{4}{>{\centering\arraybackslash}X}}
  \hline
  & TPR & FPR & MCC & F1 \\
  \hline
  \multicolumn{5}{l}{$n=50$, $\phi=0.05$} \\
  spBGC & $0.473$ (0.008) & $0.045$ (0.004) & $0.502$ (0.009) & $0.612$ (0.007) \\
  spBGCNNGP & $0.469$ (0.008) & $0.047$ (0.004) & $0.497$ (0.009) & $0.607$ (0.007) \\
  BGC & $0.508$ (0.007) & $0.062$ (0.005) & $0.511$ (0.009) & $0.637$ (0.007) \\
  \multicolumn{5}{l}{$n=50$, $\phi=0.25$} \\
  spBGC & $0.352$ (0.009) & $0.059$ (0.005) & $0.375$ (0.011) & $0.492$ (0.009) \\
  spBGCNNGP & $0.328$ (0.009) & $0.056$ (0.005) & $0.358$ (0.012) & $0.473$ (0.009) \\
  BGC & $0.527$ (0.010) & $0.329$ (0.011) & $0.209$ (0.014) & $0.547$ (0.008) \\
  \multicolumn{5}{l}{$n=50$, $\phi=0.50$} \\
  spBGC & $0.296$ (0.011) & $0.091$ (0.007) & $0.275$ (0.014) & $0.436$ (0.010) \\
  spBGCNNGP & $0.301$ (0.010) & $0.124$ (0.008) & $0.229$ (0.014) & $0.428$ (0.009) \\
  BGC & $0.559$ (0.012) & $0.471$ (0.012) & $0.096$ (0.016) & $0.524$ (0.008) \\
  \hline
  \multicolumn{5}{l}{$n=500$, $\phi=0.05$} \\
  spBGC & $0.973$ (0.003) & $0.063$ (0.005) & $0.913$ (0.006) & $0.953$ (0.003) \\
  spBGCNNGP & $0.973$ (0.003) & $0.064$ (0.005) & $0.912$ (0.006) & $0.953$ (0.003) \\
  BGC & $0.952$ (0.004) & $0.281$ (0.010) & $0.690$ (0.011) & $0.843$ (0.005) \\
  \multicolumn{5}{l}{$n=500$, $\phi=0.25$} \\
  spBGC & $0.818$ (0.006) & $0.164$ (0.008) & $0.665$ (0.010) & $0.817$ (0.005) \\
  spBGCNNGP & $0.820$ (0.006) & $0.160$ (0.008) & $0.671$ (0.010) & $0.820$ (0.005) \\
  BGC & $0.867$ (0.007) & $0.757$ (0.010) & $0.141$ (0.014) & $0.635$ (0.004) \\
  \multicolumn{5}{l}{$n=500$, $\phi=0.50$} \\
  spBGC & $0.656$ (0.008) & $0.220$ (0.009) & $0.451$ (0.012) & $0.685$ (0.007) \\
  spBGCNNGP & $0.638$ (0.009) & $0.221$ (0.009) & $0.433$ (0.012) & $0.671$ (0.007) \\
  BGC & $0.835$ (0.009) & $0.787$ (0.009) & $0.070$ (0.015) & $0.607$ (0.005) \\
  \hline
  \end{tabularx}
\end{minipage}
\hfill
\begin{minipage}{.48\linewidth}
  \centering
  \subcaption{Number of outcomes $p=9$}
  \begin{tabularx}{\linewidth}{l *{4}{>{\centering\arraybackslash}X}}
  \hline
  & TPR & FPR & MCC & F1 \\
  \hline
  \multicolumn{5}{l}{$n=50$, $\phi=0.05$} \\
  spBGC & $0.459$ (0.008) & $0.028$ (0.002) & $0.547$ (0.009) & $0.578$ (0.007) \\
  spBGCNNGP & $0.456$ (0.008) & $0.027$ (0.002) & $0.547$ (0.009) & $0.579$ (0.008) \\
  BGC & $0.473$ (0.008) & $0.047$ (0.002) & $0.511$ (0.009) & $0.567$ (0.007) \\
  \multicolumn{5}{l}{$n=50$, $\phi=0.25$} \\
  spBGC & $0.315$ (0.009) & $0.053$ (0.003) & $0.348$ (0.011) & $0.418$ (0.008) \\
  spBGCNNGP & $0.308$ (0.009) & $0.054$ (0.003) & $0.337$ (0.011) & $0.409$ (0.009) \\
  BGC & $0.516$ (0.009) & $0.329$ (0.006) & $0.159$ (0.009) & $0.362$ (0.006) \\
  \multicolumn{5}{l}{$n=50$, $\phi=0.50$} \\
  spBGC & $0.254$ (0.010) & $0.091$ (0.004) & $0.206$ (0.012) & $0.337$ (0.008) \\
  spBGCNNGP & $0.276$ (0.010) & $0.090$ (0.004) & $0.228$ (0.012) & $0.361$ (0.008) \\
  BGC & $0.542$ (0.012) & $0.479$ (0.007) & $0.051$ (0.010) & $0.304$ (0.006) \\
  \hline
  \multicolumn{5}{l}{$n=500$, $\phi=0.05$} \\
  spBGC & $0.965$ (0.004) & $0.074$ (0.003) & $0.827$ (0.006) & $0.856$ (0.005) \\
  spBGCNNGP & $0.964$ (0.004) & $0.074$ (0.003) & $0.826$ (0.006) & $0.856$ (0.005) \\
  BGC & $0.948$ (0.004) & $0.279$ (0.005) & $0.547$ (0.006) & $0.619$ (0.005) \\
  \multicolumn{5}{l}{$n=500$, $\phi=0.25$} \\
  spBGC & $0.807$ (0.007) & $0.210$ (0.005) & $0.515$ (0.007) & $0.610$ (0.005) \\
  spBGCNNGP & $0.810$ (0.007) & $0.204$ (0.005) & $0.523$ (0.008) & $0.617$ (0.006) \\
  BGC & $0.859$ (0.007) & $0.753$ (0.005) & $0.099$ (0.008) & $0.346$ (0.003) \\
  \multicolumn{5}{l}{$n=500$, $\phi=0.50$} \\
  spBGC & $0.629$ (0.009) & $0.288$ (0.005) & $0.288$ (0.008) & $0.448$ (0.006) \\
  spBGCNNGP & $0.624$ (0.009) & $0.286$ (0.005) & $0.287$ (0.009) & $0.448$ (0.006) \\
  BGC & $0.840$ (0.009) & $0.821$ (0.005) & $0.027$ (0.009) & $0.319$ (0.003) \\
  \hline
  \end{tabularx}
\end{minipage}
\end{table}

\begin{table}[H]
\centering
\small
\caption{Edge selection metrics (TPR, FPR, MCC, F1) for spBGC, spBGCNNGP, and BGC under Gaussian copula with Mat\'{e}rn 3/2 kernel for $n=1000$.
The values represent averages from 300 calculations (10 for spBGC) with standard errors in parentheses.}
\label{tb:edge_metrics_gaussian_copula_matern3_part2}
\begin{minipage}{.48\linewidth}
  \centering
  \subcaption{Number of outcomes $p=6$}
  \begin{tabularx}{\linewidth}{l *{4}{>{\centering\arraybackslash}X}}
  \hline
  & TPR & FPR & MCC & F1 \\
  \hline
  \multicolumn{5}{l}{$n=1000$, $\phi=0.05$} \\
  spBGC & $0.986$ (0.014) & $0.100$ (0.036) & $0.889$ (0.037) & $0.941$ (0.020) \\
  spBGCNNGP & $0.994$ (0.002) & $0.092$ (0.006) & $0.905$ (0.006) & $0.950$ (0.003) \\
  BGC & $0.983$ (0.003) & $0.427$ (0.011) & $0.603$ (0.010) & $0.801$ (0.004) \\
  \multicolumn{5}{l}{$n=1000$, $\phi=0.25$} \\
  spBGC & $0.914$ (0.023) & $0.287$ (0.046) & $0.637$ (0.054) & $0.818$ (0.024) \\
  spBGCNNGP & $0.899$ (0.006) & $0.237$ (0.009) & $0.673$ (0.011) & $0.832$ (0.005) \\
  BGC & $0.884$ (0.006) & $0.816$ (0.008) & $0.095$ (0.015) & $0.628$ (0.004) \\
  \multicolumn{5}{l}{$n=1000$, $\phi=0.50$} \\
  spBGC & $0.730$ (0.056) & $0.222$ (0.028) & $0.518$ (0.046) & $0.728$ (0.033) \\
  spBGCNNGP & $0.723$ (0.008) & $0.300$ (0.010) & $0.433$ (0.012) & $0.699$ (0.006) \\
  BGC & $0.885$ (0.008) & $0.860$ (0.007) & $0.043$ (0.015) & $0.615$ (0.004) \\
  \hline
  \end{tabularx}
\end{minipage}
\hfill
\begin{minipage}{.48\linewidth}
  \centering
  \subcaption{Number of outcomes $p=9$}
  \begin{tabularx}{\linewidth}{l *{4}{>{\centering\arraybackslash}X}}
  \hline
  & TPR & FPR & MCC & F1 \\
  \hline
  \multicolumn{5}{l}{$n=1000$, $\phi=0.05$} \\
  spBGC & $0.986$ (0.014) & $0.083$ (0.017) & $0.825$ (0.032) & $0.853$ (0.027) \\
  spBGCNNGP & $0.995$ (0.001) & $0.092$ (0.003) & $0.817$ (0.006) & $0.845$ (0.005) \\
  BGC & $0.981$ (0.003) & $0.421$ (0.006) & $0.450$ (0.005) & $0.534$ (0.004) \\
  \multicolumn{5}{l}{$n=1000$, $\phi=0.25$} \\
  spBGC & $0.871$ (0.026) & $0.262$ (0.034) & $0.507$ (0.032) & $0.598$ (0.025) \\
  spBGCNNGP & $0.896$ (0.006) & $0.261$ (0.005) & $0.524$ (0.007) & $0.608$ (0.005) \\
  BGC & $0.898$ (0.006) & $0.822$ (0.004) & $0.078$ (0.008) & $0.339$ (0.002) \\
  \multicolumn{5}{l}{$n=1000$, $\phi=0.50$} \\
  spBGC & $0.614$ (0.057) & $0.331$ (0.030) & $0.235$ (0.060) & $0.416$ (0.037) \\
  spBGCNNGP & $0.715$ (0.009) & $0.331$ (0.005) & $0.315$ (0.008) & $0.466$ (0.005) \\
  BGC & $0.877$ (0.008) & $0.867$ (0.004) & $0.018$ (0.009) & $0.320$ (0.003) \\
  \hline
  \end{tabularx}
\end{minipage}
\end{table}

\subsubsection{Mat\'{e}rn 5/2 Data Generating Process}

\begin{figure}[H]
\centering
\includegraphics[width=\columnwidth]{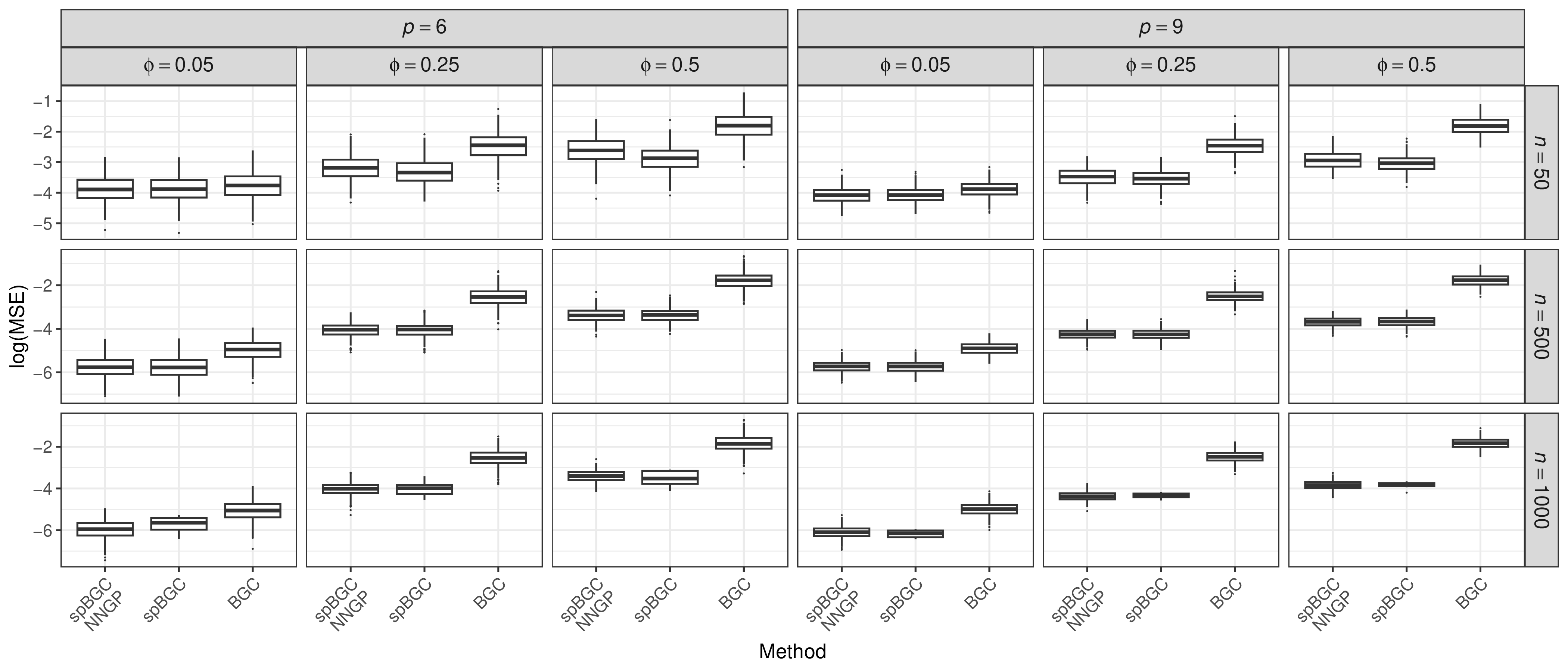}
\caption{Comparisons under Gaussian Copula with Mat\'{e}rn 5/2 kernel: logarithm of MSEs.}
\label{fig:matern5_mse}
\end{figure}

\begin{figure}[H]
\centering
\includegraphics[width=\columnwidth]{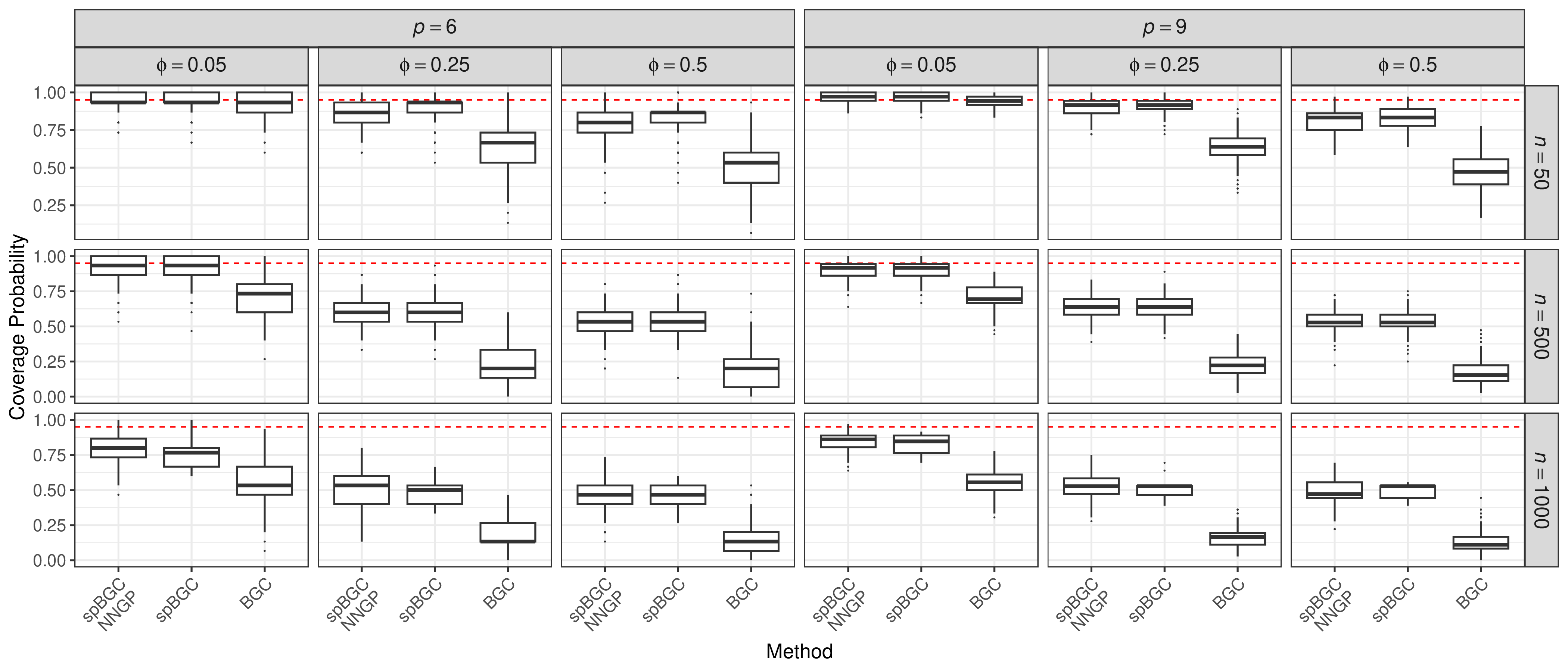}
\caption{Comparisons under Gaussian Copula with Mat\'{e}rn 5/2 kernel: coverage probabilities.}
\label{fig:matern5_cp}
\end{figure}

\begin{figure}[H]
\centering
\includegraphics[width=\columnwidth]{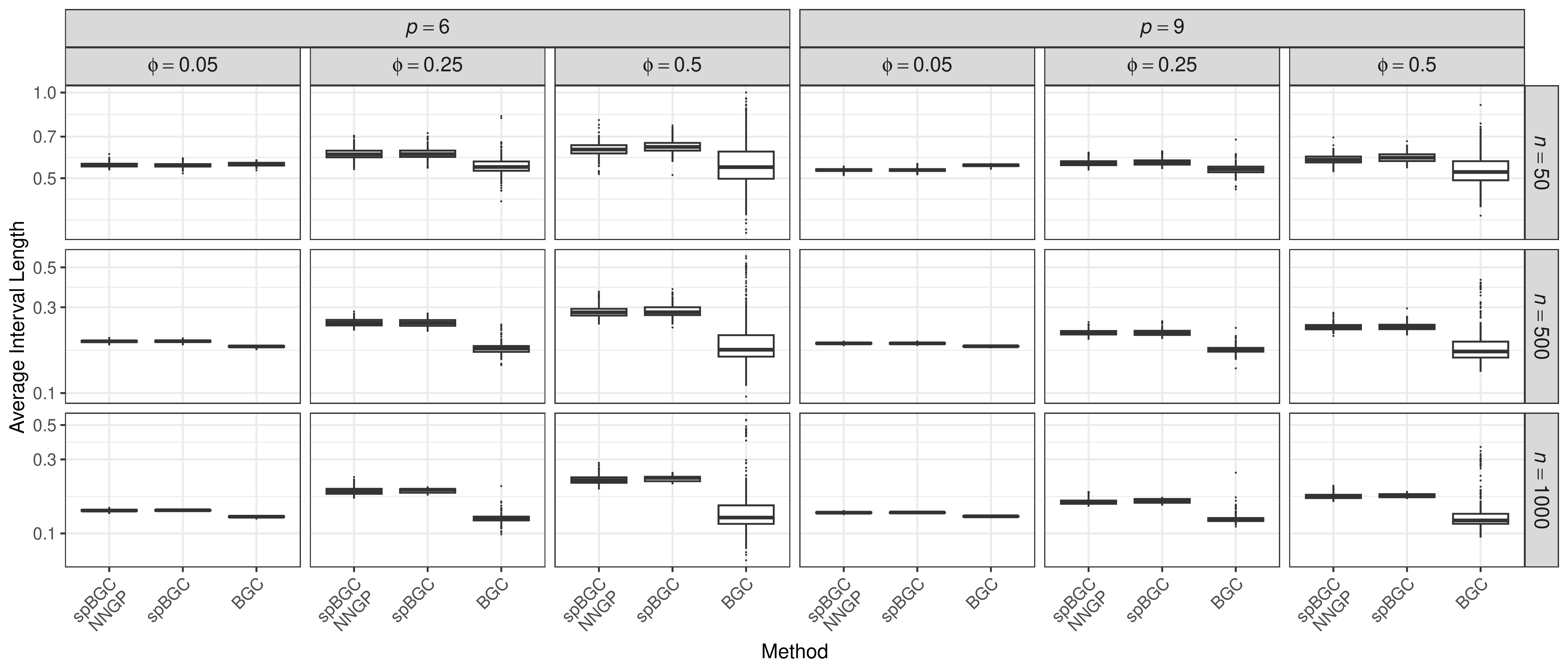}
\caption{Comparisons under Gaussian Copula with Mat\'{e}rn 5/2 kernel: average credible interval lengths.}
\label{fig:matern5_avl}
\end{figure}

Tables~\ref{tb:mse_gaussian_copula_matern5}--\ref{tb:avl_gaussian_copula_matern5} and Figures~\ref{fig:matern5_mse}--\ref{fig:matern5_avl} present the results for parameter estimation under Mat\'{e}rn 5/2.
Under Mat\'{e}rn 5/2, where the true smoothness ($\nu = 5/2$) differs most from the assumed exponential kernel ($\nu = 1/2$), the impact of misspecification becomes more pronounced.
Nonetheless, spBGC and spBGCNNGP continue to achieve substantially lower MSEs than BGC, with the logarithm of MSEs remaining reasonable across different scenarios.
The degradation is more noticeable at larger $\phi$ values, where the spatial correlation structure deviates significantly from the assumed model.
Coverage probabilities show substantial undercoverage compared to the correctly specified case, particularly at large $\phi$ (e.g., around 0.45--0.55 at $n \geq 500$ and $\phi = 0.50$), reflecting the increased severity of misspecification.
However, spBGC and spBGCNNGP still achieve considerably higher coverage probabilities than BGC (which falls below 0.20 in many scenarios), demonstrating that incorporating spatial structure provides substantial benefits even under severe model misspecification.

\begin{table}[H]
\centering
\caption{Comparisons of the logarithm of the MSEs for the spBGC, spBGCNNGP (with $m=n/10$), and BGC under Gaussian Copula with Mat\'{e}rn 5/2 kernel.
The values represent average log(MSE)s from 300 calculations (10 for spBGC at $n=1000$), with standard errors in parentheses.}
\label{tb:mse_gaussian_copula_matern5}
\begin{minipage}{.48\linewidth}
  \centering
  \subcaption{Number of outcomes $p=6$}
  \begin{tabularx}{\linewidth}{l *{3}{>{\centering\arraybackslash}X}}
  \hline
  & \multicolumn{3}{c}{$\phi$} \\
  \cline{2-4}
   & 0.05 & 0.25 & 0.50 \\
  \hline
  \multicolumn{4}{l}{$n=50$} \\
  spBGC & $-3.802$ (0.024) & $-3.246$ (0.025) & $-2.795$ (0.026) \\
  spBGCNNGP & $-3.791$ (0.025) & $-3.109$ (0.025) & $-2.521$ (0.027) \\
  BGC & $-3.680$ (0.026) & $-2.381$ (0.026) & $-1.727$ (0.025) \\
  \hline
  \multicolumn{4}{l}{$n=500$} \\
  spBGC & $-5.657$ (0.028) & $-4.002$ (0.020) & $-3.326$ (0.020) \\
  spBGCNNGP & $-5.655$ (0.028) & $-4.009$ (0.019) & $-3.335$ (0.019) \\
  BGC & $-4.869$ (0.027) & $-2.472$ (0.025) & $-1.726$ (0.022) \\
  \hline
  \multicolumn{4}{l}{$n=1000$} \\
  spBGC & $-5.654$ (0.105) & $-3.963$ (0.111) & $-3.450$ (0.115) \\
  spBGCNNGP & $-5.874$ (0.025) & $-3.970$ (0.018) & $-3.366$ (0.016) \\
  BGC & $-4.987$ (0.026) & $-2.460$ (0.024) & $-1.768$ (0.024) \\
  \hline
  \end{tabularx}
\end{minipage}
\hfill
\begin{minipage}{.48\linewidth}
  \centering
  \subcaption{Number of outcomes $p=9$}
  \begin{tabularx}{\linewidth}{l *{3}{>{\centering\arraybackslash}X}}
  \hline
  & \multicolumn{3}{c}{$\phi$} \\
  \cline{2-4}
   & 0.05 & 0.25 & 0.50 \\
  \hline
  \multicolumn{4}{l}{$n=50$} \\
  spBGC & $-4.033$ (0.015) & $-3.503$ (0.015) & $-3.014$ (0.016) \\
  spBGCNNGP & $-4.034$ (0.015) & $-3.438$ (0.018) & $-2.897$ (0.017) \\
  BGC & $-3.837$ (0.016) & $-2.418$ (0.018) & $-1.771$ (0.017) \\
  \hline
  \multicolumn{4}{l}{$n=500$} \\
  spBGC & $-5.700$ (0.016) & $-4.229$ (0.014) & $-3.657$ (0.013) \\
  spBGCNNGP & $-5.703$ (0.015) & $-4.225$ (0.014) & $-3.667$ (0.013) \\
  BGC & $-4.873$ (0.016) & $-2.451$ (0.018) & $-1.750$ (0.015) \\
  \hline
  \multicolumn{4}{l}{$n=1000$} \\
  spBGC & $-6.162$ (0.057) & $-4.334$ (0.042) & $-3.833$ (0.048) \\
  spBGCNNGP & $-6.070$ (0.016) & $-4.361$ (0.012) & $-3.818$ (0.013) \\
  BGC & $-4.956$ (0.017) & $-2.456$ (0.016) & $-1.809$ (0.015) \\
  \hline
  \end{tabularx}
\end{minipage}
\end{table}

\begin{table}[H]
\centering
\caption{Comparisons of the coverage probabilities for the spBGC, spBGCNNGP (with $m=n/10$), and BGC under Gaussian Copula with Mat\'{e}rn 5/2 kernel.
The values represent average coverage probabilities from 300 calculations (10 for spBGC at $n=1000$), with standard errors in parentheses.}
\label{tb:cp_gaussian_copula_matern5}
\begin{minipage}{.48\linewidth}
  \centering
  \subcaption{Number of outcomes $p=6$}
  \begin{tabularx}{\linewidth}{l *{3}{>{\centering\arraybackslash}X}}
  \hline
  & \multicolumn{3}{c}{$\phi$} \\
  \cline{2-4}
   & 0.05 & 0.25 & 0.50 \\
  \hline
  \multicolumn{4}{l}{$n=50$} \\
  spBGC & $0.941$ (0.004) & $0.892$ (0.005) & $0.821$ (0.007) \\
  spBGCNNGP & $0.942$ (0.004) & $0.871$ (0.005) & $0.773$ (0.008) \\
  BGC & $0.925$ (0.004) & $0.644$ (0.009) & $0.502$ (0.010) \\
  \hline
  \multicolumn{4}{l}{$n=500$} \\
  spBGC & $0.900$ (0.005) & $0.616$ (0.007) & $0.549$ (0.007) \\
  spBGCNNGP & $0.901$ (0.005) & $0.623$ (0.007) & $0.543$ (0.007) \\
  BGC & $0.706$ (0.008) & $0.241$ (0.007) & $0.191$ (0.007) \\
  \hline
  \multicolumn{4}{l}{$n=1000$} \\
  spBGC & $0.760$ (0.037) & $0.487$ (0.033) & $0.459$ (0.039) \\
  spBGCNNGP & $0.807$ (0.006) & $0.494$ (0.007) & $0.461$ (0.007) \\
  BGC & $0.556$ (0.008) & $0.177$ (0.006) & $0.146$ (0.006) \\
  \hline
  \end{tabularx}
\end{minipage}
\hfill
\begin{minipage}{.48\linewidth}
  \centering
  \subcaption{Number of outcomes $p=9$}
  \begin{tabularx}{\linewidth}{l *{3}{>{\centering\arraybackslash}X}}
  \hline
  & \multicolumn{3}{c}{$\phi$} \\
  \cline{2-4}
   & 0.05 & 0.25 & 0.50 \\
  \hline
  \multicolumn{4}{l}{$n=50$} \\
  spBGC & $0.961$ (0.002) & $0.910$ (0.003) & $0.830$ (0.004) \\
  spBGCNNGP & $0.962$ (0.002) & $0.902$ (0.003) & $0.812$ (0.005) \\
  BGC & $0.948$ (0.002) & $0.637$ (0.006) & $0.480$ (0.007) \\
  \hline
  \multicolumn{4}{l}{$n=500$} \\
  spBGC & $0.901$ (0.003) & $0.630$ (0.004) & $0.533$ (0.005) \\
  spBGCNNGP & $0.902$ (0.003) & $0.631$ (0.004) & $0.537$ (0.005) \\
  BGC & $0.706$ (0.005) & $0.231$ (0.004) & $0.172$ (0.005) \\
  \hline
  \multicolumn{4}{l}{$n=1000$} \\
  spBGC & $0.831$ (0.025) & $0.525$ (0.029) & $0.488$ (0.020) \\
  spBGCNNGP & $0.843$ (0.004) & $0.540$ (0.005) & $0.487$ (0.005) \\
  BGC & $0.560$ (0.005) & $0.165$ (0.004) & $0.131$ (0.004) \\
  \hline
  \end{tabularx}
\end{minipage}
\end{table}

\begin{table}[H]
\centering
\caption{Comparisons of the average credible interval lengths for the spBGC, spBGCNNGP (with $m=n/10$), and BGC under Gaussian Copula with Mat\'{e}rn 5/2 kernel.
The values represent average interval lengths from 300 calculations (10 for spBGC at $n=1000$), with standard errors in parentheses.}
\label{tb:avl_gaussian_copula_matern5}
\begin{minipage}{.48\linewidth}
  \centering
  \subcaption{Number of outcomes $p=6$}
  \begin{tabularx}{\linewidth}{l *{3}{>{\centering\arraybackslash}X}}
  \hline
  & \multicolumn{3}{c}{$\phi$} \\
  \cline{2-4}
   & 0.05 & 0.25 & 0.50 \\
  \hline
  \multicolumn{4}{l}{$n=50$} \\
  spBGC & $0.555$ (0.001) & $0.611$ (0.002) & $0.648$ (0.002) \\
  spBGCNNGP & $0.556$ (0.001) & $0.609$ (0.002) & $0.632$ (0.002) \\
  BGC & $0.560$ (0.001) & $0.552$ (0.002) & $0.579$ (0.008) \\
  \hline
  \multicolumn{4}{l}{$n=500$} \\
  spBGC & $0.194$ (0.000) & $0.247$ (0.001) & $0.287$ (0.001) \\
  spBGCNNGP & $0.194$ (0.000) & $0.247$ (0.001) & $0.285$ (0.001) \\
  BGC & $0.182$ (0.000) & $0.178$ (0.001) & $0.211$ (0.006) \\
  \hline
  \multicolumn{4}{l}{$n=1000$} \\
  spBGC & $0.141$ (0.000) & $0.189$ (0.003) & $0.227$ (0.004) \\
  spBGCNNGP & $0.141$ (0.000) & $0.188$ (0.001) & $0.223$ (0.001) \\
  BGC & $0.128$ (0.000) & $0.125$ (0.001) & $0.152$ (0.005) \\
  \hline
  \end{tabularx}
\end{minipage}
\hfill
\begin{minipage}{.48\linewidth}
  \centering
  \subcaption{Number of outcomes $p=9$}
  \begin{tabularx}{\linewidth}{l *{3}{>{\centering\arraybackslash}X}}
  \hline
  & \multicolumn{3}{c}{$\phi$} \\
  \cline{2-4}
   & 0.05 & 0.25 & 0.50 \\
  \hline
  \multicolumn{4}{l}{$n=50$} \\
  spBGC & $0.534$ (0.000) & $0.569$ (0.001) & $0.592$ (0.001) \\
  spBGCNNGP & $0.534$ (0.000) & $0.566$ (0.001) & $0.582$ (0.001) \\
  BGC & $0.555$ (0.000) & $0.538$ (0.001) & $0.544$ (0.005) \\
  \hline
  \multicolumn{4}{l}{$n=500$} \\
  spBGC & $0.189$ (0.000) & $0.217$ (0.001) & $0.234$ (0.001) \\
  spBGCNNGP & $0.189$ (0.000) & $0.217$ (0.000) & $0.233$ (0.001) \\
  BGC & $0.182$ (0.000) & $0.175$ (0.001) & $0.190$ (0.003) \\
  \hline
  \multicolumn{4}{l}{$n=1000$} \\
  spBGC & $0.136$ (0.000) & $0.162$ (0.002) & $0.176$ (0.002) \\
  spBGCNNGP & $0.136$ (0.000) & $0.160$ (0.000) & $0.174$ (0.000) \\
  BGC & $0.129$ (0.000) & $0.124$ (0.001) & $0.138$ (0.003) \\
  \hline
  \end{tabularx}
\end{minipage}
\end{table}

\begin{figure}[H]
\centering
\includegraphics[width=\columnwidth]{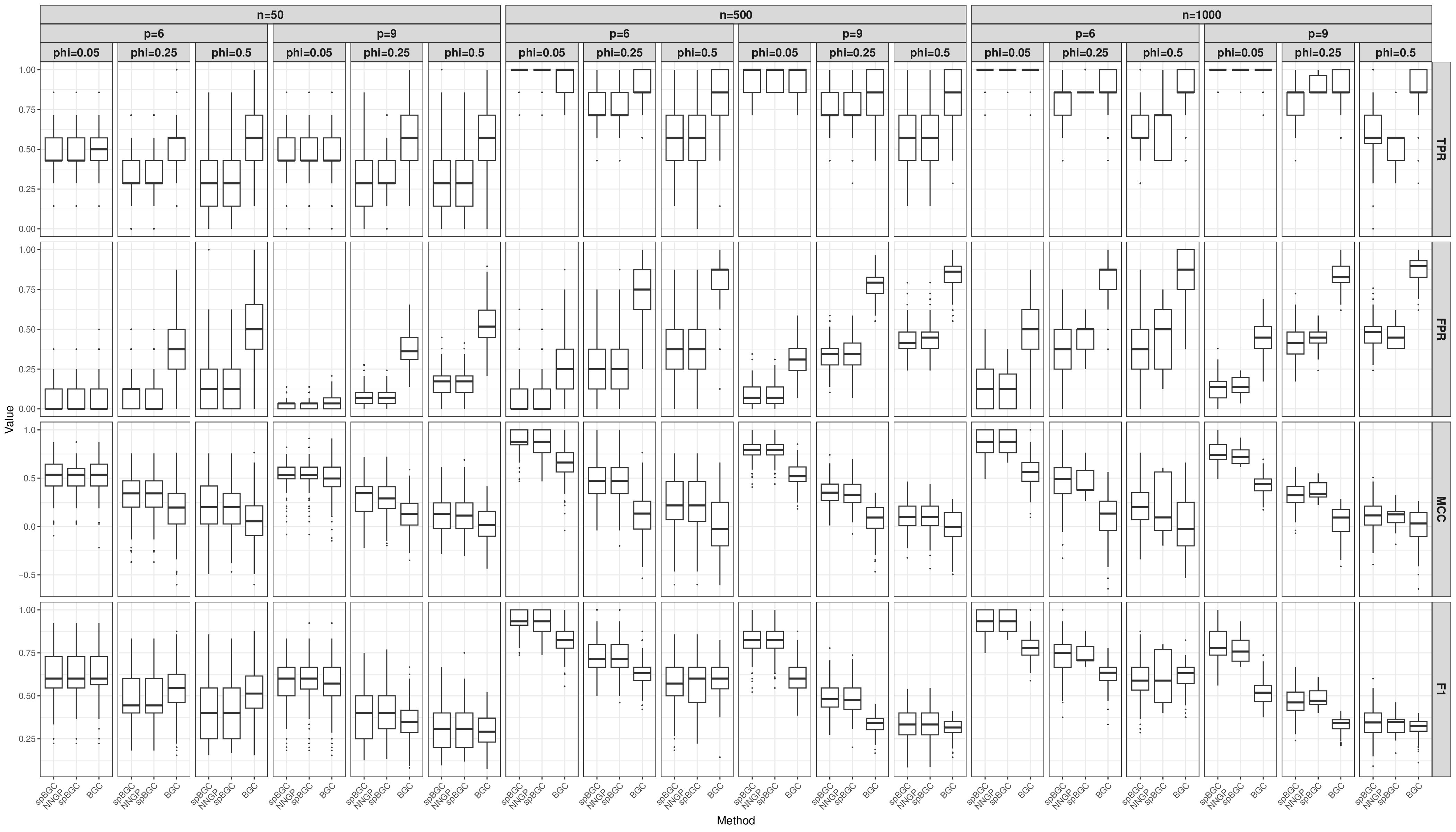}
\caption{Edge selection performance under Gaussian Copula with Mat\'{e}rn 5/2 kernel: TPR, FPR, MCC, and F1 score across different sample sizes and spatial range parameters.}
\label{fig:matern5_edge_metrics}
\end{figure}

Tables~\ref{tb:edge_metrics_gaussian_copula_matern5_part1}--\ref{tb:edge_metrics_gaussian_copula_matern5_part2} and Figure~\ref{fig:matern5_edge_metrics} present the edge selection results under Mat\'{e}rn 5/2.
For edge selection under Mat\'{e}rn 5/2, where the misspecification is most severe, the impact on performance is more pronounced than under Mat\'{e}rn 3/2.
Nonetheless, spBGC and spBGCNNGP continue to substantially outperform BGC across all scenarios.
At larger $\phi$ values, FPR increases to moderate-to-high levels (around 0.30--0.45) and MCC decreases notably (around 0.10--0.25 at $\phi = 0.50$), reflecting the challenging nature of edge selection under severe misspecification.
However, BGC's performance deteriorates even more dramatically, with very high FPR (exceeding 0.80) and near-zero MCC, making it unreliable for edge selection in spatial settings.
These results demonstrate that while severe correlation function misspecification does impact performance, incorporating spatial structure in the copula framework still provides substantial relative improvements over ignoring spatial dependence entirely.

\begin{table}[H]
\centering
\small
\caption{Edge selection metrics (TPR, FPR, MCC, F1) for spBGC, spBGCNNGP, and BGC under Gaussian copula with Mat\'{e}rn 5/2 kernel for $n=50$ and $n=500$.
The values represent averages from 300 calculations with standard errors in parentheses.}
\label{tb:edge_metrics_gaussian_copula_matern5_part1}
\begin{minipage}{.48\linewidth}
  \centering
  \subcaption{Number of outcomes $p=6$}
  \begin{tabularx}{\linewidth}{l *{4}{>{\centering\arraybackslash}X}}
  \hline
  & TPR & FPR & MCC & F1 \\
  \hline
  \multicolumn{5}{l}{$n=50$, $\phi=0.05$} \\
  spBGC & $0.463$ (0.007) & $0.047$ (0.004) & $0.492$ (0.009) & $0.602$ (0.007) \\
  spBGCNNGP & $0.465$ (0.008) & $0.045$ (0.004) & $0.495$ (0.009) & $0.603$ (0.007) \\
  BGC & $0.502$ (0.007) & $0.066$ (0.005) & $0.501$ (0.009) & $0.630$ (0.007) \\
  \multicolumn{5}{l}{$n=50$, $\phi=0.25$} \\
  spBGC & $0.346$ (0.009) & $0.085$ (0.006) & $0.335$ (0.012) & $0.478$ (0.009) \\
  spBGCNNGP & $0.333$ (0.009) & $0.094$ (0.006) & $0.307$ (0.013) & $0.465$ (0.009) \\
  BGC & $0.535$ (0.010) & $0.359$ (0.011) & $0.185$ (0.015) & $0.543$ (0.008) \\
  \multicolumn{5}{l}{$n=50$, $\phi=0.50$} \\
  spBGC & $0.292$ (0.011) & $0.157$ (0.009) & $0.179$ (0.016) & $0.412$ (0.010) \\
  spBGCNNGP & $0.305$ (0.011) & $0.172$ (0.010) & $0.167$ (0.017) & $0.420$ (0.010) \\
  BGC & $0.571$ (0.012) & $0.523$ (0.012) & $0.053$ (0.015) & $0.515$ (0.008) \\
  \hline
  \multicolumn{5}{l}{$n=500$, $\phi=0.05$} \\
  spBGC & $0.968$ (0.004) & $0.090$ (0.007) & $0.883$ (0.007) & $0.937$ (0.004) \\
  spBGCNNGP & $0.968$ (0.004) & $0.088$ (0.006) & $0.885$ (0.007) & $0.939$ (0.004) \\
  BGC & $0.947$ (0.004) & $0.299$ (0.010) & $0.668$ (0.011) & $0.833$ (0.005) \\
  \multicolumn{5}{l}{$n=500$, $\phi=0.25$} \\
  spBGC & $0.761$ (0.007) & $0.283$ (0.009) & $0.486$ (0.012) & $0.731$ (0.006) \\
  spBGCNNGP & $0.757$ (0.007) & $0.279$ (0.009) & $0.486$ (0.012) & $0.731$ (0.006) \\
  BGC & $0.872$ (0.007) & $0.772$ (0.009) & $0.130$ (0.014) & $0.633$ (0.004) \\
  \multicolumn{5}{l}{$n=500$, $\phi=0.50$} \\
  spBGC & $0.572$ (0.010) & $0.343$ (0.010) & $0.238$ (0.015) & $0.579$ (0.008) \\
  spBGCNNGP & $0.573$ (0.010) & $0.345$ (0.010) & $0.237$ (0.015) & $0.577$ (0.008) \\
  BGC & $0.825$ (0.010) & $0.817$ (0.009) & $0.020$ (0.016) & $0.593$ (0.005) \\
  \hline
  \end{tabularx}
\end{minipage}
\hfill
\begin{minipage}{.48\linewidth}
  \centering
  \subcaption{Number of outcomes $p=9$}
  \begin{tabularx}{\linewidth}{l *{4}{>{\centering\arraybackslash}X}}
  \hline
  & TPR & FPR & MCC & F1 \\
  \hline
  \multicolumn{5}{l}{$n=50$, $\phi=0.05$} \\
  spBGC & $0.461$ (0.008) & $0.029$ (0.002) & $0.547$ (0.008) & $0.579$ (0.007) \\
  spBGCNNGP & $0.451$ (0.008) & $0.028$ (0.002) & $0.540$ (0.009) & $0.570$ (0.008) \\
  BGC & $0.474$ (0.008) & $0.048$ (0.002) & $0.510$ (0.009) & $0.567$ (0.008) \\
  \multicolumn{5}{l}{$n=50$, $\phi=0.25$} \\
  spBGC & $0.329$ (0.009) & $0.078$ (0.003) & $0.306$ (0.011) & $0.409$ (0.008) \\
  spBGCNNGP & $0.321$ (0.010) & $0.079$ (0.003) & $0.298$ (0.011) & $0.395$ (0.009) \\
  BGC & $0.538$ (0.010) & $0.370$ (0.006) & $0.139$ (0.009) & $0.352$ (0.006) \\
  \multicolumn{5}{l}{$n=50$, $\phi=0.50$} \\
  spBGC & $0.275$ (0.011) & $0.163$ (0.005) & $0.116$ (0.011) & $0.309$ (0.008) \\
  spBGCNNGP & $0.290$ (0.011) & $0.166$ (0.005) & $0.130$ (0.012) & $0.318$ (0.008) \\
  BGC & $0.561$ (0.012) & $0.534$ (0.007) & $0.022$ (0.010) & $0.297$ (0.005) \\
  \hline
  \multicolumn{5}{l}{$n=500$, $\phi=0.05$} \\
  spBGC & $0.958$ (0.004) & $0.092$ (0.003) & $0.793$ (0.006) & $0.827$ (0.005) \\
  spBGCNNGP & $0.958$ (0.004) & $0.092$ (0.003) & $0.792$ (0.006) & $0.826$ (0.005) \\
  BGC & $0.948$ (0.004) & $0.300$ (0.005) & $0.527$ (0.007) & $0.603$ (0.005) \\
  \multicolumn{5}{l}{$n=500$, $\phi=0.25$} \\
  spBGC & $0.758$ (0.007) & $0.337$ (0.005) & $0.343$ (0.007) & $0.484$ (0.005) \\
  spBGCNNGP & $0.765$ (0.007) & $0.335$ (0.005) & $0.351$ (0.007) & $0.490$ (0.005) \\
  BGC & $0.851$ (0.008) & $0.774$ (0.005) & $0.075$ (0.009) & $0.337$ (0.003) \\
  \multicolumn{5}{l}{$n=500$, $\phi=0.50$} \\
  spBGC & $0.558$ (0.010) & $0.435$ (0.005) & $0.098$ (0.009) & $0.332$ (0.005) \\
  spBGCNNGP & $0.549$ (0.010) & $0.428$ (0.005) & $0.098$ (0.008) & $0.330$ (0.005) \\
  BGC & $0.831$ (0.009) & $0.834$ (0.005) & $0.004$ (0.009) & $0.313$ (0.003) \\
  \hline
  \end{tabularx}
\end{minipage}
\end{table}

\begin{table}[H]
\centering
\small
\caption{Edge selection metrics (TPR, FPR, MCC, F1) for spBGC, spBGCNNGP, and BGC under Gaussian copula with Mat\'{e}rn 5/2 kernel for $n=1000$.
The values represent averages from 300 calculations (10 for spBGC) with standard errors in parentheses.}
\label{tb:edge_metrics_gaussian_copula_matern5_part2}
\begin{minipage}{.48\linewidth}
  \centering
  \subcaption{Number of outcomes $p=6$}
  \begin{tabularx}{\linewidth}{l *{4}{>{\centering\arraybackslash}X}}
  \hline
  & TPR & FPR & MCC & F1 \\
  \hline
  \multicolumn{5}{l}{$n=1000$, $\phi=0.05$} \\
  spBGC & $0.986$ (0.014) & $0.125$ (0.042) & $0.867$ (0.041) & $0.930$ (0.022) \\
  spBGCNNGP & $0.991$ (0.002) & $0.131$ (0.007) & $0.867$ (0.007) & $0.930$ (0.004) \\
  BGC & $0.980$ (0.003) & $0.463$ (0.011) & $0.569$ (0.010) & $0.786$ (0.004) \\
  \multicolumn{5}{l}{$n=1000$, $\phi=0.25$} \\
  spBGC & $0.886$ (0.019) & $0.438$ (0.038) & $0.468$ (0.050) & $0.744$ (0.021) \\
  spBGCNNGP & $0.825$ (0.007) & $0.381$ (0.010) & $0.459$ (0.013) & $0.732$ (0.006) \\
  BGC & $0.890$ (0.007) & $0.836$ (0.008) & $0.083$ (0.015) & $0.625$ (0.004) \\
  \multicolumn{5}{l}{$n=1000$, $\phi=0.50$} \\
  spBGC & $0.667$ (0.067) & $0.444$ (0.069) & $0.231$ (0.108) & $0.611$ (0.053) \\
  spBGCNNGP & $0.633$ (0.009) & $0.420$ (0.011) & $0.220$ (0.014) & $0.597$ (0.007) \\
  BGC & $0.876$ (0.009) & $0.864$ (0.007) & $0.029$ (0.015) & $0.609$ (0.004) \\
  \hline
  \end{tabularx}
\end{minipage}
\hfill
\begin{minipage}{.48\linewidth}
  \centering
  \subcaption{Number of outcomes $p=9$}
  \begin{tabularx}{\linewidth}{l *{4}{>{\centering\arraybackslash}X}}
  \hline
  & TPR & FPR & MCC & F1 \\
  \hline
  \multicolumn{5}{l}{$n=1000$, $\phi=0.05$} \\
  spBGC & $0.986$ (0.014) & $0.141$ (0.021) & $0.735$ (0.032) & $0.774$ (0.028) \\
  spBGCNNGP & $0.993$ (0.002) & $0.131$ (0.004) & $0.756$ (0.006) & $0.792$ (0.005) \\
  BGC & $0.977$ (0.003) & $0.446$ (0.006) & $0.427$ (0.005) & $0.517$ (0.004) \\
  \multicolumn{5}{l}{$n=1000$, $\phi=0.25$} \\
  spBGC & $0.900$ (0.022) & $0.434$ (0.031) & $0.373$ (0.034) & $0.492$ (0.022) \\
  spBGCNNGP & $0.832$ (0.007) & $0.419$ (0.005) & $0.332$ (0.007) & $0.471$ (0.004) \\
  BGC & $0.897$ (0.006) & $0.839$ (0.004) & $0.064$ (0.008) & $0.334$ (0.002) \\
  \multicolumn{5}{l}{$n=1000$, $\phi=0.50$} \\
  spBGC & $0.571$ (0.063) & $0.464$ (0.028) & $0.087$ (0.050) & $0.325$ (0.029) \\
  spBGCNNGP & $0.608$ (0.010) & $0.470$ (0.006) & $0.111$ (0.009) & $0.344$ (0.005) \\
  BGC & $0.874$ (0.009) & $0.875$ (0.004) & $0.006$ (0.009) & $0.317$ (0.003) \\
  \hline
  \end{tabularx}
\end{minipage}
\end{table}

Overall, the simulation results demonstrate that the proposed spBGC and spBGCNNGP methods provide substantial improvements over BGC even under misspecification of the spatial correlation function.
Even when data are generated from smoother Mat\'{e}rn correlation functions but estimated assuming the exponential kernel, both methods achieve considerably lower MSEs and higher coverage probabilities than BGC across all scenarios.
The degradation under misspecification ranges from moderate (Mat\'{e}rn 3/2) to substantial (Mat\'{e}rn 5/2), with coverage probabilities declining to around 0.55--0.65 at large $\phi$ values; however, this is still considerably better than BGC, which falls below 0.20 in many scenarios.
For edge selection, spBGC and spBGCNNGP maintain substantially lower FPR and higher MCC than BGC, though performance degrades at larger $\phi$ values under severe misspecification.
In contrast, BGC's performance deteriorates dramatically under misspecification, with very high FPR (exceeding 0.80) and near-zero MCC, confirming that incorporating spatial structure through the GP-based copula framework provides critical benefits for reliable inference in spatial multivariate settings.

\subsection{Sensitivity to Copula Specification}
\label{sec:copula_misspec}

We evaluate the robustness of the proposed method to misspecification of the copula family.
The simulation setup follows the same configuration as Section 4 of the main manuscript.

The key difference is in the copula family used for data generation.
Latent variables $\bm{z}_{\S}$ are generated from the following spatial $t$-copula:
\begin{equation*}
\bm{z}_{\S} = (\bm{z}(\bm{s}_1)^\top, \bm{z}(\bm{s}_2)^\top, \ldots, \bm{z}(\bm{s}_n)^\top)^\top \sim t_{pn}(\bm{0}, \bm{H}(\phi)\otimes\bm{R}, \text{df}),
\end{equation*}
where $t_{pn}(\bm{\mu}, \bm{\Sigma}, \text{df})$ denotes the $pn$-dimensional multivariate $t$-distribution with location $\bm{\mu}$, scale matrix $\bm{\Sigma}$, and degrees of freedom $\text{df}$.
The density function is given by
\[
f(\bm{z}) = \frac{\Gamma((\text{df}+pn)/2)}{\Gamma(\text{df}/2)(\pi \cdot \text{df})^{pn/2}|\bm{\Sigma}|^{1/2}} \left(1 + \frac{1}{\text{df}}\bm{z}^\top\bm{\Sigma}^{-1}\bm{z}\right)^{-(\text{df}+pn)/2},
\]
with $\text{df} = 5$ for data generation.
The latent variables are then transformed to observed data via $y_j(\bm{s}_i) = F_j^{-1}[t_{\text{df}}(z_j(\bm{s}_i))]$, where $t_{\text{df}}$ is the cumulative distribution function of the standard univariate $t$-distribution with $\text{df} = 5$ degrees of freedom, and $F_j$ is the cumulative distribution function for the $j$-th marginal distribution.
The spatial $t$-copula provides heavier tails and allows for stronger dependence in the extremes compared to the Gaussian copula, making it a natural choice for investigating robustness to tail misspecification.
However, the estimation is performed assuming a spatial Gaussian copula (see Section 3 of the main manuscript).
This creates a scenario where the tail behavior of the true dependence structure is misspecified during estimation.

\begin{figure}[H]
\centering
\includegraphics[width=\columnwidth]{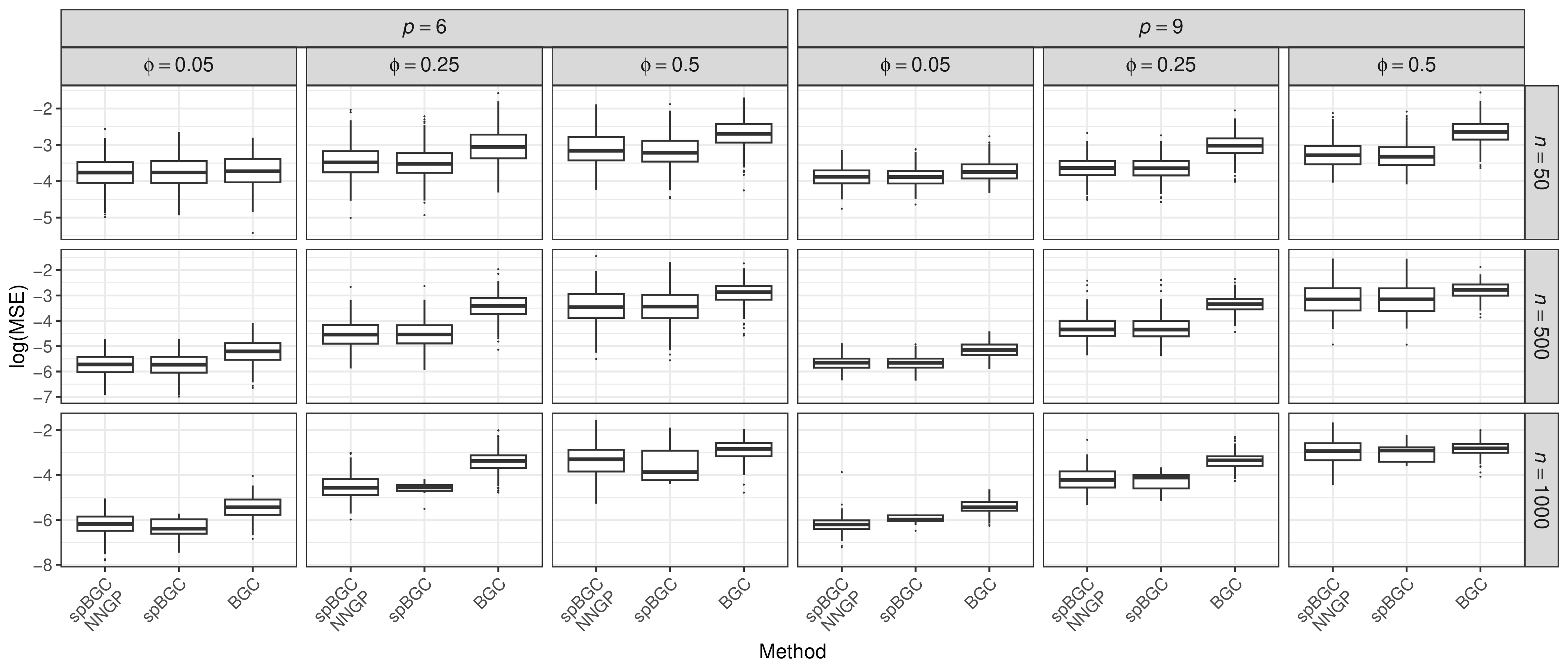}
\caption{Comparisons under $t$-Copula (df=5): logarithm of MSEs.}
\label{fig:tcopula_mse}
\end{figure}

\begin{figure}[H]
\centering
\includegraphics[width=\columnwidth]{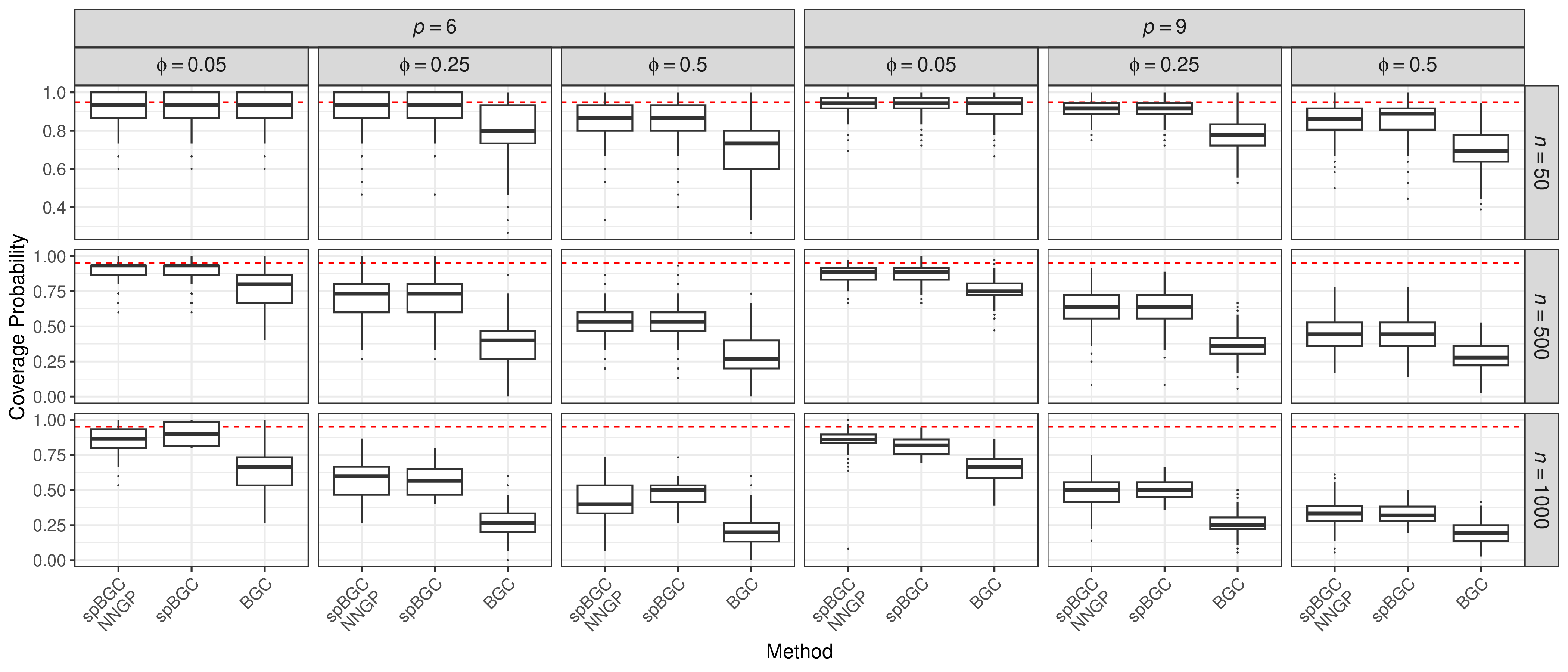}
\caption{Comparisons under $t$-Copula (df=5): coverage probabilities.}
\label{fig:tcopula_cp}
\end{figure}

\begin{figure}[H]
\centering
\includegraphics[width=\columnwidth]{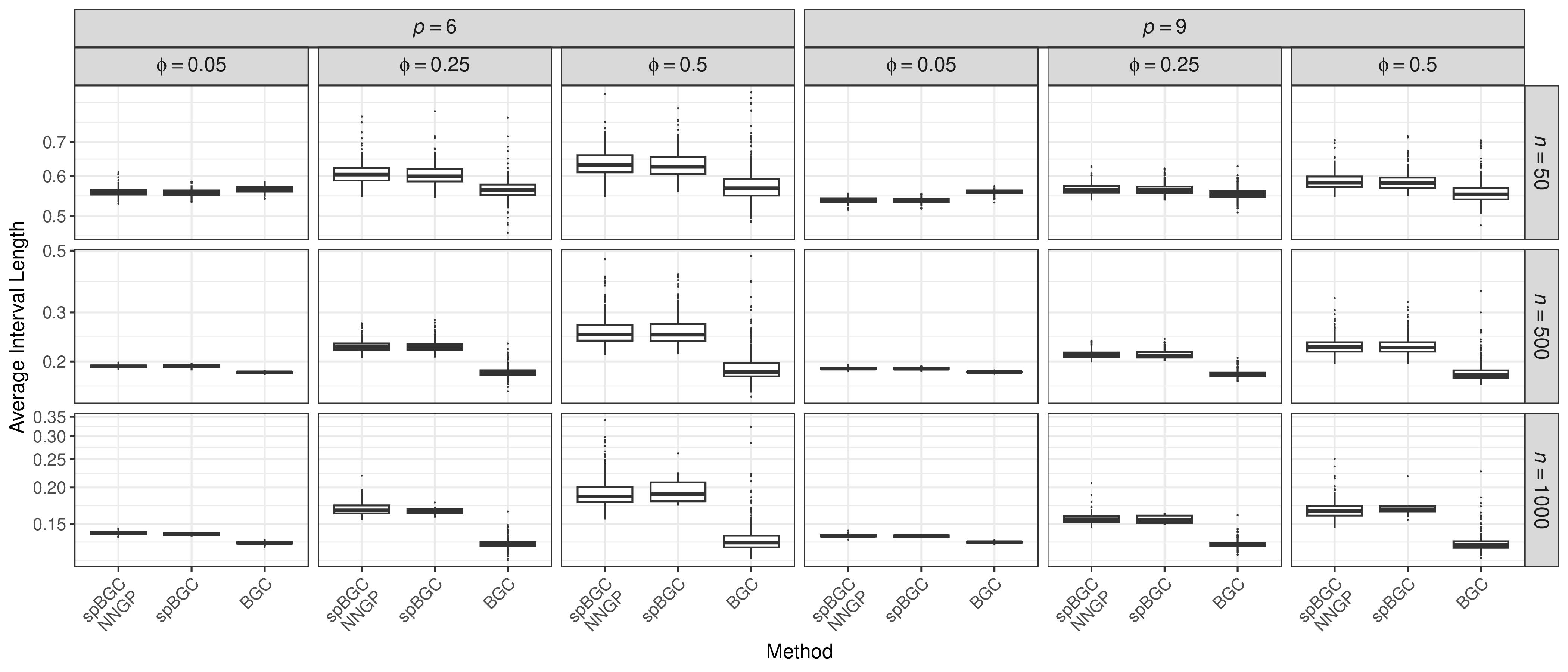}
\caption{Comparisons under $t$-Copula (df=5): average credible interval lengths.}
\label{fig:tcopula_avl}
\end{figure}

Tables~\ref{tb:mse_t_copula}--\ref{tb:avl_t_copula} and Figures~\ref{fig:tcopula_mse}--\ref{fig:tcopula_avl} present the results for parameter estimation under $t$-copula.
The results demonstrate that spBGC and spBGCNNGP achieve substantially better performance than BGC even when the true copula has heavier tails than the assumed Gaussian copula.
The logarithm of MSEs remain comparable across different spatial range parameters $\phi$, with no substantial deterioration compared to the correctly specified Gaussian copula case.
Coverage probabilities remain near the nominal 95\% level for small $\phi$ and small sample sizes; however, as $\phi$ and $n$ increase, moderate to substantial undercoverage emerges (e.g., around 0.45--0.55 at $n \geq 500$ and $\phi = 0.50$), indicating that the Gaussian copula assumption affects uncertainty quantification under tail misspecification.
Importantly, spBGC and spBGCNNGP still achieve considerably higher coverage probabilities than BGC across all scenarios.
The average credible interval lengths of spBGC and spBGCNNGP are appropriately wider than those of BGC, reflecting proper accounting for spatial dependence.

\begin{table}[H]
\centering
\caption{Comparisons of the logarithm of the MSEs for the spBGC, spBGCNNGP (with $m=n/10$), and BGC under $t$-Copula (df=5).
The values represent average log(MSE)s from 300 calculations (10 for spBGC at $n=1000$), with standard errors in parentheses.}
\label{tb:mse_t_copula}
\begin{minipage}{.48\linewidth}
  \centering
  \subcaption{Number of outcomes $p=6$}
  \begin{tabularx}{\linewidth}{l *{3}{>{\centering\arraybackslash}X}}
  \hline
  & \multicolumn{3}{c}{$\phi$} \\
  \cline{2-4}
   & 0.05 & 0.25 & 0.50 \\
  \hline
  \multicolumn{4}{l}{$n=50$} \\
  spBGC & $-3.677$ (0.025) & $-3.392$ (0.027) & $-3.085$ (0.027) \\
  spBGCNNGP & $-3.682$ (0.025) & $-3.352$ (0.029) & $-3.029$ (0.028) \\
  BGC & $-3.636$ (0.025) & $-2.937$ (0.029) & $-2.614$ (0.024) \\
  \hline
  \multicolumn{4}{l}{$n=500$} \\
  spBGC & $-5.639$ (0.026) & $-4.368$ (0.036) & $-3.235$ (0.039) \\
  spBGCNNGP & $-5.637$ (0.026) & $-4.369$ (0.035) & $-3.232$ (0.040) \\
  BGC & $-5.110$ (0.027) & $-3.338$ (0.028) & $-2.812$ (0.025) \\
  \hline
  \multicolumn{4}{l}{$n=1000$} \\
  spBGC & $-6.272$ (0.140) & $-4.560$ (0.092) & $-3.092$ (0.333) \\
  spBGCNNGP & $-6.098$ (0.026) & $-4.369$ (0.036) & $-3.084$ (0.044) \\
  BGC & $-5.343$ (0.027) & $-3.295$ (0.030) & $-2.785$ (0.025) \\
  \hline
  \end{tabularx}
\end{minipage}
\hfill
\begin{minipage}{.48\linewidth}
  \centering
  \subcaption{Number of outcomes $p=9$}
  \begin{tabularx}{\linewidth}{l *{3}{>{\centering\arraybackslash}X}}
  \hline
  & \multicolumn{3}{c}{$\phi$} \\
  \cline{2-4}
   & 0.05 & 0.25 & 0.50 \\
  \hline
  \multicolumn{4}{l}{$n=50$} \\
  spBGC & $-3.834$ (0.016) & $-3.591$ (0.018) & $-3.219$ (0.025) \\
  spBGCNNGP & $-3.835$ (0.016) & $-3.579$ (0.018) & $-3.203$ (0.024) \\
  BGC & $-3.685$ (0.017) & $-2.970$ (0.019) & $-2.599$ (0.019) \\
  \hline
  \multicolumn{4}{l}{$n=500$} \\
  spBGC & $-5.619$ (0.016) & $-4.162$ (0.036) & $-2.981$ (0.035) \\
  spBGCNNGP & $-5.618$ (0.016) & $-4.157$ (0.035) & $-2.978$ (0.035) \\
  BGC & $-5.099$ (0.018) & $-3.288$ (0.020) & $-2.747$ (0.018) \\
  \hline
  \multicolumn{4}{l}{$n=1000$} \\
  spBGC & $-5.967$ (0.067) & $-4.180$ (0.141) & $-2.920$ (0.147) \\
  spBGCNNGP & $-6.133$ (0.034) & $-4.063$ (0.033) & $-2.822$ (0.033) \\
  BGC & $-5.371$ (0.017) & $-3.312$ (0.019) & $-2.777$ (0.018) \\
  \hline
  \end{tabularx}
\end{minipage}
\end{table}

\begin{table}[H]
\centering
\caption{Comparisons of the coverage probabilities for the spBGC, spBGCNNGP (with $m=n/10$), and BGC under $t$-Copula (df=5).
The values represent average coverage probabilities from 300 calculations (10 for spBGC at $n=1000$), with standard errors in parentheses.}
\label{tb:cp_t_copula}
\begin{minipage}{.48\linewidth}
  \centering
  \subcaption{Number of outcomes $p=6$}
  \begin{tabularx}{\linewidth}{l *{3}{>{\centering\arraybackslash}X}}
  \hline
  & \multicolumn{3}{c}{$\phi$} \\
  \cline{2-4}
   & 0.05 & 0.25 & 0.50 \\
  \hline
  \multicolumn{4}{l}{$n=50$} \\
  spBGC & $0.927$ (0.004) & $0.907$ (0.005) & $0.867$ (0.006) \\
  spBGCNNGP & $0.931$ (0.004) & $0.903$ (0.005) & $0.863$ (0.006) \\
  BGC & $0.921$ (0.005) & $0.794$ (0.008) & $0.716$ (0.008) \\
  \hline
  \multicolumn{4}{l}{$n=500$} \\
  spBGC & $0.892$ (0.005) & $0.692$ (0.008) & $0.535$ (0.008) \\
  spBGCNNGP & $0.892$ (0.005) & $0.692$ (0.008) & $0.537$ (0.008) \\
  BGC & $0.757$ (0.007) & $0.388$ (0.009) & $0.304$ (0.008) \\
  \hline
  \multicolumn{4}{l}{$n=1000$} \\
  spBGC & $0.900$ (0.027) & $0.567$ (0.039) & $0.487$ (0.042) \\
  spBGCNNGP & $0.856$ (0.006) & $0.577$ (0.008) & $0.425$ (0.008) \\
  BGC & $0.649$ (0.008) & $0.260$ (0.007) & $0.214$ (0.006) \\
  \hline
  \end{tabularx}
\end{minipage}
\hfill
\begin{minipage}{.48\linewidth}
  \centering
  \subcaption{Number of outcomes $p=9$}
  \begin{tabularx}{\linewidth}{l *{3}{>{\centering\arraybackslash}X}}
  \hline
  & \multicolumn{3}{c}{$\phi$} \\
  \cline{2-4}
   & 0.05 & 0.25 & 0.50 \\
  \hline
  \multicolumn{4}{l}{$n=50$} \\
  spBGC & $0.940$ (0.003) & $0.914$ (0.003) & $0.856$ (0.005) \\
  spBGCNNGP & $0.941$ (0.003) & $0.915$ (0.003) & $0.857$ (0.005) \\
  BGC & $0.925$ (0.003) & $0.783$ (0.005) & $0.696$ (0.006) \\
  \hline
  \multicolumn{4}{l}{$n=500$} \\
  spBGC & $0.880$ (0.003) & $0.635$ (0.007) & $0.449$ (0.007) \\
  spBGCNNGP & $0.880$ (0.003) & $0.631$ (0.007) & $0.447$ (0.007) \\
  BGC & $0.761$ (0.005) & $0.368$ (0.006) & $0.284$ (0.005) \\
  \hline
  \multicolumn{4}{l}{$n=1000$} \\
  spBGC & $0.814$ (0.025) & $0.503$ (0.031) & $0.331$ (0.027) \\
  spBGCNNGP & $0.857$ (0.005) & $0.493$ (0.006) & $0.336$ (0.006) \\
  BGC & $0.659$ (0.005) & $0.260$ (0.004) & $0.208$ (0.004) \\
  \hline
  \end{tabularx}
\end{minipage}
\end{table}

\begin{table}[H]
\centering
\caption{Comparisons of the average credible interval lengths for the spBGC, spBGCNNGP (with $m=n/10$), and BGC under $t$-Copula (df=5).
The values represent average interval lengths from 300 calculations (10 for spBGC at $n=1000$), with standard errors in parentheses.}
\label{tb:avl_t_copula}
\begin{minipage}{.48\linewidth}
  \centering
  \subcaption{Number of outcomes $p=6$}
  \begin{tabularx}{\linewidth}{l *{3}{>{\centering\arraybackslash}X}}
  \hline
  & \multicolumn{3}{c}{$\phi$} \\
  \cline{2-4}
   & 0.05 & 0.25 & 0.50 \\
  \hline
  \multicolumn{4}{l}{$n=50$} \\
  spBGC & $0.557$ (0.001) & $0.603$ (0.002) & $0.634$ (0.002) \\
  spBGCNNGP & $0.558$ (0.001) & $0.607$ (0.002) & $0.637$ (0.002) \\
  BGC & $0.564$ (0.000) & $0.565$ (0.002) & $0.582$ (0.003) \\
  \hline
  \multicolumn{4}{l}{$n=500$} \\
  spBGC & $0.192$ (0.000) & $0.227$ (0.001) & $0.260$ (0.002) \\
  spBGCNNGP & $0.192$ (0.000) & $0.227$ (0.001) & $0.260$ (0.002) \\
  BGC & $0.183$ (0.000) & $0.183$ (0.000) & $0.194$ (0.002) \\
  \hline
  \multicolumn{4}{l}{$n=1000$} \\
  spBGC & $0.138$ (0.000) & $0.166$ (0.002) & $0.199$ (0.009) \\
  spBGCNNGP & $0.140$ (0.000) & $0.169$ (0.000) & $0.194$ (0.001) \\
  BGC & $0.129$ (0.000) & $0.128$ (0.000) & $0.135$ (0.001) \\
  \hline
  \end{tabularx}
\end{minipage}
\hfill
\begin{minipage}{.48\linewidth}
  \centering
  \subcaption{Number of outcomes $p=9$}
  \begin{tabularx}{\linewidth}{l *{3}{>{\centering\arraybackslash}X}}
  \hline
  & \multicolumn{3}{c}{$\phi$} \\
  \cline{2-4}
   & 0.05 & 0.25 & 0.50 \\
  \hline
  \multicolumn{4}{l}{$n=50$} \\
  spBGC & $0.537$ (0.000) & $0.565$ (0.001) & $0.586$ (0.001) \\
  spBGCNNGP & $0.537$ (0.000) & $0.565$ (0.001) & $0.587$ (0.001) \\
  BGC & $0.558$ (0.000) & $0.553$ (0.001) & $0.558$ (0.002) \\
  \hline
  \multicolumn{4}{l}{$n=500$} \\
  spBGC & $0.189$ (0.000) & $0.212$ (0.000) & $0.229$ (0.001) \\
  spBGCNNGP & $0.189$ (0.000) & $0.212$ (0.000) & $0.228$ (0.001) \\
  BGC & $0.183$ (0.000) & $0.181$ (0.000) & $0.184$ (0.001) \\
  \hline
  \multicolumn{4}{l}{$n=1000$} \\
  spBGC & $0.136$ (0.000) & $0.155$ (0.002) & $0.172$ (0.006) \\
  spBGCNNGP & $0.137$ (0.000) & $0.157$ (0.000) & $0.168$ (0.001) \\
  BGC & $0.130$ (0.000) & $0.128$ (0.000) & $0.129$ (0.001) \\
  \hline
  \end{tabularx}
\end{minipage}
\end{table}

\begin{figure}[H]
\centering
\includegraphics[width=\columnwidth]{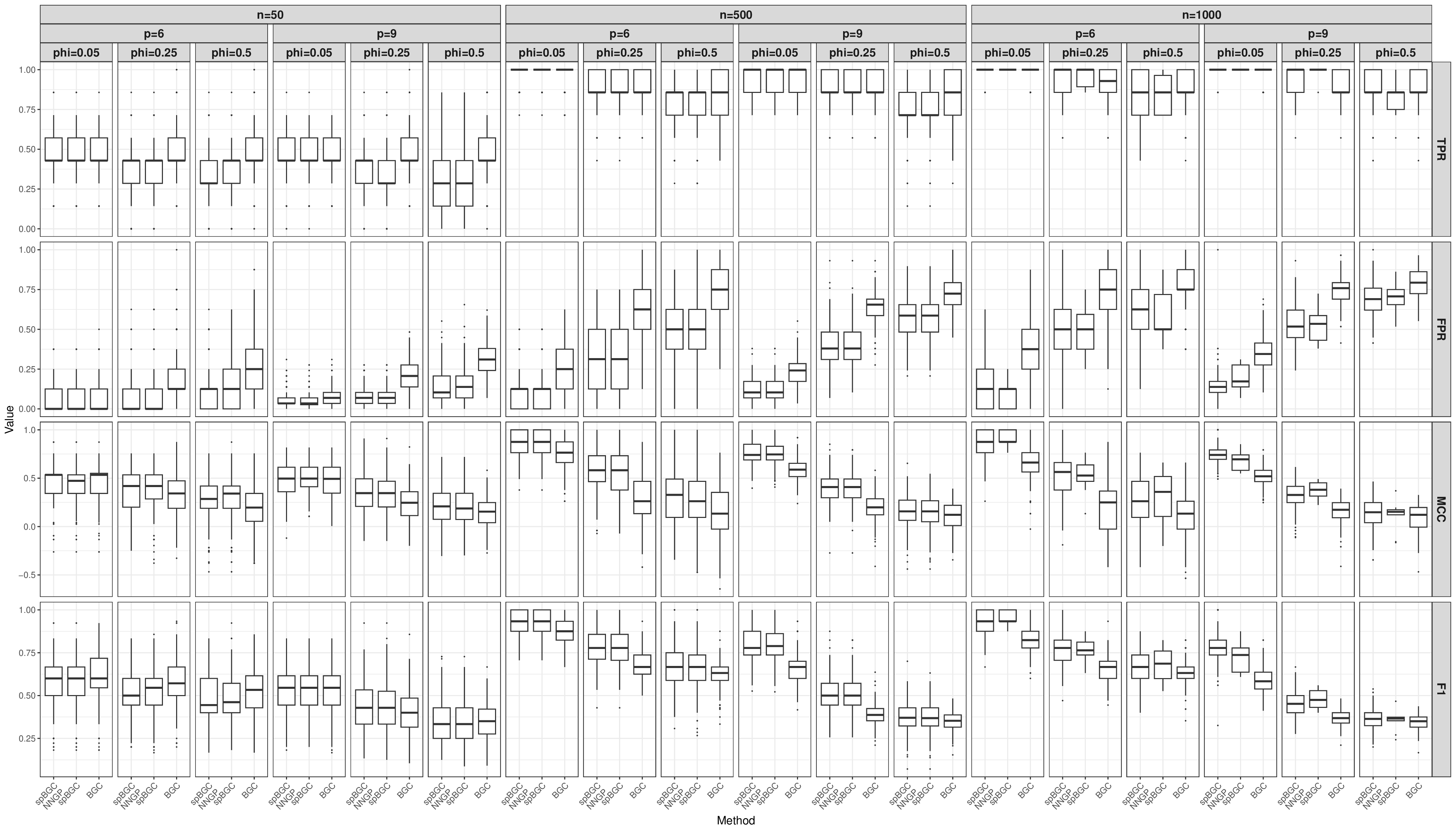}
\caption{Edge selection performance under $t$-Copula (df=5): TPR, FPR, MCC, and F1 score across different sample sizes and spatial range parameters, demonstrating robustness to copula misspecification.}
\label{fig:tcopula_edge_metrics}
\end{figure}

Tables~\ref{tb:edge_metrics_t_copula_part1}--\ref{tb:edge_metrics_t_copula_part2} and Figure~\ref{fig:tcopula_edge_metrics} present the edge selection results under $t$-copula.
For edge selection, spBGC and spBGCNNGP maintain substantially better performance than BGC despite the copula misspecification.
For small $\phi$ values, TPR remains high and FPR stays low (around 0.10--0.15); as $\phi$ increases, FPR rises to moderate-to-high levels (around 0.50--0.60 at $\phi = 0.50$) and MCC decreases, reflecting the challenging nature of edge selection under tail misspecification combined with strong spatial correlation.
Nevertheless, spBGC and spBGCNNGP still achieve substantially lower FPR and higher MCC than BGC across all scenarios.
These results demonstrate that while copula misspecification does impact edge selection performance, incorporating spatial structure in the copula framework still provides substantial relative improvements over ignoring spatial dependence entirely.

\begin{table}[H]
\centering
\small
\caption{Edge selection metrics (TPR, FPR, MCC, F1) for spBGC, spBGCNNGP, and BGC under $t$-copula (df=5) with exponential isotropic kernel for $n=50$ and $n=500$.
The values represent averages from 300 calculations with standard errors in parentheses.}
\label{tb:edge_metrics_t_copula_part1}
\begin{minipage}{.48\linewidth}
  \centering
  \subcaption{Number of outcomes $p=6$}
  \begin{tabularx}{\linewidth}{l *{4}{>{\centering\arraybackslash}X}}
  \hline
  & TPR & FPR & MCC & F1 \\
  \hline
  \multicolumn{5}{l}{$n=50$, $\phi=0.05$} \\
  spBGC & $0.456$ (0.008) & $0.066$ (0.005) & $0.460$ (0.010) & $0.587$ (0.007) \\
  spBGCNNGP & $0.451$ (0.008) & $0.060$ (0.005) & $0.465$ (0.010) & $0.584$ (0.008) \\
  BGC & $0.486$ (0.008) & $0.081$ (0.006) & $0.465$ (0.010) & $0.609$ (0.007) \\
  \multicolumn{5}{l}{$n=50$, $\phi=0.25$} \\
  spBGC & $0.374$ (0.009) & $0.085$ (0.007) & $0.369$ (0.011) & $0.507$ (0.009) \\
  spBGCNNGP & $0.366$ (0.009) & $0.086$ (0.007) & $0.361$ (0.011) & $0.501$ (0.008) \\
  BGC & $0.507$ (0.009) & $0.209$ (0.010) & $0.330$ (0.013) & $0.575$ (0.008) \\
  \multicolumn{5}{l}{$n=50$, $\phi=0.50$} \\
  spBGC & $0.357$ (0.009) & $0.130$ (0.008) & $0.280$ (0.014) & $0.478$ (0.009) \\
  spBGCNNGP & $0.345$ (0.009) & $0.130$ (0.008) & $0.267$ (0.014) & $0.465$ (0.009) \\
  BGC & $0.482$ (0.010) & $0.292$ (0.010) & $0.204$ (0.014) & $0.525$ (0.008) \\
  \hline
  \multicolumn{5}{l}{$n=500$, $\phi=0.05$} \\
  spBGC & $0.972$ (0.003) & $0.112$ (0.007) & $0.864$ (0.008) & $0.928$ (0.004) \\
  spBGCNNGP & $0.971$ (0.003) & $0.111$ (0.007) & $0.865$ (0.008) & $0.929$ (0.004) \\
  BGC & $0.963$ (0.004) & $0.251$ (0.009) & $0.729$ (0.010) & $0.861$ (0.005) \\
  \multicolumn{5}{l}{$n=500$, $\phi=0.25$} \\
  spBGC & $0.871$ (0.006) & $0.323$ (0.011) & $0.564$ (0.013) & $0.782$ (0.006) \\
  spBGCNNGP & $0.872$ (0.006) & $0.325$ (0.011) & $0.563$ (0.012) & $0.781$ (0.006) \\
  BGC & $0.877$ (0.006) & $0.617$ (0.011) & $0.297$ (0.014) & $0.682$ (0.005) \\
  \multicolumn{5}{l}{$n=500$, $\phi=0.50$} \\
  spBGC & $0.781$ (0.008) & $0.508$ (0.010) & $0.287$ (0.015) & $0.661$ (0.007) \\
  spBGCNNGP & $0.784$ (0.008) & $0.507$ (0.010) & $0.292$ (0.015) & $0.663$ (0.006) \\
  BGC & $0.836$ (0.008) & $0.715$ (0.010) & $0.147$ (0.014) & $0.629$ (0.005) \\
  \hline
  \end{tabularx}
\end{minipage}
\hfill
\begin{minipage}{.48\linewidth}
  \centering
  \subcaption{Number of outcomes $p=9$}
  \begin{tabularx}{\linewidth}{l *{4}{>{\centering\arraybackslash}X}}
  \hline
  & TPR & FPR & MCC & F1 \\
  \hline
  \multicolumn{5}{l}{$n=50$, $\phi=0.05$} \\
  spBGC & $0.457$ (0.008) & $0.054$ (0.003) & $0.484$ (0.009) & $0.539$ (0.008) \\
  spBGCNNGP & $0.454$ (0.009) & $0.054$ (0.003) & $0.481$ (0.009) & $0.537$ (0.008) \\
  BGC & $0.472$ (0.008) & $0.071$ (0.003) & $0.459$ (0.009) & $0.532$ (0.008) \\
  \multicolumn{5}{l}{$n=50$, $\phi=0.25$} \\
  spBGC & $0.351$ (0.009) & $0.080$ (0.004) & $0.332$ (0.011) & $0.424$ (0.008) \\
  spBGCNNGP & $0.355$ (0.009) & $0.078$ (0.004) & $0.340$ (0.011) & $0.427$ (0.008) \\
  BGC & $0.478$ (0.009) & $0.219$ (0.006) & $0.243$ (0.010) & $0.404$ (0.007) \\
  \multicolumn{5}{l}{$n=50$, $\phi=0.50$} \\
  spBGC & $0.308$ (0.009) & $0.143$ (0.006) & $0.189$ (0.011) & $0.344$ (0.008) \\
  spBGCNNGP & $0.314$ (0.009) & $0.139$ (0.006) & $0.202$ (0.011) & $0.352$ (0.008) \\
  BGC & $0.473$ (0.010) & $0.312$ (0.006) & $0.139$ (0.009) & $0.343$ (0.006) \\
  \hline
  \multicolumn{5}{l}{$n=500$, $\phi=0.05$} \\
  spBGC & $0.959$ (0.004) & $0.121$ (0.004) & $0.747$ (0.006) & $0.788$ (0.005) \\
  spBGCNNGP & $0.962$ (0.004) & $0.121$ (0.004) & $0.747$ (0.006) & $0.788$ (0.005) \\
  BGC & $0.951$ (0.004) & $0.239$ (0.005) & $0.592$ (0.007) & $0.656$ (0.005) \\
  \multicolumn{5}{l}{$n=500$, $\phi=0.25$} \\
  spBGC & $0.870$ (0.006) & $0.383$ (0.008) & $0.395$ (0.009) & $0.515$ (0.006) \\
  spBGCNNGP & $0.873$ (0.006) & $0.387$ (0.008) & $0.394$ (0.009) & $0.514$ (0.006) \\
  BGC & $0.869$ (0.006) & $0.637$ (0.006) & $0.197$ (0.008) & $0.388$ (0.003) \\
  \multicolumn{5}{l}{$n=500$, $\phi=0.50$} \\
  spBGC & $0.763$ (0.008) & $0.577$ (0.008) & $0.151$ (0.009) & $0.372$ (0.005) \\
  spBGCNNGP & $0.772$ (0.008) & $0.578$ (0.008) & $0.157$ (0.009) & $0.375$ (0.005) \\
  BGC & $0.846$ (0.007) & $0.723$ (0.006) & $0.109$ (0.008) & $0.350$ (0.003) \\
  \hline
  \end{tabularx}
\end{minipage}
\end{table}

\begin{table}[H]
\centering
\small
\caption{Edge selection metrics (TPR, FPR, MCC, F1) for spBGC, spBGCNNGP, and BGC under $t$-copula (df=5) with exponential isotropic kernel for $n=1000$.
The values represent averages from 300 calculations (10 for spBGC) with standard errors in parentheses.}
\label{tb:edge_metrics_t_copula_part2}
\begin{minipage}{.48\linewidth}
  \centering
  \subcaption{Number of outcomes $p=6$}
  \begin{tabularx}{\linewidth}{l *{4}{>{\centering\arraybackslash}X}}
  \hline
  & TPR & FPR & MCC & F1 \\
  \hline
  \multicolumn{5}{l}{$n=1000$, $\phi=0.05$} \\
  spBGC & $1.000$ (0.000) & $0.087$ (0.027) & $0.914$ (0.026) & $0.954$ (0.014) \\
  spBGCNNGP & $0.998$ (0.001) & $0.150$ (0.008) & $0.856$ (0.007) & $0.924$ (0.004) \\
  BGC & $0.989$ (0.002) & $0.369$ (0.011) & $0.657$ (0.010) & $0.826$ (0.005) \\
  \multicolumn{5}{l}{$n=1000$, $\phi=0.25$} \\
  spBGC & $0.957$ (0.022) & $0.475$ (0.052) & $0.526$ (0.059) & $0.769$ (0.024) \\
  spBGCNNGP & $0.949$ (0.005) & $0.469$ (0.011) & $0.523$ (0.012) & $0.768$ (0.005) \\
  BGC & $0.910$ (0.006) & $0.758$ (0.009) & $0.208$ (0.014) & $0.657$ (0.004) \\
  \multicolumn{5}{l}{$n=1000$, $\phi=0.50$} \\
  spBGC & $0.843$ (0.040) & $0.575$ (0.053) & $0.287$ (0.089) & $0.676$ (0.032) \\
  spBGCNNGP & $0.852$ (0.007) & $0.623$ (0.011) & $0.260$ (0.014) & $0.666$ (0.005) \\
  BGC & $0.884$ (0.006) & $0.796$ (0.008) & $0.118$ (0.014) & $0.633$ (0.004) \\
  \hline
  \end{tabularx}
\end{minipage}
\hfill
\begin{minipage}{.48\linewidth}
  \centering
  \subcaption{Number of outcomes $p=9$}
  \begin{tabularx}{\linewidth}{l *{4}{>{\centering\arraybackslash}X}}
  \hline
  & TPR & FPR & MCC & F1 \\
  \hline
  \multicolumn{5}{l}{$n=1000$, $\phi=0.05$} \\
  spBGC & $1.000$ (0.000) & $0.190$ (0.028) & $0.687$ (0.035) & $0.730$ (0.031) \\
  spBGCNNGP & $0.998$ (0.001) & $0.146$ (0.005) & $0.744$ (0.006) & $0.779$ (0.005) \\
  BGC & $0.991$ (0.002) & $0.342$ (0.006) & $0.523$ (0.005) & $0.591$ (0.004) \\
  \multicolumn{5}{l}{$n=1000$, $\phi=0.25$} \\
  spBGC & $0.986$ (0.014) & $0.531$ (0.035) & $0.374$ (0.028) & $0.476$ (0.018) \\
  spBGCNNGP & $0.927$ (0.005) & $0.533$ (0.007) & $0.321$ (0.007) & $0.454$ (0.004) \\
  BGC & $0.912$ (0.005) & $0.744$ (0.005) & $0.158$ (0.007) & $0.367$ (0.002) \\
  \multicolumn{5}{l}{$n=1000$, $\phi=0.50$} \\
  spBGC & $0.829$ (0.042) & $0.697$ (0.034) & $0.114$ (0.051) & $0.354$ (0.019) \\
  spBGCNNGP & $0.853$ (0.007) & $0.694$ (0.006) & $0.137$ (0.009) & $0.364$ (0.003) \\
  BGC & $0.894$ (0.006) & $0.794$ (0.005) & $0.100$ (0.008) & $0.345$ (0.002) \\
  \hline
  \end{tabularx}
\end{minipage}
\end{table}

\subsection{Sensitivity to the Structure and Magnitude of the Correlation Matrix \texorpdfstring{$\bm{R}$}{R}}
\label{sec:R_structure}

To examine how the estimator behaves when the sparse Section~4 correlation matrix is changed, we repeat the simulation of Section~4, using the same Gaussian copula, exponential-isotropic spatial kernel, mixed marginals, and $(n,p,\phi)$ grid, and altering only the true correlation matrix $\bm{R}$.
We consider (i) a dense matrix in which every off-diagonal entry is nonzero, and (ii) a small-magnitude matrix in which the nonzero entries are shrunk toward zero.
We report spBGCNNGP (with $m=n/10$) and BGC.
As in Section~4 we evaluate the MSE, coverage probability (CP) and average $95\%$ credible-interval length (AL) of the posterior-median estimates, additionally split by the truly nonzero and truly zero entries of $\bm{R}$, and we report edge-selection metrics (TPR, FPR, MCC, F1), an ``edge'' being declared when the $95\%$ credible interval of $\bm{R}_{jj'}$ excludes zero.

\subsubsection{Dense correlation matrix}
\label{sec:R_dense}

\begin{figure}[H]
\centering
\includegraphics[width=\columnwidth]{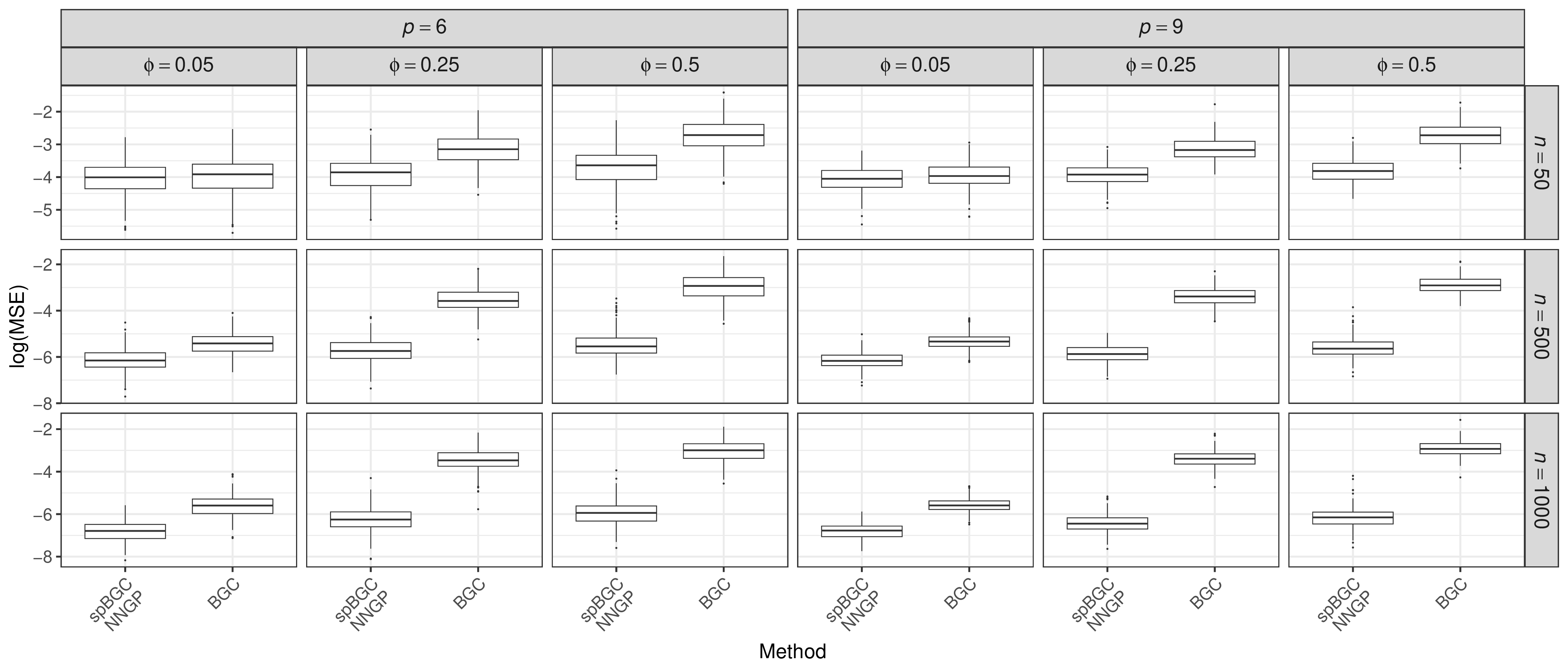}
\caption{Comparisons under a dense correlation matrix $\bm{R}$: logarithm of MSEs.}
\label{fig:A5_dense_mse}
\end{figure}

\begin{figure}[H]
\centering
\includegraphics[width=\columnwidth]{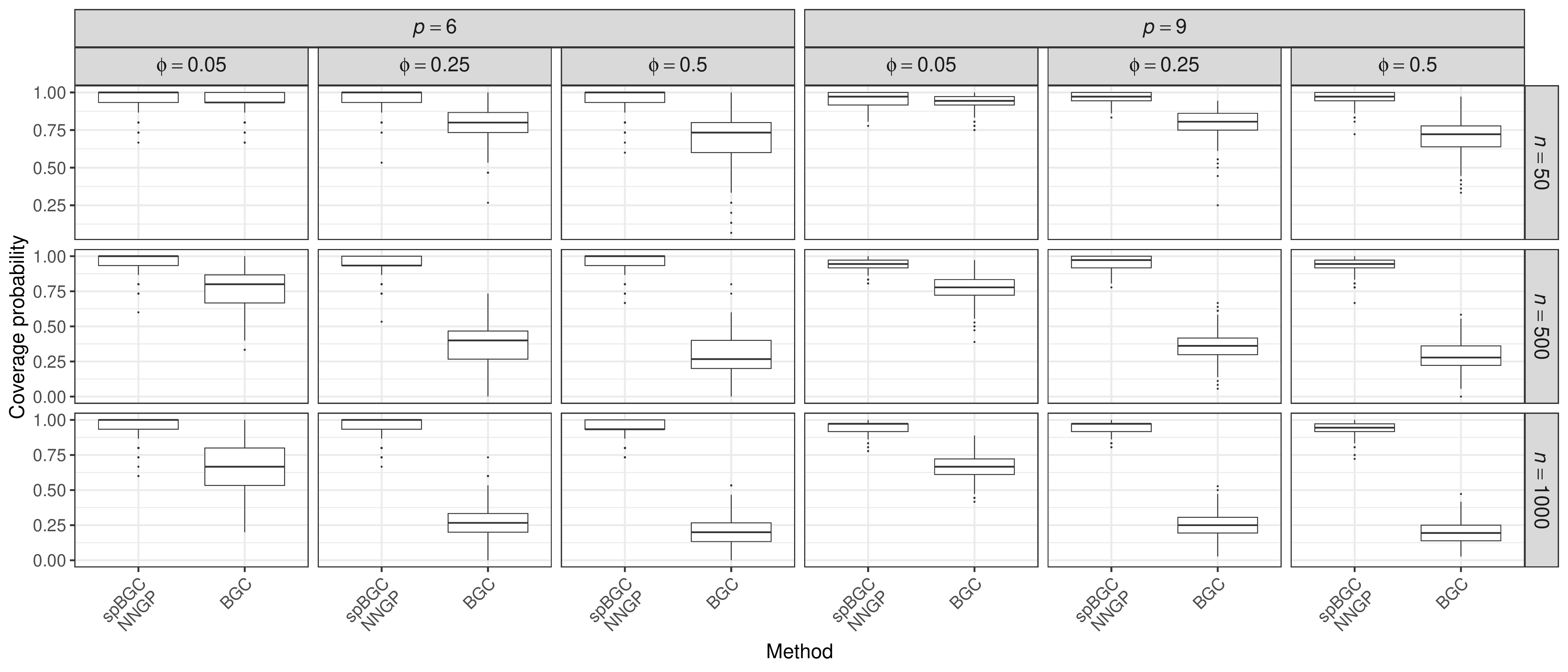}
\caption{Comparisons under a dense correlation matrix $\bm{R}$: coverage probabilities.}
\label{fig:A5_dense_cp}
\end{figure}

\begin{figure}[H]
\centering
\includegraphics[width=\columnwidth]{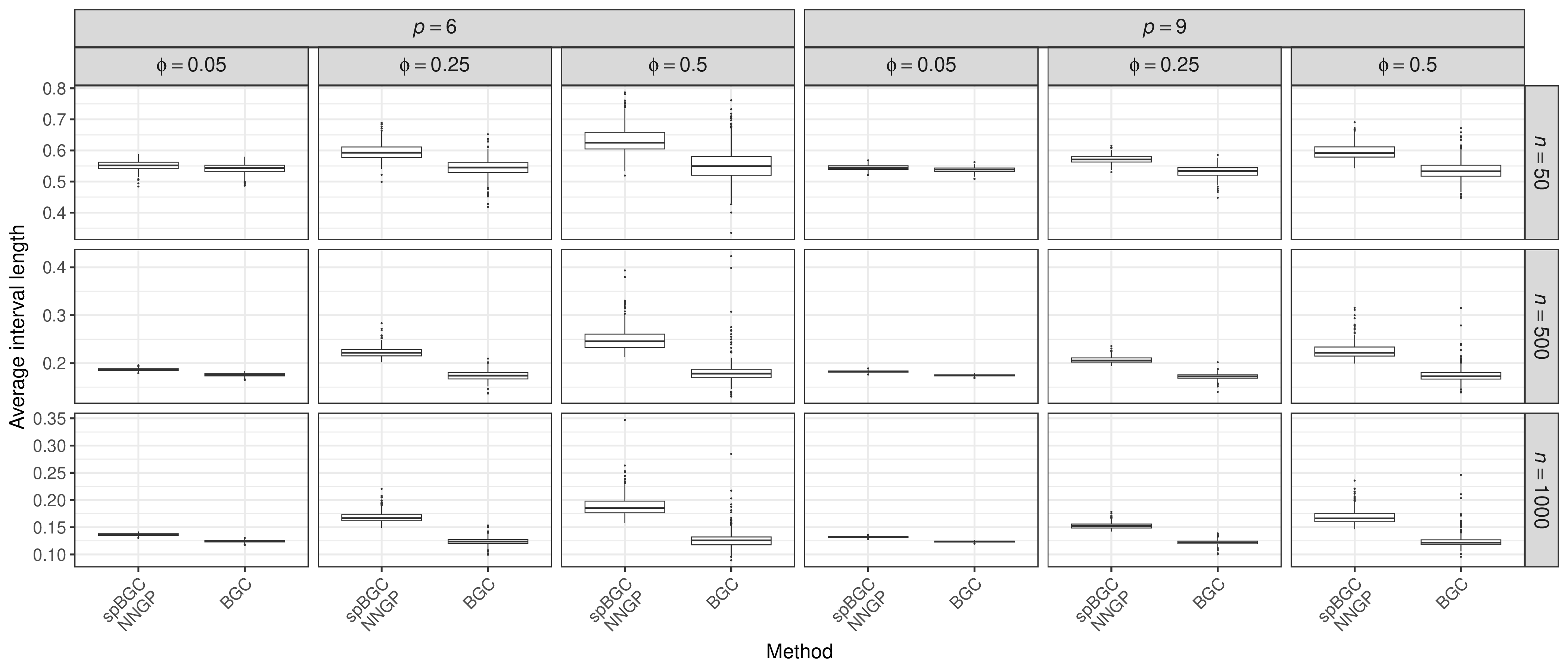}
\caption{Comparisons under a dense correlation matrix $\bm{R}$: average credible interval lengths.}
\label{fig:A5_dense_avl}
\end{figure}

We take $\bm{R}$ to be the Kac--Murdock--Szeg\H{o} matrix $\bm{R}_{jk}=0.5^{|j-k|}$, a valid positive-definite correlation matrix in which every off-diagonal entry is nonzero and the magnitudes span a wide range (there are no zero entries, so the zero/nonzero split and the false-positive rate are not defined here).
Tables~\ref{tb:A5_dense_mse}--\ref{tb:A5_dense_al} report the MSE, coverage and interval length, and Figures~\ref{fig:A5_dense_mse}--\ref{fig:A5_dense_avl} show the corresponding boxplots.
spBGCNNGP recovers the dense matrix with error and coverage of the same order as under the sparse Section~4 design. 
For example, at $n=500$, $p=6$, $\phi=0.25$ the MSE is $\approx 0.004$ and the coverage $\approx 0.94$, essentially unchanged from the sparse case, confirming that density by itself does not degrade estimation.
BGC again undercovers severely (coverage as low as $\approx 0.20$), because it treats the spatially dependent sites as independent replicates.

\begin{table}[H]
\centering
\caption{Comparisons of the logarithm of the MSEs for spBGCNNGP (with $m=n/10$) and BGC under a dense correlation matrix $\bm{R}$ (Kac--Murdock--Szeg\H{o}, $\bm{R}_{jk}=0.5^{|j-k|}$). The values represent average log(MSE)s from 300 calculations, with standard errors in parentheses.}
\label{tb:A5_dense_mse}
\begin{minipage}{.48\linewidth}
  \centering
  \subcaption{Number of outcomes $p=6$}
  \begin{tabularx}{\linewidth}{l *{3}{>{\centering\arraybackslash}X}}
  \hline
  & \multicolumn{3}{c}{$\phi$} \\
  \cline{2-4}
   & 0.05 & 0.25 & 0.50 \\
  \hline
  \multicolumn{4}{l}{$n=50$} \\
  spBGCNNGP & $-3.926$ (0.028) & $-3.785$ (0.030) & $-3.540$ (0.035) \\
  BGC & $-3.857$ (0.028) & $-3.042$ (0.028) & $-2.618$ (0.029) \\
  \hline
  \multicolumn{4}{l}{$n=500$} \\
  spBGCNNGP & $-6.013$ (0.031) & $-5.590$ (0.032) & $-5.324$ (0.042) \\
  BGC & $-5.295$ (0.030) & $-3.424$ (0.030) & $-2.840$ (0.030) \\
  \hline
  \multicolumn{4}{l}{$n=1000$} \\
  spBGCNNGP & $-6.680$ (0.030) & $-6.120$ (0.035) & $-5.788$ (0.040) \\
  BGC & $-5.497$ (0.031) & $-3.338$ (0.030) & $-2.908$ (0.029) \\
  \hline
  \end{tabularx}
\end{minipage}
\hfill
\begin{minipage}{.48\linewidth}
  \centering
  \subcaption{Number of outcomes $p=9$}
  \begin{tabularx}{\linewidth}{l *{3}{>{\centering\arraybackslash}X}}
  \hline
  & \multicolumn{3}{c}{$\phi$} \\
  \cline{2-4}
   & 0.05 & 0.25 & 0.50 \\
  \hline
  \multicolumn{4}{l}{$n=50$} \\
  spBGCNNGP & $-3.994$ (0.021) & $-3.893$ (0.018) & $-3.752$ (0.021) \\
  BGC & $-3.887$ (0.022) & $-3.085$ (0.022) & $-2.669$ (0.021) \\
  \hline
  \multicolumn{4}{l}{$n=500$} \\
  spBGCNNGP & $-6.109$ (0.019) & $-5.794$ (0.022) & $-5.509$ (0.030) \\
  BGC & $-5.274$ (0.022) & $-3.330$ (0.022) & $-2.830$ (0.022) \\
  \hline
  \multicolumn{4}{l}{$n=1000$} \\
  spBGCNNGP & $-6.741$ (0.020) & $-6.366$ (0.025) & $-6.035$ (0.034) \\
  BGC & $-5.519$ (0.020) & $-3.338$ (0.023) & $-2.852$ (0.022) \\
  \hline
  \end{tabularx}
\end{minipage}
\end{table}

\begin{table}[H]
\centering
\caption{Comparisons of the coverage probabilities for spBGCNNGP (with $m=n/10$) and BGC under a dense correlation matrix $\bm{R}$ (Kac--Murdock--Szeg\H{o}, $\bm{R}_{jk}=0.5^{|j-k|}$). The values represent average coverage probabilities from 300 calculations, with standard errors in parentheses.}
\label{tb:A5_dense_cp}
\begin{minipage}{.48\linewidth}
  \centering
  \subcaption{Number of outcomes $p=6$}
  \begin{tabularx}{\linewidth}{l *{3}{>{\centering\arraybackslash}X}}
  \hline
  & \multicolumn{3}{c}{$\phi$} \\
  \cline{2-4}
   & 0.05 & 0.25 & 0.50 \\
  \hline
  \multicolumn{4}{l}{$n=50$} \\
  spBGCNNGP & $0.954$ (0.004) & $0.960$ (0.003) & $0.952$ (0.004) \\
  BGC & $0.942$ (0.004) & $0.809$ (0.007) & $0.707$ (0.009) \\
  \hline
  \multicolumn{4}{l}{$n=500$} \\
  spBGCNNGP & $0.952$ (0.004) & $0.942$ (0.004) & $0.951$ (0.004) \\
  BGC & $0.787$ (0.008) & $0.368$ (0.008) & $0.294$ (0.008) \\
  \hline
  \multicolumn{4}{l}{$n=1000$} \\
  spBGCNNGP & $0.953$ (0.004) & $0.946$ (0.004) & $0.943$ (0.004) \\
  BGC & $0.669$ (0.009) & $0.263$ (0.008) & $0.220$ (0.007) \\
  \hline
  \end{tabularx}
\end{minipage}
\hfill
\begin{minipage}{.48\linewidth}
  \centering
  \subcaption{Number of outcomes $p=9$}
  \begin{tabularx}{\linewidth}{l *{3}{>{\centering\arraybackslash}X}}
  \hline
  & \multicolumn{3}{c}{$\phi$} \\
  \cline{2-4}
   & 0.05 & 0.25 & 0.50 \\
  \hline
  \multicolumn{4}{l}{$n=50$} \\
  spBGCNNGP & $0.959$ (0.003) & $0.963$ (0.002) & $0.959$ (0.002) \\
  BGC & $0.946$ (0.003) & $0.798$ (0.005) & $0.704$ (0.007) \\
  \hline
  \multicolumn{4}{l}{$n=500$} \\
  spBGCNNGP & $0.950$ (0.002) & $0.952$ (0.003) & $0.942$ (0.003) \\
  BGC & $0.775$ (0.006) & $0.358$ (0.006) & $0.285$ (0.006) \\
  \hline
  \multicolumn{4}{l}{$n=1000$} \\
  spBGCNNGP & $0.950$ (0.002) & $0.951$ (0.002) & $0.941$ (0.003) \\
  BGC & $0.670$ (0.006) & $0.254$ (0.005) & $0.204$ (0.004) \\
  \hline
  \end{tabularx}
\end{minipage}
\end{table}

\begin{table}[H]
\centering
\caption{Comparisons of the average credible interval lengths for spBGCNNGP (with $m=n/10$) and BGC under a dense correlation matrix $\bm{R}$ (Kac--Murdock--Szeg\H{o}, $\bm{R}_{jk}=0.5^{|j-k|}$). The values represent average lengths from 300 calculations, with standard errors in parentheses.}
\label{tb:A5_dense_al}
\begin{minipage}{.48\linewidth}
  \centering
  \subcaption{Number of outcomes $p=6$}
  \begin{tabularx}{\linewidth}{l *{3}{>{\centering\arraybackslash}X}}
  \hline
  & \multicolumn{3}{c}{$\phi$} \\
  \cline{2-4}
   & 0.05 & 0.25 & 0.50 \\
  \hline
  \multicolumn{4}{l}{$n=50$} \\
  spBGCNNGP & $0.551$ (0.001) & $0.596$ (0.002) & $0.633$ (0.003) \\
  BGC & $0.542$ (0.001) & $0.544$ (0.002) & $0.554$ (0.003) \\
  \hline
  \multicolumn{4}{l}{$n=500$} \\
  spBGCNNGP & $0.187$ (0.000) & $0.224$ (0.001) & $0.251$ (0.002) \\
  BGC & $0.176$ (0.000) & $0.174$ (0.001) & $0.182$ (0.002) \\
  \hline
  \multicolumn{4}{l}{$n=1000$} \\
  spBGCNNGP & $0.137$ (0.000) & $0.169$ (0.001) & $0.190$ (0.001) \\
  BGC & $0.124$ (0.000) & $0.124$ (0.000) & $0.127$ (0.001) \\
  \hline
  \end{tabularx}
\end{minipage}
\hfill
\begin{minipage}{.48\linewidth}
  \centering
  \subcaption{Number of outcomes $p=9$}
  \begin{tabularx}{\linewidth}{l *{3}{>{\centering\arraybackslash}X}}
  \hline
  & \multicolumn{3}{c}{$\phi$} \\
  \cline{2-4}
   & 0.05 & 0.25 & 0.50 \\
  \hline
  \multicolumn{4}{l}{$n=50$} \\
  spBGCNNGP & $0.544$ (0.000) & $0.572$ (0.001) & $0.597$ (0.001) \\
  BGC & $0.538$ (0.001) & $0.531$ (0.001) & $0.537$ (0.002) \\
  \hline
  \multicolumn{4}{l}{$n=500$} \\
  spBGCNNGP & $0.182$ (0.000) & $0.207$ (0.000) & $0.227$ (0.001) \\
  BGC & $0.175$ (0.000) & $0.172$ (0.000) & $0.176$ (0.001) \\
  \hline
  \multicolumn{4}{l}{$n=1000$} \\
  spBGCNNGP & $0.132$ (0.000) & $0.153$ (0.000) & $0.170$ (0.001) \\
  BGC & $0.124$ (0.000) & $0.122$ (0.000) & $0.125$ (0.001) \\
  \hline
  \end{tabularx}
\end{minipage}
\end{table}

\subsubsection{Small-magnitude nonzero entries}
\label{sec:R_small}

\begin{figure}[H]
\centering
\includegraphics[width=\columnwidth]{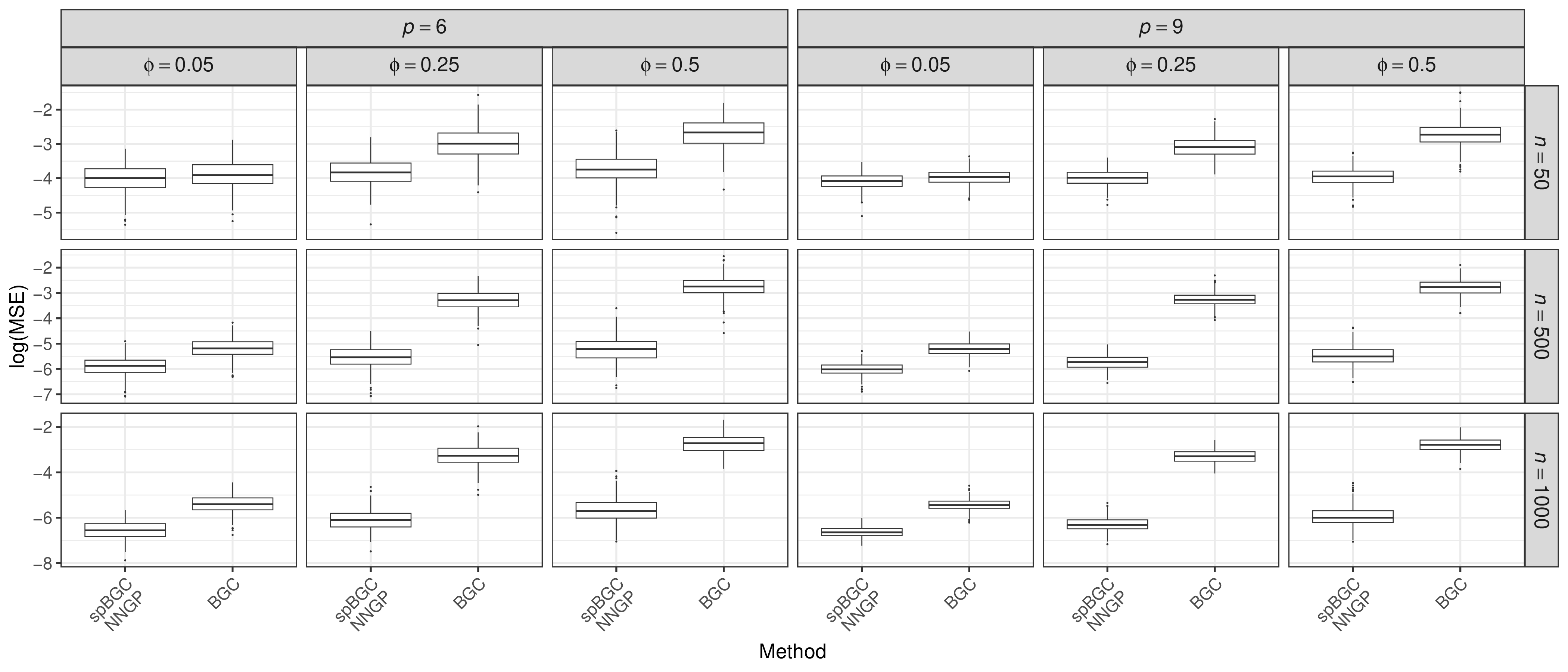}
\caption{Comparisons under a small-magnitude correlation matrix $\bm{R}$: logarithm of MSEs.}
\label{fig:A5_small_mse}
\end{figure}

\begin{figure}[H]
\centering
\includegraphics[width=\columnwidth]{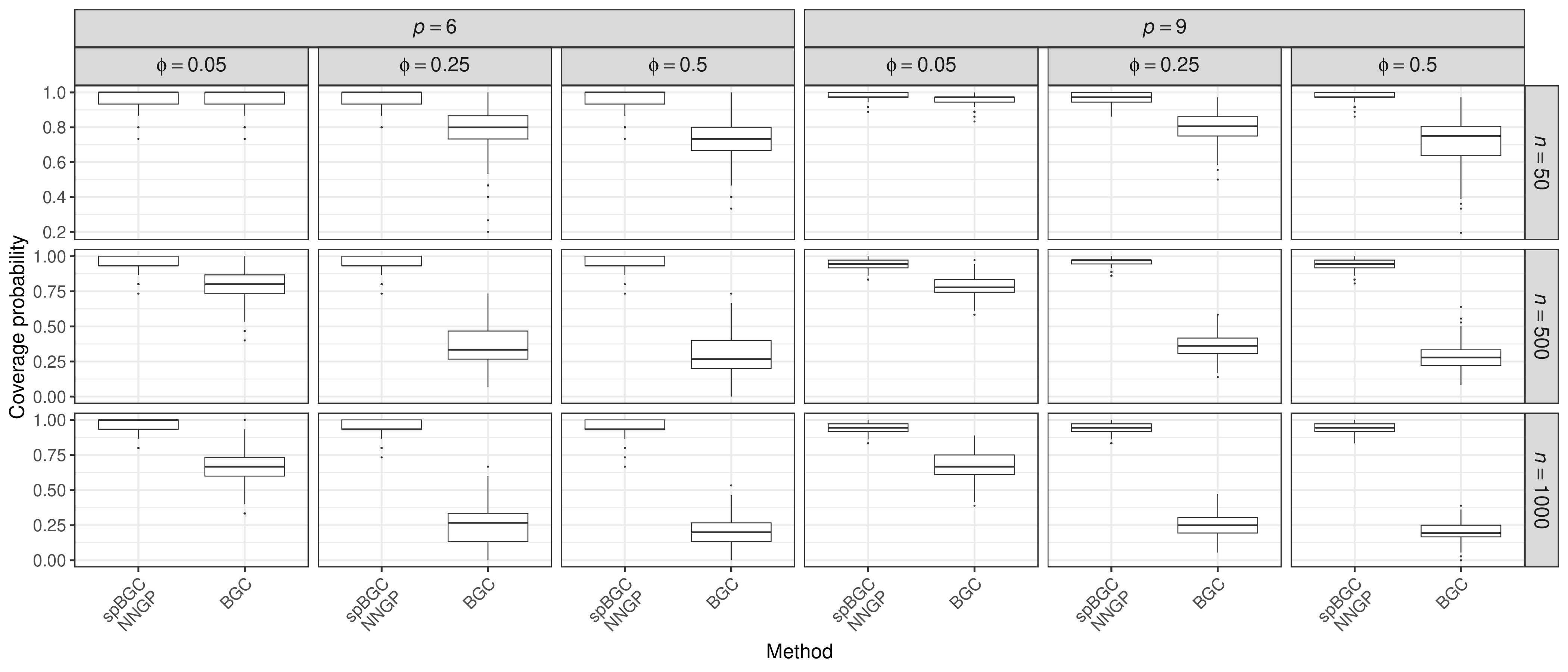}
\caption{Comparisons under a small-magnitude correlation matrix $\bm{R}$: coverage probabilities.}
\label{fig:A5_small_cp}
\end{figure}

\begin{figure}[H]
\centering
\includegraphics[width=\columnwidth]{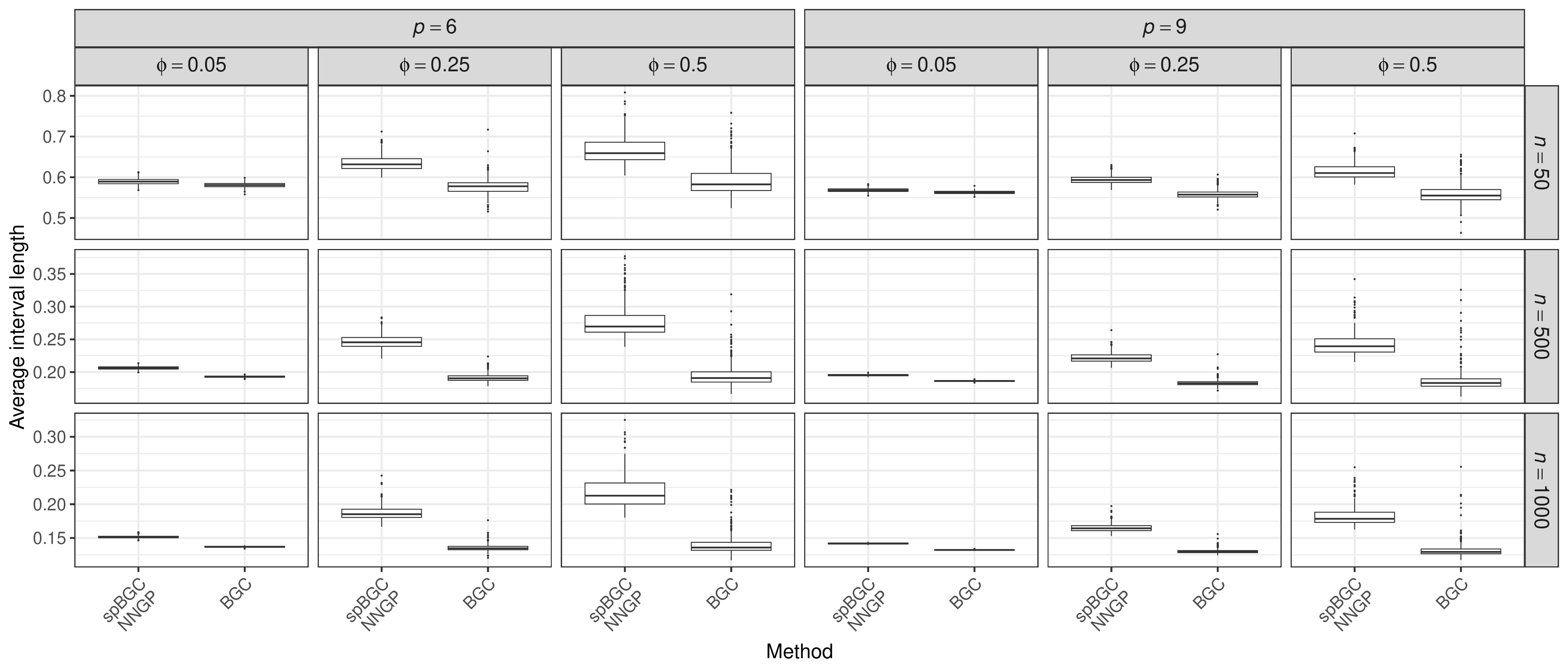}
\caption{Comparisons under a small-magnitude correlation matrix $\bm{R}$: average credible interval lengths.}
\label{fig:A5_small_avl}
\end{figure}

\begin{figure}[H]
\centering
\includegraphics[width=\columnwidth]{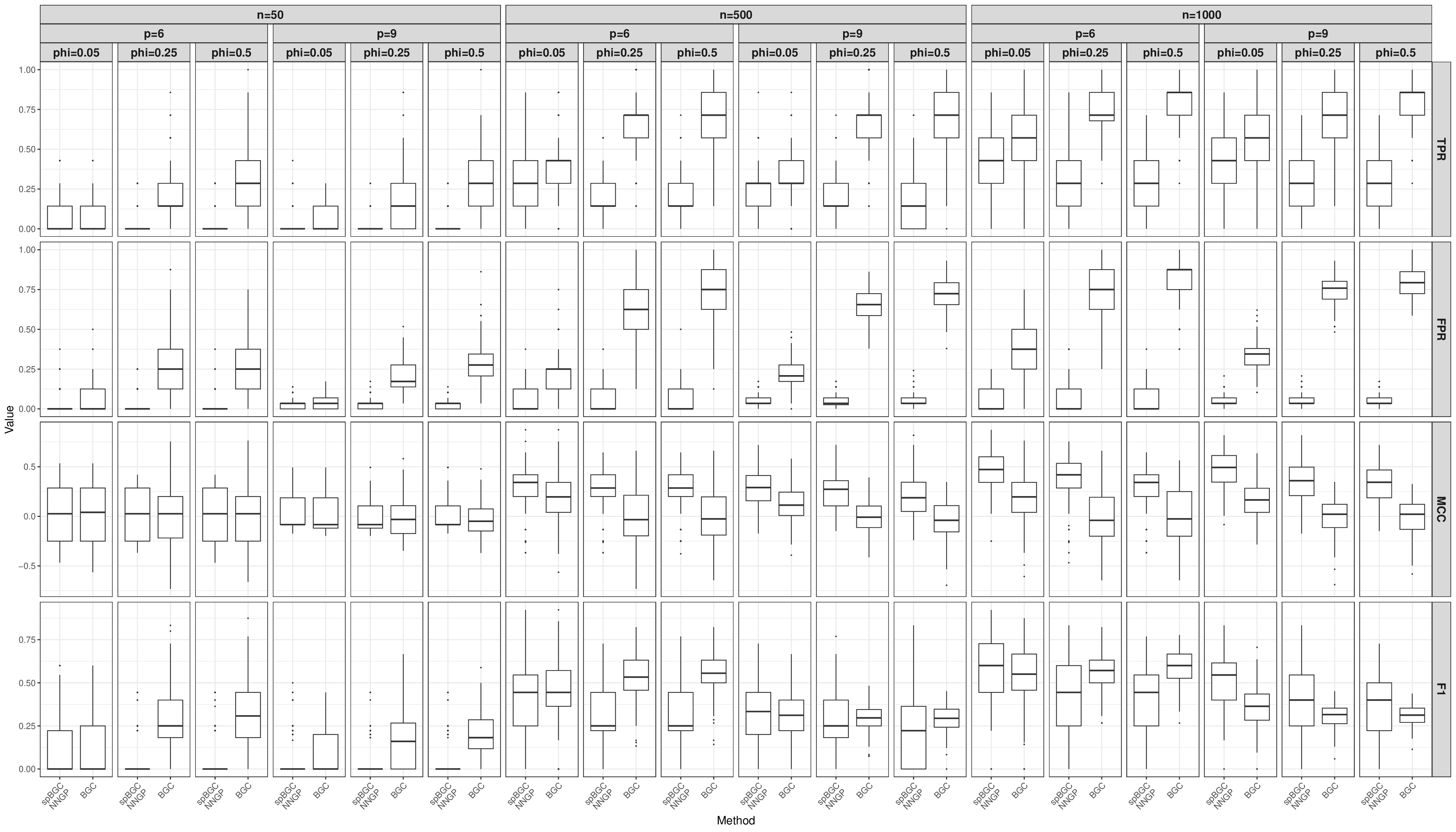}
\caption{Edge selection performance under a small-magnitude correlation matrix $\bm{R}$: TPR, FPR, MCC, and F1 score across different sample sizes and spatial range parameters.}
\label{fig:A5_small_edge}
\end{figure}

We keep the sparse Section~4 pattern of nonzero entries but scale them so that every nonzero correlation satisfies $|\bm{R}_{jj'}|\le 0.1$ (the seven nonzero values become $0.10$, $0.06$, $0.04$, $-0.04$, $-0.06$, $0.08$, and $-0.10$).
Tables~\ref{tb:A5_small_mse_all}--\ref{tb:A5_small_cp_z} report the MSE and coverage split by all, nonzero and zero entries, Table~\ref{tb:A5_small_edge} the edge-selection metrics, and Figures~\ref{fig:A5_small_mse}--\ref{fig:A5_small_edge} show the corresponding boxplots.
The estimator remains accurate and well calibrated on the small nonzero entries. 
At $n=500$, $p=6$, $\phi=0.25$ the MSE over nonzero entries is $\approx 0.005$ with coverage $\approx 0.95$, and the zero entries are likewise recovered near zero.
What changes is the power to declare a small entry nonzero. 
Because the extended rank likelihood carries little information about weak dependence, the credible intervals of small correlations usually contain zero, so the true-positive rate is low (TPR $\approx 0.21$ at $n=500$, against $\approx 0.9$ for the moderate correlations of the sparse design; Table~\ref{tb:edge_metrics_gaussian_copula_exp_iso_part1}), while the false-positive rate stays near the correctly-specified level ($\approx 0.05$).
This is the conservative, well-calibrated behaviour one wants: small true correlations are reported as small, with honest uncertainty, rather than being spuriously amplified. 
By contrast, BGC flags many spurious edges (FPR $\approx 0.6$) by mistaking spatial autocorrelation for cross-outcome dependence.
This is consistent with the near-zero real-data correlation estimates reported in the data application of the main text.

\begin{table}[H]
\centering
\caption{Comparisons of the logarithm of the MSEs (all entries) for spBGCNNGP (with $m=n/10$) and BGC under a small-magnitude correlation matrix $\bm{R}$ (the Section~4 pattern scaled so that every nonzero entry has $|\cdot|\le 0.1$). The values represent average log(MSE)s from 300 calculations, with standard errors in parentheses.}
\label{tb:A5_small_mse_all}
\begin{minipage}{.48\linewidth}
  \centering
  \subcaption{Number of outcomes $p=6$}
  \begin{tabularx}{\linewidth}{l *{3}{>{\centering\arraybackslash}X}}
  \hline
  & \multicolumn{3}{c}{$\phi$} \\
  \cline{2-4}
   & 0.05 & 0.25 & 0.50 \\
  \hline
  \multicolumn{4}{l}{$n=50$} \\
  spBGCNNGP & $-3.926$ (0.022) & $-3.739$ (0.024) & $-3.660$ (0.025) \\
  BGC & $-3.827$ (0.023) & $-2.882$ (0.027) & $-2.623$ (0.022) \\
  \hline
  \multicolumn{4}{l}{$n=500$} \\
  spBGCNNGP & $-5.850$ (0.022) & $-5.435$ (0.026) & $-5.136$ (0.031) \\
  BGC & $-5.140$ (0.021) & $-3.211$ (0.022) & $-2.662$ (0.025) \\
  \hline
  \multicolumn{4}{l}{$n=1000$} \\
  spBGCNNGP & $-6.488$ (0.022) & $-6.011$ (0.028) & $-5.525$ (0.037) \\
  BGC & $-5.325$ (0.023) & $-3.160$ (0.026) & $-2.665$ (0.024) \\
  \hline
  \end{tabularx}
\end{minipage}
\hfill
\begin{minipage}{.48\linewidth}
  \centering
  \subcaption{Number of outcomes $p=9$}
  \begin{tabularx}{\linewidth}{l *{3}{>{\centering\arraybackslash}X}}
  \hline
  & \multicolumn{3}{c}{$\phi$} \\
  \cline{2-4}
   & 0.05 & 0.25 & 0.50 \\
  \hline
  \multicolumn{4}{l}{$n=50$} \\
  spBGCNNGP & $-4.060$ (0.014) & $-3.952$ (0.014) & $-3.919$ (0.015) \\
  BGC & $-3.942$ (0.013) & $-3.052$ (0.017) & $-2.673$ (0.021) \\
  \hline
  \multicolumn{4}{l}{$n=500$} \\
  spBGCNNGP & $-5.989$ (0.014) & $-5.709$ (0.015) & $-5.410$ (0.022) \\
  BGC & $-5.175$ (0.015) & $-3.207$ (0.018) & $-2.738$ (0.018) \\
  \hline
  \multicolumn{4}{l}{$n=1000$} \\
  spBGCNNGP & $-6.608$ (0.014) & $-6.238$ (0.019) & $-5.868$ (0.028) \\
  BGC & $-5.405$ (0.015) & $-3.244$ (0.018) & $-2.742$ (0.017) \\
  \hline
  \end{tabularx}
\end{minipage}
\end{table}

\begin{table}[H]
\centering
\caption{Comparisons of the logarithm of the MSEs (nonzero entries) for spBGCNNGP (with $m=n/10$) and BGC under a small-magnitude correlation matrix $\bm{R}$ (the Section~4 pattern scaled so that every nonzero entry has $|\cdot|\le 0.1$). The values represent average log(MSE)s from 300 calculations, with standard errors in parentheses.}
\label{tb:A5_small_mse_nz}
\begin{minipage}{.48\linewidth}
  \centering
  \subcaption{Number of outcomes $p=6$}
  \begin{tabularx}{\linewidth}{l *{3}{>{\centering\arraybackslash}X}}
  \hline
  & \multicolumn{3}{c}{$\phi$} \\
  \cline{2-4}
   & 0.05 & 0.25 & 0.50 \\
  \hline
  \multicolumn{4}{l}{$n=50$} \\
  spBGCNNGP & $-3.929$ (0.033) & $-3.692$ (0.033) & $-3.551$ (0.033) \\
  BGC & $-3.801$ (0.033) & $-2.873$ (0.032) & $-2.592$ (0.030) \\
  \hline
  \multicolumn{4}{l}{$n=500$} \\
  spBGCNNGP & $-5.770$ (0.034) & $-5.267$ (0.036) & $-4.947$ (0.044) \\
  BGC & $-5.125$ (0.032) & $-3.187$ (0.032) & $-2.629$ (0.032) \\
  \hline
  \multicolumn{4}{l}{$n=1000$} \\
  spBGCNNGP & $-6.414$ (0.031) & $-5.870$ (0.041) & $-5.280$ (0.052) \\
  BGC & $-5.382$ (0.030) & $-3.124$ (0.035) & $-2.655$ (0.031) \\
  \hline
  \end{tabularx}
\end{minipage}
\hfill
\begin{minipage}{.48\linewidth}
  \centering
  \subcaption{Number of outcomes $p=9$}
  \begin{tabularx}{\linewidth}{l *{3}{>{\centering\arraybackslash}X}}
  \hline
  & \multicolumn{3}{c}{$\phi$} \\
  \cline{2-4}
   & 0.05 & 0.25 & 0.50 \\
  \hline
  \multicolumn{4}{l}{$n=50$} \\
  spBGCNNGP & $-3.960$ (0.030) & $-3.830$ (0.031) & $-3.723$ (0.032) \\
  BGC & $-3.857$ (0.028) & $-3.060$ (0.034) & $-2.630$ (0.033) \\
  \hline
  \multicolumn{4}{l}{$n=500$} \\
  spBGCNNGP & $-5.787$ (0.032) & $-5.352$ (0.033) & $-4.990$ (0.040) \\
  BGC & $-5.114$ (0.031) & $-3.174$ (0.034) & $-2.713$ (0.032) \\
  \hline
  \multicolumn{4}{l}{$n=1000$} \\
  spBGCNNGP & $-6.376$ (0.033) & $-5.816$ (0.037) & $-5.343$ (0.052) \\
  BGC & $-5.419$ (0.033) & $-3.145$ (0.032) & $-2.697$ (0.033) \\
  \hline
  \end{tabularx}
\end{minipage}
\end{table}

\begin{table}[H]
\centering
\caption{Comparisons of the logarithm of the MSEs (zero entries) for spBGCNNGP (with $m=n/10$) and BGC under a small-magnitude correlation matrix $\bm{R}$ (the Section~4 pattern scaled so that every nonzero entry has $|\cdot|\le 0.1$). The values represent average log(MSE)s from 300 calculations, with standard errors in parentheses.}
\label{tb:A5_small_mse_z}
\begin{minipage}{.48\linewidth}
  \centering
  \subcaption{Number of outcomes $p=6$}
  \begin{tabularx}{\linewidth}{l *{3}{>{\centering\arraybackslash}X}}
  \hline
  & \multicolumn{3}{c}{$\phi$} \\
  \cline{2-4}
   & 0.05 & 0.25 & 0.50 \\
  \hline
  \multicolumn{4}{l}{$n=50$} \\
  spBGCNNGP & $-3.924$ (0.030) & $-3.781$ (0.030) & $-3.767$ (0.033) \\
  BGC & $-3.851$ (0.031) & $-2.889$ (0.035) & $-2.652$ (0.029) \\
  \hline
  \multicolumn{4}{l}{$n=500$} \\
  spBGCNNGP & $-5.926$ (0.029) & $-5.610$ (0.032) & $-5.337$ (0.038) \\
  BGC & $-5.153$ (0.027) & $-3.231$ (0.027) & $-2.691$ (0.032) \\
  \hline
  \multicolumn{4}{l}{$n=1000$} \\
  spBGCNNGP & $-6.558$ (0.030) & $-6.154$ (0.033) & $-5.804$ (0.039) \\
  BGC & $-5.278$ (0.031) & $-3.193$ (0.030) & $-2.673$ (0.030) \\
  \hline
  \end{tabularx}
\end{minipage}
\hfill
\begin{minipage}{.48\linewidth}
  \centering
  \subcaption{Number of outcomes $p=9$}
  \begin{tabularx}{\linewidth}{l *{3}{>{\centering\arraybackslash}X}}
  \hline
  & \multicolumn{3}{c}{$\phi$} \\
  \cline{2-4}
   & 0.05 & 0.25 & 0.50 \\
  \hline
  \multicolumn{4}{l}{$n=50$} \\
  spBGCNNGP & $-4.085$ (0.015) & $-3.984$ (0.016) & $-3.973$ (0.017) \\
  BGC & $-3.964$ (0.015) & $-3.051$ (0.018) & $-2.684$ (0.022) \\
  \hline
  \multicolumn{4}{l}{$n=500$} \\
  spBGCNNGP & $-6.044$ (0.015) & $-5.818$ (0.017) & $-5.544$ (0.024) \\
  BGC & $-5.191$ (0.016) & $-3.215$ (0.018) & $-2.744$ (0.019) \\
  \hline
  \multicolumn{4}{l}{$n=1000$} \\
  spBGCNNGP & $-6.674$ (0.015) & $-6.374$ (0.020) & $-6.050$ (0.026) \\
  BGC & $-5.401$ (0.016) & $-3.270$ (0.019) & $-2.753$ (0.019) \\
  \hline
  \end{tabularx}
\end{minipage}
\end{table}

\begin{table}[H]
\centering
\caption{Comparisons of the coverage probabilities (all entries) for spBGCNNGP (with $m=n/10$) and BGC under a small-magnitude correlation matrix $\bm{R}$ (the Section~4 pattern scaled so that every nonzero entry has $|\cdot|\le 0.1$). The values represent average coverage probabilities from 300 calculations, with standard errors in parentheses.}
\label{tb:A5_small_cp_all}
\begin{minipage}{.48\linewidth}
  \centering
  \subcaption{Number of outcomes $p=6$}
  \begin{tabularx}{\linewidth}{l *{3}{>{\centering\arraybackslash}X}}
  \hline
  & \multicolumn{3}{c}{$\phi$} \\
  \cline{2-4}
   & 0.05 & 0.25 & 0.50 \\
  \hline
  \multicolumn{4}{l}{$n=50$} \\
  spBGCNNGP & $0.971$ (0.003) & $0.968$ (0.003) & $0.971$ (0.003) \\
  BGC & $0.958$ (0.003) & $0.794$ (0.008) & $0.737$ (0.007) \\
  \hline
  \multicolumn{4}{l}{$n=500$} \\
  spBGCNNGP & $0.948$ (0.003) & $0.950$ (0.003) & $0.956$ (0.003) \\
  BGC & $0.787$ (0.006) & $0.360$ (0.008) & $0.303$ (0.008) \\
  \hline
  \multicolumn{4}{l}{$n=1000$} \\
  spBGCNNGP & $0.956$ (0.003) & $0.952$ (0.003) & $0.946$ (0.003) \\
  BGC & $0.675$ (0.007) & $0.248$ (0.007) & $0.205$ (0.006) \\
  \hline
  \end{tabularx}
\end{minipage}
\hfill
\begin{minipage}{.48\linewidth}
  \centering
  \subcaption{Number of outcomes $p=9$}
  \begin{tabularx}{\linewidth}{l *{3}{>{\centering\arraybackslash}X}}
  \hline
  & \multicolumn{3}{c}{$\phi$} \\
  \cline{2-4}
   & 0.05 & 0.25 & 0.50 \\
  \hline
  \multicolumn{4}{l}{$n=50$} \\
  spBGCNNGP & $0.975$ (0.002) & $0.973$ (0.002) & $0.975$ (0.001) \\
  BGC & $0.963$ (0.002) & $0.810$ (0.005) & $0.720$ (0.007) \\
  \hline
  \multicolumn{4}{l}{$n=500$} \\
  spBGCNNGP & $0.950$ (0.002) & $0.956$ (0.002) & $0.950$ (0.002) \\
  BGC & $0.782$ (0.004) & $0.354$ (0.005) & $0.287$ (0.005) \\
  \hline
  \multicolumn{4}{l}{$n=1000$} \\
  spBGCNNGP & $0.949$ (0.002) & $0.951$ (0.002) & $0.944$ (0.002) \\
  BGC & $0.675$ (0.005) & $0.257$ (0.005) & $0.207$ (0.004) \\
  \hline
  \end{tabularx}
\end{minipage}
\end{table}

\begin{table}[H]
\centering
\caption{Comparisons of the coverage probabilities (nonzero entries) for spBGCNNGP (with $m=n/10$) and BGC under a small-magnitude correlation matrix $\bm{R}$ (the Section~4 pattern scaled so that every nonzero entry has $|\cdot|\le 0.1$). The values represent average coverage probabilities from 300 calculations, with standard errors in parentheses.}
\label{tb:A5_small_cp_nz}
\begin{minipage}{.48\linewidth}
  \centering
  \subcaption{Number of outcomes $p=6$}
  \begin{tabularx}{\linewidth}{l *{3}{>{\centering\arraybackslash}X}}
  \hline
  & \multicolumn{3}{c}{$\phi$} \\
  \cline{2-4}
   & 0.05 & 0.25 & 0.50 \\
  \hline
  \multicolumn{4}{l}{$n=50$} \\
  spBGCNNGP & $0.977$ (0.004) & $0.966$ (0.004) & $0.972$ (0.004) \\
  BGC & $0.957$ (0.005) & $0.803$ (0.009) & $0.756$ (0.010) \\
  \hline
  \multicolumn{4}{l}{$n=500$} \\
  spBGCNNGP & $0.950$ (0.005) & $0.946$ (0.005) & $0.955$ (0.004) \\
  BGC & $0.794$ (0.009) & $0.364$ (0.011) & $0.309$ (0.010) \\
  \hline
  \multicolumn{4}{l}{$n=1000$} \\
  spBGCNNGP & $0.957$ (0.004) & $0.955$ (0.004) & $0.942$ (0.005) \\
  BGC & $0.699$ (0.010) & $0.261$ (0.010) & $0.219$ (0.010) \\
  \hline
  \end{tabularx}
\end{minipage}
\hfill
\begin{minipage}{.48\linewidth}
  \centering
  \subcaption{Number of outcomes $p=9$}
  \begin{tabularx}{\linewidth}{l *{3}{>{\centering\arraybackslash}X}}
  \hline
  & \multicolumn{3}{c}{$\phi$} \\
  \cline{2-4}
   & 0.05 & 0.25 & 0.50 \\
  \hline
  \multicolumn{4}{l}{$n=50$} \\
  spBGCNNGP & $0.979$ (0.003) & $0.976$ (0.004) & $0.976$ (0.003) \\
  BGC & $0.965$ (0.004) & $0.843$ (0.009) & $0.745$ (0.010) \\
  \hline
  \multicolumn{4}{l}{$n=500$} \\
  spBGCNNGP & $0.949$ (0.005) & $0.952$ (0.005) & $0.944$ (0.005) \\
  BGC & $0.797$ (0.009) & $0.384$ (0.011) & $0.325$ (0.011) \\
  \hline
  \multicolumn{4}{l}{$n=1000$} \\
  spBGCNNGP & $0.950$ (0.005) & $0.945$ (0.005) & $0.936$ (0.005) \\
  BGC & $0.716$ (0.010) & $0.251$ (0.010) & $0.221$ (0.010) \\
  \hline
  \end{tabularx}
\end{minipage}
\end{table}

\begin{table}[H]
\centering
\caption{Comparisons of the coverage probabilities (zero entries) for spBGCNNGP (with $m=n/10$) and BGC under a small-magnitude correlation matrix $\bm{R}$ (the Section~4 pattern scaled so that every nonzero entry has $|\cdot|\le 0.1$). The values represent average coverage probabilities from 300 calculations, with standard errors in parentheses.}
\label{tb:A5_small_cp_z}
\begin{minipage}{.48\linewidth}
  \centering
  \subcaption{Number of outcomes $p=6$}
  \begin{tabularx}{\linewidth}{l *{3}{>{\centering\arraybackslash}X}}
  \hline
  & \multicolumn{3}{c}{$\phi$} \\
  \cline{2-4}
   & 0.05 & 0.25 & 0.50 \\
  \hline
  \multicolumn{4}{l}{$n=50$} \\
  spBGCNNGP & $0.965$ (0.004) & $0.970$ (0.003) & $0.969$ (0.004) \\
  BGC & $0.959$ (0.004) & $0.786$ (0.010) & $0.720$ (0.010) \\
  \hline
  \multicolumn{4}{l}{$n=500$} \\
  spBGCNNGP & $0.947$ (0.005) & $0.955$ (0.004) & $0.956$ (0.004) \\
  BGC & $0.781$ (0.008) & $0.358$ (0.009) & $0.298$ (0.009) \\
  \hline
  \multicolumn{4}{l}{$n=1000$} \\
  spBGCNNGP & $0.955$ (0.004) & $0.949$ (0.005) & $0.950$ (0.004) \\
  BGC & $0.654$ (0.010) & $0.238$ (0.009) & $0.193$ (0.008) \\
  \hline
  \end{tabularx}
\end{minipage}
\hfill
\begin{minipage}{.48\linewidth}
  \centering
  \subcaption{Number of outcomes $p=9$}
  \begin{tabularx}{\linewidth}{l *{3}{>{\centering\arraybackslash}X}}
  \hline
  & \multicolumn{3}{c}{$\phi$} \\
  \cline{2-4}
   & 0.05 & 0.25 & 0.50 \\
  \hline
  \multicolumn{4}{l}{$n=50$} \\
  spBGCNNGP & $0.974$ (0.002) & $0.972$ (0.002) & $0.975$ (0.002) \\
  BGC & $0.962$ (0.002) & $0.802$ (0.005) & $0.714$ (0.007) \\
  \hline
  \multicolumn{4}{l}{$n=500$} \\
  spBGCNNGP & $0.950$ (0.002) & $0.957$ (0.002) & $0.951$ (0.002) \\
  BGC & $0.778$ (0.005) & $0.347$ (0.005) & $0.278$ (0.006) \\
  \hline
  \multicolumn{4}{l}{$n=1000$} \\
  spBGCNNGP & $0.949$ (0.002) & $0.952$ (0.002) & $0.946$ (0.002) \\
  BGC & $0.666$ (0.005) & $0.258$ (0.005) & $0.204$ (0.005) \\
  \hline
  \end{tabularx}
\end{minipage}
\end{table}

\begin{table}[H]
\centering
\small
\caption{Edge selection metrics (TPR, FPR, MCC, F1) for spBGCNNGP (with $m=n/10$) and BGC under a small-magnitude correlation matrix $\bm{R}$ (the Section~4 pattern scaled so that every nonzero entry has $|\cdot|\le 0.1$). The values represent averages from 300 calculations, with standard errors in parentheses.}
\label{tb:A5_small_edge}
\begin{minipage}{.48\linewidth}
  \centering
  \subcaption{Number of outcomes $p=6$}
  \begin{tabularx}{\linewidth}{l *{4}{>{\centering\arraybackslash}X}}
  \hline
  & TPR & FPR & MCC & F1 \\
  \hline
  \multicolumn{5}{l}{$n=50$, $\phi=0.05$} \\
  spBGCNNGP & $0.046$ (0.005) & $0.035$ (0.004) & $0.044$ (0.025) & $0.075$ (0.008) \\
  BGC & $0.055$ (0.005) & $0.041$ (0.004) & $0.051$ (0.022) & $0.089$ (0.008) \\
  \multicolumn{5}{l}{$n=50$, $\phi=0.25$} \\
  spBGCNNGP & $0.036$ (0.004) & $0.030$ (0.003) & $0.028$ (0.025) & $0.060$ (0.007) \\
  BGC & $0.209$ (0.010) & $0.214$ (0.010) & $0.002$ (0.016) & $0.267$ (0.011) \\
  \multicolumn{5}{l}{$n=50$, $\phi=0.50$} \\
  spBGCNNGP & $0.032$ (0.004) & $0.031$ (0.004) & $0.007$ (0.027) & $0.053$ (0.007) \\
  BGC & $0.258$ (0.010) & $0.280$ (0.010) & $-0.021$ (0.016) & $0.310$ (0.011) \\
  \hline
  \multicolumn{5}{l}{$n=500$, $\phi=0.05$} \\
  spBGCNNGP & $0.277$ (0.009) & $0.053$ (0.005) & $0.317$ (0.012) & $0.394$ (0.011) \\
  BGC & $0.373$ (0.010) & $0.219$ (0.008) & $0.173$ (0.015) & $0.446$ (0.010) \\
  \multicolumn{5}{l}{$n=500$, $\phi=0.25$} \\
  spBGCNNGP & $0.207$ (0.008) & $0.045$ (0.004) & $0.269$ (0.013) & $0.308$ (0.011) \\
  BGC & $0.642$ (0.011) & $0.642$ (0.009) & $0.003$ (0.016) & $0.534$ (0.008) \\
  \multicolumn{5}{l}{$n=500$, $\phi=0.50$} \\
  spBGCNNGP & $0.185$ (0.008) & $0.044$ (0.004) & $0.248$ (0.013) & $0.280$ (0.011) \\
  BGC & $0.713$ (0.010) & $0.702$ (0.009) & $0.013$ (0.015) & $0.562$ (0.006) \\
  \hline
  \multicolumn{5}{l}{$n=1000$, $\phi=0.05$} \\
  spBGCNNGP & $0.431$ (0.009) & $0.045$ (0.004) & $0.465$ (0.010) & $0.566$ (0.010) \\
  BGC & $0.532$ (0.011) & $0.346$ (0.010) & $0.193$ (0.015) & $0.542$ (0.009) \\
  \multicolumn{5}{l}{$n=1000$, $\phi=0.25$} \\
  spBGCNNGP & $0.306$ (0.010) & $0.051$ (0.005) & $0.348$ (0.013) & $0.429$ (0.011) \\
  BGC & $0.754$ (0.010) & $0.762$ (0.009) & $-0.009$ (0.016) & $0.571$ (0.006) \\
  \multicolumn{5}{l}{$n=1000$, $\phi=0.50$} \\
  spBGCNNGP & $0.280$ (0.009) & $0.050$ (0.004) & $0.325$ (0.012) & $0.400$ (0.011) \\
  BGC & $0.796$ (0.009) & $0.807$ (0.008) & $-0.012$ (0.016) & $0.582$ (0.005) \\
  \hline
  \end{tabularx}
\end{minipage}
\hfill
\begin{minipage}{.48\linewidth}
  \centering
  \subcaption{Number of outcomes $p=9$}
  \begin{tabularx}{\linewidth}{l *{4}{>{\centering\arraybackslash}X}}
  \hline
  & TPR & FPR & MCC & F1 \\
  \hline
  \multicolumn{5}{l}{$n=50$, $\phi=0.05$} \\
  spBGCNNGP & $0.038$ (0.004) & $0.026$ (0.002) & $0.034$ (0.014) & $0.060$ (0.007) \\
  BGC & $0.050$ (0.004) & $0.038$ (0.002) & $0.032$ (0.012) & $0.076$ (0.007) \\
  \multicolumn{5}{l}{$n=50$, $\phi=0.25$} \\
  spBGCNNGP & $0.028$ (0.003) & $0.028$ (0.002) & $0.004$ (0.013) & $0.045$ (0.006) \\
  BGC & $0.182$ (0.009) & $0.198$ (0.005) & $-0.016$ (0.010) & $0.173$ (0.008) \\
  \multicolumn{5}{l}{$n=50$, $\phi=0.50$} \\
  spBGCNNGP & $0.030$ (0.004) & $0.025$ (0.002) & $0.009$ (0.013) & $0.047$ (0.006) \\
  BGC & $0.253$ (0.011) & $0.286$ (0.007) & $-0.032$ (0.009) & $0.196$ (0.008) \\
  \hline
  \multicolumn{5}{l}{$n=500$, $\phi=0.05$} \\
  spBGCNNGP & $0.243$ (0.009) & $0.050$ (0.002) & $0.274$ (0.012) & $0.319$ (0.010) \\
  BGC & $0.352$ (0.010) & $0.222$ (0.005) & $0.124$ (0.010) & $0.308$ (0.008) \\
  \multicolumn{5}{l}{$n=500$, $\phi=0.25$} \\
  spBGCNNGP & $0.198$ (0.009) & $0.043$ (0.002) & $0.236$ (0.013) & $0.271$ (0.011) \\
  BGC & $0.647$ (0.011) & $0.653$ (0.005) & $-0.005$ (0.010) & $0.296$ (0.004) \\
  \multicolumn{5}{l}{$n=500$, $\phi=0.50$} \\
  spBGCNNGP & $0.174$ (0.009) & $0.049$ (0.002) & $0.190$ (0.013) & $0.236$ (0.011) \\
  BGC & $0.693$ (0.011) & $0.722$ (0.006) & $-0.027$ (0.010) & $0.295$ (0.004) \\
  \hline
  \multicolumn{5}{l}{$n=1000$, $\phi=0.05$} \\
  spBGCNNGP & $0.425$ (0.009) & $0.051$ (0.002) & $0.455$ (0.010) & $0.512$ (0.009) \\
  BGC & $0.518$ (0.011) & $0.334$ (0.005) & $0.153$ (0.010) & $0.355$ (0.007) \\
  \multicolumn{5}{l}{$n=1000$, $\phi=0.25$} \\
  spBGCNNGP & $0.326$ (0.009) & $0.048$ (0.002) & $0.368$ (0.012) & $0.414$ (0.011) \\
  BGC & $0.739$ (0.010) & $0.742$ (0.005) & $-0.003$ (0.010) & $0.307$ (0.004) \\
  \multicolumn{5}{l}{$n=1000$, $\phi=0.50$} \\
  spBGCNNGP & $0.284$ (0.009) & $0.054$ (0.002) & $0.310$ (0.011) & $0.364$ (0.010) \\
  BGC & $0.788$ (0.009) & $0.796$ (0.005) & $-0.012$ (0.010) & $0.310$ (0.003) \\
  \hline
  \end{tabularx}
\end{minipage}
\end{table}

\subsection{Scalability in the Number of Outcomes \texorpdfstring{$p$}{p}}
\label{sec:p_scalability}

To examine how the cost of spBGCNNGP grows with the number of outcomes, we benchmark it for larger values of $p$ than in Section~\ref{sec:vshoff} of the main text.
So that the timings are comparable with Table~\ref{tb:time} of the main text, everything that affects the runtime is as in Section~\ref{sec:vshoff}: the implementation, the data-generating process (the first four outcomes binary, count, count and ordinal, the remaining ones continuous, with the true $\bm{R}$ of Section~\ref{sec:vshoff} so that the additional outcomes are uncorrelated), the MCMC settings ($3{,}000$ iterations with the first $1{,}000$ discarded, $m=n/10$) and the computing environment (the Apple Mac Studio with an Apple M1 Ultra chip used in Section~\ref{sec:vshoff}).
We take $\phi=0.25$, $n\in\{500,1000\}$ and $p\in\{6,9,15,30,50\}$, with $20$ replications per setting.
For $p=6$ and $9$ we reuse the first $20$ datasets of Section~\ref{sec:vshoff}; the times, measured in a separate run, are $7$--$18\%$ longer than those in Table~\ref{tb:time}, so the two sets of timings are of the same order but not identical.

Table~\ref{tb:B_time} reports the runtime and the estimation accuracy.
From $p=6$ to $p=50$ the time per iteration grows about $4.5$-fold for both $n=500$ and $n=1000$, far less than the factors $p^2\approx 69$ and $p^3\approx 579$, and a fit with $n=1000$ and $p=50$ takes about $3.5$ hours.
The mixing of $\bm{R}$ does not deteriorate with $p$: the minimum effective sample size over its entries, computed from the autocorrelation of the $2{,}000$ retained draws (cf.\ the autocorrelation plots in Section~\ref{sec:real} of the main text), averages between $85$ and $135$ in every setting, with no trend in $p$.

Table~\ref{tb:B_profile} decomposes the time per iteration by a line-level profile of one fit per setting.
The truncated-normal draws and the search for the rank-based truncation limits together take $87$--$97\%$ of the time, whereas the matrix operations, namely the conditional moments and the $\bm{R}$ and $\phi$ updates, take $2$--$13\%$; the one-off NNGP set-up takes about $35$ seconds for $n=500$ and $185$ seconds for $n=1000$, whatever the value of $p$.
Within the range examined, $p\le 50$, the computation is therefore practical.

The accuracy for the nonzero entries, however, deteriorates as $p$ grows.
The coverage of the zero entries stays between $0.94$ and $0.96$, but that of the seven nonzero entries (CP$_{\neq 0}$ in Table~\ref{tb:B_time}) falls from $0.93$ at $p=6$ to $0.59$ at $p=50$ for $n=500$, and from $0.94$ to $0.67$ for $n=1000$, while the average widths of their intervals hardly change ($0.23$--$0.24$ for $n=500$ and $0.17$--$0.18$ for $n=1000$).
Refitting the settings $p=6$ and $50$ for $n=500$ with the same data and random seeds, and keeping the estimates of the nonzero entries, shows that their point estimates move toward zero: averaged over the seven entries and the replications, the posterior mean is $91\%$ of the true value at $p=6$ and $71\%$ at $p=50$, and every one of the seven entries is closer to zero at $p=50$.
With the interval widths unchanged, this shift accounts for the loss of coverage.
It is consistent with how the inverse-Wishart prior $IW(p+2,(p+2)\bm{I})$ on $\bm{V}$ (Section~\ref{sec:vshoff}) enters the update of $\bm{V}$: its scale matrix $(p+2)\bm{I}$ is added to the cross-product of the latent variables, as if $p+2$ observations with identity correlation were added to the data, so its pull toward the identity grows with $p$ relative to the information in the $n$ spatially dependent sites; the zero entries, whose true values are those of the identity, are not affected.
Because an inverse-Wishart prior needs more than $p-1$ degrees of freedom, this pull cannot be removed by its hyperparameters alone.
Shrinkage priors on the precision matrix that shrink small entries strongly but leave large ones nearly unshrunk, such as the graphical horseshoe \citep{li2019graphical,sagar2024precision}, may reduce this attenuation; their use for the correlation matrix of the copula is left for future work.

\begin{table}[H]
\centering
\caption{Runtime and accuracy of spBGCNNGP ($m=n/10$, $\phi=0.25$) as the number of outcomes $p$ grows. Time per fit is the average over 20 replications of 3,000 MCMC iterations, with standard errors in parentheses, and time per iteration is time per fit divided by 3,000. MSE, CP and AL are as in Section~\ref{sec:vshoff}, over all entries of $\bm{R}$; CP$_{\neq 0}$ is the coverage of the seven nonzero entries.}
\label{tb:B_time}
\small
\begin{tabular}{r rr rrrr}
\hline
 & \multicolumn{2}{c}{Time (s)} & & & & \\
\cline{2-3}
$p$ & per fit & per iteration & MSE & CP & AL & CP$_{\neq 0}$ \\
\hline
\multicolumn{7}{l}{$n=500$} \\
6 & 1,243 (1.5) & 0.414 & 0.0041 & 0.940 & 0.228 & 0.929 \\
9 & 1,481 (1.7) & 0.494 & 0.0036 & 0.947 & 0.216 & 0.907 \\
15 & 1,931 (1.5) & 0.644 & 0.0028 & 0.958 & 0.198 & 0.907 \\
30 & 3,269 (3.0) & 1.090 & 0.0022 & 0.955 & 0.183 & 0.779 \\
50 & 5,642 (4.6) & 1.881 & 0.0019 & 0.960 & 0.176 & 0.586 \\
\hline
\multicolumn{7}{l}{$n=1000$} \\
6 & 2,861 (2.0) & 0.954 & 0.0021 & 0.940 & 0.167 & 0.936 \\
9 & 3,434 (3.2) & 1.145 & 0.0017 & 0.950 & 0.154 & 0.936 \\
15 & 4,460 (2.9) & 1.487 & 0.0016 & 0.935 & 0.144 & 0.900 \\
30 & 7,468 (5.4) & 2.489 & 0.0012 & 0.947 & 0.132 & 0.771 \\
50 & 12,726 (6.8) & 4.242 & 0.0010 & 0.954 & 0.127 & 0.671 \\
\hline
\end{tabular}
\end{table}

\begin{table}[H]
\centering
\caption{Share (\%) of the time per MCMC iteration spent in each step of the sampler, from a line-level profile of one fit per setting (300 iterations). TN draw is the $p$-variate truncated-normal draw by minimax tilting, truncation limits the search for the rank-based bounds, and the conditional moments and the $\bm{R}$ and $\phi$ updates the matrix operations of Section~\ref{sec:nngp} of the main text. Set-up is the one-off computation of the NNGP quantities for all candidate values of $\phi$, in seconds.}
\label{tb:B_profile}
\small
\begin{tabular}{r rrrrrr r}
\hline
 & TN & Truncation & Cond. & $\bm{R}$ & $\phi$ &  & Set-up \\
$p$ & draw & limits & moments & update & update & Other & (s) \\
\hline
\multicolumn{8}{l}{$n=500$} \\
6 & 63.6 & 33.8 & 0.8 & 0.7 & 0.9 & 0.1 & 37 \\
9 & 47.4 & 49.5 & 1.5 & 0.8 & 0.5 & 0.3 & 38 \\
15 & 44.8 & 51.2 & 2.8 & 0.8 & 0.3 & 0.1 & 32 \\
30 & 54.3 & 41.9 & 3.3 & 0.3 & 0.2 & 0.0 & 32 \\
50 & 64.1 & 30.4 & 5.1 & 0.3 & 0.2 & 0.0 & 33 \\
\hline
\multicolumn{8}{l}{$n=1000$} \\
6 & 43.1 & 53.1 & 2.8 & 0.5 & 0.4 & 0.1 & 195 \\
9 & 45.4 & 49.0 & 4.7 & 0.4 & 0.3 & 0.2 & 177 \\
15 & 31.8 & 55.2 & 12.3 & 0.4 & 0.3 & 0.0 & 178 \\
30 & 46.9 & 43.0 & 9.7 & 0.2 & 0.2 & 0.0 & 181 \\
50 & 59.0 & 31.0 & 9.7 & 0.1 & 0.2 & 0.0 & 195 \\
\hline
\end{tabular}
\end{table}

\section{Additional Results for Data Application}

This section provides additional results for the real data application presented in Section 5 of the main manuscript.
We present two supplementary analyses: sensitivity to the choice of correlation function (Section~\ref{sec:corr_func_sensitivity}) and comparison with alternative multivariate spatial models (Section~\ref{sec:inla_comparison}).

\subsection{Sensitivity to Correlation Function in Real Data Analysis}
\label{sec:corr_func_sensitivity}

We assessed the sensitivity of the real data analysis (Section 5) to the choice of spatial correlation function. 
We implemented the proposed spBGCNNGP method (Section 3) with three different correlation functions $\rho(\cdot)$ in the spatial correlation matrix $H(\phi)$: exponential, Mat\'ern with $\nu=3/2$, and Mat\'ern with $\nu=5/2$. 
The prior distributions, MCMC settings, and NNGP configuration remained identical to Section 5.

\begin{figure}[H]
\centering
\includegraphics[width=\columnwidth]{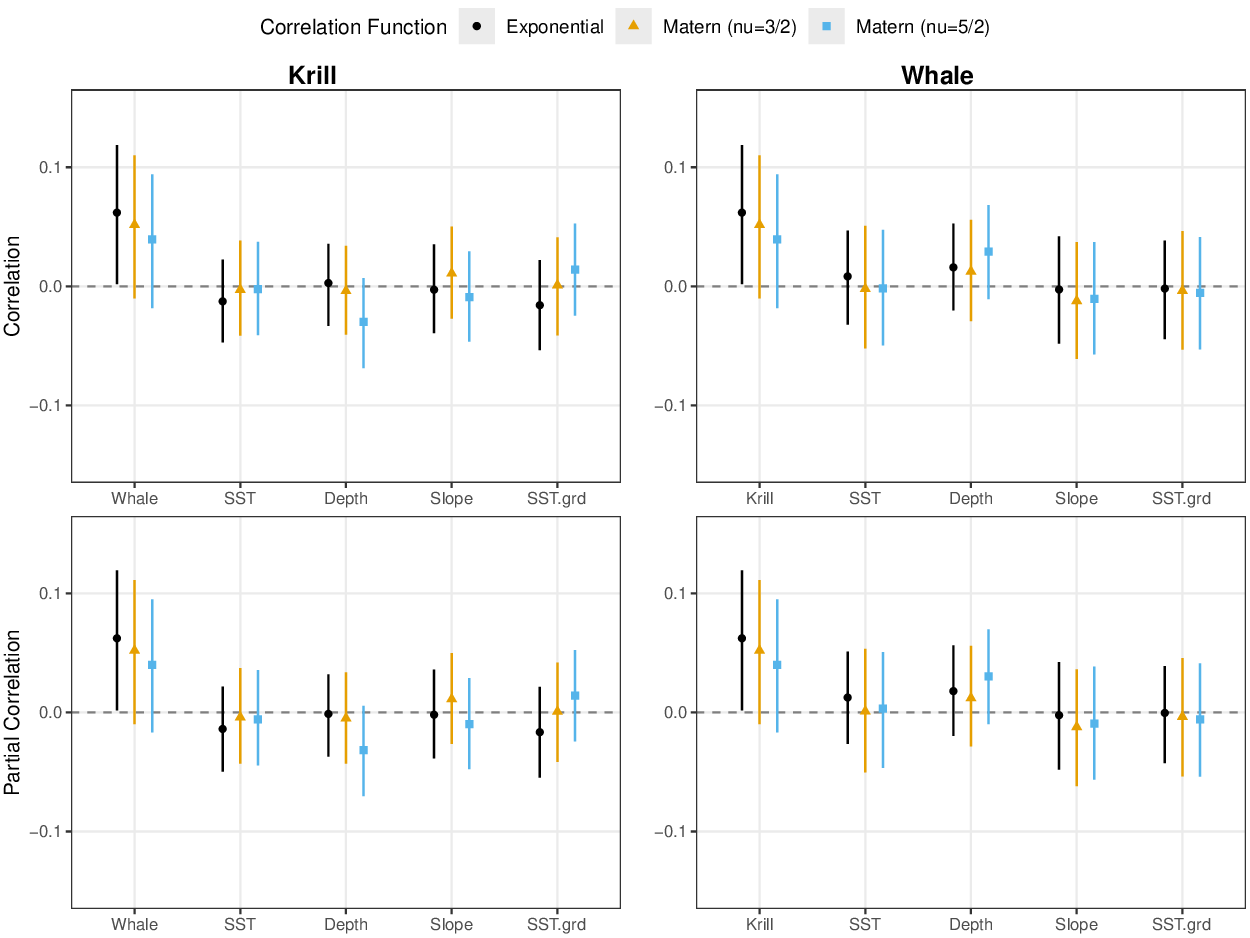}
\caption{Comparison of posterior correlation and partial correlation estimates across three spatial correlation functions (Exponential, Mat\'ern with $\nu=3/2$, and Mat\'ern with $\nu=5/2$).
Points represent posterior medians and error bars show 95\% credible intervals.}
\label{fig:corr_func_sensitivity}
\end{figure}

Figure~\ref{fig:corr_func_sensitivity} compares the posterior distributions of correlation and partial correlation coefficients across the three correlation functions.
Tables~\ref{tb:da_expo_iso}--\ref{tb:da_matern5} provide the detailed posterior quantiles (2.5\%, 50\%, and 97.5\%) for correlation and partial correlation coefficients under each correlation function.

The results demonstrate consistency in the primary finding across different correlation functions.
The correlation between Krill and Whale is consistently positive across all correlation functions, with median posterior estimates ranging from 0.039 to 0.062 for correlation and 0.040 to 0.062 for partial correlation.
While estimates for correlations and partial correlations with environmental variables show some variation across correlation functions, with point estimates occasionally differing in sign, all 95\% credible intervals overlap substantially.
This indicates that our main conclusion regarding the krill-whale relationship is robust to the choice of spatial correlation function, though some uncertainty remains in the magnitude and direction of associations with environmental covariates.

\begin{table}[H]
\caption{2.5\%, 50\%, and 97.5\% posterior quantiles of the correlation coefficients and partial correlations using exponential correlation function.}
\label{tb:da_expo_iso}
\begin{minipage}[b]{0.49\linewidth}
\centering
\subcaption{Krill}
\scalebox{0.75}{
\begin{tabular}{r rrr}
\multicolumn{1}{l}{}&\multicolumn{1}{c}{2.5\%}&\multicolumn{1}{c}{Median}&\multicolumn{1}{c}{97.5\%}\tabularnewline \hline
{\bf Correlation} & & & \\
Whale&$ 0.0019$&$ 0.0619$&$ 0.1187$\tabularnewline
SST&$-0.0472$&$-0.0126$&$ 0.0225$\tabularnewline
Depth&$-0.0332$&$ 0.0028$&$ 0.0356$\tabularnewline
Slope&$-0.0393$&$-0.0029$&$ 0.0352$\tabularnewline
SST.grd&$-0.0536$&$-0.0159$&$ 0.0219$\tabularnewline
&&& \\
{\bf Partial correlation} & & & \\
Whale&$ 0.0018$&$ 0.0623$&$ 0.1193$\tabularnewline
SST&$-0.0497$&$-0.0139$&$ 0.0217$\tabularnewline
Depth&$-0.0371$&$-0.0012$&$ 0.0320$\tabularnewline
Slope&$-0.0385$&$-0.0019$&$ 0.0359$\tabularnewline
SST.grd&$-0.0547$&$-0.0166$&$ 0.0214$\tabularnewline
\hline
\end{tabular}
}
\label{tb:krill_expo_iso}
\end{minipage}
\begin{minipage}[b]{0.49\linewidth}
\centering
\subcaption{Whale}
\scalebox{0.75}{
\begin{tabular}{r rrr}
\multicolumn{1}{l}{}&\multicolumn{1}{c}{2.5\%}&\multicolumn{1}{c}{Median}&\multicolumn{1}{c}{97.5\%}\tabularnewline \hline
{\bf Correlation} & & & \\
Krill&$ 0.0019$&$ 0.0619$&$ 0.1187$\tabularnewline
SST&$-0.0322$&$ 0.0083$&$ 0.0467$\tabularnewline
Depth&$-0.0202$&$ 0.0159$&$ 0.0526$\tabularnewline
Slope&$-0.0481$&$-0.0027$&$ 0.0420$\tabularnewline
SST.grd&$-0.0443$&$-0.0019$&$ 0.0383$\tabularnewline
&&& \\
{\bf Partial correlation} & & & \\
Krill&$ 0.0018$&$ 0.0623$&$ 0.1193$\tabularnewline
SST&$-0.0264$&$ 0.0125$&$ 0.0511$\tabularnewline
Depth&$-0.0198$&$ 0.0179$&$ 0.0563$\tabularnewline
Slope&$-0.0480$&$-0.0023$&$ 0.0424$\tabularnewline
SST.grd&$-0.0427$&$-0.0004$&$ 0.0389$\tabularnewline
\hline
\end{tabular}
}
\label{tb:whale_expo_iso}
\end{minipage}
\end{table}

\begin{table}[H]
\caption{2.5\%, 50\%, and 97.5\% posterior quantiles of the correlation coefficients and partial correlations using Mat\'ern with $\nu=3/2$ correlation function.}
\label{tb:da_matern3}
\begin{minipage}[b]{0.49\linewidth}
\centering
\subcaption{Krill}
\scalebox{0.75}{
\begin{tabular}{r rrr}
\multicolumn{1}{l}{}&\multicolumn{1}{c}{2.5\%}&\multicolumn{1}{c}{Median}&\multicolumn{1}{c}{97.5\%}\tabularnewline \hline
{\bf Correlation} & & & \\
Whale&$-0.0102$&$ 0.0517$&$ 0.1101$\tabularnewline
SST&$-0.0414$&$-0.0028$&$ 0.0384$\tabularnewline
Depth&$-0.0407$&$-0.0038$&$ 0.0340$\tabularnewline
Slope&$-0.0272$&$ 0.0109$&$ 0.0501$\tabularnewline
SST.grd&$-0.0412$&$ 0.0010$&$ 0.0412$\tabularnewline
&&& \\
{\bf Partial correlation} & & & \\
Whale&$-0.0098$&$ 0.0520$&$ 0.1112$\tabularnewline
SST&$-0.0430$&$-0.0040$&$ 0.0372$\tabularnewline
Depth&$-0.0429$&$-0.0048$&$ 0.0336$\tabularnewline
Slope&$-0.0263$&$ 0.0112$&$ 0.0499$\tabularnewline
SST.grd&$-0.0415$&$ 0.0007$&$ 0.0418$\tabularnewline
\hline
\end{tabular}
}
\label{tb:krill_matern3}
\end{minipage}
\begin{minipage}[b]{0.49\linewidth}
\centering
\subcaption{Whale}
\scalebox{0.75}{
\begin{tabular}{r rrr}
\multicolumn{1}{l}{}&\multicolumn{1}{c}{2.5\%}&\multicolumn{1}{c}{Median}&\multicolumn{1}{c}{97.5\%}\tabularnewline \hline
{\bf Correlation} & & & \\
Krill&$-0.0102$&$ 0.0517$&$ 0.1101$\tabularnewline
SST&$-0.0521$&$-0.0020$&$ 0.0508$\tabularnewline
Depth&$-0.0293$&$ 0.0125$&$ 0.0559$\tabularnewline
Slope&$-0.0609$&$-0.0123$&$ 0.0371$\tabularnewline
SST.grd&$-0.0532$&$-0.0037$&$ 0.0464$\tabularnewline
&&& \\
{\bf Partial correlation} & & & \\
Krill&$-0.0098$&$ 0.0520$&$ 0.1112$\tabularnewline
SST&$-0.0503$&$ 0.0010$&$ 0.0535$\tabularnewline
Depth&$-0.0284$&$ 0.0121$&$ 0.0559$\tabularnewline
Slope&$-0.0618$&$-0.0122$&$ 0.0363$\tabularnewline
SST.grd&$-0.0537$&$-0.0035$&$ 0.0456$\tabularnewline
\hline
\end{tabular}
}
\label{tb:whale_matern3}
\end{minipage}
\end{table}

\begin{table}[H]
\caption{2.5\%, 50\%, and 97.5\% posterior quantiles of the correlation coefficients and partial correlations using Mat\'ern with $\nu=5/2$ correlation function.}
\label{tb:da_matern5}
\begin{minipage}[b]{0.49\linewidth}
\centering
\subcaption{Krill}
\scalebox{0.75}{
\begin{tabular}{r rrr}
\multicolumn{1}{l}{}&\multicolumn{1}{c}{2.5\%}&\multicolumn{1}{c}{Median}&\multicolumn{1}{c}{97.5\%}\tabularnewline \hline
{\bf Correlation} & & & \\
Whale&$-0.0182$&$ 0.0393$&$ 0.0940$\tabularnewline
SST&$-0.0409$&$-0.0024$&$ 0.0374$\tabularnewline
Depth&$-0.0687$&$-0.0299$&$ 0.0070$\tabularnewline
Slope&$-0.0464$&$-0.0090$&$ 0.0293$\tabularnewline
SST.grd&$-0.0247$&$ 0.0141$&$ 0.0526$\tabularnewline
&&& \\
{\bf Partial correlation} & & & \\
Whale&$-0.0167$&$ 0.0399$&$ 0.0949$\tabularnewline
SST&$-0.0445$&$-0.0059$&$ 0.0356$\tabularnewline
Depth&$-0.0704$&$-0.0317$&$ 0.0055$\tabularnewline
Slope&$-0.0475$&$-0.0098$&$ 0.0288$\tabularnewline
SST.grd&$-0.0243$&$ 0.0142$&$ 0.0523$\tabularnewline
\hline
\end{tabular}
}
\label{tb:krill_matern5}
\end{minipage}
\begin{minipage}[b]{0.49\linewidth}
\centering
\subcaption{Whale}
\scalebox{0.75}{
\begin{tabular}{r rrr}
\multicolumn{1}{l}{}&\multicolumn{1}{c}{2.5\%}&\multicolumn{1}{c}{Median}&\multicolumn{1}{c}{97.5\%}\tabularnewline \hline
{\bf Correlation} & & & \\
Krill&$-0.0182$&$ 0.0393$&$ 0.0940$\tabularnewline
SST&$-0.0497$&$-0.0017$&$ 0.0473$\tabularnewline
Depth&$-0.0107$&$ 0.0292$&$ 0.0682$\tabularnewline
Slope&$-0.0573$&$-0.0106$&$ 0.0371$\tabularnewline
SST.grd&$-0.0529$&$-0.0055$&$ 0.0412$\tabularnewline
&&& \\
{\bf Partial correlation} & & & \\
Krill&$-0.0167$&$ 0.0399$&$ 0.0949$\tabularnewline
SST&$-0.0465$&$ 0.0032$&$ 0.0508$\tabularnewline
Depth&$-0.0099$&$ 0.0303$&$ 0.0699$\tabularnewline
Slope&$-0.0565$&$-0.0094$&$ 0.0386$\tabularnewline
SST.grd&$-0.0538$&$-0.0058$&$ 0.0412$\tabularnewline
\hline
\end{tabular}
}
\label{tb:whale_matern5}
\end{minipage}
\end{table}

To further assess the sensitivity of model fit to the choice of correlation function, we conducted posterior predictive checks (PPC) using Moran's $I$ statistic (see Section 1.2 of the main text), following the procedure described in Section 5.
Table~\ref{tab:ppc_moran_corr_func} presents the 95\% credible intervals of Moran's $I$ for posterior predictive samples under each correlation function.

The results show that Mat\'ern correlation functions generate data with stronger spatial correlation ($I \approx 0.999$--$1.000$) compared to the exponential function ($I \approx 0.970$--$0.996$).
For variables with very high observed spatial correlation (e.g., SST), the Mat\'ern functions provide a better fit, with the observed value falling within the 95\% credible interval.
However, for variables with moderate spatial correlation (e.g., Krill and Slope; see Table 1 in the main text for observed values), both functions overestimate spatial correlation, though the exponential function produces values closer to the observed.
This suggests that the optimal choice of correlation function may depend on the specific spatial structure of each variable, though all three functions consistently capture the presence of spatial correlation in the data.

\begin{table}[H]
\centering
\caption{Posterior predictive check using Moran's $I$ across different correlation functions.
The 95\% credible intervals are computed from 2,000 posterior predictive samples. See Table 1 in the main text for observed Moran's $I$ values.}
\label{tab:ppc_moran_corr_func}
\begin{tabular}{lccc}
\hline
 & \multicolumn{3}{c}{spBGC 95\% CI} \\
\cline{2-4}
Variable & Exponential & Mat\'ern 3/2 & Mat\'ern 5/2 \\
\hline
Krill   & [0.970, 0.995] & [0.999, 1.000] & [0.998, 1.000] \\
Whale   & [0.971, 0.995] & [0.999, 1.000] & [0.998, 1.000] \\
SST     & [0.970, 0.995] & [0.999, 1.000] & [0.998, 1.000] \\
Depth   & [0.971, 0.995] & [0.999, 1.000] & [0.998, 1.000] \\
SST.grd & [0.971, 0.995] & [0.999, 1.000] & [0.998, 1.000] \\
Slope   & [0.971, 0.996] & [0.999, 1.000] & [0.998, 1.000] \\
\hline
\end{tabular}
\end{table}

\subsection{Comparison with Multivariate Spatial Generalized Linear Model}
\label{sec:inla_comparison}

To further evaluate the proposed spBGC method, we compare it with the linear model of coregionalization (LMC) implemented via the Integrated Nested Laplace Approximation (INLA) \citep{rue2009approximate}.
The LMC, introduced by \citet{schmidt2003bayesian}, provides a flexible framework for modeling multivariate spatial data.
For $p$ response variables at location $\bm{s}$, the model can be written as:
\begin{equation}
g_j(E[Y_j(\bm{s})]) = \alpha_j + \sum_{k=1}^{j} \lambda_{jk} w_k(\bm{s}), \quad j = 1, \ldots, p,
\end{equation}
where $g_j(\cdot)$ is the link function (identity for Gaussian, log for Poisson), $\alpha_j$ is an intercept, $w_k(\bm{s})$ are independent Gaussian processes with correlation function $\rho(\cdot; \phi_k)$, and $\lambda_{jk}$ are factor loadings with $\lambda_{jj} = 1$.
The coregionalization matrix $\bm{T} = \bm{L}\bm{D}\bm{L}^\top$ captures the covariance structure of the latent Gaussian processes, where $\bm{L}$ is the lower triangular matrix of factor loadings and $\bm{D} = \text{diag}(\sigma_1^2, \ldots, \sigma_p^2)$ contains the spatial variances.
The correlation matrix $\bm{R}$ is obtained by standardizing $\bm{T}$: $R_{jk} = T_{jk} / \sqrt{T_{jj} T_{kk}}$.
This correlation matrix represents the correlations among the latent processes, which is comparable to the correlation matrix estimated by spBGC.

We fitted two versions of the LMC:
(i) a separate range model where each latent process has its own spatial range parameter $\phi_j$, and
(ii) a shared range model where all processes share a common range parameter $\phi$.
The separate range model is more flexible, allowing different spatial scales for different variables, while the shared range model corresponds to the separable covariance structure discussed in \citet{schmidt2003bayesian}.

We implemented the LMC using the R-INLA package \citep{krainski2019advanced, vanniekerk2021frontiers}, using the same exponential correlation function as in spBGC.
For fair comparison, krill biomass was modeled as Gaussian (log-transformed), whale sightings as Poisson, and the environmental variables (SST, Depth, Slope, SST.grd) as Gaussian.
Spatial coordinates were normalized to $[0, 1]^2$.

\begin{figure}[H]
\centering
\includegraphics[width=\columnwidth]{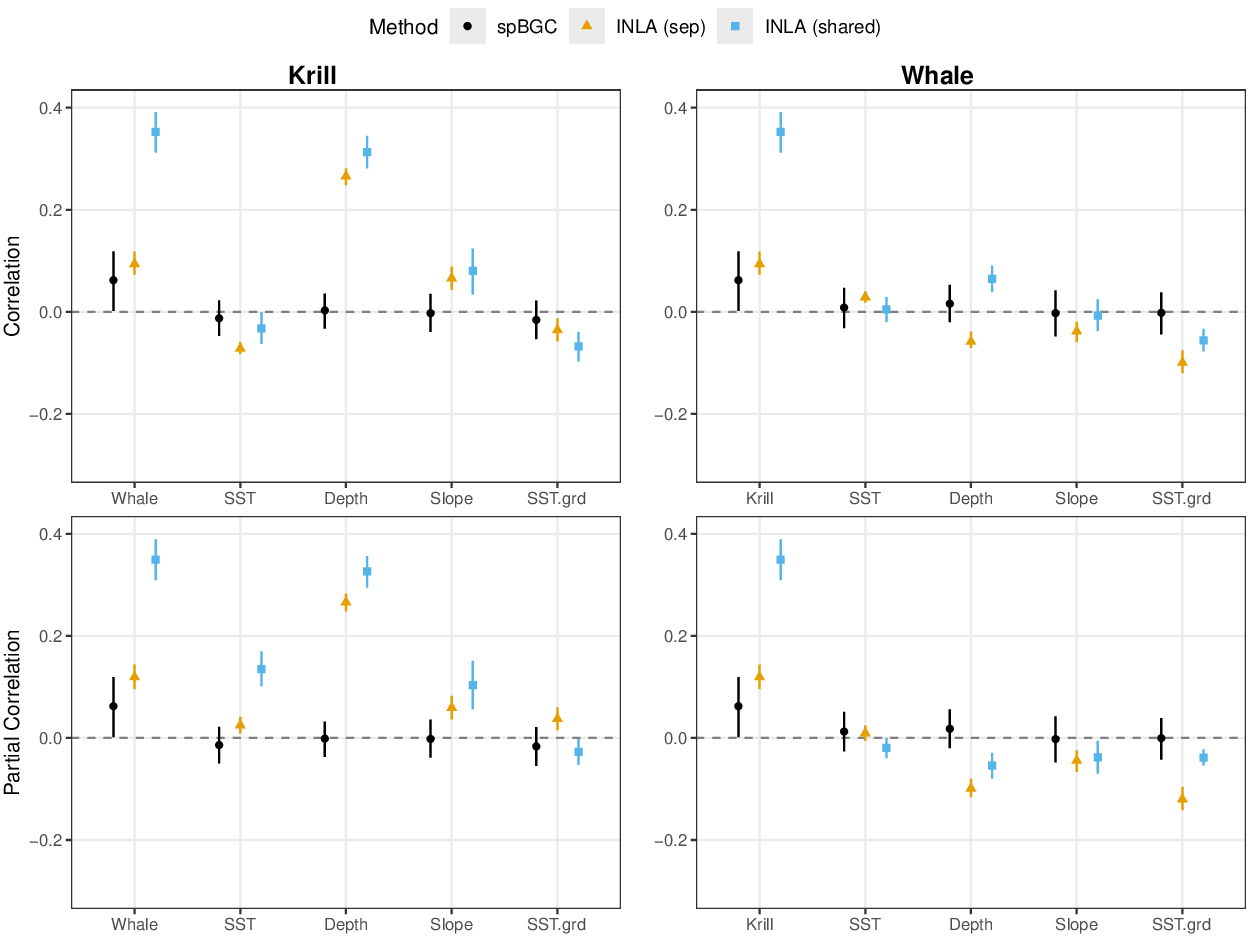}
\caption{Comparison of correlation and partial correlation estimates between krill biomass and whale sightings for spBGC and coregionalization models fitted via INLA. 
Circles: spBGC; triangles: INLA (separate range parameters); squares: INLA (shared range parameter). 
Points represent posterior medians, and error bars show 95\% credible intervals.}
\label{fig:inla_comparison}
\end{figure}

Tables~\ref{tb:inla_separate_range_comparison} and \ref{tb:inla_shared_range_comparison} compare the correlation and partial correlation estimates between spBGC and the two coregionalization models fitted via INLA.
Figure~\ref{fig:inla_comparison} visualizes the krill-whale correlation and partial correlation estimates across the three methods.
For the primary scientific question---the correlation between Krill and Whale---all three methods consistently identify a positive correlation, providing mutual validation of this ecological relationship.
Specifically, spBGC yields a median estimate of 0.062 (95\% CI: [0.002, 0.119]), the separate range model yields 0.094 (95\% CI: [0.073, 0.118]), and the shared range model yields a substantially higher estimate of 0.353 (95\% CI: [0.312, 0.392]).
In contrast, estimates for other variable pairs vary more substantially across methods.
This variation suggests that two factors may influence correlation estimates: (1) the treatment of marginal distributions (semiparametric or fully parametric specifications), and (2) the spatial range structure (common or variable-specific spatial range parameter across all variables).

Regarding the first factor, the sample skewness of variables provides evidence of potential marginal misspecification.
Among the five variables modeled as Gaussian in the LMC, Slope (skewness = 1.88), log-transformed Krill (skewness = 1.33), and SST gradient (skewness = 1.26) exhibit substantial positive skewness, indicating marked departures from the symmetric Gaussian distribution.
SST (skewness = 0.62) and Depth (skewness = $-0.53$) show moderate skewness.
For Whale modeled as Poisson in the LMC, the data exhibit substantial overdispersion with a variance-to-mean ratio of 18.2 (compared to 1 under Poisson).
Such non-normality may affect the LMC estimates, whereas spBGC is robust to marginal misspecification due to its rank-based likelihood.

Furthermore, examining the partial correlations between Whale and environmental variables reveals another distinction between spBGC and the LMC.
Both LMC models show a negative partial correlation between Whale and Slope that excludes zero: the separate range model yields $-0.044$ (95\% CI: $[-0.066, -0.024]$) and the shared range model yields $-0.038$ (95\% CI: $[-0.070, -0.006]$).
In contrast, spBGC yields $-0.002$ (95\% CI: $[-0.048, 0.042]$), with the credible interval including zero.
From an ecological perspective, a clear negative partial correlation between whale presence and slope is difficult to explain.
\cite{solvang2024} reported that krill biomass is positively associated with the continental slope, and fin whales are known to forage in areas of high krill concentration.
If whales actively seek prey aggregated along the continental slope, one would not expect a negative relationship between whale occurrence and slope after accounting for other factors.
The LMC results thus appear ecologically inconsistent, whereas spBGC does not exhibit such a contradiction.

\begin{table}[H]
\caption{Correlation and partial correlation estimates from coregionalization model with separate range parameters. 
The table shows 2.5\%, 50\%, and 97.5\% posterior quantiles.}
\label{tb:inla_separate_range_comparison}
\begin{minipage}[b]{0.49\linewidth}
\centering
\subcaption{Krill}
\scalebox{0.75}{
\begin{tabular}{r rrr}
\multicolumn{1}{l}{}&\multicolumn{1}{c}{2.5\%}&\multicolumn{1}{c}{Median}&\multicolumn{1}{c}{97.5\%}\tabularnewline \hline
{\bf Correlation} & & & \\
Whale&$ 0.0729$&$ 0.0938$&$ 0.1179$\tabularnewline
SST&$-0.0826$&$-0.0720$&$-0.0596$\tabularnewline
Depth&$ 0.2482$&$ 0.2657$&$ 0.2811$\tabularnewline
Slope&$ 0.0430$&$ 0.0658$&$ 0.0885$\tabularnewline
SST.grd&$-0.0580$&$-0.0355$&$-0.0126$\tabularnewline
&&&\\
{\bf Partial correlation} & & & \\
Whale&$ 0.0956$&$ 0.1194$&$ 0.1439$\tabularnewline
SST&$ 0.0092$&$ 0.0249$&$ 0.0412$\tabularnewline
Depth&$ 0.2480$&$ 0.2661$&$ 0.2830$\tabularnewline
Slope&$ 0.0364$&$ 0.0594$&$ 0.0826$\tabularnewline
SST.grd&$ 0.0155$&$ 0.0376$&$ 0.0600$\tabularnewline
\hline
\end{tabular}
}
\end{minipage}
\begin{minipage}[b]{0.49\linewidth}
\centering
\subcaption{Whale}
\scalebox{0.75}{
\begin{tabular}{r rrr}
\multicolumn{1}{l}{}&\multicolumn{1}{c}{2.5\%}&\multicolumn{1}{c}{Median}&\multicolumn{1}{c}{97.5\%}\tabularnewline \hline
{\bf Correlation} & & & \\
Krill&$ 0.0729$&$ 0.0938$&$ 0.1179$\tabularnewline
SST&$ 0.0178$&$ 0.0287$&$ 0.0396$\tabularnewline
Depth&$-0.0709$&$-0.0583$&$-0.0387$\tabularnewline
Slope&$-0.0588$&$-0.0383$&$-0.0197$\tabularnewline
SST.grd&$-0.1200$&$-0.0994$&$-0.0753$\tabularnewline
&&&\\
{\bf Partial correlation} & & & \\
Krill&$ 0.0956$&$ 0.1194$&$ 0.1439$\tabularnewline
SST&$-0.0059$&$ 0.0094$&$ 0.0243$\tabularnewline
Depth&$-0.1161$&$-0.0990$&$-0.0795$\tabularnewline
Slope&$-0.0663$&$-0.0443$&$-0.0242$\tabularnewline
SST.grd&$-0.1415$&$-0.1200$&$-0.0960$\tabularnewline
\hline
\end{tabular}
}
\end{minipage}
\end{table}

\begin{table}[H]
\caption{Correlation and partial correlation estimates from coregionalization model with shared range parameter.
The table shows 2.5\%, 50\%, and 97.5\% posterior quantiles.}
\label{tb:inla_shared_range_comparison}
\begin{minipage}[b]{0.49\linewidth}
\centering
\subcaption{Krill}
\scalebox{0.75}{
\begin{tabular}{r rrr}
\multicolumn{1}{l}{}&\multicolumn{1}{c}{2.5\%}&\multicolumn{1}{c}{Median}&\multicolumn{1}{c}{97.5\%}\tabularnewline \hline
{\bf Correlation} & & & \\
Whale&$ 0.3122$&$ 0.3525$&$ 0.3915$\tabularnewline
SST&$-0.0628$&$-0.0323$&$-0.0006$\tabularnewline
Depth&$ 0.2813$&$ 0.3129$&$ 0.3447$\tabularnewline
Slope&$ 0.0335$&$ 0.0802$&$ 0.1238$\tabularnewline
SST.grd&$-0.0974$&$-0.0678$&$-0.0392$\tabularnewline
&&&\\
{\bf Partial correlation} & & & \\
Whale&$ 0.3093$&$ 0.3497$&$ 0.3894$\tabularnewline
SST&$ 0.1015$&$ 0.1350$&$ 0.1699$\tabularnewline
Depth&$ 0.2946$&$ 0.3266$&$ 0.3566$\tabularnewline
Slope&$ 0.0558$&$ 0.1037$&$ 0.1511$\tabularnewline
SST.grd&$-0.0527$&$-0.0275$&$-0.0022$\tabularnewline
\hline
\end{tabular}
}
\end{minipage}
\begin{minipage}[b]{0.49\linewidth}
\centering
\subcaption{Whale}
\scalebox{0.75}{
\begin{tabular}{r rrr}
\multicolumn{1}{l}{}&\multicolumn{1}{c}{2.5\%}&\multicolumn{1}{c}{Median}&\multicolumn{1}{c}{97.5\%}\tabularnewline \hline
{\bf Correlation} & & & \\
Krill&$ 0.3122$&$ 0.3525$&$ 0.3915$\tabularnewline
SST&$-0.0199$&$ 0.0046$&$ 0.0288$\tabularnewline
Depth&$ 0.0385$&$ 0.0644$&$ 0.0905$\tabularnewline
Slope&$-0.0375$&$-0.0073$&$ 0.0242$\tabularnewline
SST.grd&$-0.0776$&$-0.0558$&$-0.0336$\tabularnewline
&&&\\
{\bf Partial correlation} & & & \\
Krill&$ 0.3093$&$ 0.3497$&$ 0.3894$\tabularnewline
SST&$-0.0393$&$-0.0196$&$ 0.0002$\tabularnewline
Depth&$-0.0796$&$-0.0542$&$-0.0293$\tabularnewline
Slope&$-0.0698$&$-0.0383$&$-0.0056$\tabularnewline
SST.grd&$-0.0539$&$-0.0384$&$-0.0222$\tabularnewline
\hline
\end{tabular}
}
\end{minipage}
\end{table}

%% file: Sp-copula_v10.bbl
\begin{thebibliography}{}

\bibitem[\protect\citeauthoryear{Apanasovich, Genton, and Sun}{Apanasovich
  et~al.}{2012}]{apanasovich2012valid}
Apanasovich, T.~V., M.~G. Genton, and Y.~Sun (2012).
\newblock A valid mat{\'e}rn class of cross-covariance functions for
  multivariate random fields with any number of components.
\newblock {\em Journal of the American Statistical Association\/}~{\em
  107\/}(497), 180--193.

\bibitem[\protect\citeauthoryear{Banerjee, Carlin, and Gelfand}{Banerjee
  et~al.}{2003}]{banerjee2003hierarchical}
Banerjee, S., B.~P. Carlin, and A.~E. Gelfand (2003).
\newblock {\em Hierarchical Modeling and Analysis for Spatial Data\/} (1 ed.).
\newblock New York: Chapman and Hall/CRC.

\bibitem[\protect\citeauthoryear{Biuw, Lindstr{\o}m, Jackson, Baines, Kelly,
  McCallum, Skaret, and Krafft}{Biuw et~al.}{2024}]{biuw2024estimated}
Biuw, M., U.~Lindstr{\o}m, J.~A. Jackson, M.~Baines, N.~Kelly, G.~McCallum,
  G.~Skaret, and B.~A. Krafft (2024).
\newblock Estimated summer abundance and krill consumption of fin whales
  throughout the scotia sea during the 2018/2019 summer season.
\newblock {\em Scientific Reports\/}~{\em 14\/}(7493), 1--12.

\bibitem[\protect\citeauthoryear{Botev}{Botev}{2017}]{botev2017normal}
Botev, Z.~I. (2017).
\newblock The normal law under linear restrictions: simulation and estimation
  via minimax tilting.
\newblock {\em Journal of the Royal Statistical Society: Series B (Statistical
  Methodology)\/}~{\em 79\/}(1), 125--148.

\bibitem[\protect\citeauthoryear{Botev and Belzile}{Botev and
  Belzile}{2021}]{botev2021truncatednormal}
Botev, Z.~I. and L.~Belzile (2021).
\newblock Truncatednormal: Truncated multivariate normal and student
  distributions, r package version 2.2.2.

\bibitem[\protect\citeauthoryear{Cressie and Huang}{Cressie and
  Huang}{1999}]{cressie1999classes}
Cressie, N. and H.-C. Huang (1999).
\newblock Classes of nonseparable, spatio-temporal stationary covariance
  functions.
\newblock {\em Journal of the American Statistical Association\/}~{\em
  94\/}(448), 1330--1339.

\bibitem[\protect\citeauthoryear{Cui, Li, Tang, Guoqing, Yumei, and Heng}{Cui
  et~al.}{2025}]{cui2025}
Cui, X., J.~Li, F.~Tang, Z.~Guoqing, W.~Yumei, and Z.~Heng (2025).
\newblock Temporal and spatial heterogeneity of antarctic krill abundance in
  relation to seawater temperature.
\newblock {\em Regional Studies in Marine Science\/}~{\em 91}, 104559.

\bibitem[\protect\citeauthoryear{Damien and Walker}{Damien and
  Walker}{2001}]{damien2001sampling}
Damien, P. and S.~G. Walker (2001).
\newblock Sampling truncated normal, beta, and gamma densities.
\newblock {\em Journal of Computational and Graphical Statistics\/}~{\em
  10\/}(2), 206--215.

\bibitem[\protect\citeauthoryear{Datta, Banerjee, Finley, and Gelfand}{Datta
  et~al.}{2016}]{datta2016hierarchical}
Datta, A., S.~Banerjee, A.~O. Finley, and A.~E. Gelfand (2016).
\newblock Hierarchical nearest-neighbor gaussian process models for large
  geostatistical datasets.
\newblock {\em Journal of the American Statistical Association\/}~{\em
  111\/}(514), 800--812.

\bibitem[\protect\citeauthoryear{Dey, Datta, and Banerjee}{Dey
  et~al.}{2022}]{dey2022graphical}
Dey, D., A.~Datta, and S.~Banerjee (2022).
\newblock Graphical gaussian process models for highly multivariate spatial
  data.
\newblock {\em Biometrika\/}~{\em 109\/}(4), 993--1014.

\bibitem[\protect\citeauthoryear{Feng and Dean}{Feng and
  Dean}{2012}]{feng2012joint}
Feng, C. and C.~Dean (2012).
\newblock Joint analysis of multivariate spatial count and zero-heavy count
  outcomes using common spatial factor models.
\newblock {\em Environmetrics\/}~{\em 23\/}(6), 493--508.

\bibitem[\protect\citeauthoryear{Finley, Datta, Cook, Morton, Andersen, and
  Banerjee}{Finley et~al.}{2019}]{finley2019efficient}
Finley, A.~O., A.~Datta, B.~D. Cook, D.~C. Morton, H.~E. Andersen, and
  S.~Banerjee (2019).
\newblock Efficient algorithms for {B}ayesian nearest neighbor {G}aussian
  processes.
\newblock {\em Journal of Computational and Graphical Statistics\/}~{\em
  28\/}(2), 401--414.

\bibitem[\protect\citeauthoryear{Geweke}{Geweke}{1991}]{geweke1991efficient}
Geweke, J. (1991).
\newblock Efficient simulation from the multivariate normal and student-t
  distributions subject to linear constraints and the evaluation of constraint
  probabilities.
\newblock In {\em Proceedings of 23rd Symposium on the Interface between
  Computing Science and Statistics}, pp.\  571--578. Interface Foundation of
  North America: Fairfax Station, VA.

\bibitem[\protect\citeauthoryear{Gneiting, Kleiber, and Schlather}{Gneiting
  et~al.}{2010}]{gneiting2010matern}
Gneiting, T., W.~Kleiber, and M.~Schlather (2010).
\newblock Mat{\'e}rn cross-covariance functions for multivariate random fields.
\newblock {\em Journal of the American Statistical Association\/}~{\em
  105\/}(491), 1167--1177.

\bibitem[\protect\citeauthoryear{Gong and Huser}{Gong and
  Huser}{2022}]{gong2022flexible}
Gong, Y. and R.~Huser (2022).
\newblock Flexible modeling of multivariate spatial extremes.
\newblock {\em Spatial Statistics\/}~{\em 52}, 100713.

\bibitem[\protect\citeauthoryear{Heaton, Datta, Finley, Furrer, Guinness,
  Guhaniyogi, Gerber, Gramacy, Hammerling, Katzfuss, Lindgren, Nychka, Sun, and
  Zammit-Mangion}{Heaton et~al.}{2019}]{heaton2019case}
Heaton, M.~J., A.~Datta, A.~O. Finley, R.~Furrer, J.~Guinness, R.~Guhaniyogi,
  F.~Gerber, R.~B. Gramacy, D.~Hammerling, M.~Katzfuss, F.~Lindgren, D.~W.
  Nychka, F.~Sun, and A.~Zammit-Mangion (2019).
\newblock A case study competition among methods for analyzing large spatial
  data.
\newblock {\em Journal of Agricultural, Biological and Environmental
  Statistics\/}~{\em 24\/}(3), 398--425.

\bibitem[\protect\citeauthoryear{Heller and Qin}{Heller and
  Qin}{2001}]{heller2001pairwise}
Heller, G. and J.~Qin (2001).
\newblock Pairwise rank-based likelihood for estimation and inference on the
  mixture proportion.
\newblock {\em Biometrics\/}~{\em 57\/}(3), 813--817.

\bibitem[\protect\citeauthoryear{Hoff}{Hoff}{2007}]{hoff2007extending}
Hoff, P.~D. (2007).
\newblock {Extending the rank likelihood for semiparametric copula estimation}.
\newblock {\em The Annals of Applied Statistics\/}~{\em 1\/}(1), 265--283.

\bibitem[\protect\citeauthoryear{Joe}{Joe}{2014}]{joe2014dependence}
Joe, H. (2014).
\newblock {\em Dependence Modeling with Copulas\/} (1 ed.).
\newblock New York: Chapman and Hall/CRC.

\bibitem[\protect\citeauthoryear{Jordan}{Jordan}{2004}]{jordan2004graphical}
Jordan, M. (2004).
\newblock Graphical models.
\newblock {\em Statistical Science\/}~{\em 19\/}(1), 140--155.

\bibitem[\protect\citeauthoryear{Katzfuss and Guinness}{Katzfuss and
  Guinness}{2021}]{katzfuss2021general}
Katzfuss, M. and J.~Guinness (2021).
\newblock A general framework for vecchia approximations of gaussian processes.
\newblock {\em Statistical Science\/}~{\em 36\/}(1), 124--141.

\bibitem[\protect\citeauthoryear{Kawaguchi, Atkinson, Bahlburg, Bernard, Cavan,
  Cox, L., Meyer, and Veytia}{Kawaguchi et~al.}{2024}]{Kawaguchi2023}
Kawaguchi, S., A.~Atkinson, D.~Bahlburg, K.~S. Bernard, E.~L. Cavan, M.~J. Cox,
  H.~S. L., B.~Meyer, and D.~Veytia (2024).
\newblock Climate change impacts on antarctic krill behaviour and population
  dynamics.
\newblock {\em Nature Reviews Earth \& Environment\/}~{\em 5\/}(1), 43--58.

\bibitem[\protect\citeauthoryear{Kotecha and Djuric}{Kotecha and
  Djuric}{1999}]{kotecha1999gibbs}
Kotecha, J.~H. and P.~M. Djuric (1999).
\newblock Gibbs sampling approach for generation of truncated multivariate
  gaussian random variables.
\newblock In {\em 1999 IEEE international conference on acoustics, speech, and
  signal processing. Proceedings. ICASSP99 (Cat. No. 99CH36258)}, Volume~3,
  pp.\  1757--1760. IEEE.

\bibitem[\protect\citeauthoryear{Krafft, Bakkeplass, Berge, Biuw, Erices,
  Jones, Knutsen, Kubilius, Kvalsund, Lindstr{\o}m, et~al.}{Krafft
  et~al.}{2019}]{krafft2019report}
Krafft, B.~A., K.~G. Bakkeplass, T.~Berge, M.~Biuw, J.~A. Erices, E.~M. Jones,
  T.~Knutsen, R.~Kubilius, M.~Kvalsund, U.~Lindstr{\o}m, et~al. (2019).
\newblock {\em Report from a krill focused survey with RV Kronprins Haakon and
  land-based predator work in Antarctica during 2018/2019}.
\newblock Havforskningsinstituttet.

\bibitem[\protect\citeauthoryear{Krafft, Macaulay, Skaret, Knutsen, Bergstad,
  Lowther, Huse, Fielding, Trathan, Murphy, Choi, Chung, Han, Lee, Zhao, Wang,
  Ying, Yu, Demianenko, Podhornyi, Vishnyakova, Pshenichnov, Chuklin, Shyshman,
  Cox, Reid, Watters, Reiss, Hinke, Arata, God{\o}, and Hoem}{Krafft
  et~al.}{2021}]{krafft2021}
Krafft, B.~A., G.~J. Macaulay, G.~Skaret, T.~Knutsen, O.~A. Bergstad,
  A.~Lowther, G.~Huse, S.~Fielding, P.~Trathan, E.~Murphy, S.-G. Choi,
  S.~Chung, I.~Han, K.~Lee, X.~Zhao, X.~Wang, Y.~Ying, X.~Yu, K.~Demianenko,
  V.~Podhornyi, K.~Vishnyakova, L.~Pshenichnov, A.~Chuklin, H.~Shyshman, M.~J.
  Cox, K.~Reid, G.~M. Watters, C.~S. Reiss, J.~T. Hinke, J.~Arata, O.~R.
  God{\o}, and N.~Hoem (2021).
\newblock Standing stock of antarctic krill (euphausia superba dana,
  1850)(euphausiacea) in the southwest atlantic sector of the southern ocean,
  2018--19.
\newblock {\em Journal of Crustacean Biology\/}~{\em 41\/}(3), 1--17.

\bibitem[\protect\citeauthoryear{Krainski, G{\'o}mez-Rubio, Bakka, Lenzi,
  Castro-Camilo, Simpson, Lindgren, and Rue}{Krainski
  et~al.}{2019}]{krainski2019advanced}
Krainski, E., V.~G{\'o}mez-Rubio, H.~Bakka, A.~Lenzi, D.~Castro-Camilo,
  D.~Simpson, F.~Lindgren, and H.~Rue (2019).
\newblock {\em Advanced Spatial Modeling with Stochastic Partial Differential
  Equations Using {R} and {INLA}\/} (1 ed.).
\newblock Boca Raton: CRC Press.

\bibitem[\protect\citeauthoryear{Krock, Kleiber, Hammerling, and Becker}{Krock
  et~al.}{2023}]{krock2023modeling}
Krock, M.~L., W.~Kleiber, D.~Hammerling, and S.~Becker (2023).
\newblock Modeling massive highly multivariate nonstationary spatial data with
  the basis graphical lasso.
\newblock {\em Journal of Computational and Graphical Statistics\/}~{\em
  32\/}(4), 1472--1487.

\bibitem[\protect\citeauthoryear{Krupskii and Genton}{Krupskii and
  Genton}{2019}]{krupskii2019copula}
Krupskii, P. and M.~G. Genton (2019).
\newblock A copula model for non-gaussian multivariate spatial data.
\newblock {\em Journal of Multivariate Analysis\/}~{\em 169}, 264--277.

\bibitem[\protect\citeauthoryear{Krupskii, Huser, and Genton}{Krupskii
  et~al.}{2018}]{krupskii2018factor}
Krupskii, P., R.~Huser, and M.~G. Genton (2018).
\newblock Factor copula models for replicated spatial data.
\newblock {\em Journal of the American Statistical Association\/}~{\em
  113\/}(521), 467--479.

\bibitem[\protect\citeauthoryear{Li, Craig, and Bhadra}{Li
  et~al.}{2019}]{li2019graphical}
Li, Y., B.~A. Craig, and A.~Bhadra (2019).
\newblock The graphical horseshoe estimator for inverse covariance matrices.
\newblock {\em Journal of Computational and Graphical Statistics\/}~{\em
  28\/}(3), 747--757.

\bibitem[\protect\citeauthoryear{Liu, Ong, Shen, and Cai}{Liu
  et~al.}{2020}]{liu2020gaussian}
Liu, H., Y.-S. Ong, X.~Shen, and J.~Cai (2020).
\newblock {When Gaussian process meets big data: A review of scalable GPs}.
\newblock {\em IEEE transactions on neural networks and learning
  systems\/}~{\em 31\/}(11), 4405--4423.

\bibitem[\protect\citeauthoryear{Liu and Wu}{Liu and
  Wu}{1999}]{liu1999parameter}
Liu, J.~S. and Y.~N. Wu (1999).
\newblock Parameter expansion for data augmentation.
\newblock {\em Journal of the American Statistical Association\/}~{\em
  94\/}(448), 1264--1274.

\bibitem[\protect\citeauthoryear{Manly and Alberto}{Manly and
  Alberto}{2016}]{manly2016multivariate}
Manly, B.~F. and J.~A.~N. Alberto (2016).
\newblock {\em Multivariate Statistical Methods: A Primer\/} (4 ed.).
\newblock New York: Chapman and Hall/CRC.

\bibitem[\protect\citeauthoryear{Mann and Lazier}{Mann and
  Lazier}{2005}]{mann2005dynamics}
Mann, K.~H. and J.~R. Lazier (2005).
\newblock {\em Dynamics of Marine Ecosystems: Biological-Physical Interactions
  in the Oceans\/} (3 ed.).
\newblock New York: Wiley.

\bibitem[\protect\citeauthoryear{Moran}{Moran}{1950}]{moran1950notes}
Moran, P. A.~P. (1950).
\newblock Notes on continuous stochastic phenomena.
\newblock {\em Biometrika\/}~{\em 37\/}(1/2), 17--23.

\bibitem[\protect\citeauthoryear{Musafer, Thompson, Wolff, and Kozan}{Musafer
  et~al.}{2017}]{musafer2017nonlinear}
Musafer, G.~N., M.~H. Thompson, R.~C. Wolff, and E.~Kozan (2017).
\newblock Nonlinear multivariate spatial modeling using nlpca and pair-copulas.
\newblock {\em Geographical Analysis\/}~{\em 49\/}(4), 409--432.

\bibitem[\protect\citeauthoryear{Paciorek and Schervish}{Paciorek and
  Schervish}{2006}]{paciorek2006spatial}
Paciorek, C.~J. and M.~J. Schervish (2006).
\newblock Spatial modelling using a new class of nonstationary covariance
  functions.
\newblock {\em Environmetrics\/}~{\em 17\/}(5), 483--506.

\bibitem[\protect\citeauthoryear{Pakman and Paninski}{Pakman and
  Paninski}{2014}]{pakman2014exact}
Pakman, A. and L.~Paninski (2014).
\newblock Exact hamiltonian monte carlo for truncated multivariate gaussians.
\newblock {\em Journal of Computational and Graphical Statistics\/}~{\em
  23\/}(2), 518--542.

\bibitem[\protect\citeauthoryear{Pettitt}{Pettitt}{1982}]{pettitt1982inference}
Pettitt, A.~N. (1982).
\newblock Inference for the linear model using a likelihood based on ranks.
\newblock {\em Journal of the Royal Statistical Society: Series B (Statistical
  Methodology)\/}~{\em 44\/}(2), 234--243.

\bibitem[\protect\citeauthoryear{{R Core Team}}{{R Core Team}}{2024}]{R2024}
{R Core Team} (2024).
\newblock {\em R: A Language and Environment for Statistical Computing}.
\newblock Vienna, Austria: R Foundation for Statistical Computing.

\bibitem[\protect\citeauthoryear{Rue, Martino, and Chopin}{Rue
  et~al.}{2009}]{rue2009approximate}
Rue, H., S.~Martino, and N.~Chopin (2009).
\newblock Approximate {B}ayesian inference for latent {G}aussian models by
  using integrated nested {L}aplace approximations.
\newblock {\em Journal of the Royal Statistical Society: Series B (Statistical
  Methodology)\/}~{\em 71\/}(2), 319--392.

\bibitem[\protect\citeauthoryear{Sagar, Banerjee, Datta, and Bhadra}{Sagar
  et~al.}{2024}]{sagar2024precision}
Sagar, K., S.~Banerjee, J.~Datta, and A.~Bhadra (2024).
\newblock Precision matrix estimation under the horseshoe-like prior--penalty
  dual.
\newblock {\em Electronic Journal of Statistics\/}~{\em 18}, 1--46.

\bibitem[\protect\citeauthoryear{Schmidt and Gelfand}{Schmidt and
  Gelfand}{2003}]{schmidt2003bayesian}
Schmidt, A.~M. and A.~E. Gelfand (2003).
\newblock A {B}ayesian coregionalization approach for multivariate pollutant
  data.
\newblock {\em Journal of Geophysical Research: Atmospheres\/}~{\em
  108\/}(D24), 8783.

\bibitem[\protect\citeauthoryear{Smith}{Smith}{2023}]{smith2021implicit}
Smith, M.~S. (2023).
\newblock Implicit copulas: An overview.
\newblock {\em Econometrics and Statistics\/}~{\em 28}, 81--104.

\bibitem[\protect\citeauthoryear{Solvang, Imori, Biuw, Lindstr{\o}m, and
  Haug}{Solvang et~al.}{2024}]{solvang2024}
Solvang, H.~K., S.~Imori, M.~Biuw, U.~Lindstr{\o}m, and T.~Haug (2024).
\newblock Categorical data analysis using discretization of continuous
  variables to investigate associations in marine ecosystems.
\newblock {\em Environmetrics\/}~{\em 35\/}(6), e2867.

\bibitem[\protect\citeauthoryear{Souris, Bhattacharya, and Pati}{Souris
  et~al.}{2018}]{souris2018soft}
Souris, A., A.~Bhattacharya, and D.~Pati (2018).
\newblock The soft multivariate truncated normal distribution with applications
  to bayesian constrained estimation.
\newblock arXiv preprint arXiv:1807.09155.

\bibitem[\protect\citeauthoryear{Stein}{Stein}{2005}]{stein2005space}
Stein, M.~L. (2005).
\newblock Space--time covariance functions.
\newblock {\em Journal of the American Statistical Association\/}~{\em
  100\/}(469), 310--321.

\bibitem[\protect\citeauthoryear{Tett, Gowen, Painting, Elliott, Forster,
  Mills, Bresnan, Capuzzo, Fernandes, Foden, Geider, Gilpin, Huxham,
  McQuatters-Gollop, Malcolm, Saux-Picart, Platt, Racault, Sathyendranath,
  van~der Molen, and Wilkinson}{Tett et~al.}{2013}]{tett2013framework}
Tett, P., R.~J. Gowen, S.~J. Painting, M.~Elliott, R.~Forster, D.~K. Mills,
  E.~Bresnan, E.~Capuzzo, T.~F. Fernandes, J.~Foden, R.~J. Geider, L.~C.
  Gilpin, M.~Huxham, A.~L. McQuatters-Gollop, S.~J. Malcolm, S.~Saux-Picart,
  T.~Platt, M.-F. Racault, S.~Sathyendranath, J.~van~der Molen, and
  M.~Wilkinson (2013).
\newblock Framework for understanding marine ecosystem health.
\newblock {\em Marine Ecology Progress Series\/}~{\em 494}, 1--27.

\bibitem[\protect\citeauthoryear{Torabi}{Torabi}{2014}]{torabi2014spatial}
Torabi, M. (2014).
\newblock Spatial generalized linear mixed models with multivariate car models
  for areal data.
\newblock {\em Spatial Statistics\/}~{\em 10}, 12--26.

\bibitem[\protect\citeauthoryear{Van~Niekerk, Bakka, Rue, and
  Schenk}{Van~Niekerk et~al.}{2021}]{vanniekerk2021frontiers}
Van~Niekerk, J., H.~Bakka, H.~Rue, and O.~Schenk (2021).
\newblock New frontiers in {B}ayesian modeling using the {INLA} package in {R}.
\newblock {\em Journal of Statistical Software\/}~{\em 100\/}(2), 1--28.

\bibitem[\protect\citeauthoryear{Vecchia}{Vecchia}{1988}]{vecchia1988estimation}
Vecchia, A.~V. (1988).
\newblock Estimation and model identification for continuous spatial processes.
\newblock {\em Journal of the Royal Statistical Society: Series B (Statistical
  Methodology)\/}~{\em 50\/}(2), 297--312.

\end{thebibliography}
